\documentclass[12pt]{report}
\usepackage{amssymb,latexsym,epsfig}

\begin{document}
\begin{titlepage}
\begin{flushright}
hep-ph/0307004
\end{flushright}
\begin{center}
\vspace{1cm} 
{\Large \bf Spontaneous and dynamical symmetry breaking 
in higher-dimensional space-time with boundary terms}

\vspace{1.5cm} 

{\large Hiroyuki~Abe\footnote{E-mail: abe@muon.kaist.ac.kr}}

\vspace{0.5cm} 

{\it Department of Physics, Korea Advanced Institute of Science and Technology, \\
Daejeon 305-701, Korea}
 
\vspace{1cm} 

{\large Doctoral thesis submitted to }

{\large Department of Physics, Hiroshima University }

{\large 18 February 2003}

\end{center}

\vspace{1cm} 
In this thesis we study physics beyond the standard model 
focusing on the quantum field theory in higher-dimensional 
space-time with some boundary terms. 
The boundary term causes nontrivial consequences 
about the vacuum structure of the higher-dimensional theory. 
We take particular note of two independent solutions to the 
weak and Planck hierarchy problem: 
``low scale supersymmetry'' and 
``dynamical electroweak symmetry breaking.'' 
From a viewpoint of the low scale supersymmetry, we study 
$F$ and $D$ term supersymmetry breaking effects on sparticle 
spectra from a boundary. 
While we also investigate a nonperturbative effect caused by a 
bulk (nonsupersymmetric) gauge dynamics on a fermion bilinear 
condensation on a boundary, and analyze the dynamical symmetry 
breaking on the brane. 
From these analyses we conclude that the field localization 
in higher-dimensional space-time involves in a nontrivial 
vacuum structure of the theory, and the resultant 
low energy four-dimensional effective theory has phenomenologically 
interesting structure. 
In a framework of purely four-dimensional theory, 
we also construct the above nontrivial effect of localization 
in the extra dimension.

\end{titlepage}

\tableofcontents

\chapter{Our world as a boundary plane}
\label{chap:intro}

The unification of all forces and matters in our world 
within a single theory is most important and challenging 
issue in modern theoretical particle physics. 
We have a quantum gauge field theory (QGFT) called standard model 
(SM) with the gauge group $SU(3)_c \times SU(2)_W \times U(1)_Y$ 
that describes strong and electroweak interaction between 
matter particles, that is quarks and leptons. 
All of the experimental results done till now assures the 
validity of SM. In a theoretical sense, however, SM has 
some serious problems. One of them is that it has so many 
free parameters which must be given from experimental 
results. For example quark and lepton masses and mixing 
angles are not predicted from SM itself. Almost all these 
parameters are related to the Higgs sector which induces 
electroweak symmetry breaking $SU(2)_W$ $\times$ $U(1)_Y$ $\to$ $U(1)_{EM}$. 
Even if we embed SM into the grand unification theory (GUT) such as 
$SU(5)$, $SO(10)$ and $E_6$, the GUT Higgs sectors also have 
many degrees of freedom and results in the same situation as SM. 
Therefore we should illuminate the origin of Higgs fields and the 
mechanism of spontaneous (dynamical) symmetry breaking in SM and GUT. 

SM also has a severe difficulty that it doesn't include 
gravity although we never doubt its existence because it is 
a most familiar force in our life. Gravity is in a sense 
very fundamental force because it originates from the structure 
of space-time as is insisted in general relativity and 
couples to all particles universally. If gravity has the same 
ultimate origin as the SM forces, it should be described by the 
quantum theory. But QGFT can not treat the gravity ably because 
of its nonrenormalizable nature. 
The most successful and interesting candidate 
of the quantum theory of gravity is string theory. String theory 
gives not only quantum gravity but also QGFT that means that it may 
include SM within itself. Unfortunately we can not explicitly show 
in a present stage that string theory gives us SM uniquely because 
it has huge number of various degenerate vacuum like or unlike SM. 

When we consider the unification of the SM forces and the gravity, 
we also have a hierarchy problem between the electroweak scale 
${\cal O}(10^2)$ GeV 
and the Planck scale ${\cal O} (10^{18})$ GeV. 
This is elegantly solved by introducing supersymmetry (SUSY) at 
the electroweak scale. 
But we have so many extra field in addition to the SM contents that 
we have another difficulty: how to realize SUSY breaking 
(dynamically) without conflicting experimental results. 
So the realistic SUSY breaking or the other mechanism which solves 
the hierarchy problem should be included in such a ultimate 
unification theory like the string theory. 

Taking above situations into account there are two kinds of positions 
for us to take to study the unification of SM and gravity. 
One is to improve string theory itself by extending it to, e.g., string 
field theory, M-theory, matrix model in order to determine the real 
vacuum that is expected to give SM in four-dimensional space-time. 
The other is to improve SM side for the purpose to find 
what kind of improvement resolves the problems in SM and likely leads 
it to the unification with gravity. In this thesis we stand on 
the latter stance. In usual bottom-to-up approaches for improving SM, 
we add new symmetry, new fields, and so on, in four-dimensional 
space-time. We should consider the origin of such new things when 
they are added. Our clue for studying beyond the SM is a fact that the 
known theories of quantum gravity such as supergravity or string theory 
are consistently defined in more than four-dimensional space-time. 
The extra dimensions should be hided somehow to give SM in 
four-dimensional space-time as an low energy effective theory 
near the electroweak scale. Inversely we should say that 
there is a possibility to have extra dimensions above the 
electroweak scale. 

Most common idea to yield four-dimensional space-time is 
to compactify extra dimensions so as not to appear in low energy 
world. Theories in higher-dimensional space-time with compact extra 
dimension generically have extra fields called moduli fields in terms 
of the four-dimensional viewpoint. 
For example, the extra dimensional components of the 
gravity or vector (gauge) fields looks like scalar fields in 
effective four-dimensional theory. These are the candidate for 
the Higgs fields that induces various symmetry breaking in 
the four-dimensional world. 

However recent discovery of D-branes~\cite{Polchinski:1995mt} 
in string theory gives new aspects to such hiding mechanism. 
D-brane is a solitonic object on which the open string endpoint 
terminates. The idea is that our four-dimensional world is realized 
on a boundary plane in higher dimensional space-time. 
We consider that our standard model matter particles are confined 
on a four-dimensional boundary. Such a picture gives many solutions 
of SM puzzles. One of the remarkable target is weak and Planck hierarchy 
problem mentioned above. If the number of space-time dimension that 
gravity propagates is lager than one of SM, our gravity scale, namely, 
four-dimensional Planck scale is suppressed by the volume of the extra 
space~\cite{Antoniadis:1990ew,Arkani-Hamed:1998rs}. 
There are some explicit models realizing SM gauge group and chiral matter 
contents directly based on such string models with D-branes, e.g., 
by intersecting D-branes~\cite{Ibanez:2001nd}. 
Some of these models can obtain weak and Planck hierarchy 
by such volume suppression mechanism. 
Of course domain walls in QGFT that is known before the discovery 
of D-brane is another candidate for such boundary plane. 
The formation of domain walls is deeply related to the spontaneous 
symmetry breaking. 

Then we are going to be interested in how to solve problems 
in SM by assuming it to be in the higher-dimensional space-time 
with such boundaries. Because almost all the ambiguities in QGFT (SM) 
originate from the sectors (e.g., Higgs sector or hidden sector 
for SUSY breaking) where spontaneous symmetry breaking occurs, 
first of all we should pay attention to such sectors. 
In this thesis we investigate spontaneous and dynamical symmetry 
breaking in quantum field theory in higher-dimensional space-time 
with some boundary terms, e.g. boundary matter field. 
In the theories of unification there are kinds of symmetries 
that should be broken spontaneously (dynamically) like 
electroweak symmetry, grand unification symmetry and SUSY. 

First we consider the SUSY gauge theories in higher-dimensional 
space-time with boundaries. Such models can give some realistic 
SUSY breaking and its mediation mechanisms. In usual scenario in 
four-dimension, it is sometimes unnatural that we don't have 
tree-level direct couplings between SUSY breaking (hidden) sector 
and SM (visible) sector. In the bulk and boundary system with extra 
dimensions, we can sequester a hidden sector (brane) from a visible 
sector (brane) by spatiality of extra dimension. Some interesting 
mechanisms of SUSY breaking communication are proposed in extra 
dimension such as gaugino, Kaluza-Klein and radion mediated scenario. 
We will briefly review them in Chapter~\ref{chap:sbmm}. 
These SUSY breaking contributions are based on the $F$ term breaking 
in some hidden supermultiplet. 
As in the case of Froggatt-Nielsen mechanism~\cite{Froggatt:1978nt} 
for realistic Yukawa couplings, if the gauge group includes $U(1)$ 
symmetry, we generally have tadpoles of auxiliary field in 
the vector multiplet on the four-dimensional boundaries. 
These tadpoles induce $D$ term SUSY 
breaking contributions by the well-known way called Fayet-Iliopoulos 
mechanism. We will also analyze SUSY breaking structure in the bulk 
with such contributions in Chapter~\ref{chap:dtermsb}. 

Next we treat the dynamical chiral symmetry breaking via fermion 
bilinear condensation, that is a basis for studying dynamical 
breaking of symmetry without large ambiguities like a scalar potential. 
The electroweak symmetry is in a sense gauged chiral symmetry, 
and SUSY breaking through gaugino condensation is similar to 
the situation of fermion bilinear condensation in dynamical 
chiral symmetry breaking. We are motivated by analyzing such 
symmetry breaking in higher-dimensional space-time with boundary 
terms. In Chapter~\ref{chap:desb} we will review scenarios 
of dynamical electroweak symmetry breaking through fermion 
bilinear condensation that is another mechanism to stabilize 
weak and Planck hierarchy without low energy SUSY. 
First we formulate the 4D effective 
Lagrangian of the gauge theory in the bulk and boundary system 
in Chapter~\ref{chap:hdgtwbm}. The most simple way to study 
dynamical symmetry breaking via fermion bilinear condensation 
is to use four-fermion approximation and follows the way by 
Nambu-Jona-Lasinio (NJL)~\cite{Nambu:tp}. 
In chapter~\ref{chap:alffim} we will analyze models with such 
a four-fermion interaction induced on a brane, 
and also the interaction which mix a bulk and a brane fermion, 
and obtain chiral phase structure in terms of the bulk configuration. 
We also take the gauge boson propagating effect into account 
by using improved ladder Schwinger-Dyson (SD) equation 
technique~\cite{Dyson:1949ha} and have nontrivial properties of 
dynamical fermion mass as a result in Chapter~\ref{chap:ailsde}. 
We conclude that the dynamical symmetry breaking in 
higher-dimensional space-time with boundary terms has rather 
rich structure compared with the usual four-dimensional one. 

In Chapter~\ref{chap:wgt} we consider the theory with product 
gauge group defined in four-dimensional space-time, which 
possesses higher-dimensional nature. This gives the new aspects 
on the difficulties in SM in terms of the higher-dimensional 
theories, but just in four-dimensional space-time. 
Finally in Chapter~\ref{chap:cd} we conclude this thesis.

\section{Brane world picture}

The known theories of quantum gravity like supergravity or 
superstring theory are consistently defined in more than 
four-dimensional space-time. 
The extra dimensions should be hided somehow to give SM in 
four-dimensional space-time as an low energy effective theory. 
Most common idea to yield four-dimensional space-time is 
to compactify extra dimensions so as not to appear in low energy 
world. However recent discovery of D-branes~\cite{Polchinski:1995mt} 
in string theory gives new aspects to such hiding mechanism. 
That is the idea that our four-dimensional world is realized 
on a boundary plane in higher dimensional space-time. 
We consider, for example, that our standard model matter particles 
are confined on a four-dimensional boundary by certain mechanism 
like D-brane or domain wall, while the gravity (gauge) fields propagate 
whole bulk space-time. Such a picture gives many solutions to 
SM puzzles like hierarchy problem. 

The domain wall that is known before the D-brane appearance is 
a candidate for such boundary plane. The formation of 
domain walls is deeply related to the spontaneous or dynamical 
symmetry breaking. For instance we consider the next action 
in five-dimensional space-time: 
\begin{eqnarray}
{\cal L} = \int d^4x\, dy \left[ \bar\psi iD \!\!\!\!/ \psi 
+ Y \varphi \bar\psi \psi + \varphi D^2 \varphi + V(\varphi) \right], 
\nonumber 
\end{eqnarray}
where $D \!\!\!\!/ = \gamma^M D_M$ and $D^2 = D^M D_M$. 
$Y$ is a Yukawa coupling and $D_M$ is the covariant derivative 
in terms of gauge and general coordinate invariance. 
The appropriate potential $V(\varphi)$ can give a kink solution, 
$\langle \varphi \rangle = \varphi_{\rm kink}(y)$ 
where $\varphi_{\rm kink}(\infty)=-\varphi_{\rm kink}(-\infty)$. 
In such kink background the profile of fermion zero mode is given by 
$\gamma_5 \partial_y \psi_{L,R} = Y \varphi_{\rm kink}(y) \psi_{L,R}$, 
which results in 
\begin{eqnarray}
\psi^{(0)}_{L,R}(y) 
= \psi^{(0)}_{L,R}(0) 
  \exp \left[ \mp Y \int_0^y dy'\, \varphi_{\rm kink}(y') \right], 
\nonumber 
\end{eqnarray}
where $\mp$ in the exponent originates from the eigenvalue of $\gamma_5$. 
Thus we understand that $\psi^{(0)}_L(y)$ localizes at $y^\ast$ where 
$\varphi_{\rm kink}(y^\ast)=0$. If we consider ultimate kink solution 
$\varphi_{\rm kink}(y) \sim \textrm{sgn} (y)$, we have delta-function 
type localization $\psi^{(0)}_L(y) \sim \delta (y)$. We note that 
massless vector fields can not localized on the kink backgrounds. 
Therefore we know that such kink localization mechanism gives us 
higher-dimensional gauge theory with boundary matters as an effective 
theory. The kink solution divides two different vacua {\it spatially} from 
each other, and we call the boundary between them `domain wall' here. 

Such domain wall can be induced in point particle field theories 
in higher-dimensional space-time, and is an interesting candidate 
for our four-dimensional world. 
In string theory we consequentially have such wall-like objects. 
For example if we compactify string theory into orbifold~\cite{Dixon:jw}, 
the twisted sector, that comes from the winding modes around the orbifold 
fixed points, looks like a field theory on the orbifold fixed planes in low energy. 
Such a situation also gives us bulk and boundary field theories. 
Furthermore in 1995 J.~Polchinski~\cite{Polchinski:1995mt} suggested that 
in string theory there should exist Dirichlet membrane (D-brane), on which 
the open string endpoint terminates, because of a T-duality nature 
of the string theory. 
E.~Witten also showed~\cite{Witten:1995ex} that we have supersymmetric 
Yang-Milles theory on such D-branes. Here we don't show details of D-brane 
because we need the knowledge of string theory which is beyond the 
purpose of this thesis. But the low energy properties of D-brane is well 
approximated by the boundary action in higher-dimensional space-time. 

\begin{figure}[t]
\centerline{\epsfig{figure=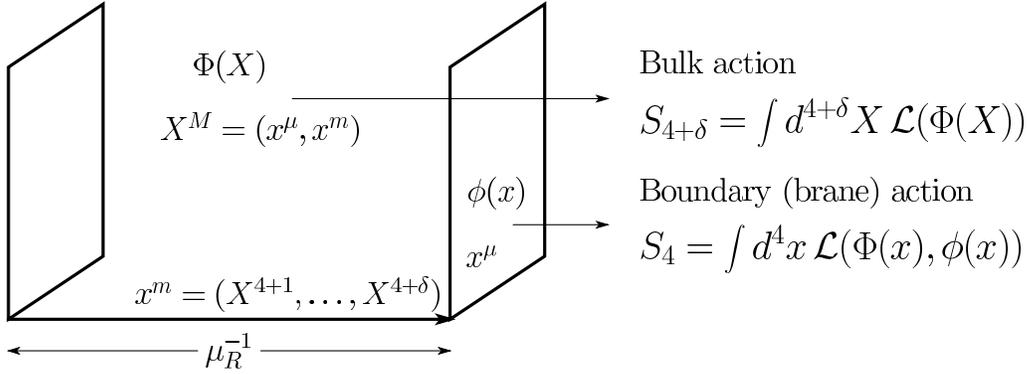,width=0.8\linewidth}}
\caption{Schematic view of the brane world.}
\label{fig:braneworld}
\end{figure}
As above we can consider a possibility that our four-dimensional 
world is realized on a boundary plane in higher-dimensional space-time. 
Though the origin of such boundary plane may be domain wall, orbifold fixed 
plane, D-brane, and something like that, we approximate them by the 
delta-function type boundary plane for simplicity and generality, and we 
don't specify the concrete origin of it in this thesis. We generically call 
such scenarios `brane world' models in this thesis. The brane world is 
schematized in Fig.~\ref{fig:braneworld}. 

\section{Hierarchy problem and brane world picture}

As is also described at the exordium, 
up to the energy scale we can test experimentally, SM well describes 
the forces between matter particles except for the gravity. 
Almost all the candidates for the theory of gravity at the Planck scale 
$M_{\rm Pl}$ are defined with SUSY in more than four-dimensional space-time, 
such as supergravity or superstring theory. It is considered that the 
SM particles lie in higher-dimensional space-time and extra dimensions 
are compactified smaller than the size we can detect in the low energy 
experiments, or lie on the 4D subspace such as a boundary plane 
in higher dimensional space-time. 
As we discussed above, recent interesting suggestion is that SM may be 
realized on a boundary plane in the higher dimensional space-time. 
One of the candidates of such object is D-brane, orientifold plane, 
orbifold fixed plane, or domain wall. 
For example recent studies in the superstring theory provide us some 
concrete examples realizing SM like structure in the D-brane systems 
\cite{Ibanez:2001nd}. 
Such models now casting new ideas on physics beyond the SM. 

Basis of the SM is a spontaneous gauge symmetry breaking 
that results in the existence of massive gauge bosons, W and Z. 
We need at least one (doublet) scalar field, called Higgs field 
which develops a vacuum expectation value (VEV) to give masses 
to the W and Z gauge bosons and breaks the SM gauge symmetry. 
These gauge boson masses are related to the electroweak scale 
$M_{\rm EW}$ at which the SM gauge symmetry is broken as 
$SU(3)_c \times SU(2)_W \times U(1)_Y \to SU(3)_c \times U(1)_{EM}$. 
The mass scale of the W and Z gauge bosons shows that $M_{\rm EW}$ is 
of the order TeV scale, while the gravitational scale, namely Planck 
scale $M_{\rm Pl}$ is of the order $10^{19}$ GeV. 
If we consider the unification of SM and gravity, we should 
explain the hierarchical structure between $M_{\rm Pl}$ and $M_{\rm EW}$. 
Since the mass of the scalar field is not protected by any symmetries,
the radiative correction brings the Higgs mass to the fundamental scale 
$M_{\rm Pl}$ without a fine tuning. This is so called `hierarchy' or 
`fine tuning' problem which is briefly mentioned at the exordium. 
There exist several interesting ideas solving weak and Planck hierarchy 
problem. Here we list some of them:
\begin{description}
\item[(a)]
Low scale supersymmetry breaking, 
\item[(b)] 
Fermion bilinear condensation (composite Higgs), 
\item[(c)] 
Higgs field as a pseudo Nambu-Goldstone (NG) boson, 
\item[(d)] 
Higgs field originating in a gauge field, 
\item[(e)] 
Reduction of the fundamental scale 
due to a compactification. 
\end{description}

Although they are coming out a lot of successful results based on the mechanisms 
(c) and (d), in this thesis we concentrate on the first two items, 
(a) low scale supersymmetry breaking, and 
(b) fermion bilinear condensation. 
In the brane world, each mechanisms of (a)-(d) can be mixed with (e). 
So first of all we review the mechanism of (e), the reduction of 
the fundamental scale through  a compactification of extra dimensions. 

\subsection{Large and warped extra dimension}

There are two remarkable brane world models 
solving this hierarchy problem. One is the large extra dimension 
scenario~\cite{Antoniadis:1990ew,Arkani-Hamed:1998rs} and the other is 
the warped extra dimension 
scenario in Randall-Sundrum (RS) brane world~\cite{Randall:1999ee}.
The idea of these two models is that we have weak gravity, namely large 
Planck scale, if only the gravity propagates some extra dimensions and 
SM fields are confined on a four-dimensional brane. 

For example we consider $(4+\delta)$-dimensional space-time with the metric 
\[ G_{MN}(X)dX^MdX^N = g_{\mu \nu}(x) dx^\mu dx^\nu + d\mathbf{y}^2 \] 
where $x^\mu$ and $\mathbf{y}$ is the four- and $\delta$-dimensional coordinates 
respectively. Only the gravity can propagate $\delta$-dimensional extra space. 
If each $\mathbf{y}$-direction is compactified to a circle 
with radius $1/\mu_c$, the gravity and matter action roughly becomes 
\begin{eqnarray}
\int d^4x \int d^\delta \mathbf{y} \left[ M_\ast^{2+\delta} \sqrt{-G} R(G) 
+ \delta^\delta (\mathbf{y}-\mathbf{y}_\ast) 
\sqrt{-g}{\cal L}_{\rm matter}(g_{\mu\nu},\psi,\phi,\ldots;M_\ast) \right] 
\nonumber \\
\sim \int d^4x \sqrt{-g} \left[ \mu_c^{-\delta} M_\ast^{2+\delta} R(g) 
+ {\cal L}_{\rm matter}(g_{\mu\nu},\psi,\phi,\ldots;M_\ast) \right],
\nonumber 
\end{eqnarray}
where $\mathbf{y}_\ast$ is the position of the brane and 
$M_\ast$ is the fundamental scale of the theory. 
Thus the four-dimensional Planck scale is given by 
\[ M_{\rm Pl}^2 \sim (M_\ast/\mu_c)^{\delta} M_\ast^2. \]
So If $\mu_c \ll M_\ast$ we have a large hierarchy between $M_{\rm Pl}$ 
and $M_\ast$. In the scenario of large extra dimension we take 
$M_\ast \sim {\cal O}(1)$ TeV and we know $M_{\rm Pl} \sim 10^{16}$ TeV. 
Thus the compactification scale should be $\mu_c \sim 10^{-32/\delta}$ TeV 
and we have another hierarchy problem between $\mu_c$ and $M_\ast$ 
though the hierarchical structure can be reduced for large $\delta$. 

Furthermore we consider a warped background metric 
\[ G_{MN}(X)dX^MdX^N = e^{-2ky} g_{\mu \nu}(x) dx^\mu dx^\nu + dy^2, \] 
where $k$ is a AdS curvature and $y$ is a compact dimension with its 
radius $1/\mu_c$. In such case roughly again we have 
\begin{eqnarray}
\int d^4x \int dy \left[ M_\ast^3 \sqrt{-G} R(G) 
+ \delta (y-y_\ast) 
\sqrt{-g}{\cal L}_{\rm matter}(g_{\mu\nu},\psi,\phi,\ldots;M_\ast) \right] 
\nonumber \\
\sim \int d^4x \sqrt{-g} \left[ (1-e^{-2\pi k/\mu_c}) k^{-1} M_\ast^{3} R(g) 
+ {\cal L}_{\rm matter}(g_{\mu\nu},\psi,\phi,\ldots;e^{-2ky_\ast}M_\ast) \right].
\nonumber 
\end{eqnarray}
Thus the four-dimensional Planck scale is given by 
\[ M_{\rm Pl}^2 \sim (1-e^{-2\pi k/\mu_c}) (M_\ast/k) M_\ast^2. \]
L.~Randall and R.~Sundrum found~\cite{Randall:1999ee} that the above warped 
metric is a static solution of the Einstein equation in five-dimensional 
space-time if certain relation exists between the bulk cosmological constant 
and brane tensions. And they proposed that by choosing 
\[ 10 \mu_c \sim k \sim M_\ast \quad (y_\ast = \pi/\mu_c), \]
which is ${\cal O} (10)$ tuning, we have 
$M_{\rm Pl}/M_\ast^{\rm brane} \sim 10^{16}$ 
where $M_\ast^{\rm brane} \equiv e^{-2ky_\ast}M_\ast$ is a {\it reduced} 
fundamental mass scale on the $y_\ast =\pi/\mu_c$ brane. 

The scenarios of large and warped extra dimension provide new solution to 
the weak and Planck hierarchy problem. 
The remarkable consequence of these solutions is that 
the fundamental scale of the theory can be lowered, e.g., extremely to 
just above the electroweak scale, while we {\it believed} that the 
fundamental scale of the theory should be near the four-dimensional gravity 
scale, namely Planck scale $M_{\rm Pl}$. We also note that the same mechanism 
of volume suppression works for realizing hierarchical structures of the 
other couplings though it is applied to the gravity coupling here. 
In addition, in this thesis, we will see that another structure is also 
important in the brane world scenarios such as wave function localization 
in the extra dimension. For instance the wave function localization can 
give not only the source of hierarchical structure between some couplings 
at the tree level but also different phase structures in a single theory. 
The wave function localization of the gauge-boson higher modes can change 
the nonperturbative dynamics on a boundary plane where they are localized.

\subsection{TeV or fundamental scale supersymmetry breaking}

SUSY also enables to stabilize the scalar mass against the radiative 
correction of ${\cal O}(M_{\rm Pl})$ because it is related to 
the mass of the fermionic superpartner that can be protected 
by appropriate symmetry. SUSY permits the existence of light scalar 
fields like Higgs field, then we introduce it around the TeV scale 
in order to keep the Higgs mass $\sim$ TeV, which also leads into a 
consequence that all the fermions in the SM have their scalar partners 
with their masses around TeV. 
These extra scalar fields have not been observed in any experiments yet. 
Thus we need some mechanisms in the SUSY breaking process to control 
these extra light scalars not to appear inside the region of the present 
experimental observation. Because of such reason, in general the theory 
must have `hidden' sector where SUSY is broken which doesn't couple 
directly to the visible sector. 
A certain mediation mechanism to communicate it to the 
`visible sector' is required at the loop level through renormalizable 
interactions or at the tree level through nonrenormalizable interactions. 
But how to hide the hidden sector is also problem, 
especially in four-dimensional space-time. 
Recently some people consider that the hidden sector setup merges into 
the brane world picture, that separates the hidden and visible sector 
by spatiality in the extra dimension. They use the bulk (or moduli) fields 
to communicate the SUSY breaking signals from the hidden to the visible 
sector~\cite{Randall:1998uk,Kaplan:1999ac,Kobayashi:2000ak,Chacko:2000rm}. 

The hidden sector is the minimal requirement from the supertrace 
theorem and we need more severe conditions to the SUSY breaking from the 
experiment. One of them is from the supersymmetric flavor problem, that is 
the masses of the SM fermions are disjointed each other while the masses 
of their scalar partners should be almost degenerate. 
With regards to such considerations, in Part~\ref{part:ssbb} of this thesis, 
we consider the SUSY breaking effect in our stand point, that is, 
in higher-dimensional space-time with four-dimensional boundary plane. 
We will also know that if we have $U(1)$ gauge symmetry even in the bulk 
we should be careful about the $D$ term contribution to the SUSY flavor problem. 

Aside from the above low scale SUSY breaking scenario, 
we can also consider the stability of the electroweak scale and the flavor 
problem as follows. As we see above, almost all the problems in the TeV scale 
SUSY come from too much light extra fields in addition to the SM one. 
We need complicated setup about SUSY breaking and its mediation mechanism 
to avoid these problems. Remember that SUSY itself is needed for the 
consistency of the quantum theory of gravity, while `TeV scale SUSY' is 
required in order to stabilize the electroweak scale $M_{\rm EW}$, 
and to realize gauge coupling unification in minimal SUSY SM (MSSM). 
However, in a recent brane world picture, there is a possibility that the 
strong and electroweak gauge group come from the different 
brane~\cite{Ibanez:2001nd}. In this case the unification of SM and gravity is 
able to occur directly without grand unification and the gauge coupling 
unification in MSSM may be accidental. We can also consider the case that 
gauge coupling unification is not depend on the TeV scale SUSY (MSSM) and 
it happens via the other mechanism, e.g. extra dimensional 
effect~\cite{Dienes:1998vh,Randall:2001gc}. 

If there is another mechanism to stabilize mass of the Higgs field, 
it is possible to break SUSY with a higher breaking scale, even around 
the Planck scale. Such a mechanism can set SUSY free from any 
problems at TeV scale. We consider one of such stabilization mechanism 
in Chapter~\ref{chap:desb}. 
The idea is that we use fermion bilinear condensation instead of 
elementary scalar condensation. This enables us to obtain light Higgs 
scalar as in the case of pion in QCD. This is well known as dynamical 
electroweak symmetry breaking. The most interesting candidate for such 
fermion to be condensed is top quark in SM~\cite{Miransky:1988xi}. But 
within theories in four-dimensional space-time, we need extra unknown 
force that makes top quark bilinear condensed successfully. 
Here we remember our stand point that the gauge field propagates in 
extra dimensions. The Kaluza-Klein modes of the gauge field can be the 
candidate for such unknown force. Therefore in Part~\ref{part:dsbb} of 
this thesis we give an analysis of dynamical symmetry breaking in 
(nonSUSY) gauge theory in higher-dimensional space-time with 
four-dimensional boundary plane on which matter fermion lives. 

As we discussed, there are the other mechanisms that stabilize 
the light Higgs mass such as the Higgs field as a pseudo NG boson, 
and the Higgs field from a gauge field. These are quite interesting 
scenarios, however, we will not treat these models in this thesis 
because of the limitation of the space.

%%%%%%%%%%%%%%%%%%%%%%%%%%%%%%%%%%%%%%%%%%%%%%%%%%%%%%%%%%%%%%%%%%%%%
%%%%%%%%%%%%%%%%%%%%%%%%%%% Part I %%%%%%%%%%%%%%%%%%%%%%%%%%%%%%%%%%
%%%%%%%%%%%%%%%%%%%%%%%%%%%%%%%%%%%%%%%%%%%%%%%%%%%%%%%%%%%%%%%%%%%%%
\part{Supersymmetry breaking with boundaries}
\label{part:ssbb}

\chapter{$F$ term supersymmetry breaking and its mediation mechanisms}
\label{chap:sbmm}

If we have supersymmetry just above the electroweak scale 
the structure (mechanism) of supersymmetry breaking itself should 
give serious effects on low energy physics and experiments. 
In this chapter we investigate how supersymmetry breaking effects 
appear at low energy. It is known that there are two kind of 
contributions to such effects. One is called $F$ term contribution 
that originates in the VEV of an auxiliary field in certain chiral 
multiplet, and the other is $D$ term contribution which comes from 
the VEV of an auxiliary field in certain $U(1)$ vector multiplet. 
In part~\ref{part:ssbb} of this thesis we study such supersymmetry 
breaking especially in higher-dimensional space-time with boundaries. 
Following~\cite{Martin:1997ns} first we review SUSY breaking effects 
in usual 4D ${\cal N}=1$ theory caused by nonvanishing $F$ term 
from the SUSY breaking sector~\cite{Brignole:1993dj}. 

\section{Soft supersymmetry breaking terms in MSSM}
In four-dimensional space-time, the most general ${\cal N}=1$ globally 
supersymmetric Lagrangian for vector and chiral supermultiplet, 
$V=\{V_a\}$ and $\Phi=\{\Phi_{I}\}$ respectively, is given by 
\begin{eqnarray}
{\cal L} = 
  \int d^2 \theta d^2 \bar\theta K(\Phi^\dagger e^{2gV},\Phi)
+ \int d^2 \theta \left[ W(\Phi) + \textrm{h.c.} \right] 
+ \int d^2 \theta \left[ f_{ab}(\Phi) W^\alpha_{\ a} W_{\alpha b} 
                                 + \textrm{h.c.} \right]. 
\label{eq:mgengsusy}
\end{eqnarray}
and the renormalizability restricts the functions 
$K(\Phi^\dagger,e^{2gV},\Phi)$, $W(\Phi)$ and $f_{ab}(\Phi)$ 
respectively to the forms 
\begin{eqnarray}
K(\Phi^\dagger e^{2gV},\Phi) 
  &=& \Phi^\dagger e^{2gV} \Phi, \label{eq:Krncond} \\
W(\Phi) 
  &=& \frac{1}{2} m_{ij} \Phi_i \Phi_j 
     +\frac{1}{6} y_{ijk} \Phi_i \Phi_j \Phi_k, \label{eq:Wrncond} \\
f_{ab}(\Phi) &=& \delta_{ab}, \label{eq:frncond}
\end{eqnarray}
where the parameters $m_{ij}$ and $y_{ijk}$ are determined 
by the charge of $\Phi_i$. 
The minimal supersymmetric SM (MSSM) has the superpotential 
\begin{eqnarray}
W_{\rm MSSM} 
=  \bar{u} {\bf y}_u QH_u - \bar{d} {\bf y}_d QH_d 
 - \bar{e} {\bf y}_e LH_d + \mu H_u H_d, 
\label{eq:mssmsp}
\end{eqnarray}
where $\mu$ corresponds to the Higgs boson mass and dimensionless 
Yukawa coupling parameters ${\bf y}_u$, ${\bf y}_d$ and ${\bf y}_e$ 
are $3 \times 3$ matrices in family space. 
The $SU(2)_W$ weak isospin indices are contracted by anti-symmetric 
tensor $\epsilon^{\alpha \beta}$. 

A realistic phenomenological model must contain SUSY breaking, 
but we should preserve SUSY below the cut-off scale of the theory 
in order to avoid the quadratic divergences. Thus the supersymmetry 
should spontaneously broken at some scale. In addition if we consider 
that SUSY assures electroweak stability, it should be broken just above 
the electroweak scale. In the context of a general renormalizable theory 
(\ref{eq:mgengsusy}) with (\ref{eq:Krncond})-(\ref{eq:frncond}), the 
possible soft supersymmetry breaking terms are 
\begin{eqnarray}
{\cal L}_{\rm soft} 
&=& -\frac{1}{2} (M_\lambda \lambda^a \lambda^a + \textrm{c.c.})
    -(m^2)^i_j (\phi^j)^\ast \phi_i \nonumber \\ && \quad 
    -\left[  \frac{1}{2} b^{ij} \phi_i \phi_j 
           + \frac{1}{6} a^{ijk} \phi_i \phi_j \phi_k 
           + \textrm{c.c.} \right], \nonumber \\
{\cal L}'_{\rm soft} 
&=&  -\frac{1}{2} c_i^{jk} (\phi^i)^\ast \phi_j \phi_k + \textrm{c.c.}, 
\nonumber 
\end{eqnarray}
where $\lambda^a$ is the gaugino field in the vector multiplet $V_a$, 
$\phi_i$ is the scalar component of the chiral multiplet $\Phi_I$, and 
$\textrm{c.c.}$ stands for the charge conjugated terms. 
In MSSM, ${\cal L}'_{\rm soft}$ contribution is usually negligibly small 
and we consider only ${\cal L}_{\rm soft}$. 
MSSM given by (\ref{eq:mssmsp}) has soft breaking terms written as 
\begin{eqnarray}
{\cal L}_{\rm soft}^{\rm MSSM} &=& 
-\frac{1}{2} \left( 
     M_3 \tilde{g} \tilde{g} + M_2 \widetilde{W} \widetilde{W}
    +M_1 \widetilde{B} \widetilde{B} \right) + \textrm{c.c.} 
\nonumber \\ && 
-\left( \tilde{\bar{u}} {\bf a}_u \widetilde{Q} H_u 
       -\tilde{\bar{d}} {\bf a}_d \widetilde{Q} H_d 
       -\tilde{\bar{e}} {\bf a}_e \widetilde{L} H_d \right)
    \textrm{c.c.} 
\nonumber \\ && 
-\widetilde{Q}^\dagger {\bf m}_Q^2 \widetilde{Q} 
-\widetilde{L}^\dagger {\bf m}_L^2 \widetilde{L} 
-\tilde{\bar{u}} {\bf m}_{\bar{u}}^2 \tilde{\bar{u}}^\dagger 
-\tilde{\bar{d}} {\bf m}_{\bar{d}}^2 \tilde{\bar{d}}^\dagger 
-\tilde{\bar{e}} {\bf m}_{\bar{e}}^2 \tilde{\bar{e}}^\dagger 
\nonumber \\ && 
-m_{H_u}^2 H_u^\ast H_u 
-m_{H_d}^2 H_d^\ast H_d 
-(b H_u H_d +\textrm{c.c.}). 
\nonumber 
\end{eqnarray}

The constraints on flavor changing neutral current (FCNC) and 
CP-violation from current experiments suggest that some soft 
breaking parameters take forms as 
\begin{eqnarray}
&
{\bf m}_Q^2 = m_Q^2 {\bf 1}, \ 
{\bf m}_{\bar{u}}^2 = m_{\bar{u}}^2 {\bf 1}, \ 
{\bf m}_{\bar{d}}^2 = m_{\bar{d}}^2 {\bf 1}, \ 
{\bf m}_L^2 = m_L^2 {\bf 1}, \ 
{\bf m}_{\bar{e}}^2 = m_{\bar{e}}^2 {\bf 1},& 
\nonumber \\ 
&
{\bf a}_u = A_{u0} {\bf y}_u, \ 
{\bf a}_d = A_{d0} {\bf y}_d, \ 
{\bf a}_e = A_{e0} {\bf y}_e,& 
\nonumber \\ 
&
\arg (M_1), \arg (M_2), \arg (M_3), 
\arg (A_{u0}), \arg (A_{d0}), \arg (A_{e0}) 
=0 \textrm{ or } \pi,& 
\nonumber 
\end{eqnarray}
where 
\begin{eqnarray}
{\bf y}_u \sim \pmatrix{0 & 0 & 0 \cr 0 & 0 & 0 \cr 0 & 0 & y_t}, \ 
{\bf y}_d \sim \pmatrix{0 & 0 & 0 \cr 0 & 0 & 0 \cr 0 & 0 & y_b}, \ 
{\bf y}_e \sim \pmatrix{0 & 0 & 0 \cr 0 & 0 & 0 \cr 0 & 0 & y_\tau}. 
\nonumber 
\end{eqnarray}
If these conditions are satisfied at the high energy scale, the 
renormalization group (RG) evolution does not introduce new CP-phases, 
and supersymmetric contributions to FCNC and CP-violating observables 
can be acceptably small at low energy scale. 

\section{Communicating supersymmetry breaking from hidden sector}

In the previous section we have introduced possible soft breaking 
parameters in MSSM and some constraints on them. Now we consider 
the origin of the soft breaking parameters. Unfortunately it is known that 
a $D$ term VEV for $U(1)_Y$ does not lead to an acceptable spectrum, 
and there is no gauge-singlet candidate in MSSM contents whose $F$ term 
could develop VEV. That is, the ultimate SUSY-breaking order parameter can't 
belong to any of the supermultiplet in MSSM, so one needs a hidden sector 
where supersymmetry breaks and a SUSY-breaking communication mechanism 
from the hidden to the visible MSSM sector. 

Furthermore we have another difficulty if we consider the SUSY-breaking 
mediation mechanism through renormalizable interaction at tree-level 
because of the supertrace theorem. In theories with spontaneous SUSY 
breaking, we have general sum rule between the tree-level squared masses: 
\begin{eqnarray}
\textrm{STr}{\cal M}^2 = \sum_J (-1)^{2J} (2J+1) {\cal M}_J^2 =0, 
\nonumber 
\end{eqnarray}
where ${\cal M}_J$ denotes the tree level mass of a particle with spin $J$. 
This implies that if the chiral fermions are light as in SM 
we should have light scalar fields which has never been discovered. 

Therefore we need a SUSY-breaking communication mechanism 
through tree-level nonrenormalizable or loop-level renormalizable 
interaction. A lot of proposals for such mechanism are given 
and we will see them in the following. 

\subsection{Gravity mediation}
First we show a case with tree-level nonrenormalizable 
interaction. A typical example of such interaction is gravity. 
The gravity couples to all particles universally and it communicate 
the SUSY breaking effect from the hidden to the MSSM sector. 
In the supergravity Lagrangian, the relevant terms to the 
soft SUSY-breaking are written as\footnote{
Even these soft breaking terms are restricted version, for simplicity, 
in a general ${\cal N}=1$ 4D supergravity Lagrangian. For a most general 
form we should refer \cite{Nilles:1983ge}, for example.} 
\begin{eqnarray}
{\cal L}  &=& 
-\frac{1}{M_{\rm Pl}} F_X \sum_a \frac{1}{2} f_{a} \lambda^a \lambda^a 
+ \textrm{c.c.} \nonumber \\ && 
-\frac{1}{M_{\rm Pl}^2} F_X F_X^\ast k^i_j \phi_i (\phi^j)^\ast \nonumber \\ && 
-\frac{1}{M_{\rm Pl}} F_X \left(
         \frac{1}{2} \mu'^{ij} \phi_i \phi_j 
       + \frac{1}{6} y'^{ijk} \phi_i \phi_j \phi_k \right) 
+\textrm{c.c.} \nonumber \\ && + \ldots, \nonumber 
\end{eqnarray}
where $F_X$ is the auxiliary field for a chiral supermultiplet $X$ in 
the hidden sector, and $\phi_i$ and $\lambda^a$ is the scalar and 
gaugino field respectively in the MSSM. 
The parameters $f_a$, $k^i_j$, $y'^{i,j,k}$, $\mu'^{ij}$ and the order 
parameter $F_X$ are given by the K\"ahler function $K$, superpotential 
$W$ and gauge kinetic function $f_{ab}$ in Eq.~(\ref{eq:mgengsusy}), 
and their derivatives. 

For example if we assume the diagonal form: 
\begin{eqnarray}
K &=& \widetilde{K}(X) 
     +k\delta_{\bar{i}j} 
      (\Phi^{\bar{i}})^\dagger e^{2g\sum_a V_a} \Phi^j, 
      \label{eq:diagK} \\ 
W &=& \widetilde{W}(X) 
     +\frac{1}{2} \beta \mu^{ij} \Phi_i \Phi_j 
     +\frac{1}{6} \alpha y^{ijk} \Phi_i \Phi_j \Phi_k, 
      \label{eq:diagW} \\
f_{ab} &=& f\delta_{ab}, 
      \label{eq:diagf}
\end{eqnarray}
the parameters are given as 
$f_a =f$, $k^i_j=k \delta^i_j$, $y'^{i,j,k}=\alpha y^{i,j,k}$, 
$\mu'^{ij}=\beta \mu^{ij}$ and 
$F_X = M_{\rm Pl}^3 \widetilde{W} \partial_{X} \widetilde{K} 
+ M_{\rm Pl} \partial_X \widetilde{W} $. 
The VEV of the scalar potential is 
$V_0 = \langle V \rangle 
= M_{\rm Pl}^2 \left( |F_X|^2 - 3m_{3/2}^2 \right)$ where 
$m_{3/2}^2=M_{\rm Pl}^{-4}e^{\widetilde{K}}|\widetilde{W}|^2$ 
provides the gravitino mass. In the case of MSSM with 
$\Phi^i = \{ Q,\bar{u},\bar{d},L,\bar{e},H_u,H_d \}$ and 
superpotential (\ref{eq:mssmsp}), 
we have the soft SUSY-breaking parameters as 
\begin{eqnarray}
&& M_3 = M_2 = M_1 = m_{1/2}, \nonumber \\
&& {\bf m}_Q^2 = {\bf m}_{\bar u}^2 = {\bf m}_{\bar d}^2 = 
{\bf m}_L^2 = {\bf m}_{\bar e}^2 = m_0^2 {\bf 1}, \ 
m_{H_u}^2 = m_{H_d}^2 = m_0^2 \nonumber \\
&& {\bf a}_u = A_0 {\bf y}_u,\ {\bf a}_d = A_0 {\bf y}_d,\ 
{\bf a}_e = A_0 {\bf y}_e, \nonumber \\
&& b = B_0 \mu, 
\nonumber 
\end{eqnarray}
where 
\begin{eqnarray}
m_{1/2} = f \frac{\langle F_X \rangle}{M_{\rm Pl}},\ 
m_0^2 = k \frac{|\langle F_X \rangle|^2}{M_{\rm Pl}^2},\ 
A_0 = \alpha \frac{\langle F_X \rangle}{M_{\rm Pl}},\ 
B_0 = \beta \frac{\langle F_X \rangle}{M_{\rm Pl}}.\ 
\label{eq:diaggravmedpara}
\end{eqnarray}

Further particular models of gravity-mediated SUSY-breaking 
are more predictive, relating some of the parameters in 
Eq.~(\ref{eq:diaggravmedpara}) to each other and to the 
mass of the gravitino $m_{3/2}$ like 
\begin{eqnarray}
&& m_{1/2}=m_0=A_0 \quad 
\textrm{(minimal)}, 
\label{eq:minimalsugra} \\
&& m_0^2=m_{3/2}^2,\ m_{1/2}=-A_0=\sqrt{3}m_{3/2} \quad 
\textrm{(dilaton-dominated)}, 
\label{eq:dilatondomsugra} \\ 
&& m_{1/2} \gg m_0,\ A_0,\ m_{3/2} \quad 
\textrm{(no-scale)}. 
\label{eq:noscalesugra}
\end{eqnarray}

In Eqs.~(\ref{eq:diagK})-(\ref{eq:diagf}) we have only assumed 
diagonal forms. Generically there is no reason to appear in 
such simple forms. From general form of $K$, $W$ and $f_{ab}$ 
we have large dependence on flavor of soft terms even at the 
cut-off scale (as an initial condition of RG flow) 
which would be dangerous from the viewpoint of FCNCs. 
Also in Eqs.~(\ref{eq:diagK})-(\ref{eq:diagf}) we have assumed 
the form of $K$, $W$ and $f_{ab}$ in which the hidden sector ($X$ field) 
is completely separated from the visible sector ($\Phi_i$ fields). 
This assumption will become more natural in the brane world picture 
that will be discussed latter.

\subsection{Gauge mediation}
Next we show a case with renormalizable interaction. 
The typical example is gauge 
interaction~\cite{Dine:1982gm,Dine:1993yw,Giudice:1998bp}. 
We assume the existence of the MSSM sector $\{ \Phi_i \}$, 
the messenger sector $\{ \widetilde\Phi_a \}$ and 
SUSY-breaking sector $\{ X \}$. 
The messenger sector fields $\widetilde\Phi_a$ are charged under 
MSSM gauge interactions and $X$ is MSSM gauge singlet. 
We assume constant and diagonal K\"ahler and gauge kinetic functions, 
and a superpotential 
\begin{eqnarray}
W &=& W_{\rm MSSM}(\Phi_i)+W_{\rm mess}(\widetilde\Phi_a,X), 
      \label{eq:gaugemedW} \\
W_{\rm mess} 
&=& X \widetilde{\Phi}_a^\dagger \widetilde{\Phi}_a. 
\nonumber 
\end{eqnarray}
We also assume the VEV of the superfield X as 
\begin{eqnarray}
\langle X \rangle = M_X + F_X \theta^2, 
\nonumber 
\end{eqnarray}
where $M_X$ gives a mass of the messenger field and 
$F_X$ is an order parameter of the SUSY breaking. 
Phenomenologically we should take $M_X^2 \gg F_X$ 
in order to decouple the messenger fields from 
low energy MSSM with soft SUSY breaking terms. 

In such situation, we can explicitly calculate 
the soft parameters in MSSM up to its running effect. 
As is shown in appendix~\ref{app:wrsp} these are extracted from 
the wave function renormalization~\cite{Giudice:1997ni} as 
\begin{eqnarray}
M_a(\mu) &=& -\frac{1}{2} \left. 
\frac{\partial \ln S_a(X,t)}{\partial \ln X} 
\right|_{X=M_X} \frac{F_X}{M_X}, 
\label{eq:gaugemedMa} \\
m_i^2(\mu) &=& -\left. 
\frac{\partial^2 \ln Z_i(X^\dagger,X,t)}
{\partial \ln X^\dagger \partial \ln X} 
\right|_{X=M_X} \frac{F_X^\dagger F_X}{M_X^\dagger M_X}, 
\label{eq:gaugemedmi} \\
a_{ijk}(\mu) &=& \left. \frac{\partial 
\ln \left[ Z_i(X^\dagger,X,t)Z_j(X^\dagger,X,t)
Z_k(X^\dagger,X,t) \right]}{\partial \ln X} 
\right|_{X=M_X} \frac{F_X}{M_X}, 
\label{eq:gaugemedAi}
\end{eqnarray}
where $t=\ln (M_X^2/\mu^2)$ and $\mu$ $(= e^{-t/2}M_X < M_X)$ 
is the low energy scale at which 
the soft terms are defined, and $S_a(M_X,t)$ and $Z_i(M_X^\dagger,M_X,t)$ 
are the wave function renormalization factor of the MSSM vector multiplet $V^a$ 
and chiral multiplet $\Phi_i$ respectively. Therefore these soft parameters 
can be given by the well-known MSSM gauge couplings $g_a(\mu)$ 
of the vector multiplets $V_a$ as are shown in appendix~\ref{app:wrsp}. 
So they are quite predictive. 

In this gauge mediated scenario, the soft terms are generated 
at the messenger scale $M_X$. It is natural to assume 
$M_X \ll M_F \sim M_{\rm Pl}$, where $M_F$ is the flavor breaking scale 
at which the flavor dynamics is frozen and the flavor dependence is 
condensed into only Yukawa couplings in MSSM. 
In such situation the soft terms receive flavor breaking effects 
through only Yukawa couplings in $Z_i$ because the gauge interactions 
are flavor universal. This is the advantage of gauge mediation 
comparing with the gravity mediated case. 

\subsection{Anomaly mediation}
In this section we consider the extreme situation that MSSM 
sector and SUSY-breaking sector are completely decouple to 
each other. Even in such case the SUSY breaking mediation is 
possible if the superconformal anomaly exists 
in the system~\cite{Randall:1998uk}.

In the superconformal formulation we introduce the 
Weyl compensator field $\varphi$. In such formulation 
the above visible and hidden decoupling are written by 
\begin{eqnarray}
{\cal L} 
&=& \int d^4 \theta \varphi^\dagger \varphi 
    K_{\rm visible}(\Phi^\dagger,\Phi) 
   +\int d^2 \theta \varphi^3 W_{\rm visible}(\Phi) \nonumber \\ && 
   +\int d^4 \theta \varphi^\dagger \varphi 
    K_{\rm hidden}(X^\dagger,X) 
   +\int d^2 \theta \varphi^3 W_{\rm hidden}(X), 
\nonumber 
\end{eqnarray}
and the supersymmetric world is described by 
$\langle \varphi \rangle =1$. If SUSY breaking occurs in the 
hidden sector by the nonzero $F_X$, we also obtain the 
nonzero $F_\varphi$ by integrating out the hidden sector. 
We then have $\langle \varphi \rangle =1+\theta^2 F_\varphi/\Lambda$ 
and the SUSY breaking is communicated to the visible sector 
through the Weyl compensator $\varphi$. 

The soft parameters in such case are given by 
Eqs.~(\ref{eq:gaugemedMa})-(\ref{eq:gaugemedAi}) 
in which we replace as $M_X \to \Lambda$ and 
$F_X \to F_\varphi$, where $\Lambda$ is the cut off parameter 
of the theory. This can be understood as follows. 
We perform a scale transformation $\varphi \mu \to \mu$ and 
$\varphi \Phi \to \Phi$ in the theory. 
(We should note that $\varphi$ is not a dynamical field.) 
At the tree level the bilinear part of 
$K_{\rm visible}(\Phi^\dagger,\Phi)$ becomes as 
\begin{eqnarray}
\int d^4 \theta \varphi^\dagger \varphi 
\Phi^\dagger \Phi \to 
\int d^4 \theta \Phi^\dagger \Phi. 
\nonumber 
\end{eqnarray}
Therefore there is no SUSY breaking at the tree level because 
the tree level Lagrangian is (assumed to be) scale invariant. 
However at the loop level, we need a regularization and 
we introduce a cut off scale $\Lambda$. The effective Lagrangian 
transforms, under the above transformation, as 
\begin{eqnarray}
\int d^4 \theta 
Z(\Lambda/\mu, \Lambda^\dagger/\mu^\dagger) 
\varphi^\dagger \varphi \Phi^\dagger \Phi 
\to \int d^4 \theta 
Z(\varphi \Lambda/\mu, 
\varphi^\dagger \Lambda^\dagger/\mu^\dagger) 
\Phi^\dagger \Phi, 
\nonumber 
\end{eqnarray}
where $Z$ is a wave-function renormalization of $\Phi$. 
Then we have SUSY breaking at the loop level. 
This comes from the scale (super Weyl) anomaly 
of the theory. 

Following the similar argument to the above gauge-mediated 
case (see also appendix~\ref{app:wrsp}) 
except that the messenger scale is now the cut off scale, 
the wave function renormalization leads us to the explicit form 
of the soft parameters. We can write them as 
\begin{eqnarray}
M_a(\mu) &=& 
-\frac{\beta_{g_a^2}}{2g_a^2} \frac{F_\varphi}{\Lambda}, 
\label{eq:am1} \\ 
m_i^2(\mu) &=& 
-\frac{1}{4}\frac{\partial \gamma_i}{\partial \ln \mu}
 \left| \frac{F_\varphi}{\Lambda} \right|^2, \label{eq:am2} \\
a_{ijk}(\mu) &=& 
-\frac{1}{2}(\gamma_i+\gamma_j+\gamma_k)
\frac{F_\varphi}{\Lambda}, \label{eq:am3}
\end{eqnarray}
where $\beta_{g_a^2} = 2g_a \beta_{g_a}$ is a beta function 
of the vector multiplet $V_a$, and $\gamma_i$ is an 
anomalous dimension of the chiral multiplet $\Phi_i$. 
These beta functions and anomalous dimensions are the 
usual MSSM ones, because extra messenger fields don't exist here, 
and the soft parameters are generated at the cut off scale. 
So the result is very simple. 
We also note that the Weyl compensator couplings are flavor universal, 
we have flavor independent soft parameters at the cut off scale. 
However in a minimal anomaly mediated model it is known that we 
have negative slepton masses at the low energy due to the RG flow. 
This is the serious problem of minimal scenario of anomaly mediation. 

\section{Spatially sequestering hidden sector in extra dimensions}

In the previous sections we have only assumed that the 
hidden (SUSY breaking) sector and MSSM sector don't couple directly 
to each other. The assumption is, however, sometimes unnatural 
because there frequently exist cases that the direct couplings 
between hidden and visible sector, that keeps respecting all 
symmetries, can be written. 

\begin{figure}[t]
\centerline{\epsfig{figure=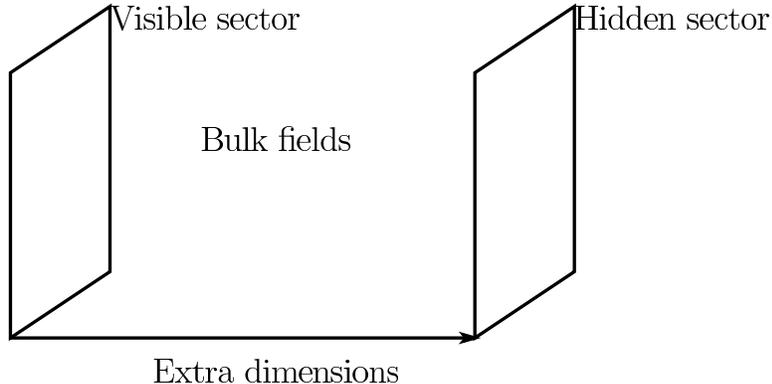,width=0.6\linewidth}}
\caption{Schematic view of the hidden sector in brane world.}
\label{fig:hiddensector}
\end{figure}

Recent elegant solution to such a problem is obtained by 
sequestering the hidden sector by the spatiality in extra 
dimensions. The brane world picture naturally gives such 
sequestering mechanism as is shown in Fig~\ref{fig:hiddensector}. 
It is very interesting idea and here we review SUSY-breaking 
mediation mechanisms in such situation in this section\footnote{
If we consider SUSY in more than four-dimension, we need 
e.g. orbifold compactification to obtain realistic (chiral) 
4D ${\cal N}=1$ SUSY. So here we assume such nontrivial 
compactification.}. 
Even if the hidden and visible sector are separated spatially 
in extra dimension, the anomaly-mediated SUSY breaking 
contribution always exists when the system possesses 
superconformal anomaly. So we should take 
Eqs.~(\ref{eq:am1})-(\ref{eq:am2}) into account in any case. 
Unfortunately this simple setting of pure anomaly mediation 
results in the negative slepton masses. We need some 
correction in order to obtain realistic spectrum. 
In addition to these anomaly mediated contributions, 
we have additional contributions mediated by some bulk 
fields if they exist. We show several typical examples 
of such SUSY-breaking contribution in the following. 

\subsection{Gaugino mediation}

First we consider the equivalent setup to the gauge mediation 
except for the assumption that `bulk' SM gauge multiplet communicate 
the SUSY-breaking effect from messenger (hidden) sector to the SM 
matter chiral multiplets spatially separated from the former 
in extra dimensions. In such case, the zero mode gaugino obtains 
the same soft mass (\ref{eq:gaugemedMa}) as the gauge-mediated case, 
but the other soft parameters related to scalars are suppressed as 
\begin{eqnarray}
M_a(M_X) &\sim& \frac{F_X}{M_X}, \nonumber \\
m_i^2(M_X) &\sim& \frac{1}{(R \Lambda)^2} 
\left| \frac{F_X}{M_X} \right|^2 \ll M_a^2(M_X), \nonumber \\ 
A_{i0}(M_X) &\sim& \frac{1}{R \Lambda} \frac{F_X}{M_X} 
\ll M_a(M_X), \nonumber 
\end{eqnarray}
for $1/R < M_X < \Lambda$. 
This is because the scalar soft terms are induced through the loop effects of 
bulk SM gauge multiplet {\it without zero mode}, namely loop effects 
from KK higher modes only, because the loop should be attached both 
sectors (both boundary planes) separated by the spatial distance $\pi R$. 
Due to this fact we obtain a no-scale type initial conditions 
(\ref{eq:noscalesugra}) for the soft parameters at the compactification scale. 
Because the SUSY breaking masses for all chiral matter 
fields other than the third generation squarks are dominated by the gaugino 
loop in this type of mediation, this scenario is known as `gaugino mediated' 
SUSY breaking~\cite{Kaplan:1999ac}. 
If the gaugino-mediated contribution dominates the anomaly-mediated one 
the problem of negative slepton mass in latter case is dissolved. 
Moreover the gaugino-mediated SUSY-breaking scenario gives 
necessity of the no-scale type initial condition which is only assumed 
in the gravity-mediated scenario.

\subsection{Kaluza-Klein and radion mediation}

We can also consider the case that SUSY-breaking contribution 
is dominated by the radion chiral multiplet $T$ with a non vanishing 
VEV of its zero mode, $\langle T_0 \rangle =1/R+\theta^2F_{T_0}$. 
The radion is referred as a 
field that represents physical degrees of freedom for oscillating the size 
of the extra dimensions. This extremely originates from the gravity multiplet. 
In the low energy effective theory below the compactification scale, $1/R = M_c$, 
the radion only couples to the Kaluza-Klein (KK)~\cite{Kaluza:kk} higher modes 
of the other field. Thus the KK zero modes or fields on a brane obtain 
the soft SUSY breaking terms through the loop effects of the higher 
modes if the SUSY-breaking effect is expressed by the VEV of the auxiliary 
component in the radion chiral multiplet in the low energy effective theory. 
This situation is known as `Kaluza-Klein (KK) mediated' SUSY-breaking 
scenario~\cite{Kobayashi:2000ak}. The soft terms are given by 
\begin{eqnarray}
M_a (\mu) &=& \frac{\beta_{\alpha_a}}{2\alpha_a}\frac{F_{T_0}}{M_c}, \nonumber \\
m_i^2 (\mu) &=& -\frac{\delta}{2}\gamma_i
          \left| \frac{F_{T_0}}{M_c} \right|^2, \nonumber \\
A_{i0} (\mu) &=& -\gamma_i\frac{F_{T_0}}{M_c}, 
\nonumber 
\end{eqnarray}
where
\begin{eqnarray}
\beta_\alpha &=& \frac{d \alpha}{dt} 
              =  \frac{b_{\rm KK}}{2\pi}N(\mu)\alpha^2, \nonumber \\ 
\gamma_i &=& \frac{d \ln Z_i}{dt} 
          =  \frac{c_{\rm KK}}{2\pi}N(\mu)\alpha, 
\nonumber 
\end{eqnarray}
$\alpha = g^2/(4\pi)$ and 
$b_{\rm KK}$ is the beta function coefficient determined by 
KK mode contributions and $C_{\rm KK}=4C_2(R_{\phi_i})$. 
$N(\mu)$ is the number of the KK modes below the scale $\mu$ 
and is approximated by the volume of a sphere with radius $\Lambda/\mu$, 
which is given by 
$N(\mu) \simeq \pi^{\delta/2}(\mu/\Lambda)^\delta/\Gamma(1+\delta/2)$, 
where $\delta$ is the number of the extra dimensions. 

If the gaugino condensation occurs in a bulk and a brane super Yang-Mills 
theories, a dynamical superpotential $W_{\rm eff} = ae^{-bT}+c$ can be 
generated and it is known that this superpotential can stabilize the VEV 
of the radion $1/R$, and simultaneously give nonzero $F_T$. So this is an 
interesting example of dynamical supersymmetry breaking and its mediation 
by the radion (radion-mediated SUSY breaking~\cite{Chacko:2000rm}).  
In the radion-mediated scenario the supersymmetry-breaking order parameter 
$F_T$ is a VEV of the full five-dimensional auxiliary field in a radion 
multiplet, while in the KK-mediated scenario $F_{T_0}$ is a VEV of the zero 
mode of it.

\section{Fermion flavor structure and $U(1)$ symmetry}
In the above section phenomenologically we have noticed 
the SUSY flavor problem, and we saw various SUSY-breaking 
mediation mechanisms which can give degenerate sfermion spectrum 
as an initial condition of RG flow. 
However as we mentioned in Chapter~\ref{chap:intro}, 
on the other hand, we should have hierarchical structure of 
mass and mixing angle in the fermion sector in MSSM. 
For example the CKM matrix is approximately given as 
\begin{eqnarray}
V_{\rm CKM} = 
\pmatrix{1 & \lambda & \lambda^3 \cr 
         \lambda & 1 & \lambda^2 \cr 
         \lambda^3 & \lambda^2 & 1}, 
\label{eq:Vckm}
\end{eqnarray}
where $\lambda \sim 0.22$ is the Cabibbo angle. 
Usually we don't treat the explicit mechanism of flavor symmetry 
breaking and only assume flavor-breaking Yukawa couplings in MSSM. 
In the limit where Yukawa couplings vanish, we have global $U(3)^5$ 
flavor symmetry in MSSM that originates from flavor independence of 
the MSSM gauge interactions. In a case that we consider flavor dynamics 
we assume that the subgroup $G_{\rm f}$ of $U(3)^5$ is a fundamental 
(global or local) symmetry. We sometimes consider that $G_{\rm f}$ is 
broken spontaneously by the VEV of some scalar fields. 
In such scenario hierarchical structure of Yukawa coupling is 
generated at each stage of $G_{\rm f}$ symmetry breaking, or generated 
by the small value of a ratio between the VEV and the scale of 
flavor dynamics. 

As a typical example we consider the latter case. 
One of the way deriving realistic CKM matrix and fermion masses 
is known as Froggatt-Nielsen mechanism~\cite{Froggatt:1978nt}. 
We assume ($G_{\rm f}=$) $U(1)$ symmetry and extra field $\chi$ with 
its $U(1)$ charge $q_\chi =-1$. We also assume that 
the MSSM field $\Phi_i$ has a charge $q_{\Phi_i}$. We have 
up-sector couplings as 
\begin{eqnarray}
( \chi/M_{\rm Pl} )^{n_{ij}} u_i Q_j H_u, 
\label{eq:fnhdc}
\end{eqnarray}
where $n_{ij}= q_{u_i}+q_{Q_j}+q_{H_u}$. 
Similarly the down and the lepton sector couplings are 
given. After the $U(1)$ breaking by the VEV, 
$\langle \chi \rangle \ne 0$, we have hierarchical Yukawa 
couplings $({\bf y}_u)_{ij} \sim \epsilon^{n_{ij}}$ for the up-sector 
if $\epsilon= \langle \chi \rangle /M_{\rm Pl} <1$, 
and similarly hierarchical ${\bf y}_d$ and ${\bf y}_e$ are given. 

We should note that if such $U(1)$ symmetry is gauged we generally 
have contributions to soft SUSY breaking parameters 
from the Fayet-Iliopoulos (FI) $D$ term of the $U(1)$ 
vector multiplet. In such case the flavor breaking 
and the SUSY breaking can be deeply connected to each other. 
Furthermore, if the $U(1)$ is anomalous, it is known that 
the FI term is induced through the one-loop effects 
even though there is no such FI term in the tree level.

\subsection{Anomalous $U(1)$ symmetry from 4D string models}
\label{ssec:anomalousu1}

It is also known that such anomalous $U(1)$ gauge symmetries appear 
in 4D string models~\cite{Witten:dg,Kobayashi:1996pb,Ibanez:1998qp}.
Its anomaly is canceled by the 4D Green-Schwarz mechanism. 
For example, in weakly coupled 4D heterotic string models, the 4D dilaton 
field $S$ is required to transform nontrivially at one-loop level,
\begin{equation}
S \rightarrow S + {i \over 2 }\delta_{GS}\Lambda_X ,
\label{GS-S}
\end{equation}
under the $U(1)$ transformation with the transformation 
parameter $\Lambda_X$, where the Green-Schwarz coefficient 
is assigned as $\delta_{GS} = {\rm tr} Q_X /(48 \pi^2)$.
Then, the FI term $\xi$ is induced by the vacuum expectation value (VEV) 
of the dilaton field $S$,
\begin{equation}
\xi_S = {1 \over 2} \delta_{GS} 
\langle K'(S + \bar S) \rangle  M_{\rm Pl}^2,
\label{GS-FI}
\end{equation}
where $K'(S + \bar S)$ is the first derivative of K\"ahler 
potential of the dilaton field. At the tree level, we have 
$K(S + \bar S)=- \ln (S + \bar S)$, and the VEV of the dilaton field 
provides the 4D gauge coupling $g_{\rm 4D}^2 = 1/\langle {\rm Re} S\rangle $. 
The FI term is suppressed by one-loop factor unless ${\rm tr} Q_X$ 
is quite large. In type I models, other singlet fields, i.e. moduli fields, 
also contribute to the 4D Green-Schwarz mechanism and the FI term is 
induced by their VEVs~\cite{Ibanez:1998qp}. Therefore, the total magnitude 
of FI term is arbitrary in type I models, while its value of heterotic models 
is one-loop suppressed compared with $M_{\rm Pl}$. 

Some fields develop their VEVs along $D$-flat directions 
to cancel the FI term, and the anomalous $U(1)$ gauge symmetry is broken 
around the energy scale $\xi^{1/2}$, e.g., just below $M_{\rm Pl}$ in 
heterotic models. Hence this anomalous $U(1)$ gauge symmetry does not 
remain at low energy scale, although its discrete subsymmetry or 
the corresponding global symmetry may remain even at low energy scale.
However, it is known that the anomalous $U(1)$ symmetry with nonvanishing 
FI term would provide phenomenologically interesting effects. 
Actually, several applications of the anomalous $U(1)$ was investigated. 
As is mentioned above, one of interesting applications is generation of 
hierarchical Yukawa couplings by the Froggatt-Nielsen 
mechanism~\cite{Froggatt:1978nt}. We can use higher dimensional couplings 
(\ref{eq:fnhdc}) as the origin of hierarchical fermion masses and mixing 
angles, where the ratio 
$\langle \chi \rangle / M_{\rm Pl} \sim \xi^{1/2}/M_{\rm Pl}$ 
plays a important role for deriving fermion masses and mixing 
angles~\cite{Ibanez:ig}. 

The anomalous $U(1)$ symmetry breaking induces $D$ term contributions 
to soft SUSY breaking scalar masses~\cite{Nakano:1994sw} as is also 
mentioned above. These $D$ term contributions are, in general, 
proportional to $U(1)$ charges and their overall magnitude 
is of the order of the other soft SUSY breaking scalar masses. 
Thus, these $D$ term contributions become another source of 
nonuniversal sfermion masses. That would be dangerous from 
the viewpoint of FCNCs. 

On the other hand, there is a model of SUSY breaking and its mediation 
mechanism via the anomalous $U(1)$ and FI term~\cite{Dvali:1996rj}. 
(See also Ref.~\cite{Kageyama:vf}.) In this type of models, $D$ term 
contributions can be much larger than other soft scalar masses and 
gaugino masses. For example, in the case that only the top quark has 
vanishing $U(1)$ charge, stop and gaugino masses can be ${\cal O}(100)$ 
GeV and the other sfermion masses can be ${\cal O}(10)$ TeV. This type 
of sfermion spectra can satisfy FCNC constraints as well as the 
naturalness problem. Actually that corresponds to the decoupling 
solution\footnote{
With this type of spectra, we must be careful about 
two-loop renormalization group effects on stop masses, 
which could become tachyonic~\cite{Arkani-Hamed:1997ab}.} 
for the SUSY flavor problem~\cite{Dimopoulos:1995mi}. 

In this section we have reviewed a flavor structure derived 
from $U(1)$ symmetry. We have also remarked that the 
$D$ term contributions to the sfermion masses can be important 
if we have such $U(1)$ symmetry in the theory. 
Then we will study such $D$ term effects in brane world scenarios 
in the next chapter.

\chapter{$D$ term supersymmetry breaking from boundary}
\label{chap:dtermsb}

In the previous chapter we reviewed some SUSY breaking mediation 
mechanisms. Various mechanisms are proposed to solve the 
SUSY flavor problem. We have also seen that, in the brane world framework, 
we find a solution to the difficulty how to separate hidden sector 
from the visible sector. In such spatially sequestered hidden-sector 
setup, we can have some natural soft SUSY breaking parameters as an 
initial condition of renormalization group running. 

We have also noticed a $U(1)$ flavor symmetry and hierarchical structure 
in the fermion sector. 
An important mechanism suggested by Froggatt-Nielsen is introduced in 
order to generate hierarchical fermion masses and mixings in (MS)SM. 
The FN mechanism requires extra $U(1)$ symmetry. If it is anomalous 
gauge symmetry, even in a four-dimensional supergravity framework, 
we have FI term of the vector multiplet. 
This also gives SUSY breaking contributions to the soft parameters. 

Now we are interested in gauged (e.g., flavor) $U(1)$ dynamics 
in the brane world. 
If we introduce $U(1)$ gauge symmetry with FI term in the bulk, 
how large are the $D$ term contributions to bulk and brane sfermion masses? 
In addition it is known that the FI term can exist on fixed points in generic 
orbifold theories~\cite{GrootNibbelink:2002qp,GrootNibbelink:2002wv,
Ghilencea:2001bw,Barbieri:2001cz,Scrucca:2001eb,Barbieri:2002ic} 
(see also appendix~\ref{app:olidfi}) even in the local SUSY (supergravity) 
framework~\cite{Barbieri:2001cz} and even without anomalous symmetry. 
Such FI term can also be {\it dynamically} generated on the 
four-dimensional boundaries~\cite{GrootNibbelink:2002qp,GrootNibbelink:2002wv,
Ghilencea:2001bw,Barbieri:2001cz,Scrucca:2001eb,Barbieri:2002ic} 
because of the orbifolding as is shown in appendix~\ref{app:olidfi}. 

Therefore, in this chapter\footnote{This chapter is based on Ref.~\cite{Abe:2002ps} 
with T.~Higaki and T.~Kobayashi}, we study SUSY $U(1)$ gauge theory 
in higher-dimensional (especially five-dimensional, for simplicity) 
space-time with four-dimensional boundaries, and the extra dimension 
is compactified by $S_1/Z^2$ (as a concrete example to obtain chiral theory). 
The SUSY is ${\cal N}=1$ in terminology of four-dimensional space-time 
due to the orbifolding. 
In such circumstance we study the SUSY breaking structures, 
especially caused by $U(1)$ FI terms on the boundary~\cite{Abe:2002ps}.

\section{5D $S^1/Z_2$ model with generic FI terms}
\label{sec:2}

We consider 5D SUSY model on $M^4 \times S^1/Z_2$.
Because of orbifolding, SUSY is broken into ${\cal N}=1$ 
SUSY in terminology of 4D theory.
We concentrate on a $U(1)$ vector multiplet 
and charged hypermultiplets in bulk.
The 5D vector multiplet consists of a 5D vector field 
$A_M$, gaugino fields $\lambda_{\pm}$, a real scalar field 
$\Sigma$ and a triplet of auxiliary fields $D_a$ ($a=1,2,3$). 
Their $Z_2$ parities are assigned in a way consistent 
with 4D ${\cal N}=1$ SUSY and $U(1)$ symmetry~\cite{Mirabelli:1997aj}. 
For example, the $Z_2$ parity of 
$\Sigma$ field as well as $A_5$ is assigned as odd. 
A 5D hypermultiplet consists of 
two complex scalar fields $\Phi_{\pm}$ and their 
4D superpartners as well as their auxiliary fields.
The fields $\Phi_{\pm}$ have $Z_2$ parities $\pm$ and 
$U(1)$ charge of the $Z_2$ even field $\Phi_+$ is 
denoted by $q$, while the corresponding $Z_2$ odd field 
has the $U(1)$ charge $-q$.
$Z_2$ odd fields should vanish at the fixed points, e.g.
\begin{equation}
\Sigma(0)=\Sigma(\pi R) = 0,\qquad 
\Phi_-(0) = \Phi_-(\pi R)=0.
\end{equation}
The first derivatives of $Z_2$ even fields 
along $y$ should vanish at the fixed points, e.g.
\begin{equation}
\partial_y \Phi_+(0)= \partial_y \Phi_+(\pi R) =0.
\end{equation}
In addition to these bulk fields, 
our model includes brane fields, i.e. 
4D $N=1$ chiral multiplets on the two fixed points, 
and their scalar components are written as  
$\phi_I$ ($I=0,\pi$), where $I$ denotes two 
fixed points.
Their $U(1)$ charges are denoted by $q_I$.
The detailed formulation is shown in appendix~\ref{app:osform}. 
At the first stage, we do not consider any 
superpotential on the fixed points.

In this section, following \cite{Abe:2002ps} we study the VEV of $\Sigma$ 
with generic values of $\xi_0$ and $\xi_\pi$, and consider zero mode 
wave function profiles under nontrivial VEV of $\Sigma$ in bulk. 
We will also study mass eigenvalues and profiles of higher modes. 
In addition, we will consider an application on SUSY breaking and examine 
implication of $D$ term contributions to SUSY breaking scalar masses in 
the next section. 

We consider the following generic FI terms on the fixed points: 
\begin{equation}
\xi (y) = \xi_0 \delta (y) +\xi_\pi \delta (y - \pi R).
\label{5D-FI}
\end{equation}
Here we just assume these generic FI terms.\footnote{
If they are generated at one-loop level, the sum of 
FI coefficients $\xi_0 + \xi_\pi$ is 
proportional to the sum of $U(1)$ charges 
as is shown in Eqs.~(\ref{eq:dfibulk}) and 
(\ref{eq:dfibrane}). Thus, in the case with 
$\xi_0 + \xi_\pi \neq 0$ we have anomaly. 
It is not clear that some singlet fields 
(brane/bulk moduli fields) may transform nontrivially 
under $U(1)$ transformation to cancel anomaly and 
may generate FI terms like Eqs.~(\ref{GS-S}) and 
(\ref{GS-FI}).  Moreover, in what follows 
we will discuss nontrivial profiles of bulk fields 
and such profiles might change one-loop 
calculations of FI term from the case with trivial 
profiles. Thus, let us just assume the above FI terms 
and study their aspects classically.} 
This model corresponds to the higher-dimensional (actually 
five-dimensional) extension of 4D models mentioned 
in~\ref{ssec:anomalousu1} and we are interested 
in such situation. 
The scalar potential relevant to our study is 
written in the form~\cite{GrootNibbelink:2002qp,Mirabelli:1997aj,
Marti:2002ar,Arkani-Hamed:2001tb}
\begin{eqnarray}
V &=& \frac{1}{2} \left[
-\partial_y \Sigma + \xi(y) +gq|\Phi_+|^2 - 
gq|\Phi_-|^2 + \sum_{I=0,\pi} gq_I|\phi_I|^2 
\delta (y-IR) \right]^2 \nonumber \\
&+& |\partial_y \Phi_+ -gq \Sigma \Phi_+|^2 
+  |\partial_y \Phi_- +gq \Sigma \Phi_-|^2 
+ g^2q^2|\Phi_+ \Phi_-|^2 +\cdots ,
\label{potential}
\end{eqnarray}
where $g$ is the five-dimensional gauge coupling 
related to four-dimensional gauge coupling $g_{\rm 4D}$ 
as $g^2 = 2 \pi R g_{\rm 4D}^2$. 
(See Eq.~(\ref{eq:appsp}) in appendix~\ref{app:osform}.) 

The first term comes from the $D$ term and 
the other terms originate from the $F$ terms. 
So the $D$-flat direction is written as 
\begin{equation}
-\partial_y \Sigma + \xi(y) +gq|\Phi_+|^2 - 
gq|\Phi_-|^2+  \sum_{I=0,\pi} gq_I|\phi_I|^2 
\delta (y-IR) =0,
\label{D-flat}
\end{equation}
and the $F$-flat direction satisfies
\begin{eqnarray}
\partial_y \Phi_{\pm} \mp gq \Sigma \Phi_{\pm} &=& 0,  
\label{F-flat1} \\ 
\Phi_+ \Phi_- &=& 0.\label{F-flat2}
\end{eqnarray}
It is convenient that we integrate Eq.~(\ref{D-flat}) 
and the $D$-flat condition becomes 
\begin{equation}
\frac{1}{2}(\xi_0 +\xi_\pi) +\int^{\pi R}_0 dy 
(gq|\Phi_+|^2 - gq|\Phi_-|^2) +
{1 \over 2}  \sum_{I=0,\pi} gq_I|\phi_I|^2 =0, 
\label{D-flat2}
\end{equation}
where we have applied the boundary conditions 
$\Sigma(0)=\Sigma(\pi R) =0$. 
Hereafter we refer the sign of the $U(1)$ charges 
$q$ and $q_I$ such that $\xi_0+\xi_\pi \ge 0$, 
namely we redefine $(q,q_I)$ as 
$[2\theta(\xi_0+\xi_\pi)-1](q,q_I)$ $\to$ $(q,q_I)$. 
We also denote the ratio of $\xi_\pi$ to $\xi_0$ as 
$r^\pi_0 = \xi_\pi / \xi_0$.

One of important aspects is that when $\Sigma$ 
develops its VEV, that generates masses of bulk matter fields.
When SUSY is not broken, the profiles of zero modes of $Z_2$ even 
fields $\Phi_+$ satisfy the following equation,
\begin{equation}
\partial_y \Phi^{(0)}_{+} - gq\langle \Sigma\rangle \Phi^{(0)}_{+} =0,
\label{phizeroee}
\end{equation}
that is, the profiles are written as 
\begin{equation}
\Phi^{(0)}_{+ }(y) =\Phi^{(0)}_{+ }(0)
\exp \left[gq \int^y_0 dy' \langle \Sigma (y')\rangle \right].
\label{profile-1}
\end{equation}
Fermionic superpartners also have the same profiles as their 
scalar fields (\ref{profile-1}).
If SUSY is broken and SUSY breaking scalar masses 
are induced, zero mode profiles of bulk scalar 
components would change, but fermionic components 
do not change their profiles from 
Eq.~(\ref{profile-1}).
Furthermore, if SUSY breaking scalar masses 
are constant in bulk, zero modes profiles of 
scalar components are not changed, but scalar mass eigenvalues  
are shifted from zero by SUSY breaking scalar masses.
Thus, the VEV of $\Sigma$ is important.
In the following subsections, we study 
the VEV of $\Sigma$ and zero mode profiles 
as well as higher modes systematically.

We also consider the VEV of $Z_2$ odd bulk fields $\Phi_-$ 
in the SUSY vacuum.
The $F$-flat condition requires the VEV of $\Phi_-$ 
satisfy the following equation,
\begin{equation}
\langle \Phi_-(y)\rangle  =\langle \Phi_{-}(0)\rangle 
\exp \left[-gq \int^y_0 dy' \langle \Sigma (y')\rangle \right], 
\label{vev-odd}
\end{equation}
and the $Z_2$ odd bulk fields should satisfy the boundary condition, 
$\Phi_{-}(0) =0$.
This implies that $\langle \Phi_-(y)\rangle $ vanishes everywhere 
unless $ \langle \Sigma\rangle $ has a singularity. 
Thus $\langle \Phi_-(y)\rangle =0$ gives real vacuum in the unbroken 
SUSY case. Also for the broken SUSY case it gives at least a local 
minimum of the potential if there are no tachyonic modes on this vacuum. 
Therefore we will set $\langle \Phi_-(y)\rangle =0$ in all the cases in 
the following.\footnote{Actually we will see that no tachyonic mode appears 
in all the cases that we analyze here.} 

In the following we will see that the charged hyper multiplet 
is localized in the extra direction, 
depending on the charge and the value of the FI term. 
We call FI term with a condition $\int dy\, \xi = \xi_0 + \xi_\pi = 0$ 
`integrable', and $\int dy\, \xi = \xi_0 + \xi_\pi \ne 0$ `nonintegrable.' 

\subsection{Zero mode localization: integrable FI term}
\label{sec:lifi}

First we consider the integrable FI term, i.e. $\xi_0 + \xi_\pi = 0$. 
In this case, from the scalar potential (\ref{potential}) and the $D$-flat 
condition (\ref{D-flat2}), we easily understand that only $\Sigma$ 
can develops nonvanishing VEV, and 
$\langle \Phi_+(y) \rangle = \langle \phi_I \rangle = 0$.
By putting this into Eq.~(\ref{potential}) and varying $\Sigma$ 
we obtain the minimization condition,
$\partial^2_y \Sigma - \partial_y \xi(y) =0$.
Its solution is obtained as 
\begin{equation}
\partial_y \langle \Sigma \rangle = \xi(y) + C .
\label{vevphi}
\end{equation}
where the boundary condition $\Sigma(0)=\Sigma(\pi R ) =0$ 
determines the integral constant $C$ as 
$C=-(\xi_0 + \xi_\pi)/(2\pi R)$. 
Thus we obtain 
\begin{equation}
\langle \Sigma(y) \rangle = {1 \over 2 }\xi_0 {\rm sgn}(y) + Cy 
+{1 \over 2} \xi_\pi ({\rm sgn}(y-\pi R) +1),
\label{VEVSigma-1}
\end{equation}
in the region $0 \leq y \leq \pi R$.\footnote{The case with 
$\xi_0 \neq 0$ and $\xi_\pi =0$ was discussed  in 
Ref.~\cite{Mirabelli:1997aj}, and the case with $\xi_0+\xi_\pi =0$ 
was studied in Ref.~\cite{GrootNibbelink:2002qp}. The above general 
case includes these specific models.} 
The vacuum energy turn out to be $V_4 = \int dy C^2/2$. Therefore SUSY 
is broken generally. Only in the case with $\xi_0 + \xi_\pi = 0$, we 
have $C=0$ and SUSY is unbroken. Even in the broken SUSY case the 
vacuum energy is independent to $y$ (constant). 
%%%%%%%%%%%%%%%%%%%%%%%%%%%%%%%%%%
\begin{figure}[t]
\begin{center}
\begin{minipage}{0.47\linewidth}
   \centerline{\epsfig{figure=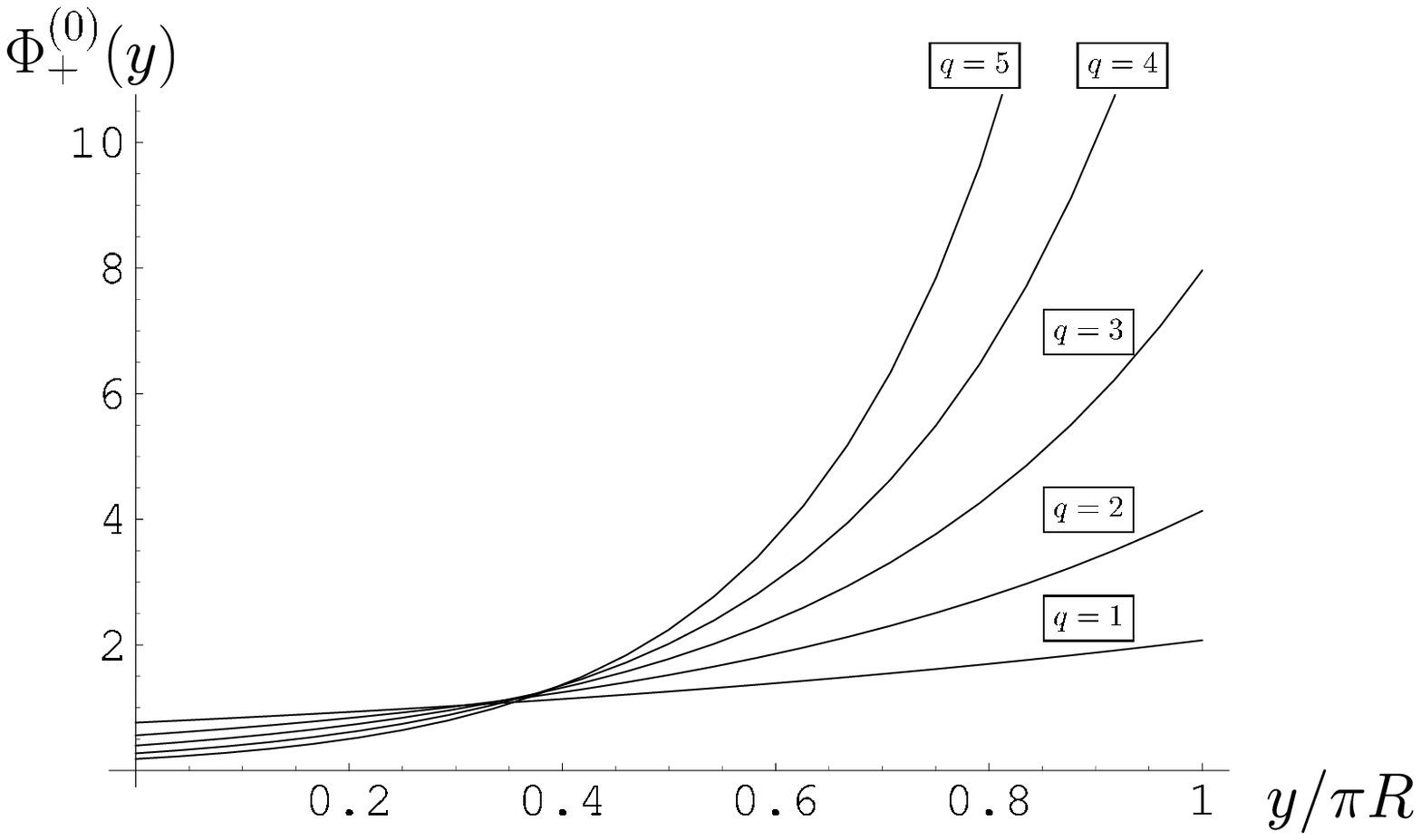,width=\linewidth}}
   \centerline{(a) $C=0$}
\end{minipage}
\end{center}
\begin{center}
\begin{minipage}{0.47\linewidth}
   \centerline{\epsfig{figure=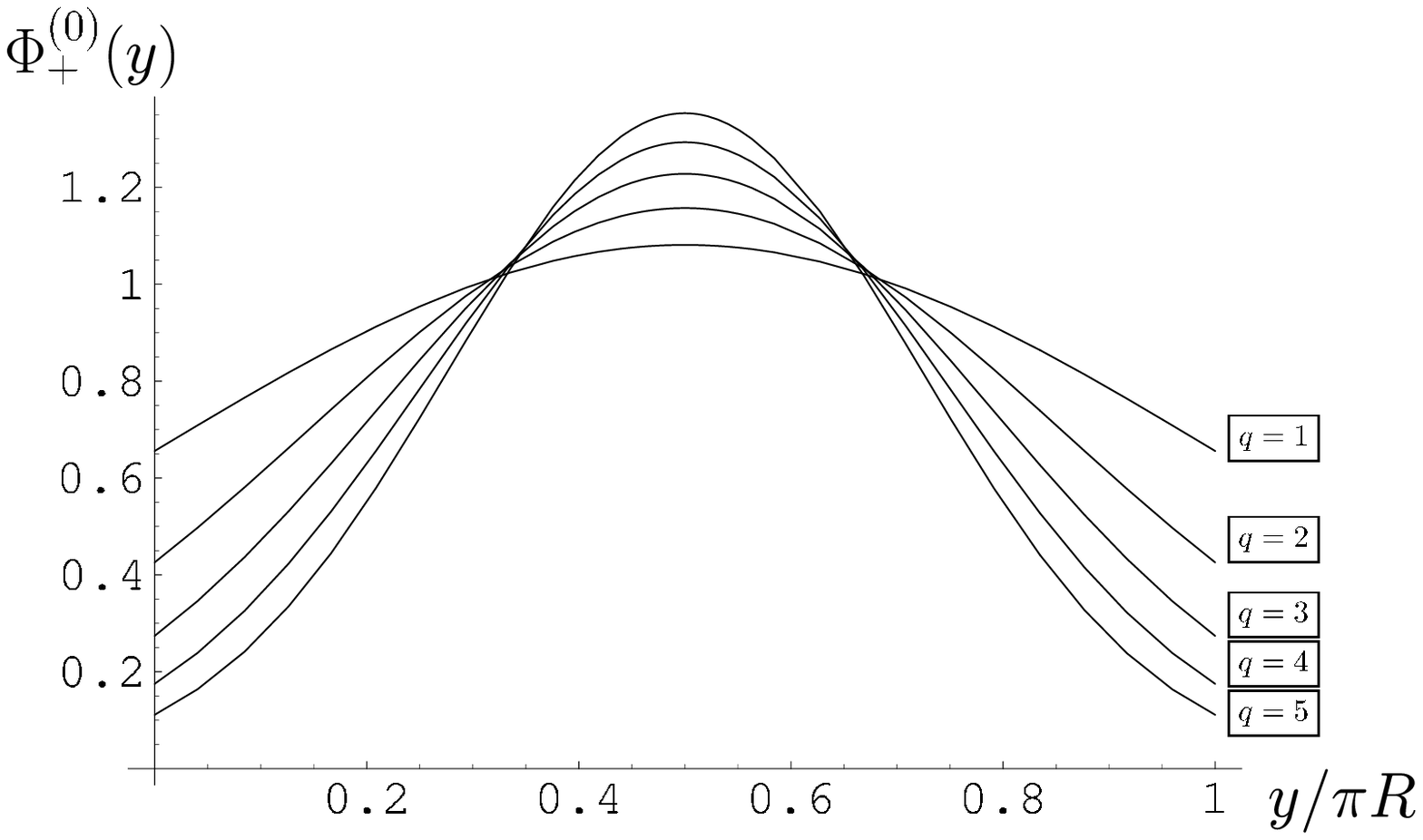,width=\linewidth}}
   \centerline{(b) $C \ne 0$ ($r^\pi_0=1$)}
\end{minipage}
\hfill 
\begin{minipage}{0.47\linewidth}
   \centerline{\epsfig{figure=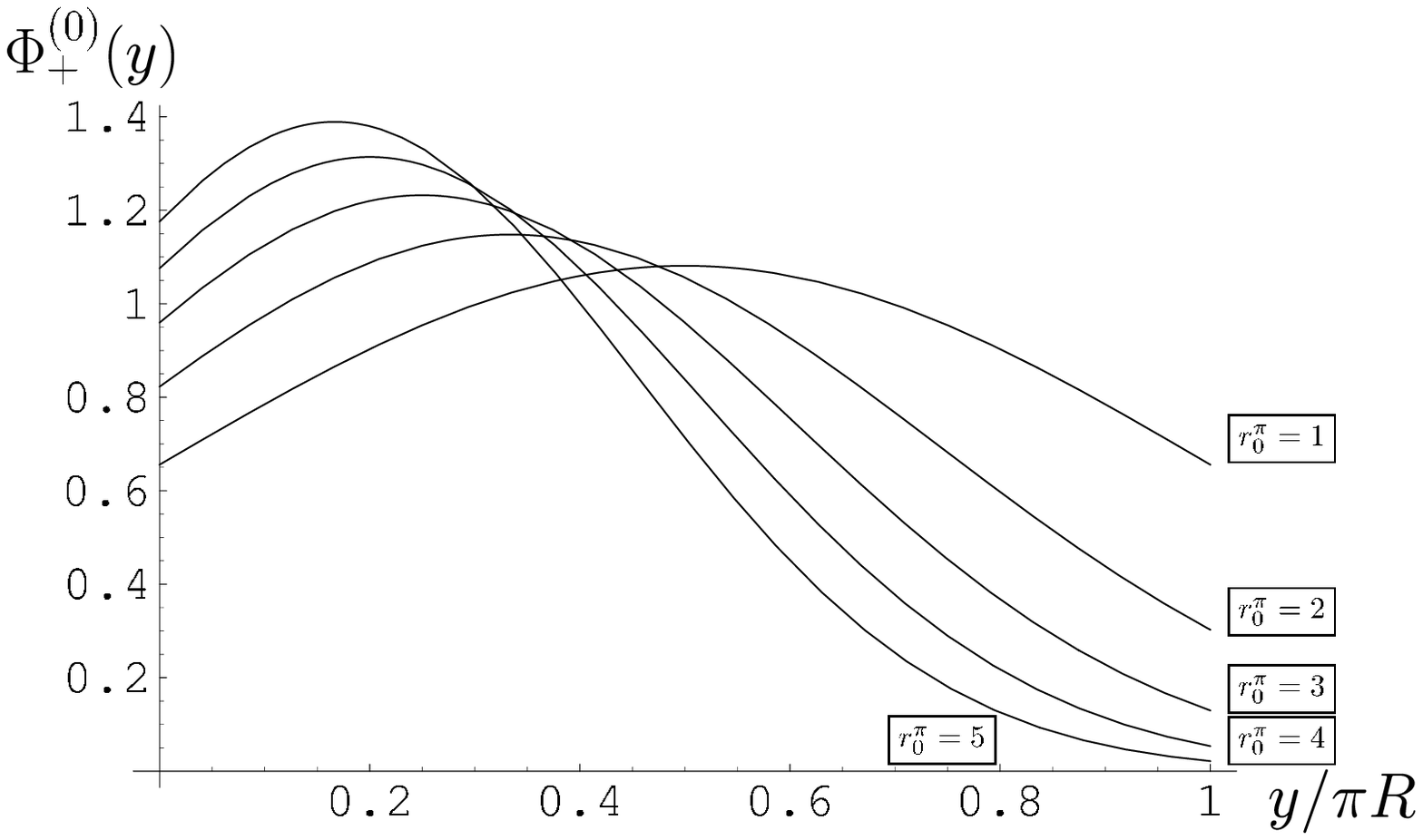,width=\linewidth}}
   \centerline{(c) $C \ne 0$ ($q=1$)}
\end{minipage}
\end{center}
\caption{Wave-function profile of the zero mode of 
$\Phi_+$. These are plotted by setting 
$g\pi R \xi_0/2 \equiv 1$ 
and varying $q$ or $r^\pi_0 = \xi_\pi/\xi_0$.}
\label{fig:gaussian}
\end{figure}
%%%%%%%%%%%%%%%%%%%%%%%%%%%%%%%%%%

With the above VEV (\ref{VEVSigma-1}), we consider 
the profile of bulk fields. 
Using Eq.~(\ref{profile-1}), we can write the profiles of 
both bosonic and fermionic zero modes of $Z_2$ even bulk field 
in the region $0 \leq y \leq \pi R$ as 
\begin{eqnarray}
\Phi_+(y) &=& \Phi_+(0) \exp \left[{gq \over 2}\xi_0 y \right], 
\label{profile-2}
\end{eqnarray}
for $C=0$ 
($\xi_0+\xi_\pi = 0$)~\cite{Arkani-Hamed:2001tb,GrootNibbelink:2002qp}.

\subsection{Zero mode localization: nonintegrable FI term}

In this section we proceed to the case of the nonintegrable FI term. 
In this case we have different results depending on the sign of 
the charge of hyper and chiral multiplets. 

\subsubsection{Case with $^\forall q, q_I>0$}
First we consider the case that all of bulk and brane fields 
except $Z_2$ odd bulk fields have positive charges, i.e. 
$q > 0$ and $q_I > 0$. 
In this case, from the scalar potential (\ref{potential}) 
and the $D$-flat condition (\ref{D-flat2}), we also understand 
that only $\Sigma$ can develops nonvanishing VEV, and 
$\langle \Phi_+(y) \rangle = \langle \phi_I \rangle = 0$.
As we will show in Eq.~(\ref{Dterm-1}), 
even for the case with $C \neq 0$ ($\xi_0+\xi_\pi \ne 0$), 
SUSY breaking scalar masses are constant along the $y$ direction. 
Thus for $C \ne 0$, Using Eq.~(\ref{profile-1}), we can write the 
profiles of both bosonic and fermionic zero modes of $Z_2$ even bulk 
field in the region $0 \leq y \leq \pi R$ as~\cite{Abe:2002ps} 
\begin{eqnarray}
\Phi_+(y) &=& \Phi_+\left( \frac{\xi_0}{\xi_0 + \xi_\pi}\pi R \right) 
\exp \left[-{gq \over 2\pi R}(\xi_0 +\xi_\pi)
\left(y-\frac{\xi_0}{\xi_0 + \xi_\pi}\pi R \right)^2 \right].  
\label{profile-3}
\end{eqnarray}
In this case the scalar masses are shifted from zero by $D$ term 
contributions. Actually this profile is interesting. 
This form is the Gaussian profile which is localized 
at $y = \pi R \xi_0/(\xi_0 + \xi_\pi)$. 
These profiles are shown in Fig.~\ref{fig:gaussian}. 
Note that we are considering the case with $q > 0$ 
for all of $Z_2$ even bulk fields.
This localization point is independent of $U(1)$ charges, 
that is, all of zero modes are localized at the same point.
For example, if $\xi_0 = \xi_\pi$, all of zero modes are 
localized at the point $y= (\pi R)/2$.
In some models the localization point may be out of the region 
$[0,\pi R]$ and in such case we see only a tail part of 
the Gaussian profile.
Even in the very large FI limit, i.e. $\xi_0, \xi_\pi \rightarrow 
\infty$, their ratio ${\xi_0 / (\xi_0 + \xi_\pi)}$ can be finite.
On the other hand, the width of the Gaussian profile 
depends on $U(1)$ charges $q$.
That would be useful e.g. to derive hierarchical couplings 
by wave function overlapping.

The scalar zero mode has SUSY breaking scalar mass term 
due to nonvanishing $D$ term, $-gqC|\Phi_+|^2$, in bulk, 
and that is constant along the $y$ direction as said above.
Namely, the scalar zero mode has the following mass 
\begin{equation}
m^2_D = g q (\xi_0 + \xi_\pi)/(2 \pi R) .
\label{Dterm-1}
\end{equation}
Note that this $D$ term contribution is positive for the 
present case with $q>0$.
Similarly, the brane fields on both branes have 
the $D$ term contribution to SUSY breaking scalar mass  
\[ m^2_{D,I} = g q_I (\xi_0 + \xi_\pi)/2 \pi R. \]
This $D$ term contribution is also positive 
for the present case with $q_I>0$.
There is no zero mode with tachyonic mass.
Note that the overall magnitude of $D$ term 
contributions to scalar masses are universal up to 
$U(1)$ charges in bulk and both branes.
This point will be considered in section~\ref{sec:3}, again.
The profiles and mass eigenvalues of the scalar higher modes 
$\Phi_\pm$ are analyzed in appendix~\ref{app:higher1}. 

\subsubsection{Case with $^\exists q_I<0$} \label{sec:2.2.2}
Here we consider another case that there is a brane field with 
the charge satisfying $q_I < 0$.
In this case, such brane field develops its VEV 
along the $D$-flat direction and $U(1)$ is broken. 
For concreteness, we assign such brane field lives on the $y=0$ brane.
The $D$-flat condition is satisfied with the VEV,
$gq_0 \langle |\phi_0| \rangle ^2 = -(\xi_0 + \xi_\pi)$. 
Now it is convenient to define the effective FI term~\cite{Abe:2002ps}, 
$\xi'(y) = \xi'_0 \delta (y) + \xi'_\pi \delta (y-\pi R)$, 
where $\xi'_0 = \xi_0 + gq_0\langle |\phi_0|^2 \rangle$ and 
$\xi'_\pi = \xi_\pi$. 
Note that for these effective FI coefficients we have 
$\xi'_0 + \xi'_\pi =0$. 
Therefore the $D$-flat direction for $\langle \Sigma \rangle$ 
is obtained by replacing $\xi(y)$ by $\xi'(y)$ and putting 
$C=0$ in Eqs.(\ref{vevphi}) and (\ref{VEVSigma-1}), i.e., 
$\langle \Sigma(y) \rangle = (\xi'_0/2) {\rm sgn}(y) 
+(\xi'_\pi/2) \left({\rm sgn}(y-\pi R) +1\right)$.
Since we assume no superpotential for $\phi_0$, 
SUSY is unbroken along this direction.

The zero mode profile of $Z_2$ even bulk field is 
obtained by replacing $\xi(y)$ by $\xi'(y)$ in 
Eq.~(\ref{profile-2}), i.e. 
\[ \Phi_+^{(0)}(y) = \Phi_+^{(0)}(0) 
\exp \left[\frac{1}{2}gq \xi'_0 y \right].\] 
This is effectively the same as the case that was studied 
systematically in Ref.~\cite{GrootNibbelink:2002qp}. 
Mass eigenvalues and wave function profiles 
of higher modes are the same as Eqs.~(\ref{eq:czerokk}) and 
(\ref{eq:czerowf}) respectively in appendix~\ref{app:higher1} 
except replacing $\xi$ by $\xi'$. 

\subsubsection{Case with $^\exists q<0$} \label{sec:2.2.3}
%%%%%%%%%%%%%%%%%%%%%%%%%%%%%%%%%%
\begin{figure}[t]
\begin{center}
\begin{minipage}{0.48\linewidth}
   \centerline{\epsfig{figure=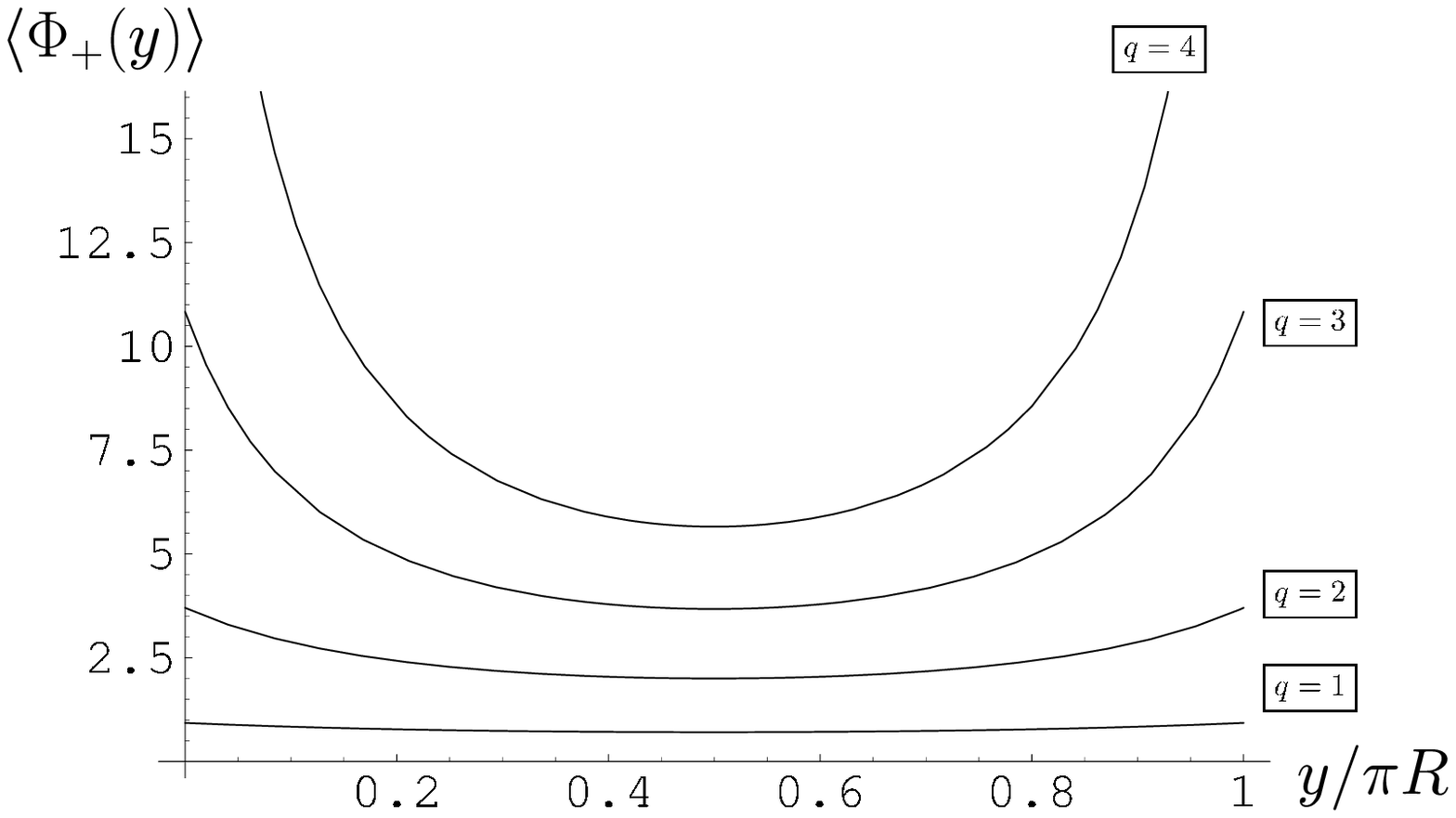,width=\linewidth}}
   \centerline{(a) $\langle \Phi_+ (y) \rangle$ $(r^\pi_0=1)$}
\end{minipage}
\hfill
\begin{minipage}{0.48\linewidth}
   \centerline{\epsfig{figure=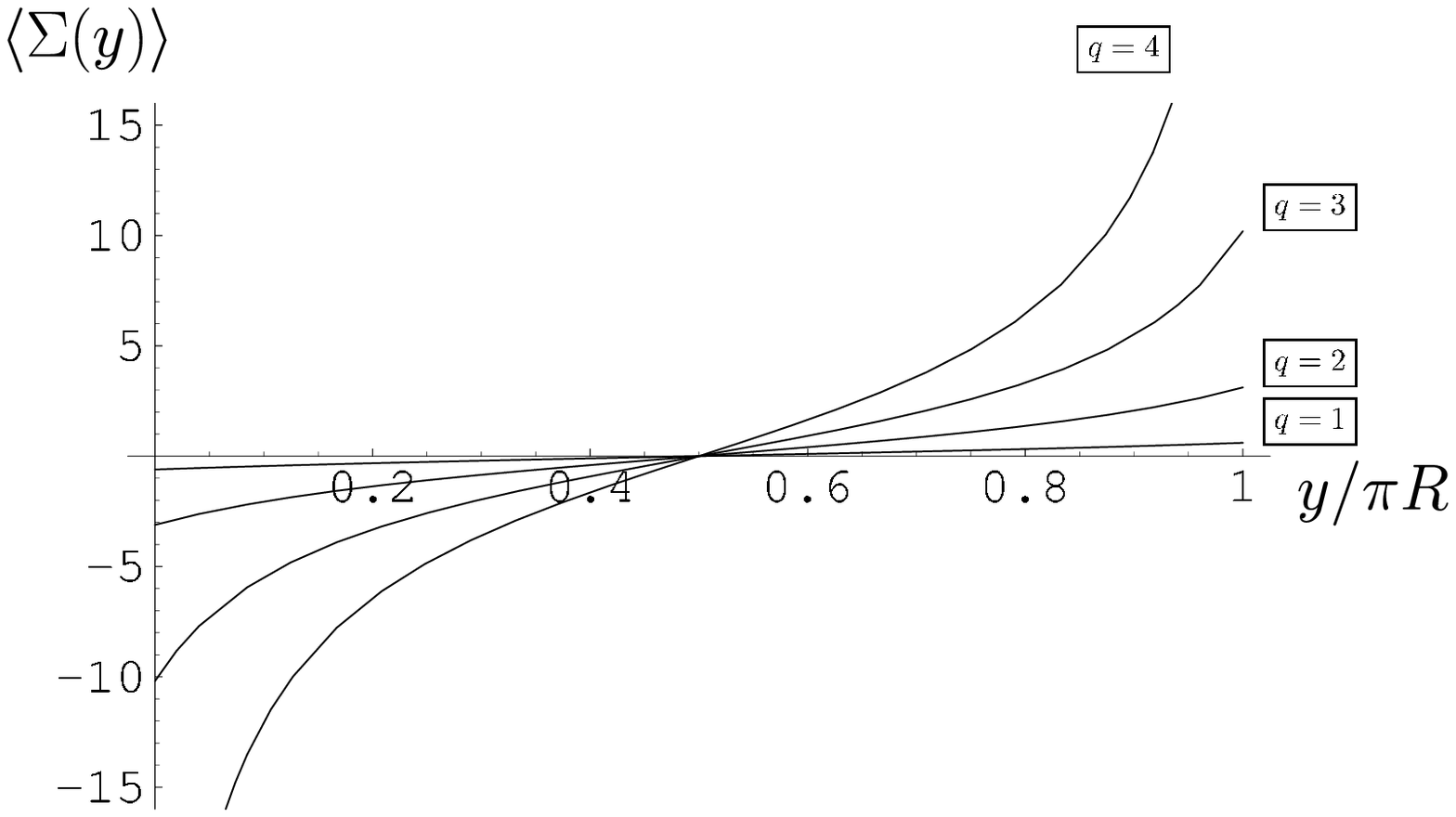,width=\linewidth}}
   \centerline{(b) $\langle \Sigma (y) \rangle$ $(r^\pi_0=1)$}
\end{minipage}
\end{center}
\caption{$y$-dependence of the VEVs $\langle \Phi_+(y) \rangle$ 
and $\langle \Sigma (y) \rangle$. 
These are plotted by setting $r^\pi_0\equiv 1$, $g\pi R \xi_0/2 \equiv -1$, 
$a=a_+$ ($m=0$) and varying $q$.}
\label{fig:susybv}
\end{figure}
%%%%%%%%%%%%%%%%%%%%%%%%%%%%%%%%%%
If there is a $Z_2$ even bulk field with the charge 
satisfying $q < 0$, such bulk field 
can also develop its VEV along flat direction.
Here we consider the case that $\Sigma$ and the bulk field 
$\Phi_+$ with the charge  $q < 0$ 
develop their VEVs.
In this case, the $D$-flat and $F$-flat conditions 
are written as
\begin{eqnarray}
-\partial_y \langle \Sigma \rangle + \xi (y) 
+gq \langle |\Phi_+| \rangle ^2 = 0, 
\label{D-flat3}
\\
\partial_y \langle \Phi_+ \rangle -gq 
\langle \Sigma \rangle \langle \Phi_+ \rangle =0 .
\label{F-flat3}
\end{eqnarray}
We find the following form of function is 
the solution~\cite{Kaplan:2001ga,Abe:2002ps}: 
\begin{eqnarray}
\langle |\Phi_+ (y)| \rangle &=& 
{A \over \cos [ay + b + c_0 {\rm sgn}(y) 
+ c_\pi {\rm sgn }(y-\pi R)]}, 
\label{vevphi-3} \\
\langle \Sigma (y) \rangle &=& {a \over gq}
\tan[ay + b+ c_0 {\rm sgn}(y) 
+ c_\pi {\rm sgn }(y-\pi R)], 
\label{vevSigma-3}
\end{eqnarray}
where $a^2 = g^2 q^2 A^2$ and $(2a/gq) c_I = \xi_I$, and 
along this direction SUSY is unbroken but $U(1)$ is broken. 
The boundary condition, $\partial_y \Phi_+(0) = \Sigma(0) =0$ 
requires $b = c_\pi $ and the same boundary condition for $y= \pi R$ 
requires $a\pi R + c_0 + c_\pi = m \pi$, where $m$ is an integer. 
These result in a solution 
$a_{\pm} = (m \pi \pm \sqrt{m^2\pi^2 - 2\pi R gq 
(\xi_0 + \xi_\pi)})/2 \pi R$. 
Recall that only the bulk field with the charge 
$q <0$ can develop the above VEV.
Thus, the inside of the square root is always positive.
Moreover we have to require the VEVs 
$\langle \Phi_+ \rangle $ and $\langle \Sigma \rangle$ 
be not singular in the region $0 < y < \pi R$.
That implies that the singular point 
$y=(n\pi +\pi/2-c_0)/a$,  where $n$ is any integer, 
should not be in the region 
$0 < y < \pi R$.
The VEVs of Eqs.~(\ref{vevphi-3}) and (\ref{vevSigma-3}) 
are shown in Fig.~\ref{fig:susybv}.

The zero mode profile of a bulk field with the charge $q'$ 
is given as follows. Such profile is obtained by substituting 
this VEV of $\Sigma$ into Eq.~(\ref{profile-1}). 
It results in~\cite{Abe:2002ps} 
\begin{eqnarray}
\Phi_+'^{(0)} (y) = 
\Phi_+'^{(0)} (0) \cos^{-q'/q} [ay + b + c_0 {\rm sgn}(y) 
+ c_\pi {\rm sgn }(y-\pi R)].
\label{zerophi}
\end{eqnarray}
Regarding the profiles and mass eigenvalues of higher modes 
some analyses are given in appendix~\ref{app:higher2}.

\section{Supersymmetry breaking from boundary}
\label{sec:3}
So far, we have studied VEVs and zero mode profiles 
systematically in the cases with generic values 
of $\xi_0$ and $\xi_\pi$.
In any case, bulk fields have nontrivial profiles depending 
on their $U(1)$ charges. This means that the coupling between 
fields with different charge can be suppressed due to the 
small wave function overlapping. 
That would be useful e.g. to explain some hierarchical 
couplings~\cite{Marti:2002ar,Kaplan:2001ga}.\footnote{
For example, if bulk fermionic fields have 
their profiles as Eq.~(\ref{profile-3}) and 
the Higgs field live on the $y=0$ brane, 
then the suppression factors to Yukawa couplings 
due to wave function overlapping are obtained as 
$Y_{ij} \sim \exp \left[ -g\pi R(q_i + q_j)/2(\xi_0 +\xi_\pi) \right]$  
between the bulk fields with charge $q_i$ and $q_j$.} 
In such localized system, we should consider SUSY breaking 
and its effects on the soft parameters. In section~\ref{sec:3.1} 
we consider a toy model for SUSY breaking. This is the 5D simple 
extension of the 4D models studied in Ref.~\cite{Dvali:1996rj}. 
In section~\ref{sec:3.2}, we give an analysis of $D$ term 
contributions~\cite{Nakano:1994sw}.

\subsection{Toy model for supersymmetry breaking} 
\label{sec:3.1}
Our starting point is almost the same as section~\ref{sec:2.2.2}.
We consider the case that 
all the bulk fields have positive $U(1)$ charges and 
all of the brane fields also have positive charges, 
but only one brane field $\chi_0$ at $y=0$ has 
the negative $U(1)$ charge.
Here we normalize such field has the $U(1)$ charge, 
$q_{\chi_0} = -1$.
If this brane field has no superpotential 
on the brane, there is a SUSY flat direction 
studied in section~\ref{sec:2.2.2}, i.e. 
$\partial_y \langle \Sigma \rangle = 
(\xi_0 - g\langle |\chi_0|^2\rangle )\delta (y) + 
\xi_\pi \delta (y-\pi R)$, and 
$\langle |\chi_0|^2\rangle   = (\xi_0 + \xi_\pi)/g$. 
Along this direction $U(1)$ is broken although SUSY is unbroken.

Here we assume the mass term of $\chi_0$ 
in the superpotential as\footnote{
This type of mass terms can be generated 
by dynamics on the brane as discussed for 4D models 
in Ref.~\cite{Dvali:1996rj,Kageyama:vf}.
In this case, the dynamical generated mass can 
also have a $U(1)$ charge and other types of 
$U(1)$ charge assignment for $\chi'_0$ are 
possible. Alternatively, we could write a mass term 
only by $\chi_0$ itself. For simplicity of our toy model, 
we take the above mass term and charge assignment.} 
\begin{equation}
W = m\chi_0 \chi'_0 \delta (y),
\label{mass-term}
\end{equation}
where $\chi'_0$ is the brane field on $y=0$ 
and has the $U(1)$ charge $q_{\chi'_0} =1$. 
With this mass term, SUSY is broken.
Suppose that the field $\chi_0$ develops its VEV 
$\langle \chi_0 \rangle$. 
Then the analysis in section~\ref{sec:lifi} implies that 
the scalar potential $V$ is minimized by 
the VEV $\partial_y \langle \Sigma \rangle = \xi(y) 
-g\langle \chi_0 \rangle^2 \delta(y) +C$ that results 
in $\langle \Sigma \rangle = (\xi_0 - g\langle \chi_0 \rangle^2)
{\rm sgn} (y)/2 + Cy + \xi_\pi({\rm sgn} \left(y-\pi R)+1\right)/2$  
where $C = - ({\xi_0 -g\langle \chi_0 \rangle^2 + 
\xi_\pi)/(2 \pi R)}$~\cite{Abe:2002ps}. The 4D scalar potential 
with the above brane mass term (\ref{mass-term}) 
is written as $V_4 = m^2\langle \chi_0 \rangle^2 + 
(\xi_0 +\xi_\pi - g\langle \chi_0 \rangle^2)^2/4\pi R$. 
Minimizing this potential, we obtain 
$\langle \chi_0 \rangle^2 = (\xi_0 + \xi_\pi)/g - 
2\pi R m^2/g^2$. 
With this VEV, SUSY breaking parameters are obtained 
\begin{eqnarray}
F_{\chi'_0} &=& m \langle \chi_0 \rangle  = 
m\sqrt{{\xi_0 + \xi_\pi \over g} - 
{ 2 \pi R \over g^2}m^2 }, \nonumber \\
D &=& -\partial_y \langle \Sigma \rangle + \xi (y) 
-g \langle \chi_0 \rangle^2 \delta (y) = -C = 
\frac{m^2}{g}. 
\label{vevSigma-5}
\end{eqnarray}
Note that this $F$ term is induced only on the 
$y=0$ brane.

With this VEV of $\Sigma$, the fermionic 
zero mode as well as the corresponding bosonic mode 
has the following profile,
\begin{equation}
\Phi_+^{(0)}(y) = 
A \exp \left[ -\frac{1}{2}qm^2 
\left(y-\frac{g\xi_0}{2m^2}\right)^2 \right]. 
\label{profile-5}
\end{equation}
That is the Gaussian profile.
In a way similar to section~\ref{sec:lifi},
 profiles of higher modes as well as 
their mass eigenvalues are obtained.

Suppose again that the Higgs field lives on the $y=0$ brane.
Then the suppression factors to Yukawa couplings 
due to wave function overlapping are obtained as 
$ Y_{ij} \sim \exp \left[- g^2 \xi_0^2(q_i + q_j)/4m^2 \right]$ 
between the bulk fields with charge $q_i$ and $q_j$.
This form is the same as one obtained from the Froggatt-Nielsen 
mechanism, although the mechanism is totally different.
For example, we assume that the top quark lives on the 
$y=0$ brane and has no $U(1)$ charge.
Then the top Yukawa coupling would be large.

\subsection{Gaugino and sfermion masses}

In this vacuum, SUSY is broken and 
SUSY breaking terms are induced by 
nonvanishing $D$ term and $F$ term.
All of the bulk gaugino masses are induced by 
$ \left[ \chi_0 \chi'_0 W^\alpha W_\alpha/ \Lambda^3)\delta(y) \right]_D$ 
where $\Lambda$ is the cut-off scale, 
that is, the zero mode gaugino mass $M_{\lambda_0}$ is induced as 
\begin{equation}
M_{\lambda_0} \sim {F_{\chi'_0} \chi_0 \over 2 \pi R \Lambda^3} =
{m v^2 \over 2 \pi R \Lambda^3} .
\end{equation}

On the other hand, the bulk field with charge $q$ 
has SUSY breaking scalar mass due to nonvanishing 
$D$ term as 
\begin{equation}
m^2_q  = qm^2. 
\end{equation}
The brane field also has the same $D$ term contribution 
to the SUSY breaking scalar mass $m^2_I = q_I m^2 $ .
Here the VEV of $\Sigma$ field plays a role to 
mediate SUSY breaking to bulk fields and brane fields 
on the other brane as $D$ term contributions to 
SUSY breaking scalar masses~\cite{Abe:2002ps}. 
In addition, the brane field on the $y=0$ brane 
has the contact term $\left[ |\chi_0|^2 |\phi_0|^2/\Lambda^2\right]_F$ 
which induces the following $F$ term contribution to scalar masses 
\begin{equation}
m^2_{\phi_0} \sim {m^2 v^2 \over \Lambda^2}  .
\end{equation}
Similarly, the bulk scalar fields with vanishing 
$U(1)$ charge has the scalar mass $m^2_{\Phi} \sim
m^2 v^2 / (R \Lambda^3)$ from the $y=0$ brane. 
In the case with $R = O(1/\Lambda)$, we have the following ratio,
\begin{equation}
m_q : m_{\phi_0} : |M_\lambda| = m : m\varepsilon : 
m \varepsilon^2,
\end{equation}
where $\varepsilon = v/\Lambda$.

Then we know that we have $D$ term contribution to scalar masses 
like 4D theory. One of important points is that $D$ term is constant 
along the $y$-direction in our toy model.
Therefore, {\it bulk fields and brane fields on both branes have 
the same effects}.
For this point, the field $\Sigma$ plays a role. 
Its VEV, in particular $\partial_y \langle \Sigma \rangle$, 
cancels the effect of $D$ term localization in bulk. 
That is the reason why $D$ term contribution appears everywhere in bulk. 
Then, $D$ term contributions, in general, generate 
flavor-dependency in sfermion masses everywhere 
in bulk and on both branes when 
matter fields have different $U(1)$ charges. 
That is dangerous from the viewpoint of FCNC experiments 
like 4D models unless $D$ term contributions are 
quite large or suppressed compared with gaugino masses.

We can change the brane field $\chi'_0$ by a bulk field.
Then we have the same result.
As another scenario, we can use the bulk field instead of 
$\chi_0$, assuming the mass term superpotential 
only on the $y=0$ brane.
We can consider more complicated models, e.g. the model 
with two $U(1)$s and their FI terms.
Such cases would be studied elsewhere in order to 
derive interesting aspects, e.g. realistic Yukawa couplings 
and sparticle spectra.

\subsection{More generic case of supersymmetry breaking}
\label{sec:3.2}
So far, we have assumed that the superpotential (\ref{mass-term})
breaks SUSY with a combination effect of the FI terms.
Here we give a comment on other cases without 
such specific superpotential.
Under the same setup except the superpotential (\ref{mass-term}), 
we assume that SUSY is softly broken and 
SUSY breaking scalar mass of $\chi_0$ is induced as 
$m_{\chi_0}$.
Through this SUSY breaking, gaugino masses and 
soft scalar masses of other fields can be induced.
In this case, the previous discussion holds same except replacing 
$m$ by $m_{\chi_0}$.
The $\Sigma$ field has a nontrivial VEV as Eq.~(\ref{vevSigma-5}), 
and the fermionic zero mode of bulk field 
has a nontrivial profile as Eq.~(\ref{profile-5}).
Furthermore, bulk fields with charge $q$ and 
brane fields with charge $q_I$ on both branes 
have $D$ term contributions to scalar masses as 
\begin{equation}
m^2_q = q m^2_{\chi_0},\qquad m^2_I = q_I m^2_{\chi_0},
\end{equation}
respectively.
Again, the VEV of $\Sigma$ field plays a role to 
induce $D$ term contributions to 
SUSY breaking scalar masses
for bulk fields and brane fields on the other brane.
If $m_{\chi_0}$ is of order of gaugino masses, 
we have the SUSY flavor problem.
Such $D$ term contributions must be suppressed 
compared with gaugino masses~\cite{Kobayashi:2002mx}, 
or alternatively those must be quite larger to realize 
the decoupling solution.

\section{Discussions: supersymmetry breaking with boundaries}
\label{sec:4}

In Chapter~\ref{chap:sbmm} we reviewed supersymmetry breaking 
mediation mechanisms in a phenomenological viewpoint, 
especially FCNC constraints. 
Current experimental results imply 
almost degenerate sfermion masses although there 
exist hierarchical fermion masses and mixings. 
Thus the way communicating supersymmetry breaking should be 
almost flavor blind. We saw many viable mediation mechanisms 
in the braneworld framework for such purpose, 
like anormaly, gaugino, Kaluza-Klein and 
radion mediated scenario. In addition to such communications 
we should also have a mechanism to generate hierarchical 
fermion mixings without any conflicts with the former success. 
In such mechanism, $U(1)$ symmetry sometimes plays an 
important role for realizing hierarchical structure 
in the fermion sector, and would give $D$ term contribution 
to the soft SUSY breaking scalar masses. 

For example, as is mentioned in the last part of Chapter~\ref{chap:sbmm}, 
four-dimensional ${\cal N}=1$ supersymmetric models with 
an anomalous $U(1)$ gauge symmetry and nonvanishing FI term 
can give many phenomenologically interesting aspects. 
In addition to these four-dimensional models, recently 
five-dimensional $S^1/Z_2$ models with the FI term were 
studied and interesting aspects were 
shown~\cite{GrootNibbelink:2002qp,GrootNibbelink:2002wv,
Ghilencea:2001bw,Barbieri:2001cz,Scrucca:2001eb,Barbieri:2002ic} 
as is partly described in the previous sections. 
The $S^1/Z_2$ extra dimension with the radius $R$ 
has two fixed points, which we denote $y=0$ and $\pi R$.
In general (rigid SUSY framework), we can put two 
{\it independent} FI terms on these two fixed points, i.e. 
$\xi_0$ and $\xi_\pi$. 

In this chapter we have considered VEVs of $\Sigma$ 
and brane/bulk fields in the 5D model with generic FI terms.
We have studied systematically the VEV of $\Sigma$ 
in generic case that $\xi_0$ and $\xi_\pi$ are 
independent each other~\cite{Abe:2002ps}. 
The nontrivial VEV of $\Sigma$ generates bulk mass terms for 
$U(1)$ charged fields and their zero modes have 
nontrivial profiles.
In particular, in the SUSY breaking case, 
fermionic zero modes as well as corresponding bosonic modes 
have Gaussian profiles. 
Also in other cases, bulk fields have nontrivial profiles.
The profiles and mass eigenvalues of higher modes are shown 
in appendix~\ref{app:higher1}.

Such nontrivial profiles would be useful to explain hierarchical couplings. 
A toy model for SUSY breaking has been studied. Sizable $D$ term contributions 
to scalar masses have obtained.
The overall magnitude of $D$ term contributions are same everywhere in bulk 
and also on both branes~\cite{Abe:2002ps}. We have to take into account these 
$D$ term contributions and other SUSY breaking terms for a realistic scenario 
to explain Yukawa hierarchy.

Indeed, there are a lot of related studies with a special value of FI terms. 
In Ref.~\cite{Mirabelli:1997aj}, SUSY breaking in bulk was discussed 
by the FI term. In Ref.~\cite{Marti:2002ar,Arkani-Hamed:2001tb,Kaplan:2001ga}, 
zero mode wave function profiles have been studied and their profiles depend 
on $U(1)$ charges. Then, hierarchical Yukawa couplings have been 
derived~\cite{Marti:2002ar,Kaplan:2001ga}.
In particular, bulk field profiles have been studied systematically in 
Ref.~\cite{GrootNibbelink:2002qp}, which considered the models that the sum of the 
FI coefficients vanishes, i.e. $\xi_0 + \xi_\pi = 0$. 
That corresponds to vanishing FI term in four-dimensional effective theory.
Actually, the SUSY breaking model of Ref.~\cite{Mirabelli:1997aj} and 
the Yukawa hierarchy model of Ref.~\cite{Kaplan:2001ga} have 
the FI terms like $\xi_0 \neq 0$ and $\xi_\pi = 0$.
The analysis here includes all those results.

The analysis here is purely classical.
Actually one-loop calculations were done to 
show the FI term is generated at the loop-level.
In generic case, $U(1)$ is anomalous.
For quantum consistency, we would need 
anomaly cancellation by the Green-Schwarz mechanism and/or 
the Chern-Simons term. 
For the former, brane/bulk singlet (moduli) fields 
might be transformed as Eq.~(\ref{GS-S}) 
such that the Green-Schwarz mechanism work.
However, it is not clear.
For the latter, the SUSY Chern-Simons term includes 
the term $\Sigma F^{MN} F_{MN}$~\cite{Arkani-Hamed:2001tb}.
(See also Refs.~\cite{Arkani-Hamed:2001is,Pilo:2002hu}.)
Thus, a nontrivial VEV of $\Sigma$ might change 
the gauge coupling in bulk and/or brane, although in most of case 
$U(1)$ is broken. It is also noted that, in the supergravity framework,
the Chern-Simons term with a parity odd coefficient which mixes the $U(1)$ 
gauge field and a graviphoton is relevant to the FI term 
on the boundaries~\cite{Barbieri:2002ic}.

%%%%%%%%%%%%%%%%%%%%%%%%%%%%%%%%%%%%%%%%%%%%%%%%%%%%%%%%%%%%%%%%%%%%%
%%%%%%%%%%%%%%%%%%%%%%%%%%% Part II %%%%%%%%%%%%%%%%%%%%%%%%%%%%%%%%%
%%%%%%%%%%%%%%%%%%%%%%%%%%%%%%%%%%%%%%%%%%%%%%%%%%%%%%%%%%%%%%%%%%%%%
\part{Dynamical symmetry breaking on a boundary}
\label{part:dsbb}

\chapter{Dynamical electroweak symmetry breaking}
\label{chap:desb}

The quantum theory of gravity likely has SUSY in 
higher-dimensional space-time. 
Low energy SUSY can be the most elegant and popular candidate 
for stabilizing the electroweak scale. 
The experimental prediction of coupling unification 
derived from MSSM renormalization group analysis 
also favors the existence of SUSY. In the SUSY theory 
the problems related to the electroweak symmetry breaking 
and flavor physics are replaced to the problems of SUSY 
breaking and its communication to the visible sector. Therefore 
in Part~\ref{part:ssbb} of this thesis, we have reviewed 
various SUSY-breaking mediation mechanisms. 
Especially we have analyzed the properties of a concrete 
model in five-dimensional space-time with FI terms on 
the four-dimensional boundaries in relation to the 
FN mechanism of flavor symmetry breaking. 

We can also consider the stability of the electroweak scale and the 
SUSY flavor problem as follows. Almost all the problems in the 
low energy (TeV scale) SUSY come from too much light 
extra fields in addition to the SM one. We need complicated 
setup about SUSY breaking and its mediation mechanism to avoid these 
problems as shown in Part~\ref{part:ssbb} of this thesis. 
Remember that SUSY itself is needed for the consistency of the quantum 
theory of gravity, while `TeV scale SUSY' is required in order to stabilize 
the electroweak scale $M_{\rm EW}$, and to realize gauge coupling 
unification. However, in a recent brane world picture, 
there is a possibility that the strong and electroweak gauge group 
come from the different brane~\cite{Ibanez:2001nd}. In this case the 
unification of SM and gravity is able to occur directly without grand 
unification and the gauge coupling unification in MSSM may be accidental. 
We can also consider the case that gauge coupling unification 
is not depend on the TeV scale SUSY (MSSM) and it happens by the 
other mechanism, e.g. extra dimensional effect~\cite{Dienes:1998vh,Randall:2001gc}. 
If there is another mechanism to stabilize mass of the Higgs field, 
it is possible to break SUSY at higher energy scale, even around 
the Planck scale. Such a mechanism can set SUSY free from any 
problems at TeV scale. 

\section{Dynamical electroweak symmetry breaking}

The idea that the electroweak symmetry is broken 
dynamically gives another solution to the stabilization 
problem of the electroweak Higgs mass scale. 
By replacing elementary Higgs scalar to the fermion bilinear, 
we obtain its stabilized mass in analog of small pion mass 
derived from dynamical chiral symmetry breaking in QCD, 
or technically of Cooper pairing in superconductivity. 
The motivation of such composite Higgs scenario is not 
only the stabilization of the light scalar masses but also 
the fact described below. 

Higgs scalars are used in gauge theories to break symmetries 
spontaneously. In SM they generate masses for $W$ and $Z$ gauge 
bosons and also for quarks and leptons through Yukawa interaction. 
The different sizes of the fermion masses and mixings can be 
accommodated by having different sizes of the couplings. 
The complex Yukawa couplings can give rise to CP-violation 
through the diagonalization of the fermion mass matrices. 
In terms of the {\it elementary} Higgs scalars, their self 
couplings and Yukawa couplings are quite unconstrained so long as 
they satisfy the requirements of gauge invariance. 
Thus gauge theories with elementary Higgs scalars have many 
arbitrary parameters associated with the Higgs fields. 
(Of course this problem exists in GUT and also in SUSY 
gauge theories.) 
This results in the fact that the masses and mixings can't 
be predicted and must be introduced as parameters into the 
theory. 

Now we have enough motivation to study composite Higgs 
scenario, and we review such models and their problems 
in the following in this chapter. 

\section{Electroweak symmetry breaking by QCD}
First we see what happens in SM without elementary Higgs fields 
following the textbook by Cheng and Li~\cite{Cheng:bj}. 
For simplicity, let us restrict ourselves to one family of fermions 
\begin{eqnarray}
q_L=\pmatrix{u \cr d}_L, \ l_L=\pmatrix{\nu \cr e}_L.
\nonumber 
\end{eqnarray}
The Lagrangian is given as 
\begin{eqnarray}
{\cal L}
&=& -\frac{1}{4}G^{\alpha \mu \nu}G^{\alpha}_{\mu \nu}
    -\frac{1}{4}W^{i \mu \nu}W^{i}_{\mu \nu}
   -\frac{1}{4}B^{\mu \nu}B_{\mu \nu} \nonumber \\ && 
+i(\bar{q}\gamma_\mu D^\mu q + \bar{l}\gamma_\mu D^\mu l), 
\nonumber 
\end{eqnarray}
where $\alpha=1,2,\ldots,8$ and $i=1,2,3$ are the indices of 
adjoint representation of color and weak isospin, respectively. 
All gauge bosons and all fermions likely remain massless, 
but this is not the case due to the existence of QCD. 
The fact that the $u$ and $d$ quark are massless implies 
that we have the flavor symmetry $SU(2)_W$ $\times$ $SU(2)_R$. 
It is a famous result that the QCD strong dynamics spontaneously 
break the above symmetry to the diagonal subgroup $SU(2)_{L+R}$ 
due to the fermion bilinear condensation 
\begin{eqnarray}
\langle \bar{u}u \rangle_0 = \langle \bar{d}d \rangle_0 \ne 0. 
\nonumber 
\end{eqnarray}
Such condensation results in the existence of three massless 
Goldstone bosons $\pi^{\pm,0}$. The effective scalar 
and pseudo-scalar fields $(\sigma,\vec{\pi})$ $\sim$ 
$(\bar{q}q,i\bar{q}\vec{\tau}\gamma_5 q)$ transform as 
\begin{eqnarray}
H_{\rm eff} = \pmatrix{\pi_1+i\pi_2 \cr \sigma+i\pi_3}, 
\nonumber 
\end{eqnarray}
in the weak isospin basis. Thus the VEV $\langle \sigma \rangle =v$ 
breaks the $SU(2)_W$ $\times$ $U(1)_Y$ symmetry down to the $U(1)_{\rm em}$ 
with $\vec{\pi}$ being eaten by the three gauge bosons to become 
$W^{\pm}$ and $Z$. We however obtain $v \sim f_\pi = 95$ MeV from the QCD 
dynamics with its dynamical scale $\Lambda_{\rm QCD} \simeq 200$ MeV. 
This results in 
\begin{eqnarray}
M_W = \frac{1}{2}gv \simeq 30 \textrm{ MeV }, 
\nonumber 
\end{eqnarray}
that is about three orders of magnitude smaller than the experimental 
value of the mass of $W$ boson ($80$ GeV). 
But this simple mechanism of dynamical symmetry breaking gives 
the correct relation $M_W = M_Z \cos \theta_W$ where $\theta_W$ 
is the Weinberg angle, because we have an $SU(2)_{L+R}$ symmetry 
remaining. 

We see that QCD itself breaks down the electroweak gauge group 
in just right pattern. It has, however, following two difficulties: 
\begin{enumerate}
\item $W$ boson mass scale is too small compared with the 
experimental value, \label{enum:diff1}
\item fermions remain massless. \label{enum:diff2}
\end{enumerate}

\section{Technicolor model}

It is straightforward to overcome problem~\ref{enum:diff1}. 
We assume the existence of another QCD-like strong interaction, 
called technicolor interaction (TC), which has a scale parameter 
$\Lambda_{\rm TC}$ such that it produces the phenomenologically 
correct mass for $W$ as 
\begin{eqnarray}
M_W = \frac{1}{2}g_{\rm TC}f_{\pi}^{\rm TC} \simeq 80 \textrm{ GeV }. 
\nonumber 
\end{eqnarray}
From the analysis analogous to QCD we should have 
$f_\pi^{\rm TC} \simeq 250$ GeV and $\Lambda_{\rm TC}\simeq {\cal O}(1)$ 
TeV. Namely the technicolor interaction (e.g. $SU(3)_{\rm TC}$) is 
similar to QCD except that the fermion bilinear condensate at energy 
three orders of magnitude lager than QCD. We also need fermions 
$q_{\rm TC}$ called technifermions that carry technicolors, 
such that their bilinear condenses resulting in correct mass of 
the $W$ boson. 

The picture as described above still does not give solution 
to difficulty~\ref{enum:diff2}, namely massless fermions. 
Quarks and leptons have separate chiral symmetries which remain 
unbroken. So we need to find ways to obtain effective Yukawa coupling 
between SM fermions and technimesons. One possible way to do this 
is to extend the technicolor gauge group to extended technicolor 
(ETC) gauge group. It is achieved by putting SM fermions $f$ and 
technifermions $f_{\rm TC}$ in a single irreducible representation 
of the ETC group. We assume that ETC breaks down to TC at some energy 
scale $\mu_{\rm ETC}$. The gauge boson $V_\mu^{\rm ETC/TC}$ exists 
in ETC but not in TC acquires mass 
$M_{\rm ETC/TC} \simeq g_{\rm ETC}\mu_{\rm ETC}$, and couples to current 
$\bar{f}_{\rm TC}\gamma_\mu f$. Thus we obtain the effective four-fermion 
interactions mediated by $V_\mu^{\rm ETC/TC}$ in the form 
\begin{eqnarray}
\frac{1}{2} \left( \frac{g_{\rm ETC}}{M_{\rm ETC/TC}} \right) 
(\bar{f}_{{\rm TC},L}\gamma_\mu f_L)(\bar{f}_R \gamma^\mu \bar{f}_{{\rm TC},R}), 
\nonumber 
\end{eqnarray}
that transforms to 
\begin{eqnarray}
-\frac{1}{2\mu_{\rm ETC/TC}^2}\left[
(\bar{f}_{\rm TC}f_{\rm TC})(\bar{f}f)-
(\bar{f}_{\rm TC}\gamma_5f_{\rm TC})(\bar{f}\gamma_5f)+\ldots \right]. 
\nonumber 
\end{eqnarray}
The technifermion bilinear condensation then produces a mass for SM fermions, 
\begin{eqnarray}
m_f = \frac{1}{2\mu_{\rm ETC/TC}^2}\langle \bar{f}_{\rm TC}f_{\rm TC} \rangle. 
\nonumber 
\end{eqnarray}
Since $\langle \bar{f}_{\rm TC}f_{\rm TC} \rangle \simeq f_\pi^{\rm TC}$ $\sim$ 
$(1 \textrm{ TeV})^3$, one needs $\mu_{\rm ETC/TC} \simeq 30$ TeV 
to obtain $m_f \sim 1$ GeV. 

Because $V_\mu^{\rm ETC/TC}$ is SM singlet, we need a set of technifermions 
for each SM fermion in order to give all SM fermions masses. For SM one family, 
we need eight sets of technifermions 
$u_{\rm TC}^a$, $d_{\rm TC}^a$, $\nu_{\rm TC}$ and $e_{\rm TC}$, where 
$a=1,2,3$ is the color index. Thus we have $SU(8)_L$ $\times$ $SU(8)_R$ 
flavor symmetry under the technicolor interaction. When this chiral symmetry 
is spontaneously broken, three NG bosons are absorbed in $W^\pm$ and 
$Z$ gauge bosons, remaining huge number (sixty) of light pseudo-NG bosons. 

We see that the ETC scenario can solve both of difficulties~\ref{enum:diff1} 
and~\ref{enum:diff2} in dynamical electroweak symmetry breaking. 
However it needs huge number of unobserved fermions resulting huge number of 
light pseudo-NG bosons. The current FCNC experiment excludes simple ETC 
which reads to too light quarks and unobserved light pseudo-NG bosons. 
It is known that the large contribution to FCNC in ETC is originated in 
relatively small anomalous dimension of the technifermion bilinears. 
Thus we should find some mechanism to give large anomalous dimension of 
fermion bilinear operators. One of such mechanisms is realized in 
gauged NJL model.

\section{Top mode standard model}
As we saw above one of difficulties in ETC is too many pseudo-NG bosons 
that is originated in large global flavor symmetry. We need gauged flavor 
symmetry to absorb such pseudo-NG bosons, that implies the existence of 
flavor changing gauge bosons. Inversely if we integrate out such 
(heavy) flavor gauge bosons at the beginning, we would have 
flavor-breaking and inter-generational four-fermion interactions. 
And as mentioned above the large anomalous dimension can be derived 
from gauged NJL model which contains four-fermion interactions 
as fundamental ones. These facts lead us to study TC with 
flavor-mixing (effective) four-fermion interactions. 
The most economic one is obtained by identifying the technifermion 
to a quark in SM. Most appropriate candidate for such SM quark is 
top quark. It is also consistent with the fact that top quark mass 
is near the electroweak scale while the others are sufficiently small. 

From these reasons SM without elementary Higgs fields but with 
flavor-breaking and inter-generational four-fermion interactions 
is very interesting. Such model has Lagrangian 
\begin{eqnarray}
{\cal L} &=& 
  {\cal L}^{\rm SM}_{\rm gauge} 
+ {\cal L}^{\rm SM}_{\rm fermion} 
+ {\cal L}_{\rm 4f}, \nonumber \\ 
{\cal L}_{\rm 4f} &=& 
\frac{1}{M_\ast^2} \Bigg[
 \kappa^{(1)}_{ii';jj'} \left( \bar\psi^{i\hat{a}}_L \psi^{i'\hat{b}}_R \right) 
                        \left( \bar\psi^{j\hat{b}}_L \psi^{j'\hat{a}}_R \right) 
+\kappa^{(2)}_{ii';jj'} \left( \bar\psi^{i\hat{a}}_L \psi^{i'\hat{b}}_R \right) 
                             (i\tau_2)^{\hat{a}\hat{c}}(i\tau_2)^{\hat{b}\hat{d}}
                        \left( \bar\psi^{j\hat{c}}_L \psi^{j'\hat{d}}_R \right) 
\nonumber \\ && \hspace{3cm}
+\kappa^{(3)}_{ii';jj'} \left( \bar\psi^{i\hat{a}}_L \psi^{i'\hat{b}}_R \right) 
                             (i\tau_3)^{\hat{b}\hat{c}}
                        \left( \bar\psi^{j\hat{c}}_L \psi^{j'\hat{a}}_R \right) 
\Bigg]+\textrm{h.c.}, 
\label{eq:gen4f}
\end{eqnarray}
where $N_c$ is the number of colors ($N_c=3$ in reality), 
$\kappa^{(1)}_{ii';jj'}$, $\kappa^{(2)}_{ii';jj'}$ and 
$\kappa^{(3)}_{ii';jj'}$ are the dimensionless four-fermion couplings, 
$i,i',j,j'=1,2,\ldots,6$ and $\hat{a},\hat{b},\hat{c},\hat{d}=1,2$ are 
the family and the isospin indices, respectively, 
and $M_\ast$ is a scale of the new physics. 
The family index $i$ in $\psi^{ia}$ runs from $1$ to $6$ where each $i$ 
corresponds to $\psi^{1a}=\pmatrix{u \cr d}$, $\psi^{2a}=\pmatrix{c \cr s}$, 
$\psi^{3a}=\pmatrix{t \cr b}$, $\psi^{4a}=\pmatrix{\nu_e \cr e}$, 
$\psi^{5a}=\pmatrix{\nu_\mu \cr \mu}$ and $\psi^{6a}=\pmatrix{\nu_\tau \cr \tau}$. 
In order to realize the dynamical symmetry breaking by top condensation, 
we need an assumption in this most general four-fermion interaction that 
the top quark four-fermion coupling is supercritical, 
$\kappa_t$ $\equiv$ $\kappa^{(1)}_{33;33}+\kappa^{(3)}_{33;33}$ $\gtrsim$ 
$\kappa_c \simeq 1$, where $\kappa_c$ is the critical coupling. 
Because of this assumption we call this model `top mode SM'. 

Also we assume no extra gauge or fermion fields in the top mode SM. 
It is known that the model can have large anomalous dimension and 
few light pseudo-NG bosons. This is very successful results. 
However only the problem resulting from the detailed analysis is 
that the top quark mass is slightly large ($m_t \sim 250$ GeV) 
if we require natural cut-off scale ($\Lambda \sim 10^{19}$ GeV). 
And as the cut-off decreases the top mass increases from the value. 
Some mechanisms reducing top mass are proposed, e.g. 
top quark seesaw mechanism~\cite{Dobrescu:1997nm}.

\section{Electroweak symmetry breaking and compact extra dimensions}
In the previous section we have resigned renormalizable frame work 
of dynamical electroweak symmetry breaking to introducing 
four-fermion interactions. But we have also seen that this works 
quite well. Therefore the next task is to find what is the origin 
of the nonrenormalizable four-fermion interactions. 
One candidate is the spontaneously broken gauge theories 
that gives such interactions after it is integrated out. 
One of them is known as the top color scenario~\cite{Hill:1991at}. 

Recent interesting possibility is that some parts of such four-fermion 
interactions can be provided by gauge theories with compact 
extra dimensions~\cite{Dobrescu:1998dg}. 
In the top mode SM the other gauge bosons than the SM ones are assumed 
to be integrated out at the higher energy scale which results in 
four-fermion interactions. In the extra dimensional scenario we also assume 
the situation that only SM gauge fields have zero modes at the electroweak 
scale. 

In the case that SM gauge fields propagate in bulk with compact 
extra dimensions and SM matter fields is confined on a four-dimensional 
brane, we have massive KK tower of only gauge fields. If the cut-off 
scale $\Lambda$ is near the compactification scale $M_c$, 
the extra dimensional kinetic energy of the KK modes can be neglected 
and we have four-fermion interactions in the effective theory 
after integrating out such heavy KK higher modes of SM gauge fields. 
For instance by integrating out KK gluons and KK hypercharge gauge boson, 
we have respectively four-fermion interactions~\cite{Dobrescu:1998dg} 
\begin{eqnarray}
&& -\frac{g_c^2}{M_c^2}
 \left( \sum_{i=1}^3 c_c^i \, 
 \bar\psi^{i\hat{a}} \gamma_\mu T^a \psi^{i\hat{a}} \right)^2, 
\label{eq:kkgl4f} \\
&& -\frac{g_Y^2}{M_c^2} 
\left[ \frac{1}{3} \sum_{i=1}^6 c_Y^i \left( 
       \bar\psi^{i\hat{a}}_L \gamma_\mu \psi^{i\hat{a}}_L 
    +4 \bar\psi^{i1}_R \gamma_\mu \psi^{i1}_R 
    -2 \bar\psi^{i2}_R \gamma_\mu \psi^{i2}_R  \right) \right]^2, 
\label{eq:kkhy4f} 
\end{eqnarray}
where $g_c$ and $g_Y$ are the color and hypercharge gauge coupling respectively, 
$c_c^i$ and $c_Y^i$ are the dimensionless parameters defined by the 
bulk and boundary structure in compact dimensions. Here most simple case with 
$c_c^i \equiv c_c$ and $c_Y^i \equiv c_Y$ is chosen. 
The four-fermion interaction (\ref{eq:kkgl4f}) have a possibility to induce 
electroweak symmetry breaking. For super-critical $c_c$, Eq.~(\ref{eq:kkgl4f}) 
would produce an $SU(N_q)$ symmetric condensate and an $SU(N_q)$ adjoint 
of NG bosons ($N_q$ is the number of quark flavors). All the quarks would 
acquire the same dynamical mass. It is thus necessary to identify a source 
of (quark) flavor symmetry breaking. By considering four-fermion interaction 
(\ref{eq:kkhy4f}) in addition, (\ref{eq:kkgl4f}) $+$ (\ref{eq:kkhy4f}) 
may induces VEVs only for the Higgs fields made up of the $u$, $c$ and $t$, 
namely up-type quarks because (\ref{eq:kkhy4f}) is attractive for the up-type 
quarks and repulsive for the down-type quarks. We need, however, 
inter-generational flavor symmetry breaking in order to achieve top 
condensation only. One of the idea is to produce the inequality 
$c_c^3 > c_c^1$, $c_c^2$ by changing the extra dimensional configuration.
 We will see an example for such configuration in a warped extra dimension. 

Up to here we have only consider to break electroweak symmetry dynamically, 
i.e. to obtain $W$ and $Z$ masses. However we should also realize fermion 
masses and mixings. The four-fermion interactions (\ref{eq:kkgl4f}) $+$ 
(\ref{eq:kkhy4f}) does not generate such terms and the exchange of KK massive 
modes of the other SM gauge bosons than gluons or hypercharge gauge bosons 
also do not. Thus we conclude that we need more structure than SM gauge fields 
in the bulk with compact extra dimensions. Indeed we need flavor-changing 
massive fields to generate fermion masses and mixings under the above 
situation of the electroweak symmetry breaking by top condensation. 
As we have seen in ETC theories in the previous sections it seems to be 
difficult to complete such scenario in the statement of renormalizable 
interaction. It may be necessary that the quantum-gravitational effects 
supply with such mechanism, e.g. exchange of the massive modes in the 
string states may provide it. However these topics are outside the range 
of this thesis. 

In Part~\ref{part:dsbb} of this thesis we assume that the quantum 
gravitational effects gives us compact extra dimensions and general 
{\it perturbative}  four-fermion interactions like Eq.~(\ref{eq:gen4f}) 
($\kappa^{(1)}_{ii';jj'}$, $\kappa^{(2)}_{ii';jj'}$, 
$\kappa^{(3)}_{ii';jj'}$ $\ll 1$). 
We concentrate on the systematical studies on how the fermion bilinear 
condensation occurs in the context of gauge theories in the bulk space-time 
with compact extra dimensions. In the case $M_c \simeq \Lambda$ the vacuum 
structure of such systems can be analyzed by using the above local 
four-fermion approximation. We can apply the way developed by NJL and 
its large following. In the case $M_c \ll \Lambda$ the kinetic (derivative) 
term can't be neglected and we should include such effects. 
In the next chapter we analyze the general structure of higher-dimensional 
gauge theory with compact extra dimensions and boundary matters.

\chapter{Higher-dimensional gauge theory with boundary matter}
\label{chap:hdgtwbm}

Motivated by the facts shown in the previous chapter, in this chapter 
we give an analysis of dynamical symmetry breaking in (nonSUSY) gauge 
theory in higher-dimensional space-time with four-dimensional boundary 
planes. 

As is shown in the previous chapter in detail, we notice the recent 
suggestions that the electroweak (chiral) symmetry is broken down 
dynamically by the gauge interaction in more than 4D 
space-time ~\cite{Dobrescu:1998dg,Cheng:1999bg,Arkani-Hamed:2000hv,
Hashimoto:2000uk,Rius:2001dd,Abe:2002yb,Abe:2001yi,Abe:2001ax,Abe:2000ny}. 
The idea is that we use fermion bilinear condensation instead of 
elementary scalar condensation. This enables us to obtain light Higgs 
scalar as in the case of pion in QCD. This is well-known as dynamical 
electroweak symmetry breaking. The most interesting candidate of such 
fermion to condensate is top quark in SM~\cite{Miransky:1988xi}. 
But within theories in four-dimensional space-time, we don't know 
what is the force that makes top quark bilinear condensed. 
Here we remember our stand point that the gauge field propagating 
extra dimensions. 
The higher dimensional gauge theory may realize the stabilization of 
a light Higgs field through the composite Higgs scenario. This is based 
on the idea that the gauge theory in compact extra dimension has 
Kaluza-Klein (KK) massive gauge bosons and they act as a binding force 
between fermion and anti-fermion that results in the composite scalar 
field. 

If SM is embedded in higher-dimensional space-time with boundaries, 
the dynamical mechanism of electroweak symmetry breaking can be 
realized in a certain bulk gauge theory (or SM itself in the bulk). 
SM gauge fields in the bulk has a 
possibility~\cite{Arkani-Hamed:2000hv,Hashimoto:2000uk} 
to realize top quark condensation~\cite{Miransky:1988xi} on the boundary 
which results in electroweak symmetry breaking. 
It is very interesting because it may have no extra field (elementary 
Higgs, technicolor gauge boson, technifermion etc.) except for the 
SM contents (without elementary Higgs) in order to break the
electroweak symmetry. Therefore we launched a plan to study the 
dynamical symmetry breaking in the brane world models in detail.
 
There is no enough knowledge about the (nonperturbative) 
dynamics of the higher dimensional gauge theory with boundary terms. 
In the remaining part of Part~\ref{part:dsbb} we study the basic 
properties of dynamical symmetry breaking in the bulk gauge theory 
couples to a fermion on a boundary. 
Based on the effective theory of the bulk Yang-Mills theory that 
is defined below the {\it reduced} cutoff scale on the boundary plane, 
we analyze dynamical chiral symmetry breaking on it.
 First we approximate the gauge boson exchange effects by four-fermion 
interactions and study the chiral phase structure in terms of the bulk 
configurations. Next we formulate the ladder Schwinger-Dyson (SD) 
equation~\cite{Dyson:1949ha} of the fermion propagator. 
It is numerically solved by using the improved ladder approximation 
with power low running of the effective gauge coupling. 
We show how the dynamics of the bulk gauge field affects
the chiral symmetry breaking on the boundary.

\section{Lagrangian of the system} \label{sec:lag}

In the following we analyze dynamical symmetry breaking on 
the four dimensional brane with bulk gauge theory 
in various brane world models. The bulk gauge theory is, of course, 
more than four dimensional theory and ill defined in ultraviolet (UV) 
region and we need some regularization about it. 
In this thesis we use effective Lagrangian of it defined below the 
{\it reduced} cutoff scale on the brane. 

In this section we derive the KK reduced 4D Lagrangian of the bulk 
gauge theory in the brane worlds, $M_4 \times S^1$ type space-time. 
We can easily extend it to the case of higher dimensional 
bulk space-time, $M_4 \times T^\delta$, it is shown in the next section. 
Because one of our goal is to analyze the dynamically induced mass 
scale on the brane in the Randall-Sundrum space-time, 
i.e. the slice of AdS$_5$, 
we take account for the Lagrangian and mode function in a curved 
extra dimension. The original motivation to introduce a curved extra 
dimension by Randall and Sundrum is to produce the weak and Planck hierarchy 
from the exponential factor in the space-time metric. 
The factor is called `warp factor', and the extra space `warped 
extra dimension'. A nonfactorizable geometry with the warp factor 
distinguishes the RS brane world from the others. 
In the RS model we consider the fifth dimension $y$ which is compactified 
on an orbifold, $S^1/Z_2$ of radius $R$ and two three-branes at the orbifold 
fixed points, $y^\ast=0$ and $\pi R$. 
Requiring the bulk and boundary cosmological constants to be related, 
Einstein equation in five-dimension leads to the solution 
\cite{Randall:1999ee}, 
\begin{equation}
ds^2=G_{MN}dx^Mdx^N=e^{-2k|y|}\eta_{\mu\nu}dx^\mu dx^\nu - dy^2, 
\label{eq:RSmetric}
\end{equation}
where $M,N$ = $0,1,2,3,4$, $\mu,\nu=0,1,2,3$ and 
$\eta_{\mu\nu}={\rm diag}(1,-1,-1,-1)$. 
$k$ is the AdS curvature scale with the mass dimension one, 
and $k=0$ describes the flat $M_4 \times S^1$ space-time. 
In the following we study a bulk vector field, $A_M$ 
in this background.  
\begin{figure}[t]
\centerline{\epsfig{figure=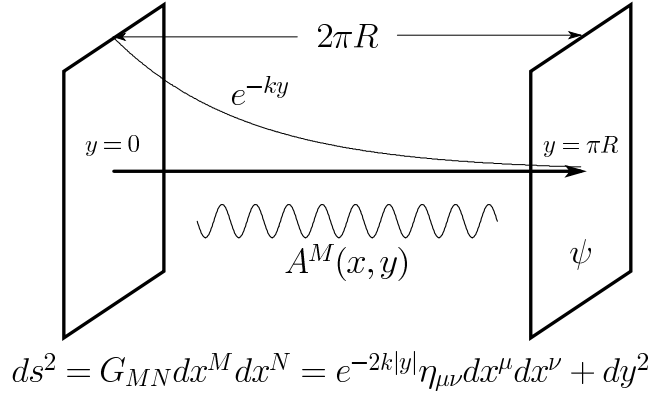,width=0.5\linewidth}}
\caption{Schematic view of the RS warped brane world. 
If we introduce certain bulk and boundary cosmological constant, 
the extra dimension becomes curved by a warp factor $e^{-ky}$. 
$k=0$ corresponds to a flat brane world.}
\label{fig:warpbw}
\end{figure}

\subsection{KK reduced Lagrangian of bulk gauge theory}
Substituting the metric $G_{MN}$ defined in Eq.~(\ref{eq:RSmetric}) 
and taking $A_4=0$ gauge, the Lagrangian of a bulk gauge field 
in the RS brane world is written by 
\begin{eqnarray}
{\cal L}_{\rm gauge}^{(5)}
= -\frac{1}{4} \sqrt{-G} F^{MN}F_{MN} 
= -\frac{1}{4} F^{\mu \nu}F_{\mu \nu}
    +\frac{1}{2} e^{-2k|y|} (\partial_4 A_\mu)^2
    +{\cal L}_{\rm SI}^{(5)},
\label{eq:bulkgauge}
\end{eqnarray}
where ${\cal L}_{\rm SI}^{(5)}$ includes the gluon three and four point 
self interaction in the bulk. 
We drop the explicit symbol of the trace operation in terms of the 
Yang-Mills index in the Lagrangian throughout this thesis.   
We perform the KK mode expansion as 
\begin{eqnarray}
A_\mu (x,y) 
= \frac{1}{\sqrt{2 \pi R}}
  \left(
  \sum_{n=0}^\infty A_\mu^{(n)}(x) \chi_n (y)
 +  \sum_{n=1}^\infty \tilde{A}_\mu^{(n)}(x) \tilde{\chi}_n (y)
  \right), 
\nonumber 
\end{eqnarray}
where $\chi_n (y)$ and $\tilde{\chi}_n (y)$ are the $Z_2$ even and odd 
mode function of $n$-th KK excited mode $A_\mu^{(n)}(x)$ and 
$\tilde{A}_\mu^{(n)}(x)$, respectively. 
The $Z_2$ orbifold condition thus means $\tilde{A}_\mu^{(n)}(x) \equiv 0$ 
for $^\forall n$, i.e. dropping all the odd modes. After KK expansion, 
we integrate the Lagrangian (\ref{eq:bulkgauge}) over the y-direction and obtain 
\begin{eqnarray}
{\cal L}_{\rm gauge}
&=& \sum_{n=0}^\infty \left[ 
    -\frac{1}{4}F^{(n)\mu \nu}(x)F_{\mu \nu}^{(n)}(x)
    +\frac{1}{2} M_n^2 A^{(n)\mu} A_{\mu}^{(n)} \right] \nonumber \\ &&
   +\sum_{n=1}^\infty \left[ 
    -\frac{1}{4}\tilde{F}^{(n)\mu \nu}(x)\tilde{F}_{\mu \nu}^{(n)}(x)
    +\frac{1}{2} M_n^2 \tilde{A}^{(n)\mu} \tilde{A}_{\mu}^{(n)} \right], 
\label{eq:bulkgaugekinmass}
\end{eqnarray}
where the KK mode function $\chi_n(y)$ and mass eigenvalue $M_n$ satisfy 
\begin{eqnarray}
-\partial_y \left( e^{-2k|y|} \partial_y \chi_n(y) \right)
  &=& M_n^2 \chi_n(y), \label{eq:KKeigeneq} \\
\frac{1}{2\pi R} \int_{-\pi R}^{\pi R} dy\, \chi_n(y) \chi_m(y) 
  &=& \delta_{nm}, \label{eq:normcond}
\end{eqnarray}
and $\tilde{\chi}_n(y)$ obeys the same equations. 
The boundary conditions are given by 
\begin{eqnarray}
\partial_y \chi_n(y^\ast) = \tilde{\chi}_n(y^\ast) = 0 
\qquad (y^\ast=0,\pi R). 
\label{eq:boundarycond}
\end{eqnarray}
Eq.~(\ref{eq:bulkgaugekinmass}) is the 4D effective Lagrangian 
of the bulk gauge theory and 
Eqs.~(\ref{eq:KKeigeneq})-(\ref{eq:boundarycond})
determine the form of the mode function and the eigenvalue $M_n$. 
Below we solve Eq.~(\ref{eq:KKeigeneq}) with the boundary 
condition (\ref{eq:boundarycond}) and the normalization 
(\ref{eq:normcond}) for the flat ($k=0$) and warped ($k \ne 0$) extra
dimensions. 

We call the case with $k=0$ `flat brane world'. In this case the solution 
of Eqs.~(\ref{eq:KKeigeneq})-(\ref{eq:boundarycond}) is given by
\begin{eqnarray}
\chi_0 (y) &=& 1, \nonumber \\
\chi_n (y) &=& \sqrt{2} \cos (n y/R) 
  \quad (n =1,2,\ldots), \label{eq:KKwf} \\
\tilde{\chi}_n (y) &=& \sqrt{2} \sin (n y/R) 
  \quad (n =1,2,\ldots), \label{eq:oKKwf} \\
M_n &=& n\mu_R, \label{eq:KKmasstorus}
\end{eqnarray}
where $\mu_R=1/R$. 
These are the KK mode function and mass eigenvalue of the bulk gauge field 
in the flat brane world with $M_4 \times S^1$. 
We can easily extend it to higher dimensional bulk space-time,
$M_4 \times T^\delta$, that is shown latter. 

$k \ne 0$ with orbifold condition, $\tilde{A}_\mu^{(n)}(x) \equiv 0$ 
for $^\forall n$, gives the RS brane world. 
The solution of Eqs.~(\ref{eq:KKeigeneq})-(\ref{eq:boundarycond}) 
reads~\cite{Chang:2000nh,Gherghetta:2000qt}
\begin{eqnarray}
\chi_0(y) &=& 1, \nonumber \\
\chi_n(y) &=& \frac{e^{k|y|}}{N_n} 
\left[ J_1 \left( \frac{M_n}{k} e^{k|y|} \right) 
 + c_n Y_1 \left( \frac{M_n}{k} e^{k|y|} \right) \right] 
  \quad (n=1,2,\ldots), \label{eq:rsKKwf} 
\end{eqnarray}
where the coefficient and the normalization factor are 
given by 
$c_n = -J_0 \left( \frac{M_n}{k} \right)/Y_0 \left( \frac{M_n}{k} \right)$ 
and $N_n^2 = \frac{1}{2 \pi R} \int_{-\pi R}^{\pi R} dy \, 
e^{2k|y|} \left[ J_1 \left( \frac{M_n}{k} e^{k|y|} \right) 
 + c_n Y_1 \left( \frac{M_n}{k} e^{k|y|} \right) \right]^2$ respectively 
and $J$ and $Y$ are the Bessel functions. 
The KK mass eigenvalue $M_n$ is obtained as the solution of 
$J_0 \left( \frac{M_n}{k} \right) 
Y_0 \left( \frac{M_n}{k} e^{\pi kR} \right)
= Y_0 \left( \frac{M_n}{k} \right) 
J_0 \left( \frac{M_n}{k} e^{\pi kR} \right)$. 
For the case with $kR \gg 1$ and $M_n \ll k$ the asymptotic form 
of the KK mass eigenvalue is simplified to
\begin{eqnarray}
M_0 &=& 0, \nonumber \\
M_n &\simeq& (n-1/4)\mu_{kR} \quad (n=1,2,\ldots), 
\label{eq:KKmassasym}
\end{eqnarray}
where $\mu_{kR} \equiv \pi e^{-\pi k R} k$. 
We find that the spectrum of the excited mode is shifted by the 
factor $1/4$ and the mass difference between neighbor modes 
is suppressed by the warp factor $e^{-\pi k R}$. 
In the same asymptotic limit $n$-dependence of the mode function 
vanishes on the $y=\pi R$ brane. It reduces to 
\begin{eqnarray}
\chi_n(\pi R) \simeq \sqrt{2\pi kR}. 
\nonumber 
\end{eqnarray}

\begin{figure}[t]
\begin{center}
\begin{minipage}{0.48\linewidth}
   \centerline{\epsfig{figure=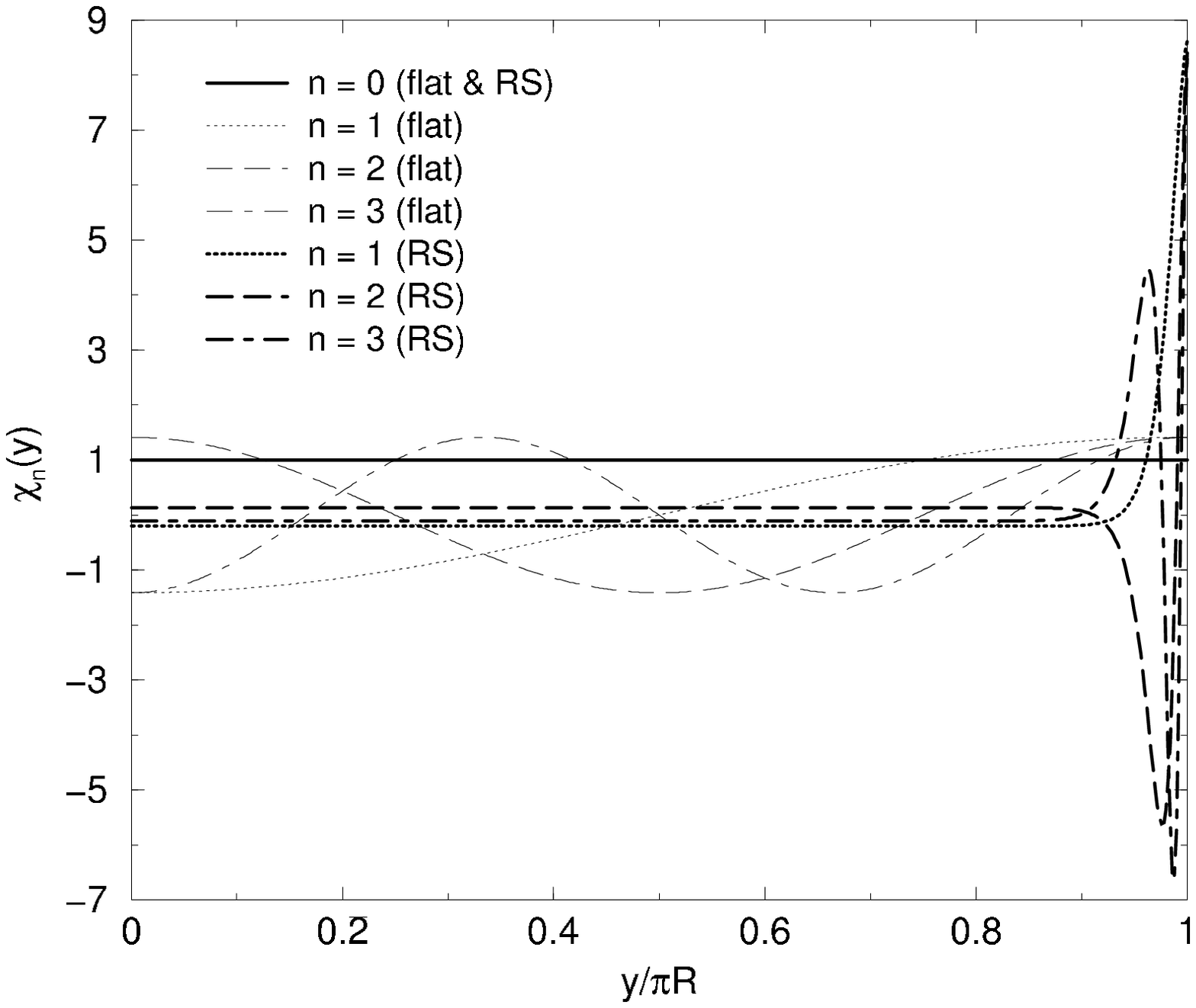,width=\linewidth}}
\end{minipage}
\hfill
\begin{minipage}{0.48\linewidth}
   \centerline{\epsfig{figure=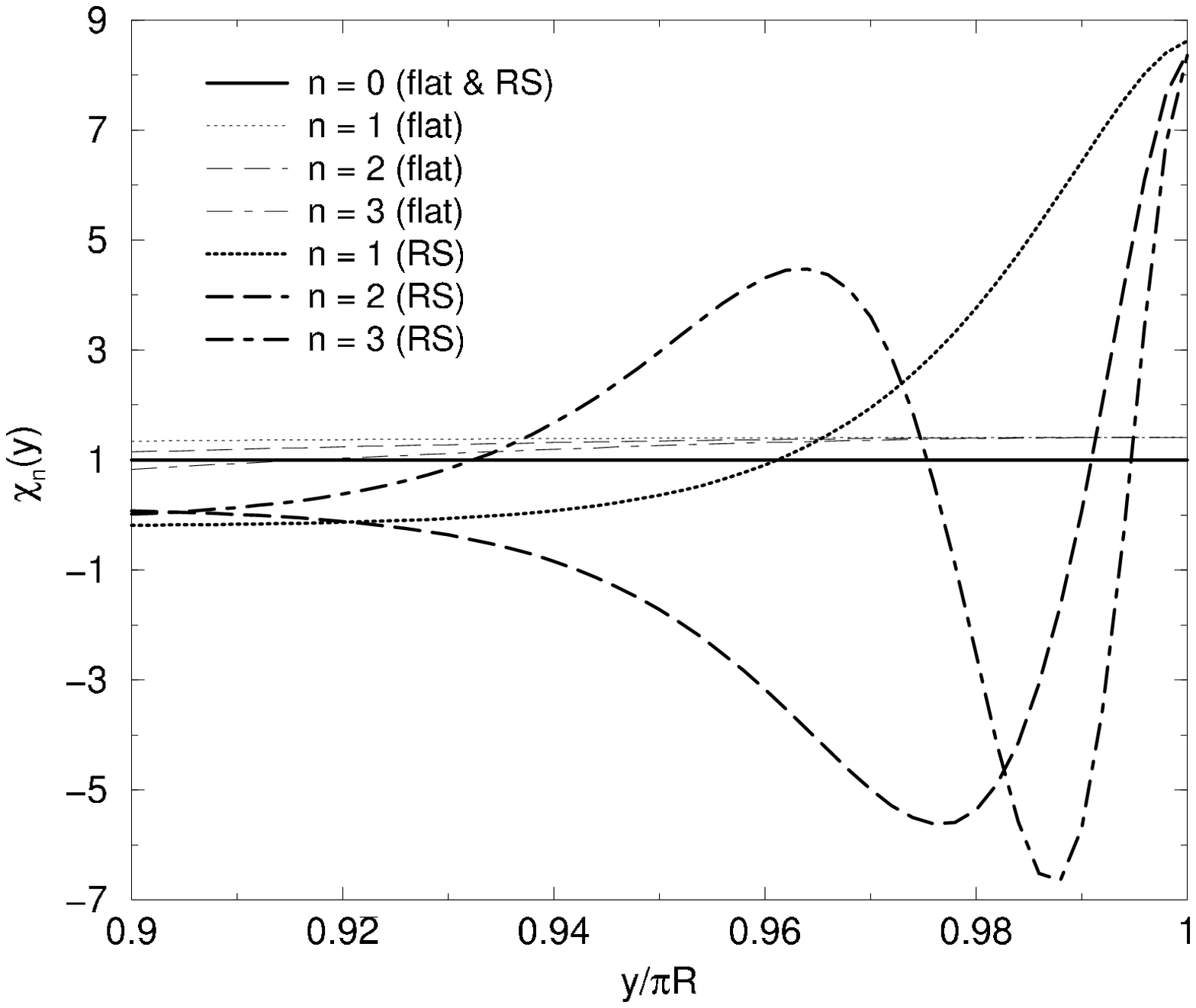,width=\linewidth}}
\end{minipage}
\end{center}
\caption{The profile of the mode functions $\chi_n(y)$ 
for $kR=11$, 
as a function of the coordinate of the extra dimension $y$. 
$\chi_n (y)$ are plotted so as to take a positive value at 
the $y/\pi R =1$. }
\label{fig:WF}
\end{figure}

The profile of the mode functions $\chi_n(y)$ 
are shown in Fig.~\ref{fig:WF}. 
From the figure we know that the 
KK excited gauge bosons are localized in the vicinity of 
the $y=\pi R$ brane in the RS brane world due to the 
finite curvature scale $k$ of the extra dimension, while 
the zero mode is flat in the direction of the extra dimension. 
We will see that the localized KK excited modes 
enhance (suppress) the dynamical symmetry breaking on the $y=\pi R$ 
($y=0$) brane.
%%%%%%%%%%%%%%%%%%%%%%%%%%%%%%%%%%
%%%%%%%%%%%%%%%%%%%%%%%%%%%%%%%%%%

\subsection{Coupling to brane fermion}

Next we define the coupling of the bulk gauge field 
to a fermion on the brane. By decomposing the metric as 
$G_{MN}=E_{\ M}^{\bar{M}} \eta_{\bar{M}\bar{N}} E_{\ N}^{\bar{N}}$
where 
$\eta_{\bar{M}\bar{N}}={\rm diag}(1,-1,-1,-1,-1)$, 
we define 
\begin{eqnarray}
E_{\bar{M}}^{\ M} = \left(
\begin{array}{cc}
e^{+k|y|}\delta_{\bar{\mu}}^{\ \mu} & 0 \\
0 & 1 
\end{array}
\right), \quad
E_{\ M}^{\bar{M}} = \left(
\begin{array}{cc}
e^{-k|y|}\delta^{\bar{\mu}}_{\ \mu} & 0 \\
0 & 1 
\end{array}
\right), 
\nonumber 
\end{eqnarray}
where the row and column of the matrix correspond to the indices 
$M$ and $\bar{M}$ respectively. $E_{\bar{M}}^{M}$ gives the coupling 
of a fermion on the brane and the bulk gauge field. 
That is described as 
\begin{eqnarray}
{\cal L}_{\rm coupling}^{(5)} 
&=& {(\det E_{\ \mu}^{\bar{\mu}})}^{-1} g_{\rm 5D} \bar{\psi}(x) 
    i\Gamma^{\bar{M}} E_{\bar{M}}^{\ M}(y) A_M(x,y) \psi(x) 
    \delta (y-y^\ast) \nonumber \\
&=& g_{\rm 5D} \bar{\psi}(x)
    iA\!\!\!/(x,y) \psi(x) \delta (y-y^\ast). 
\label{eq:bbcoup}
\end{eqnarray}
where $\Gamma_{\bar{M}}=(\gamma_\mu,i\gamma_5)$, 
$A\!\!\!/ = \gamma^\mu A_\mu$, 
$g_{\rm 5D}$ is the five dimensional gauge coupling and 
$y^\ast$ is the position of the brane. 
In the second line we take $A_4=0$ gauge and 
rescale $\psi$ field as $\psi \to e^{3k|y^\ast|/2}\psi$ 
in order to canonically normalize the kinetic term of $\psi$. 
By integrating (\ref{eq:bbcoup}) over the extra dimensional 
coordinate we obtain the 4D effective gauge coupling written by 
\begin{eqnarray}
{\cal L}_{\rm coupling} 
&=& \sum_{n=0}^\infty g_n 
    \bar{\psi}(x) i{A\!\!\!/}^{(n)}(x) \psi(x) 
   +\sum_{n=1}^\infty \tilde{g}_n 
    \bar{\psi}(x) i{\tilde{A}\!\!\!/}^{(n)}(x) \psi(x), 
    \label{eq:4Dgcterm}
\end{eqnarray}
where 
\begin{eqnarray}
g_n = \frac{\chi_n(y^\ast)}{\chi_0(y^\ast)} g_0, \quad 
g_0 = \frac{\chi_0(y^\ast)}{\sqrt{2\pi R}} g_{\rm 5D} 
    \equiv g, 
\label{eq:gn}
\end{eqnarray}
where $g$ is the ordinary four dimensional gauge coupling. 
$\tilde{g}_n$ is defined by replacing $\chi$ with $\tilde{\chi}$ 
in Eq.~(\ref{eq:gn}). 
In the case $y^\ast =0,\pi R$, i.e. $\tilde{g}_n=0$  
(see Eq.~(\ref{eq:boundarycond})), 
$\tilde{A}_\mu^{(n)}(x)$ decouples in Eq.~(\ref{eq:4Dgcterm}). 
We notice that the coupling of the gauge boson 
excited modes $g_n$ are enhanced/suppressed by the value of the 
mode function $\chi_n$ on the $y=y^\ast$ brane, compared with it of 
zero mode. Thus we obtain larger effect from the KK excited gauge 
boson on the brane they localize.

\section{Effective theory}
In the previous section we have derived the Lagrangian of the 
bulk gauge field and its coupling to a fermion on the brane. 
Putting them together we obtain the effective Lagrangian for the system 
of the bulk gauge theory couples to fermion on the $y=y^\ast(=0,\pi R)$ 
brane as 
\begin{eqnarray}
{\cal L} 
&=& \frac{1}{2} \sum_{n=0}^{N_{\rm KK}} A_\mu^{(n)} \left[ 
    \eta^{\mu \nu} \left( \partial^2 + M_n^2 \right)
   -(1-\xi) \partial_\mu \partial_\nu \right] A_\nu^{(n)}
    \nonumber \\ &&
   +\frac{1}{2} \sum_{n=1}^{N_{\rm KK}} \tilde{A}_\mu^{(n)} \left[ 
    \eta^{\mu \nu} \left( \partial^2 + M_n^2 \right)
   -(1-\xi) \partial_\mu \partial_\nu \right] \tilde{A}_\nu^{(n)}
   +{\cal L}_{\rm SI} \nonumber \\ &&
   +\bar{\psi} \left( i\partial_\mu + g A_\mu^{(0)} \right) 
    \gamma^\mu \psi
   + g \sum_{n =1}^{N_{\rm KK}} \frac{\chi_n(y^\ast)}{\chi_0(y^\ast)} 
    \bar{\psi} A_\mu^{(n)} \gamma^\mu \psi, 
\label{eq:efflag}
\end{eqnarray}
where ${\cal L}_{\rm SI}$ represents the self interaction (and mixing) 
terms of $A_\mu^{(n)}$ and $\tilde{A}_\mu^{(n)}$, and 
$N_{\rm KK}$ stands for the cutoff of the KK summation. 
It should be noted that we introduce the gauge fixing parameter 
$\xi$ on the brane. 
(See appendix in Ref.~\cite{Abe:2001yi} for more details.) 
$\xi$ is chosen so that a QED like Ward-Takahashi identity 
is hold in the latter analysis. 
We can easily extend the above system to 
higher dimensional bulk space-time with $M_4 \times T^\delta$ 
by replacing $M_n$ and $\chi_n$ as 
\begin{eqnarray}
M_n&\to&\sqrt{M_{n_1}^2+M_{n_2}^2+\ldots+M_{n_\delta}^2},\nonumber \\
\chi_n(y)&\to&\chi_{n_1}(y_1)\chi_{n_2}(y_2) \cdots \chi_{n_\delta}(y_\delta),
\nonumber 
\end{eqnarray}
where $n_1, n_2, \ldots, n_\delta = 0,1,2,\ldots$ 
correspond to the indices of the KK excited mode for each extra dimension 
$(y_1, y_2, \ldots, y_\delta)$ respectively. 
$M_{n_i}$ and $\chi_{n_i}(y_i)$ $(i=1,...,\delta)$ are given 
in Eqs.~(\ref{eq:KKwf}) and (\ref{eq:KKmasstorus}) respectively. 
In addition we should interpret $\sum_n$ as $\sum_{n_1} \sum_{n_2} 
\cdots \sum_{n_3}$ with a cut-off condition, 
$n^2 = n_1^2+n_2^2+\ldots+n_\delta^2 \le N_{\rm KK}^2$. 

In the following discussion we regard the KK reduced Lagrangian 
(\ref{eq:efflag}) as an effective one below the bulk fundamental scale 
$\Lambda_{\rm 5D}$. Thus we cutoff the momentum of the loop integral,
including only the propagator of bulk fields, at the scale 
$\Lambda_{\rm 5D}$. On the other hand we cutoff the momentum
at the {\it reduced} cutoff scale $\Lambda$ on the brane in 
the case that the loop integral includes the brane fields.
Loop integrals in the ladder Schwinger-Dyson equation is only
the latter case. In the flat brane world $\Lambda$ should be 
the same order of $\Lambda_{\rm 5D}$, while in the RS brane world 
we have an large difference between $\Lambda$ and $\Lambda_{\rm 5D}$
due to the warp factor. 
We also define $N_{\rm KK}$ as the number of the KK modes with their mass 
below the {\it reduced} cutoff scale $\Lambda$. 
In our numerical analyses, however, we cutoff the KK mode summation 
at $N_{\rm KK}+10$ in order to avoid the sharp threshold effect.

\subsection{Local four-fermion approximation}

As is mentioned in the previous chapter, we assume that the
KK mode mass $M_n$ is on the order of the cutoff scale $\Lambda$ 
to deal with chiral symmetry breaking analytically as a first step 
in the investigation of the phase structure. 
If all the KK excited modes are sufficiently heavy, terms with KK modes 
in the Lagrangian (\ref{eq:efflag}) reduce to four-fermion interactions. 
After integrating out the KK excited modes, we obtain
\begin{equation}
{\cal L}_{\rm eff} 
    = -\frac{1}{4} F^{(0)\mu \nu}F^{(0)}_{\mu \nu} 
      +\bar{\psi} \left( i\partial_\mu +gA_\mu^{(0)} \right) \gamma^\mu \psi
      -\frac{1}{2} 
       G \left( \bar{\psi} \gamma^\mu \psi \right) 
         \left( \bar{\psi} \gamma_\mu \psi \right), 
\label{lag:ff}
\end{equation}
where we neglect more-than four-fermion interactions 
induced by the gauge boson self couplings if the gauge group is non-Abelian. 
The four-fermion coupling constant $G$ is given by
\begin{eqnarray}
G = g^2 T^a T^a \sum_{n \ne 0}^{N_{\rm KK}} 
    \frac{\chi_n^2 (y^\ast)}{M_n^2}. 
\label{eq:GNJL}
\end{eqnarray}
Using the Fierz transformation, the four-fermion interaction terms 
in (\ref{lag:ff}) can be rewritten as
\begin{eqnarray}
\cal{L}_{\rm eff} 
  &=& -\frac{1}{4} F^{(0)\mu \nu}F^{(0)}_{\mu \nu} 
      +\bar{\psi} \left( i\partial_\mu +gA_\mu^{(0)} \right) \gamma^\mu \psi
      +\frac{1}{2} 
       G \left[ \left( \bar{\psi} \psi \right)^2 + 
                \left( \bar{\psi} i\gamma_5 \psi \right)^2 \right],\qquad
\label{GNJLLag}
\end{eqnarray}
where we ignore the tensor-type interactions. 

For instance, the case of Abelian gauge theory with a flat compact 
extra dimension with its radius $R$, i.e. $M_n=n/R$ and $\chi_n=1$, we have 
\begin{eqnarray}
G = \sum_{n \ne 0}^{N_{\rm KK}}\frac{8\pi\alpha R^2}{n^2} 
\ \stackrel{N_{\rm KK}\rightarrow \infty}{\longrightarrow}\  
\frac{4 \pi^2 \alpha R^2}{3}, \label{G1:GNJL}
\end{eqnarray}
If the gauge field is confined to the brane, the four-fermion interaction
term is not generated, i.e. $G=0$.

\chapter{Analysis of local four-fermion interaction models}
\label{chap:alffim}

In this chapter\footnote{This chapter is based on Refs.~\cite{Abe:2001yi,
Abe:2001ax, Abe:2000ny} with K.~Fukazawa, T.~Inagaki, H.~Miguchi and T.~Muta.} 
we analyze four-fermion interaction models as a simple 
approximation of the bulk gauge theory with fermions on branes.

\section{Dynamical chiral symmetry breaking on a boundary}

Equation~(\ref{GNJLLag}) is identical to the  gauged 
Nambu-Jona-Lasinio (NJL) model. 
A variety of methods have been developed to evaluate the phase structure 
in the gauged NJL model (see Refs.~\cite{Kondo:1991mq} and~\cite{Miransky:vk}). 
Using the bifurcation method, the critical line in the $\alpha$-$G$ plane 
is analytically obtained as~\cite{Appelquist:1988fm,Kondo:1988qd} 
\begin{equation}
\frac{G_{\rm critical} \Lambda^2}{4\pi^2}
=\frac{1}{4}\left(1+\sqrt{1-\frac{3\alpha_{\rm critical}}{\pi}}\right)^2\ ,\ \ \  
\left(\alpha < \frac{\pi}{3}\right)
\label{cl:GNJL}
\end{equation}
where $G_{\rm critical}$ and $\alpha_{\rm critical}$ denote the critical values 
of $G$ and $\alpha$ respectively. For $\alpha < \pi /3$ and $G > G_{\rm critical}$, 
the fermion mass function $B(p^2)$ behaves as
\begin{equation}
  B(p^2)=\frac{\sqrt{1-\omega^2}}{\omega}\frac{\mu^2}{\sqrt{p^2}} 
  \mbox{sinh}\left(
    \frac{\omega}{2}\mbox{ln}\frac{p^2}{\mu^2}+\mbox{tanh}^{-1}\omega
  \right) ,
\label{mass:GNJL}
\end{equation}
where $\omega=\sqrt{1-3\alpha/\pi}$ and $\mu$ is given by
\begin{equation}
  \mu=\Lambda \exp \left[
    -\frac{1}{\omega}\mbox{tanh}^{-1}\left(
      \frac{\omega/2}{-1/4-\omega^2/4+G \Lambda^2/(4\pi^2)}
    \right)
  \right] .
\end{equation}
\begin{table}[t]
\begin{center}
\begin{tabular}{l|c|c|c|c|c}
\hline \hline
\ & $N_{\rm KK}$ & Radius & $M_1$ & $\alpha_1$ & $\alpha_{\rm critical}$ \\
\hline 
QED & 0 & \ & \ & 0 & $\pi/3 \sim 1.05$ \\
Flat & 1 & $R \Lambda =1$ & $\sim\Lambda$ & $2\alpha$ & $0.83$ \\
Flat & $\infty$ & $R \Lambda =1$ & $\sim\Lambda$ & $2\alpha$ & $0.63$ \\
Warped ($y=\pi R$) & 1 & $2kR=1$ & $\sim0.80k$  & $\sim 4.06\alpha$ & $0.40$ \\
Warped ($y=0$) & 1 & $2kR=1$ & $\sim0.80k$ & $\sim 0.92\alpha$ & $0.94$ \\ 
\hline
\end{tabular}
\end{center}
\caption{The critical couplings $\alpha_{\rm critical}$ evaluated from 
the result of the four-fermion approximation for $N_{\rm KK}=1$. 
$R$ is the radius of the flat extra dimension. 
The AdS curvature is taken as $k=\Lambda$. 
In the third line, $N_{\rm KK} \to \infty$ in the flat 
extra dimension means that the summation of the KK modes is taken 
to be infinity, but $\Lambda$ is fixed to a finite value. 
This corresponds to the existence of an anisotropy between the cutoff 
of the bulk and of the brane.} 
\label{GNJLTab}
\end{table}
The coupling constants $\alpha$ and $G$ are not independent
parameters for the brane world models considered here. 
Substituting Eq.~(\ref{eq:GNJL}) into Eq.~(\ref{cl:GNJL}), 
we find the critical coupling constant $\alpha_{\rm critical}$. 
In Table~\ref{GNJLTab} we list values of the 
critical coupling constant for some characteristic cases~\cite{Abe:2001yi}. 
Since the local four-fermion approximation must be valid for heavy KK 
excitations, we consider the case $R \sim 1/\Lambda$ for the flat 
extra dimension and $2kR\sim 1$ for the warped extra dimension, 
where only the lightest KK mode has mass below the cutoff scale. 
In the flat extra dimension, the lightest KK excitation mode is 
almost decoupled. Thus, the KK modes
have only a small correction to enhance chiral symmetry
breaking. This result is similar to that without KK modes, i.e. QED. 
However, the critical coupling in the warped extra dimension is less 
than half the value of that in QED in the $y=\pi R$ brane, 
because the coupling constants between the fermion and the KK modes 
are enhanced by the mode expansion function $\chi_n(y^\ast)$ in 
Eq.~(\ref{eq:gn}).

\begin{figure}[t]
\begin{center}
\epsfig{figure=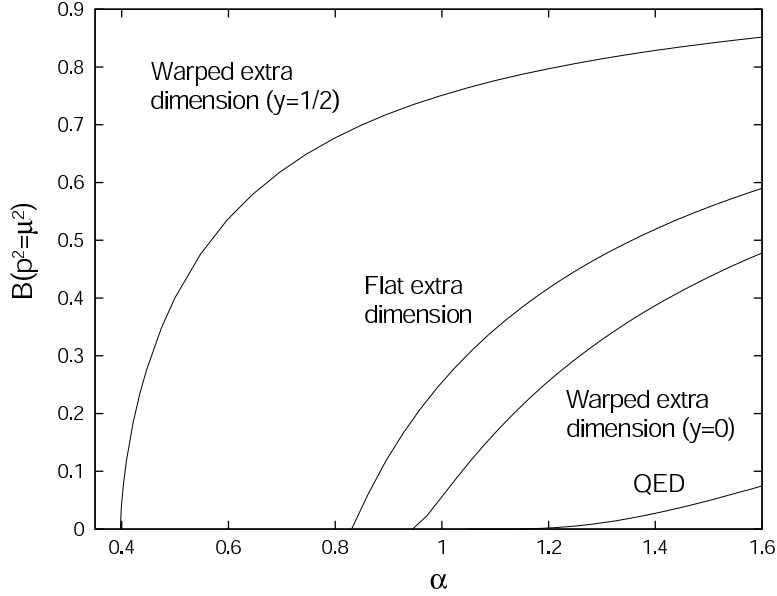,width=0.6\linewidth}
\caption{Behavior of the mass function $B(p^2=\mu^2)$ as a function
 of $\alpha$ for the cases of QED, a flat extra dimension 
($R\sim 1/\Lambda$), and a warped extra dimension 
($2kR\sim 1$) in the gauged NJL model. (Here $y$ is denoted as a 
dimensionless coordinate $y/(2\pi R) \to y$.)}
\label{Ba0}
\end{center}
\end{figure} 
In Fig.~\ref{Ba0} we plot the mass function $B(p^2)$ as
a function of the coupling constant $\alpha$ for fixed $p$. 
It is seen that the mass function smoothly
decreases to zero as the coupling constant $\alpha$ decreases. 
In other words, no mass gap is observed. 
Thus, a second order phase transition is realized. 
Such behavior is considered as a generalization of Miransky scaling. 
On the $y=0$ brane in the warped extra dimension, the effective coupling
constant of the lightest KK mode is nearly equal to $\alpha$, and the situation
is similar to that for the brane in the flat extra dimension. Therefore, 
in the following calculations, we focus on the $y = 1/2$ 
brane for the warped extra dimension, and the $y = 0$ brane is ignored.

\section{Bulk and boundary mixing}
\label{sec2}

In the above we only consider the case that all the matter fields 
are confined on four-dimensional boundaries. 
In this section we treat rather complicated situation. 
The fermion on a brane can interact with a fermion in the bulk 
through the common gauge interaction in the bulk. 
We analyze the bulk and boundary mixed four-fermion interaction models 
instead of such complicated system of gauge theory by following 
the previous aspects. 
The bulk fermions interact with themselves as well as with fermions 
on the four-dimensional branes through the exchange of KK modes of the 
gauge bosons which may be assumed to exist in the 
bulk~\cite{Chang:2000nh,Gherghetta:2000qt}, or through the exchange of 
the graviton and its Kaluza-Klein excited modes. 
As discussed above the interactions among fermions generated as a 
result of the exchange of all the Kaluza-Klein excited modes of the 
graviton or gauge bosons may be expressed as effective four-fermion 
interactions~\cite{Dobrescu:1998dg,Cheng:1999bg,Han:1998sg}. 
According to the four-fermion interactions we expect that 
the dynamical generation of fermion masses will take place.
In the present communication we look for a possibility of the dynamical 
fermion mass generation under the influence of the bulk fermions through 
the effective four-fermion interactions.
The possible source of fermion masses on the branes is two-fold, i. e. 
the dynamically generated fermion masses and masses of the Kaluza-Klein 
excited modes of the bulk fermions. 

We introduce a bulk fermion $\Psi$ in addition to a brane one $\psi$, 
and assume that $\Psi$ and $\psi$ possess same representation 
(same charge) under a bulk gauge group ($U(1)$ for simplicity). 
We integrate out heavy KK gauge bosons by assuming that the 
compactification scale of the extra dimensions is near the cut-off 
scale. Furthermore we neglect the effect of gauge boson zero mode 
as a zeroth approximation for analyzing phase structure. 
In such case we have a Lagrangian in five-dimensional space-time: 
\begin{eqnarray}
{\cal L}^{(5)} &\simeq& 
\bar{\Psi} i \gamma^M \partial_M \Psi 
- (\bar{G}/2) (\bar\Psi \gamma^\mu \Psi) (\bar\Psi \gamma_\mu \Psi)
  \nonumber \\ && 
- \big[ \bar\psi i \gamma^\mu \partial_\mu \psi 
- (\bar{G}/2) (\bar\psi \gamma^\mu \psi) (\bar\psi \gamma_\mu \psi) 
  \nonumber \\ && \hspace{3cm}
- \bar{G} (\bar\Psi \gamma^\mu \Psi) (\bar\psi \gamma_\mu \psi) 
\big] \delta(y). 
\nonumber 
\end{eqnarray}
The four-fermion interaction $\bar{G}$ is given by $\bar{G}=2\pi R G$ 
where $G$ is given in Eq.~(\ref{eq:GNJL}). 
Because of a factor $1/2$ in the first and second line we assume 
that a four-fermion interaction in the third line that mixes bulk 
$\Psi$ and brane $\psi$ is the most attractive channel. 
Accordingly we start with the following Lagrangian for our 
analysis~\cite{Abe:2000ny}:
\begin{equation}
{\cal L\/}^{(5)}  =  
  \bar{\Psi} i \gamma^M \partial_M \Psi  
+ [ \bar{\psi} i \gamma^\mu \partial_\mu \psi 
- \bar{G} (\bar{\Psi} \gamma^\mu \Psi) (\bar{\psi} \gamma_\mu \psi) 
  ] \,\delta(y),    
\label{eq:sLag}
\end{equation}
$\bar{G}$ has mass dimension $-3$, and fermions $\Psi$ and $\psi$ 
are assumed to be of $N_f$ components.

\subsection{Flat bulk}

\begin{figure}[t]
\begin{minipage}{0.48\linewidth}
   \centerline{\epsfig{figure=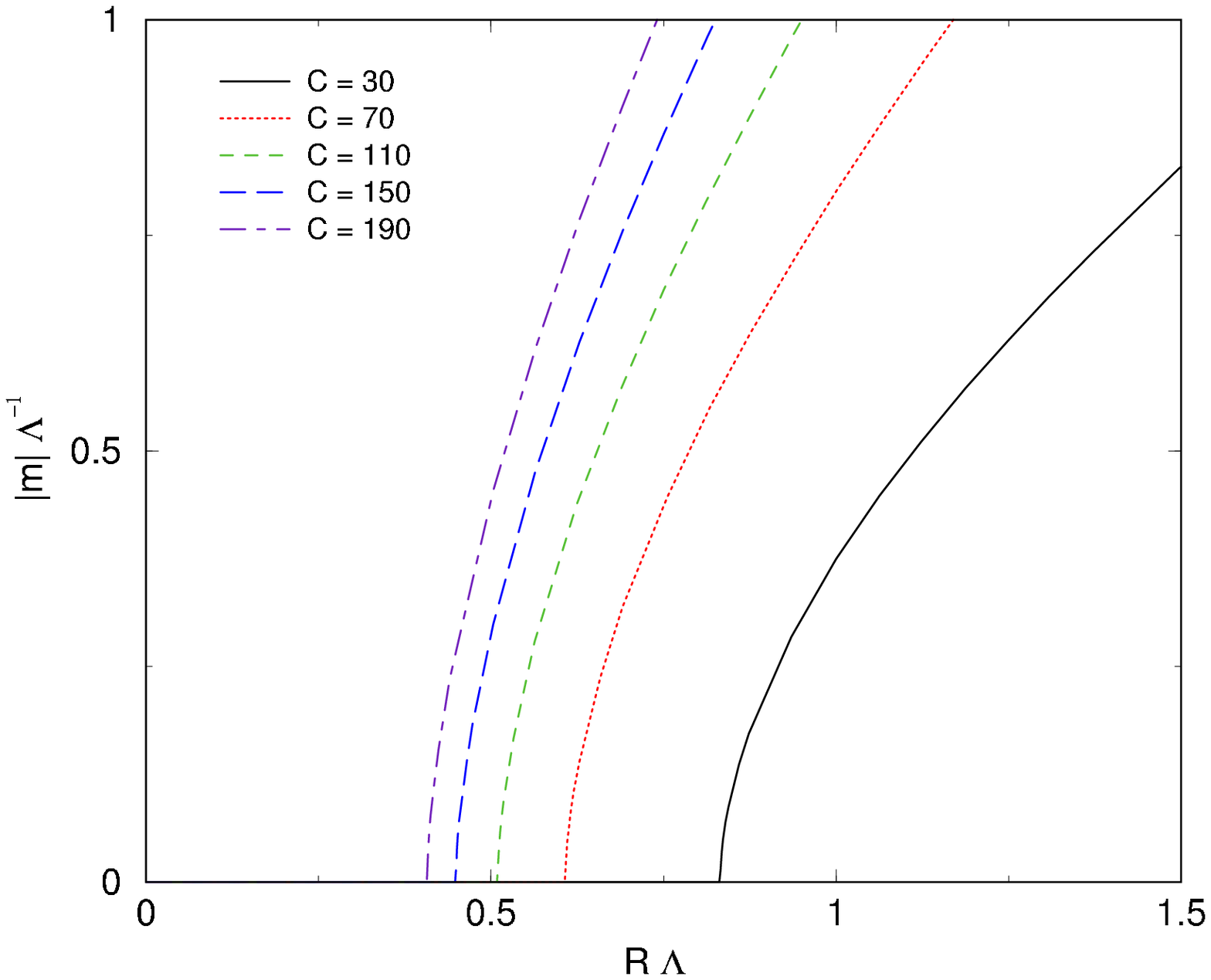,width=\linewidth}}
   \centerline{(a)}
\end{minipage}
\hfill
\begin{minipage}{0.48\linewidth}
   \centerline{\epsfig{figure=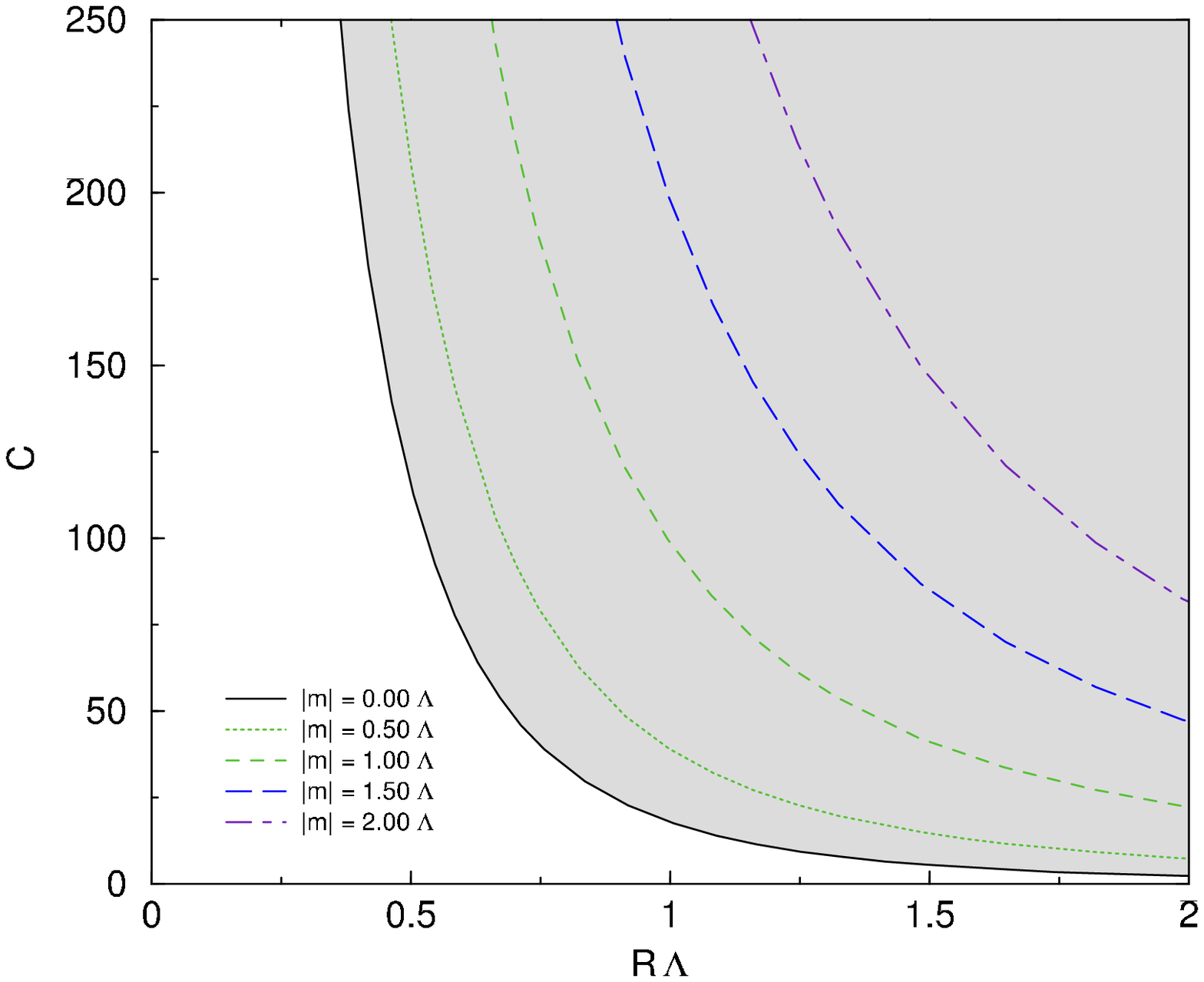,width=\linewidth}}
   \centerline{(b)}
\end{minipage}
\caption{(a) Dynamical fermion mass $m$ as a function of the 
compactification radius $R$ with the dimensionless coupling 
$C \equiv \bar{G}/(2\pi R^3) = G/R^2$ fixed. The fermion mass is generated 
dynamically beyond a critical $R$. 
(b) Contour curve of $m$. The critical curve is drawn 
by real line and the massive region is filled.}
\label{fig:c-R}
\end{figure}

First we assume that $\Psi$ and $\psi$ are in the flat brane world, 
namely with vanishing AdS curvature scale ($k=0$). 
In this case the bulk fermion $\Psi$ has usual $\cos$ ($\sin$) type 
KK mode expansion and a $y$ integration can be performed easily. 
After Fierz transformation and chiral rotation 
$\Psi \to e^{i\frac{\pi}{4}\gamma^5}\Psi$, 
$\psi \to e^{i\frac{\pi}{4}\gamma^5}\psi$ we introduce scalar 
and pseudo-scalar auxiliary fields $\sigma \sim \bar{\Psi}\psi$ and 
$\pi \sim \bar\Psi i\gamma_5 \psi$. 
Here we assume (or from the beginning consider such a four-fermion model) 
that $\langle \pi \rangle =0$ because of e.g. the relict from 
five-dimensional Lorentz invariance. 
Then our four-dimensional Lagrangian is rewritten as 
\begin{eqnarray}
  {\cal L\/}^{(4)} 
      =   \bar{\Psi} (M + i /\!\!\!\partial) \Psi 
                  \ - \  \left| \sigma \right|^2,
\nonumber 
\end{eqnarray}
where 
$\Psi^t \equiv 
       \left( \psi \, , \, \Psi_{0} \, , \, \Psi_{1} \, , \,
           \Psi_{-1} \, , \, \Psi_{2} \, , \, \Psi_{-2} 
           \cdots
        \right)$
and 
\begin{eqnarray}
  M \equiv      
       \left(
         \matrix{
            0 & m^{*} & m^{*} & m^{*} & m^{*} & m^{*} & \cdots \cr
            m & 0     & 0     & 0     & 0     & 0     & \cdots \cr
            m & 0     & \frac{1}{R}   & 0     & 0     & 0 & \cdots   \cr
            m & 0     & 0     & - \frac{1}{R} & 0     & 0 & \cdots   \cr
            m & 0     & 0     & 0     & \frac{2}{R}   & 0 & \cdots   \cr
            m & 0     & 0     & 0     & 0     & - \frac{2}{R} & \cdots   \cr
            \vdots & \vdots & \vdots & \vdots & \vdots & \vdots & \ddots  
         }
       \right), \qquad (m=\sqrt{\bar{G}/2\pi R}\,\sigma). \label{M} 
\end{eqnarray}
If $\sigma$ acquires a nonvanishing vacuum expectation value,
we replace $\sigma$ in $m$ by its vacuum expectation value 
$\left\langle \sigma \right\rangle$, 
i.e. $m = \sqrt{\bar{G}/2\pi R} \left\langle \sigma \right\rangle$. 
The eigenvalues of matrix $M$ determine the masses of four-dimensional fermions. 
Obviously we find 
that the lightest eigenvalue is given by
\begin{eqnarray}
\lambda_{\pm 0} = \pm \ |m| \quad {\textrm{for}} \quad |m| \ll 1/R. 
\label{Mass}
\end{eqnarray}
Thus we conclude that within our scheme there is a possibility of having the 
light fermion masses which is much smaller than the mass of the Kaluza-Klein
modes of the bulk fermion. 
By performing the path-integration for fermion field $\Psi$ 
we find the effective potential for $\sigma$ 
in the leading order of the $1/N_f$ expansion:
\begin{eqnarray}
V(\sigma)  =  |\sigma|^2 &-& \frac{1}{2 \pi^2}
                \int_0^\Lambda dx \, x^3
                \ln \left[x^2 + |m|^2 (\pi xR) \coth(\pi x R)\right] 
                \nonumber \\
          &-& \frac{1}{2 \pi^2} \sum_{j=1}^{\infty}\, 
                \int_0^\Lambda dx \, x^3
                \ln \left[x^2 + \left( \frac{j}{R} \right)^2\right].
\nonumber 
\end{eqnarray}
The gap equation to determine the vacuum expectation value 
$\left<\sigma\right>$ reads
\begin{eqnarray}
\frac{\partial V(\sigma)}{\partial |\left<\sigma \right>|} 
   =  2 |\left< \sigma \right>| \, \left\{ 1- \frac{g^2}{2 \pi^2}\,
      \int_0^\Lambda dx \, \frac{x^3}{2 x \tanh(\pi x R) 
      + g^2 |\left< \sigma \right>|^2}
      \right\} = 0.
\label{Gap}
\end{eqnarray}
By numerical observation of Eq. (\ref{Gap}) we find that there exists 
a nontrivial solution for $|\sigma|$ for a suitable range of parameters 
$g$ and $R$ and the solution corresponds to the true minimum of the 
effective potential. That is shown in Fig.~\ref{fig:c-R}. 
Accordingly the fermion mass is generated dynamically. 
Here the auxiliary field $\sigma$ (or the composite field $\bar{\psi}\Psi$) 
acquires a vacuum expectation value. Moreover it is easily confirmed that 
the phase transition associated with this symmetry breaking is of 
second order. 

As was shown in Eq. (\ref{Mass}) the lowest fermion mass on the 
four-dimensional brane is $m=Ng \left<|\sigma|\right>$ 
where $\left<|\sigma|\right>$ is determined by solving Eq. (\ref{Gap}). 
The critical curve which represents the critical radius as a function of the 
coupling constant is shown in Fig.~\ref{fig:c-R} as the curve for $m=0$. 
If we assume $\Lambda \sim 1/R \sim$ TeV like in the large extra dimension 
scenario~\cite{Antoniadis:1990ew,Arkani-Hamed:1998rs}, light ($\ll$ TeV) 
fermions are obtained in the region between the solid line and the 
dot-dashed line in the Fig.~\ref{fig:c-R}, i.e. just above the critical 
line of the second phase transition. 

\subsection{Warped bulk}

In this section we introduce a bulk fermion $\Psi$ and two fermions 
$\psi_1$ and $\psi_2$ on a $y=0$ and a $y=\pi R \equiv L/2$ brane 
respectively in the RS warped brane 
world~\cite{Chang:2000nh,Gherghetta:2000qt}. 
We consider such two kind of brane fermion $\psi_1$ and $\psi_2$ 
because the invariance of replacing two branes each other in the warped 
dimension is violated, i.e. $\psi_1$ has different property from $\psi_2$. 

Following Chang {\it et al.}\cite{Chang:2000nh,Gherghetta:2000qt} 
we derive the 
mode expansion of the bulk fermion in the RS space-time such that 
\begin{eqnarray}
\Psi(x,y) = 
  \frac{e^{3k|y|}}{\sqrt{L}}
  \sum_n \left[ \Psi_L^{(n)}(x){\xi (y)} 
              + \Psi_R^{(n)}(x){\eta(y)} \right],
\label{modeex}
\end{eqnarray}
\begin{eqnarray}
\left\{
\matrix{
{\xi (y)}
  = \sqrt{\frac{2\pi kR}{1-e^{-\pi kR}}} e^{-k|\pi R-y|}
    \sin \displaystyle{ \frac{m_n}{k} } (1-e^{2k|y|}) \cr \cr
{\eta(y)} 
  = \sqrt{\frac{2\pi kR}{1-e^{-\pi kR}}} e^{-k|\pi R-y|}
    \cos \displaystyle{ \frac{m_n}{k} } (1-e^{2k|y|}) } 
\right.,
\nonumber 
\end{eqnarray}
where 
$m_n \equiv {n\pi k}/{(e^{\pi kR}-1)}$. 
With this expansion the kinetic terms of the KK modes become 
\begin{eqnarray}
{\cal L\/}^{(4)}_{\rm bluk} 
  = \int dy\, E \bar{\Psi} i \gamma^M D_M \Psi
  = \sum_{n} \left[ \bar{\Psi}^{(n)}i /\!\!\!\partial \Psi^{(n)}
      - \frac{m_n}{2 \pi R} \bar{\Psi}^{(n)} \Psi^{(n)} \right].
\nonumber 
\end{eqnarray}
In the following we adopt the similar analysis to the previous 
flat case to the RS background.

Now we extend the previous Lagrangian (\ref{eq:sLag}) to the case 
in the RS warped background with a vielvein 
$E_{\bar{M}}^{\ \ M} = \left( 
\matrix{ e^{2k|y|}\eta_{\bar{\mu}}^{\ \mu} & 0 \cr 0 & 1 } \right)$ 
whose square becomes the metric $G_{MN}$. 
Then we start with the following five-dimensional Lagrangian: 
\begin{eqnarray}
{\cal L\/}^{(5)}
     &=&  E \bar\Psi i \gamma^M D_M \Psi 
      +   E_{(1)} \left[ {
          \bar\psi_1 i \gamma^\mu \partial_\mu \psi_1 
          - \bar{G}_1 (\bar\Psi \gamma^\mu \Psi) 
                    (\bar\psi_1 \gamma_\mu \psi_1) \, } \right]
          \delta(y)  \nonumber \\ && \hspace{1cm}
      +   E_{(2)} \left[ {
          \bar\psi_2 i \gamma^\mu \partial_\mu \psi_2 
          - \bar{G}_2 (\bar\Psi \gamma^\mu \Psi) 
                    (\bar\psi_2 \gamma_\mu \psi_2) \, } \right]
          \delta(y-\pi R), 
\label{eq:5Lag}
\end{eqnarray}
where $\bar{G}_1$ and $\bar{G}_2$ is the four-fermion coupling on $y=0$ and 
$y=\pi R$ brane, respectively. 
They should satisfy $\bar{G}_1=2\pi R G|_{y^\ast=0}$ and 
$\bar{G}_2=2\pi R G|_{y^\ast=\pi R}$ in our case, i.e. they are induced 
via the bulk gauge interaction. 
After chiral rotation $\Psi \to e^{i \frac{\pi}{4} \gamma^5} \Psi$ 
and $\psi_i \to e^{i \frac{\pi}{4} \gamma^5}\psi_i$, 
we introduce scalar and pseudo-scalar auxiliary fields 
$\sigma_i \sim \bar\Psi \psi_i$ and $\pi_i \sim \bar\Psi i\gamma_5 \psi_i$ 
where $i=1,2$. Here we assume (or from the beginning consider such model) 
that $\langle \pi_i \rangle =0$ because of e.g. the relict from five-dimensional 
Lorentz invariance. Thus our four-dimensional Lagrangian is rewritten as 
\begin{eqnarray} 
   {\cal L\/}^{(5)}
       &=&  E \bar{\Psi} i \gamma^M \partial_M \Psi
         +  \left[ \bar{\psi_1} i /\!\!\!\partial \psi_1 
         -  N_f \left| \sigma_1 \right|^2 
         +  (\sqrt{\bar{G}_1} \sigma_1 \bar{\Psi}  \psi_1  +  h.c.) \right] 
            \delta(y) \nonumber \\ && 
         +  e^{-4 \pi kR}\left[ e^{\pi kR} 
            \bar{\psi_2} i /\!\!\!\partial \psi_2 
         -  N_f \left| \sigma_2 \right|^2 
         +  (\sqrt{\bar{G}_2} \sigma_2 \bar{\Psi}  \psi_2  +  h.c.) \right] 
            \delta(y-\pi R).
\nonumber 
\end{eqnarray}
Now we substitute the mode expansion of $\Psi$ of Eq. (\ref{modeex}) 
into the Lagrangian and integrate it over the extra dimension. Taking 
$e^{-\frac{3}{2}\pi kR}\psi_2 \to \psi_2$ and $e^{-2\pi kR}\sigma_2 \to \sigma_2$,
Lagrangian in four-dimension becomes 
\begin{eqnarray}
  {\cal L\/}^{(4)} 
&\equiv& \int dy \,{\cal L\/}^{(5)} \nonumber \\ 
&=& \sum_{n} \left[ \bar{\Psi}^{(n)}i /\!\!\!\partial \Psi^{(n)}
 -  \frac{n}{R} \bar{\Psi}^{(n)} \Psi^{(n)} \right] 
 +  \bar{\psi_1} i /\!\!\!\partial \psi_1 
 +  \bar{\psi_2} i /\!\!\!\partial \psi_2 
 -  N_f (\left| \sigma_1 \right|^2 + \left| \sigma_2 \right|^2) 
    \nonumber \\ &&
 +  \sum_n \left( \mu_1 \bar{\Psi}_R^{(n)} \psi_1     
 +  h.c.\right) 
 +  \sum_n \left( (-1)^n \mu_2 \bar{\Psi}_R^{(n)} \psi_2     
 +  h.c.\right),
\label{4LagRS}
\end{eqnarray}
where
\begin{eqnarray}
\mu_1 \equiv \displaystyle{\sqrt{\bar{G}_1/(2\pi R)}\,\sigma_1},\ \ 
\mu_2 \equiv \displaystyle{\sqrt{\bar{G}_2/(2\pi R)}\,\sigma_2},\ \ 
    R_{\rm w}   \equiv \frac{1}{m_1} = \frac{e^{\pi kR}-1}{\pi k} 
         \sim    ({\rm TeV})^{-1}.
\nonumber 
\end{eqnarray}
Here we note that even though $\Lambda \sim M_{\rm Pl}$
the mass of the 1st KK excited mode $1/R_{\rm w}$ can be of the 
order of TeV without fine-tuning because of the warp factor 
(for $kR \sim 11$) in the RS metric.

\begin{figure}[t]
\centerline{\epsfig{figure=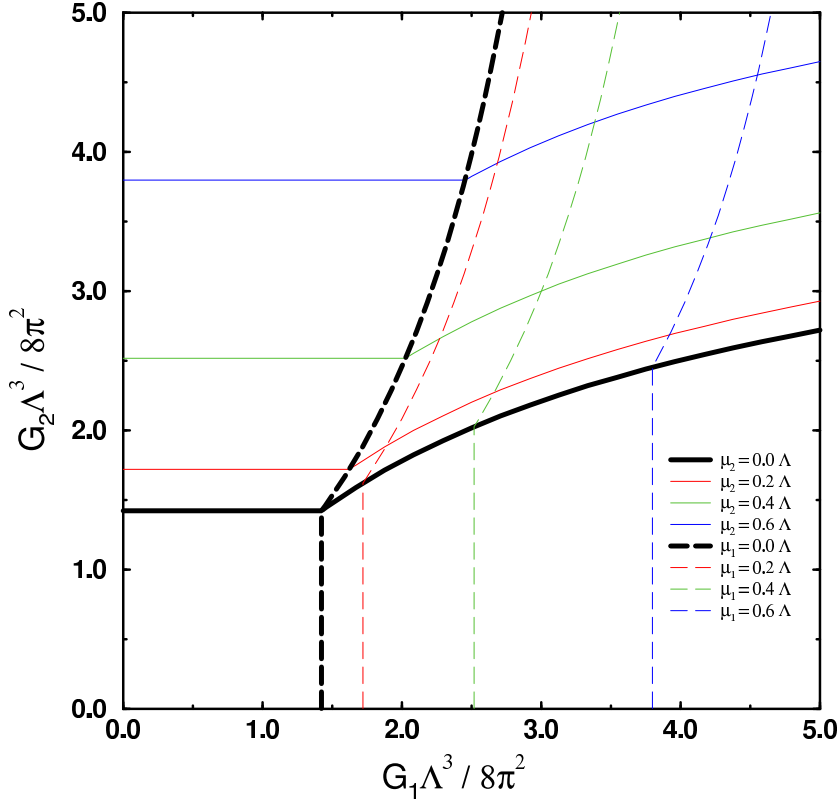,width=0.65\linewidth}}
\caption{The phase structure of the vacuum.}
\label{phase}
\end{figure}

After integrating out all fermionic degrees of freedom, 
we obtain the effective potential for $\sigma_i (\mu_i)$ 
as follows~\cite{Abe:2001ax}: 
\begin{eqnarray}
\lefteqn{\bar{V}({\mu_1}/\Lambda,{\rm 5D}/\Lambda)} 
\hspace{0cm} \nonumber \\
&=& [ \,V({\mu_1}/\Lambda,
      {\mu_2}/\Lambda) - V(0,0)\,] / \Lambda^4 
     \nonumber \\
&=& 2\pi R\Lambda
    \left[ \frac{({\mu_1}/\Lambda)^2}{\bar{G}_1 \Lambda^3} + 
           \frac{({\mu_2}/\Lambda)^2}{\bar{G}_2 \Lambda^3} 
           \right] \nonumber \\ &&
   -\frac{1}{2\pi^2} \int_0^1 dz\, z^3 
     \ln \Bigg[ 1 + 
                    \frac{\pi R\Lambda}{z} \left\{ 
                    \left( \frac{{\mu_1}}{\Lambda} \right)^2 
                  + \left( \frac{{\mu_2}}{\Lambda} \right)^2 
                    \right\} 
                    \coth (\pi z R\Lambda) 
                    \nonumber \\ && \hspace{4cm}
                  + \frac{\pi^2}{2} (R\Lambda)^2 
                    \left(\frac{{\mu_1}}{\Lambda}\right)^2 
                    \left(\frac{{\mu_2}}{\Lambda}\right)^2  
                    \Bigg]. 
\nonumber 
\end{eqnarray} 
The behavior of the vacuum is determined by solving the gap equations 
($i,j=1,2$;\ $i\ne j$):
\begin{eqnarray}
\lefteqn{\frac{\partial \bar{V}}{\partial (\mu_i/\Lambda)}}
\hspace{0cm} \nonumber \\ 
 &=&   R\Lambda \frac{\mu_i}{\Lambda} \, 
      \Bigg\{ \frac{4\pi}{\bar{G}_i \Lambda^3} 
    - \frac{1}{2\pi}\, 
      \int_0^1 dz 
         \frac{z^3 (\pi R\Lambda \mu_j^2 \tanh(\pi z R\Lambda)+2)}
         {z \left\{ 1+\frac{\pi^2}{2}(R\Lambda)^2 
                    \left( \frac{\mu_1}{\Lambda} \right) 
                    \left( \frac{\mu_2}{\Lambda} \right) \right\} 
          \tanh(\pi z R\Lambda) + \pi R\Lambda 
            \left\{ \left( \frac{\mu_1}{\Lambda} \right) + 
                    \left( \frac{\mu_2}{\Lambda} \right) \right\}}
      \Bigg\} \nonumber \\ 
 &=&   0.
\nonumber 
\end{eqnarray}
The system has a second order phase transition. 
The phase structure of this system is summarized in the Fig.~\ref{phase}.
If $\bar{G}_i \gtrsim \bar{G}_i^{\rm critical}$, we see $\left< \mu_i \right> \gtrsim 0$. 
Therefore we can have three mass scale $\Lambda \sim M_{\rm Pl}$, 
$R_{\rm w}^{-1} \sim$ TeV and $\left< \mu_2 \right> \ll$ TeV in the RS brane world 
if we can realize coupling just above its critical value. 

Now we realize that $\mu_1$ and $\mu_2$ can have nonzero vacuum expectation 
value and Lagrangian (\ref{4LagRS}) has mixing term 
$\bar{\Psi}_R\psi_i + {\rm h.c.}$.
The physics on the $y=\pi R$ brane is examined by 
integrating out the invisible field $\psi_1$. Setting 
$\sqrt{2}\Psi^{(n)}_L \equiv N^{(n)}-M^{(n)}$ and 
$\sqrt{2}\Psi^{(n)}_R \equiv N^{(n)}+M^{(n)}$\ ($n\ne 0$), 
the effective Lagrangian on the $y=\pi R$ brane is obtained as follows: 
\begin{eqnarray}
{\cal L\/}^{(4)}_{\rm eff}
  =  \bar{\Theta}_2 [\, i\partial\!\!\!/ + M_2 + |\mu_1|^2 P \,] \Theta_2,
\label{eq:eff4Lag}
\end{eqnarray}
where 
$\Theta_2^T 
\equiv  \left( \psi_2 \, , \, \Psi^{(0)}_R \, , \, N^{(1)} \, , \,
           M^{(1)} \, , \, N^{(2)} \, , \, M^{(2)} 
           \cdots \right)$ and 
\begin{eqnarray}
  M_2  \equiv  \left(
       \matrix{
       0 & \mu_2^{*} & -\mu_2^{*} & -\mu_2^{*} & \mu_2^{*} & \mu_2^{*} & \cdots \cr
       \mu_2 & 0 & 0 & 0 & 0 & 0 & \cdots \cr
      -\mu_2 & 0 & \frac{1}{R_{\rm w}} & 0 & 0 & 0 & \cdots   \cr
      -\mu_2 & 0 & 0 & - \frac{1}{R_{\rm w}} & 0 & 0 & \cdots   \cr
       \mu_2 & 0 & 0 & 0 & \frac{2}{R_{\rm w}} & 0 & \cdots   \cr
       \mu_2 & 0 & 0 & 0 & 0 & - \frac{2}{R_{\rm w}} & \cdots   \cr
       \vdots & \vdots & \vdots & \vdots & \vdots & \vdots & \ddots  
       }
               \right), \quad 
P  \equiv      \left(
         \matrix{
            0 & 0 & 0 & 
            \cdots \cr
            0 & (i\partial\!\!\!/)^{-1} & (i\partial\!\!\!/)^{-1} & 
            \cdots \cr 
            0 & (i\partial\!\!\!/)^{-1} & (i\partial\!\!\!/)^{-1} & 
             \cdots   \cr          
            \vdots & \vdots & \vdots & \ddots  
         }
        \right).
\label{eq:m2matrix}
\end{eqnarray}  
We have seen above that the dynamical fermion mass generation 
is of the second order phase transition, and we find the parameter 
region of $(\bar{G}_1,\bar{G}_2)$ for 
$0 < \left< \mu_2 \right> \ll \Lambda$, $\left< \mu_1 \right> =0$. 
Thus in Eq. (\ref{eq:eff4Lag}) we have a possibility to get 
a light (Dirac) fermion by diagonalizing Eq.~(\ref{eq:m2matrix}) 
on the $y=\pi R$ brane. We are now treating the case that the four-fermion 
interactions are induced by a common gauge theory. From Fig.~\ref{phase} 
we know that we might obtain such result because $\bar{G}_1 \ll \bar{G}_2$ 
due to the localization effect of the gauge boson KK modes 
(see Fig.~\ref{fig:WF}). 

\section{Discussions: four-fermion approximation}
\noindent
In this chapter we analyzed the effect of heavy KK gauge bosons 
on the dynamical symmetry breaking using local four-fermion 
approximation. The chiral phase transition on the brane with [$U(1)$] 
gauge theory in the bulk is revealed to be of the second order 
and the critical coupling is less than that of the usual 4D theory. 
We obtained the basic structure of dynamical symmetry breaking on a 
brane induced by bulk gauge theory. 

We also analyzed the case that there exists a bulk fermion\footnote{ 
It is an interesting scenario that the radius of the extra dimension 
can be stabilized by the Casimir energy of the bulk fields~\cite{
Goldberger:1999uk}. The Casimir force can be easily calculated by 
using path integral formulation~\cite{Abe:1999ts} even in the SUSY case.} 
in addition to the brane fermions. In such case it is difficult to analyze 
the exact gauge NJL approximation and we adopt further approximations. 
We found that the mixing of the brane fermions with the bulk fermions 
does not lead to the lightest fermion masses of order $1/R_{\rm w}$ 
and the dynamically generated fermion masses are not of order 
$1/R_{\rm w}$ but of order $\left< \mu_2 \right>$. 
If we assume that the coupling or compactification radius is near its 
critical value, $\left< \mu_2 \right>$ can be much smaller than $\Lambda$ 
because the dynamical fermion mass generated under the second-order phase 
transition is small near the critical parameters irrespectively of 
$\Lambda$ and $1/R_{\rm w}$. In this case the possibility of having 
low mass fermions resulted from the dynamical origin that is quite different 
from the ones in other approaches in which low mass fermions are expected to 
show up as a result of the kinematical origins~\cite{Arkani-Hamed:1998vp,
Dienes:1998sb,Mohapatra:1999zd,Das:1999dx,Yoshioka:1999ds}. 
In order to apply this dynamical mechanism to realistic problems, 
however, it seems that we need certain mechanism to realize such 
a critical value of parameter because the four-fermion coupling 
$\bar{G}$, compactification radius $R$ and gauge coupling $g$ is 
related by Eq.~(\ref{eq:GNJL}). 

In the four-fermion approximation we neglect the kinetic energy of 
KK gauge bosons that becomes more important as the compactification 
scale $1/R$ decreases compared with the cut-off scale $\Lambda$. 
Therefore in the next chapter we will construct another way to treat 
directly the bulk gauge boson effect on the dynamical symmetry breaking 
on a brane.

\chapter{Analysis of improved ladder Schwinger-Dyson equation}
\label{chap:ailsde}

As was shown in the previous chapter, 
in higher-dimensional gauge theories with compact extra dimensions, 
the effect of gauge boson KK mode is well approximated 
by four-fermion interactions between charged matter fermions 
if the scale of the compactification is less than the cut-off 
scale of the theory. When we treat the case that the compactification 
scale is sufficiently small compared with the cut-off scale 
we can not neglect the effect from the gauge boson kinetic term. 

\begin{figure}[t]
\centerline{\epsfig{figure=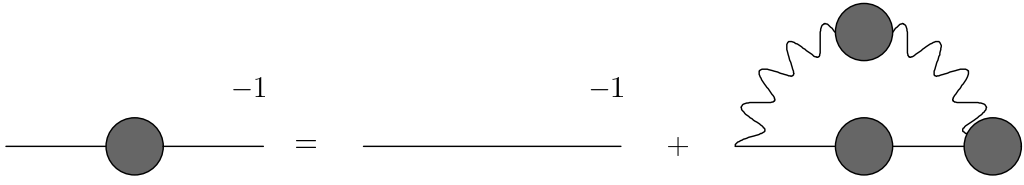,width=0.8\linewidth}}
\caption{Schwinger-Dyson equation for a fermion propagator 
in gauge theory. The filled circles mean that it includes all 
quantum corrections.}
\label{fig:sdeq}
\end{figure}

\begin{figure}[t]
\centerline{\epsfig{figure=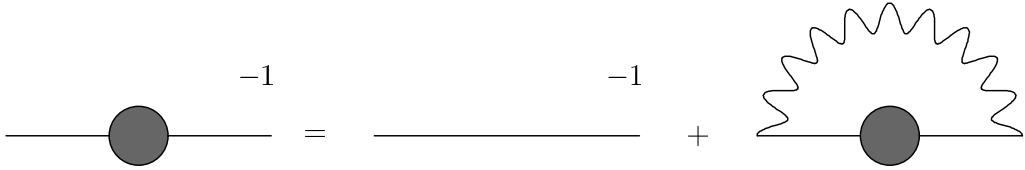,width=0.8\linewidth}}
\caption{Ladder Schwinger-Dyson equation for a fermion propagator 
in gauge theory. The filled circles mean that it includes all 
quantum corrections.}
\label{fig:ldsdeq}
\end{figure}

In this chapter\footnote{This chapter is based on Refs.~\cite{Abe:2002yb,
Abe:2001yi} with K.~Fukazawa and T.~Inagaki} 
we utilize Schwinger-Dyson (SD)~\cite{Dyson:1949ha} 
equation for analyzing such case. It is equivalent to a self-consistent 
equation shown in Fig.~\ref{fig:sdeq} 
that is derived from an extremum condition of the effective action 
against the variation by a two-point function (propagator). 
Actually we can not solve the exact SD equation, and in most case 
we solve it by adopting ladder approximation shown in 
Fig~\ref{fig:ldsdeq} that is obtained by replacing a propagator 
and vertex function with a tree one except for the target propagator. 
In the following we apply SD equation to the analyses of 
a dynamical fermion mass on a boundary in higher-dimensional 
space-time that is induced by a bulk gauge theory. 

\section{Schwinger-Dyson equation of boundary propagator}
\label{sec:effsd}

From the Lagrangian (\ref{eq:efflag}) we construct a SD equation 
for the fermion propagator, 
\begin{eqnarray}
iS^{-1}(p) 
= iS^{-1}_0(p) + \sum_{n=0}^{N_{\rm KK}} \int \frac{d^4 q}{i(2\pi)^4}
  \left[ -ig_n T^a \Gamma^M \right] S(q) 
  \left[ -ig_n T^a \Gamma^N \right] D_{MN}^{(n)} (p-q), 
\label{eq:orgSDeq}
\end{eqnarray}
where $T^a$ is the gauge group representation matrix of the fermion, and 
$S_0(p) = i/p\!\!\!/$, $S(p)$ and $D_{MN}^{(n)} (p-q)$ are the free fermion 
propagator, full fermion propagator and full $n$-th excited KK gauge boson 
propagator respectively. In this thesis we use the ladder approximation 
in which $D_{MN}^{(n)}$ is replaced by the free propagator, 
\begin{eqnarray}
D_{MN}^{(n)}(k) &\equiv& \frac{-i}{k^2-M_n^2}
\left[ \eta_{MN} -(1-\xi)\frac{k_Mk_N}{k^2-\xi M_n^2} \right]. 
\nonumber 
\end{eqnarray}

By writing the full fermion propagator as 
\begin{eqnarray}
iS^{-1} (p) &\equiv& A(-p^2) p\!\!\!/ -B(-p^2), 
\nonumber 
\end{eqnarray}
the SD equation (\ref{eq:orgSDeq}) becomes the 
simultaneous integral equation of $A$ and $B$,
\begin{eqnarray}
A(p^2) &=& 1+ \int_0^{\Lambda^2} dq^2\, 
            \frac{q^2A(q^2)}{A^2(q^2)q^2 +B^2(q^2)} 
            \sum_{n=0}^{N_{\rm KK}} L_\xi (p^2,q^2;M_n,\alpha_n),  
\label{eq:SDA} \\
B(p^2) &=& \int_0^{\Lambda^2} dq^2\, 
         \frac{q^2 B(q^2)}{A^2(q^2)q^2 +B^2(q^2)} 
         \sum_{n=0}^{N_{\rm KK}} K_\xi (p^2,q^2;M_n,\alpha_n), 
\label{eq:SDB}
\end{eqnarray}
where~\cite{Bando:qd} 
\begin{eqnarray}
L_\xi (p^2,q^2;M_n,\alpha_n)
&=& \frac{\alpha_n}{4\pi} 
    \Bigg[ q^2f_{M_n^2}^2 (p^2,q^2) 
         + \frac{2q^2}{M_n^2} \left\{ f_{M_n^2} (p^2,q^2)-
                             f_{\xi M_n^2} (p^2,q^2) \right\} 
           \nonumber \\  && \hspace{1cm} 
         - \frac{(q^2)^2+p^2q^2}{2M_n^2} \left\{ f_{M_n^2}^2 (p^2,q^2)-
                             f_{\xi M_n^2}^2 (p^2,q^2) \right\} 
    \Bigg], \label{eq:kerL} \\
K_\xi (p^2,q^2;M_n,\alpha_n)
&=& \frac{\alpha_n}{4\pi} 
    \bigg[ 4f_{M_n^2} (p^2,q^2) 
         + \frac{p^2+q^2}{M_n^2} \left\{ f_{M_n^2} (p^2,q^2)-
                             f_{\xi M_n^2} (p^2,q^2) \right\} 
           \nonumber \\  && \hspace{1cm} 
         - \frac{p^2q^2}{M_n^2} \left\{ f_{M_n^2}^2 (p^2,q^2)-
                             f_{\xi M_n^2}^2 (p^2,q^2) \right\} 
    \Bigg], \label{eq:kerK}
\end{eqnarray}
\begin{eqnarray}
f_M (p^2,q^2) 
&=& \frac{2}{p^2+q^2+M+\sqrt{(p^2+q^2+M)^2 -4p^2q^2}}, 
\nonumber 
\end{eqnarray}
\begin{eqnarray}
\alpha_n &\equiv& \alpha \chi_n^2(y^\ast)
\qquad (n \ne 0), \nonumber \\ 
\alpha_0 &=& \alpha = g^2/4\pi. 
\nonumber 
\end{eqnarray}
Since our analysis is based on the effective theory on the 
four-dimensional brane below the {\it reduced} cutoff,
the explicit UV cutoff $\Lambda$ in terms of the 
loop momentum {\it on the brane} appears. 
$A(x)$ and $B(x)$ are the wave-function normalization 
factor and the mass function of the fermion respectively. 
The mass function $B(x)$ is the oder parameter 
of chiral symmetry breaking on the brane. 
The chiral symmetry is broken down for $B(x) \ne 0$. 
By solving Eqs.~(\ref{eq:SDA}) and (\ref{eq:SDB}) 
we obtain the behavior of the dynamical fermion 
mass on the brane.

\section{Improved ladder analysis of dynamical mass on the boundary}
\label{sec:ila}

In this section we numerically solve the SD equation 
and analyze dynamical brane fermion mass induced by 
the bulk gauge theory~\cite{Abe:2002yb,Abe:2001yi}. 

\begin{figure}[t]
\centerline{\epsfig{figure=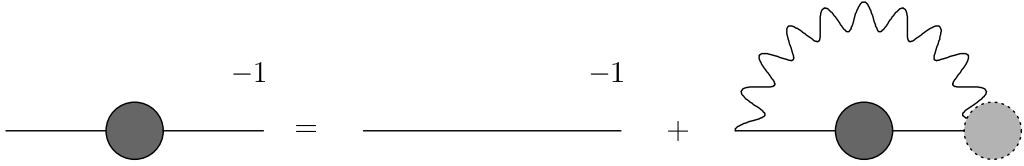,width=0.8\linewidth}}
\caption{Improved ladder Schwinger-Dyson equation for a 
fermion propagator in gauge theory. 
The darkly filled circles mean that it includes all 
quantum corrections. The lightly filled circle means that 
we put (perturbative) running coupling there. }
\label{fig:imldsdeq}
\end{figure}

First we analyze the case of the bulk Abelian gauge theory (QED) 
with $N_{\rm KK} \lesssim 10$~\cite{Abe:2001yi}, 
and the results are shown in Sec.~\ref{sec:bqed}. 
In such case we can use the point vertex approximation 
depicted in Fig~\ref{fig:ldsdeq}. 
Then we proceed to the Yang-Mills case~\cite{Abe:2002yb}. 
To solve the SD equation of Yang-Mills theory we use the 
improved ladder approximation shown in Fig.~\ref{fig:imldsdeq} 
where the running coupling is imposed to the vertex function. 
The running coupling is derived from the truncated KK approach 
proposed in Ref.~\cite{Dienes:1998vh}. 
We use the asymptotic form (\ref{eq:KKmassasym}) of the KK 
spectrum to evaluate the running coupling in RS brane world. 

In Sec.~\ref{sec:qcd} we review the process of analyzing SD equation in 
usual 4D QCD~\cite{Aoki:1990eq}, and then we extend it to the brane world bulk 
Yang-Mills theory in Sec.~\ref{sec:bqcdf} and \ref{sec:bqcdw}. 
For convenience we choose QCD ($SU(3)_c$ Yang-Mills theory with 
$N_{\rm f}=3$ flavor fermion on the brane) as 
a concrete example of the bulk Yang-Mills theory, 
because the bulk results can be compared with 
the usual 4D QCD. We expect that the essential point of
the results is not changed in the other bulk Yang-Mills 
theory such as technicolor. 

To solve Eqs.~(\ref{eq:SDA}) and (\ref{eq:SDB}) numerically, we
employ the iteration method. Starting from suitable trial functions for
$A(p^2)$ and $B(p^2)$, we numerically evaluate the right-hand sides of
Eqs.~(\ref{eq:SDA}) and (\ref{eq:SDB}). We next use the resulting
functions, $A(p^2)$ and $B(p^2)$, to calculate the right-hand sides of
Eqs.~(\ref{eq:SDA}) and (\ref{eq:SDB}), and we iterate the calculational 
procedure until stable solutions were obtained. At each iteration, 
the integration is performed using the Monte Carlo method, 
and it is cut off at the mass scale $\Lambda$.

\subsection{QED in bulk} \label{sec:bqed}
\begin{figure}[h]
\begin{center}
\begin{minipage}{0.48\linewidth}
\epsfig{figure=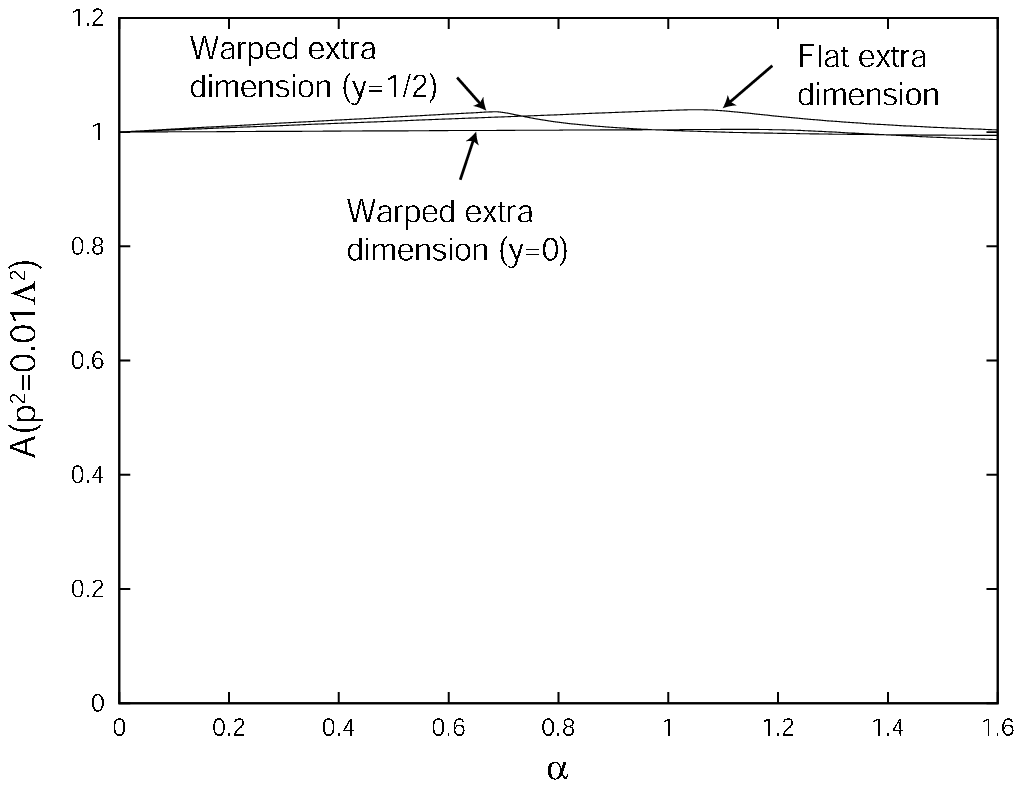,width=\linewidth}
\caption{Behavior of the wave-function $A(p^2=0.01\Lambda^2)$ as a function 
of the coupling constant $\alpha$ for the cases of a flat extra dimension 
($R\sim 1/\Lambda$) with $\xi\sim 0.21$ and a 
warped extra dimension ($2kR\sim 1$) with $\xi\sim 0.23(y=\pi R)$ 
and $\xi\sim 0.04(y=0)$. (Here $y$ is denoted as a 
dimensionless coordinate $y/(2\pi R) \to y$.)}
\label{Aa}
\end{minipage}
\hfill
\begin{minipage}{0.48\linewidth}
\epsfig{figure=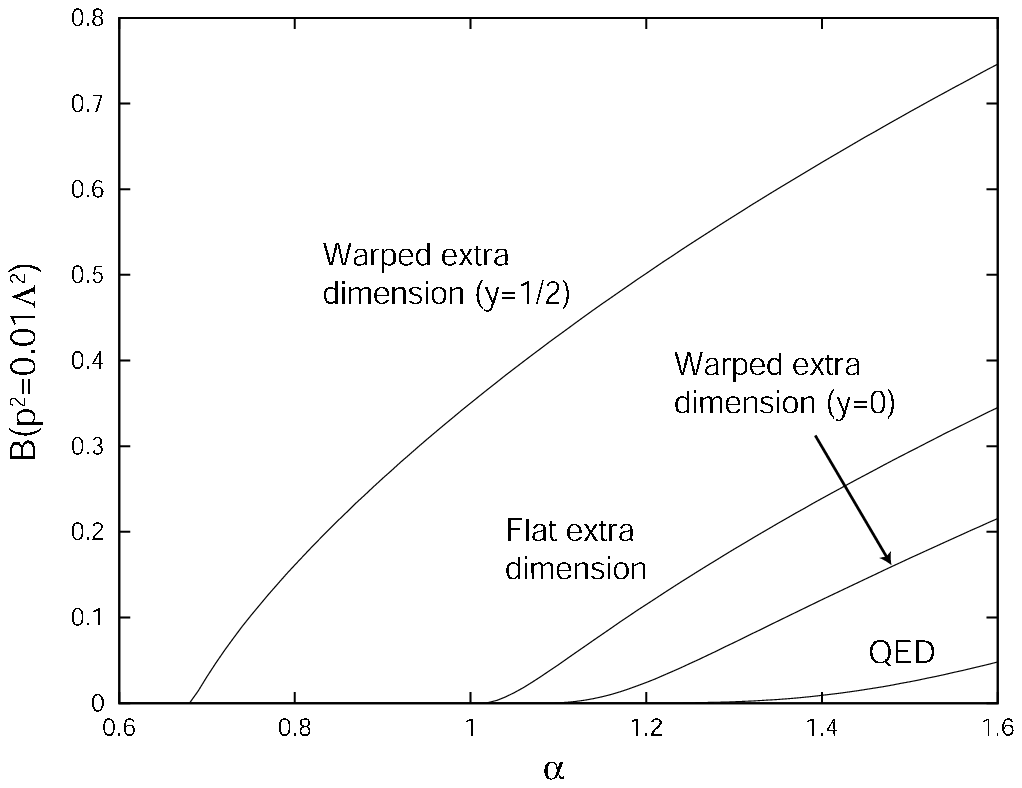,width=\linewidth}
\caption{Behavior of the mass function $B(p^2=0.01\Lambda^2)$ as a function 
of $\alpha$ for the cases of QED, a flat extra dimension ($R\sim 1/\Lambda$) 
and a warped extra dimension ($2kR\sim 1$) in the ladder SD equation. 
(Here $y$ is denoted as a 
dimensionless coordinate $y/(2\pi R) \to y$.)}
\label{Ba}
\end{minipage}
\end{center}
\end{figure}

\begin{figure}[h]
\begin{center}
\begin{minipage}{0.48\linewidth}
\epsfig{figure=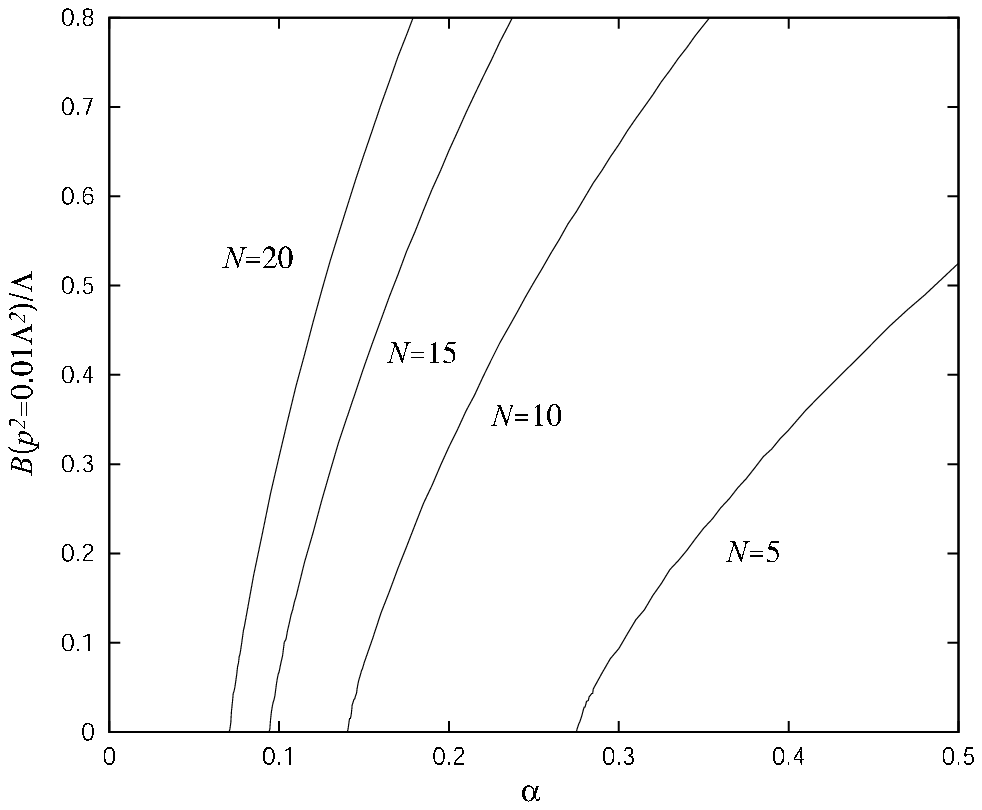,width=\linewidth}
\caption{Behavior of the mass function $B(p^2=0.01\Lambda^2)$ as a function 
of $\alpha$ for a flat extra dimension with 
$R=5/\Lambda (N_{\rm KK}=5)$, $R=10/\Lambda (N_{\rm KK}=10)$, 
$R=15/\Lambda (N_{\rm KK}=15)$ and $R=20/\Lambda (N_{\rm KK}=20)$ 
in the ladder SD equation.}
\label{ADDNkk}
\end{minipage}
\hfill
\begin{minipage}{0.5\linewidth}
\epsfig{figure=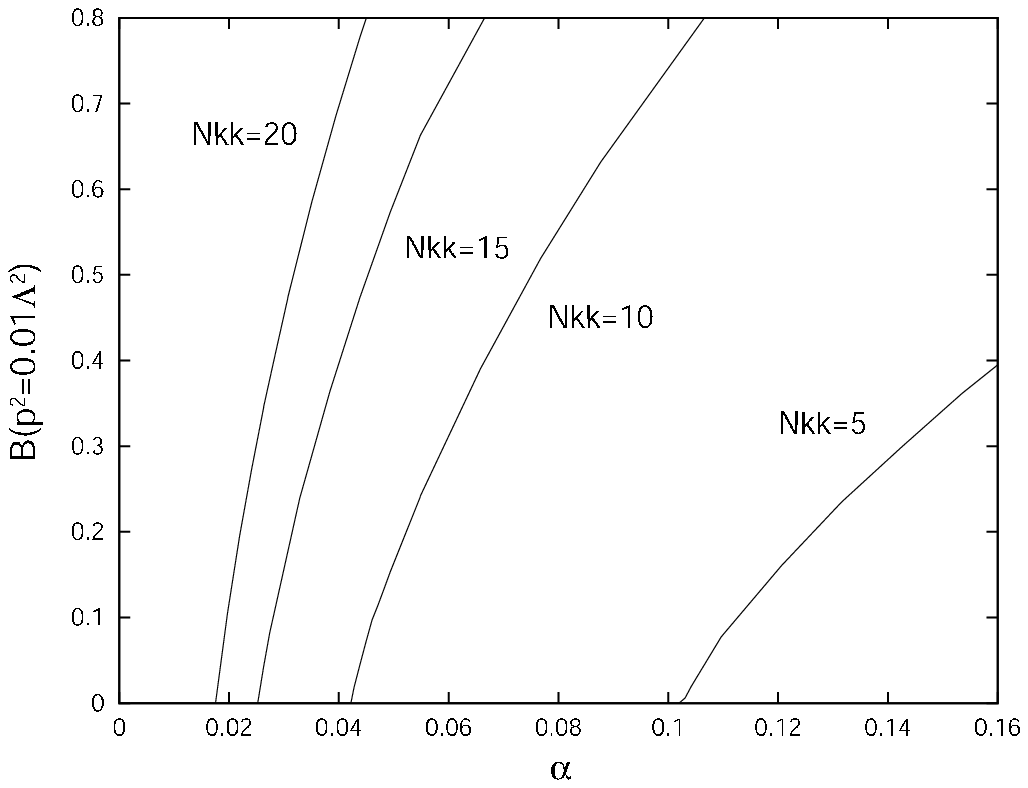,width=\linewidth}
\caption{Behavior of the mass function $B(p^2=0.01\Lambda^2)$ as a function 
of $\alpha$ for a warped extra dimension $(y=\pi R)$ 
with $1kR=1.80 (N_{\rm KK}=5)$, $2kR=2.22 (N_{\rm KK}=10)$, 
$2kR=2.47 (N_{\rm KK}=15)$ and $2kR=2.65 (N_{\rm KK}=20)$ 
in the ladder SD equation.}
\label{RS05Nkk}
\end{minipage}
\end{center}
\end{figure} 

First, we consider the flat extra dimension with $R\sim 1/\Lambda$ and the 
warped extra dimension with $2kR \sim 1$. 
In this case only the lightest KK mass is less than the cutoff scale, 
i.e., $N_{\rm KK}=1$.

If the gauge fields do not propagate in the extra dimension, i.e. if no KK
modes exist, the Landau gauge, $\xi=0$, is consistent with the
Ward-Takahashi identity for the ladder SD equation. In the bulk gauge
theories, massive KK modes appear in the brane, and the Ward-Takahashi
identity is not guaranteed within the ladder approximation in the Landau
gauge. We attempted to solve the ladder SD equation using various values of 
the gauge parameter $\xi$. As illustrated in Fig.~\ref{Aa}, it is possible to
choose the gauge parameter $\xi=\xi_{WT}$ for which the fermion wave-function
$A(p^2)$ is almost unity, and the Ward-Takahashi identity nearly holds
on the brane~\cite{Abe:2001yi}. 
In the models considered here, $\xi_{WT}$ is given by
$\xi_{WT} \sim 0.21$ (flat,$R\sim1/\Lambda$), 
$\xi_{WT} \sim 0.31$ (warped,$2kR\sim 1;\ y=\pi R$), and 
$\xi_{WT} \sim 0.11$ (warped,$2kR\sim 1;\ y=0$). 
In the following we consider the gauge parameter $\xi=\xi_{WT}$. 

To study the structure of the chiral phase transition, we observe the
behavior of the mass function $B(p^2)$ at some fixed value of $p$. 
The $\alpha$ dependence of the mass function $B(p^2)$ with 
$p^2=0.01\Lambda^2$ is plotted in Fig.~\ref{Ba}. 
As seen there, the chiral phase transition is of the second order 
(or continuous), since a fermion mass is generated at a critical 
value of the coupling constant $\alpha$ without any discontinuity. 
In the $y=0$ brane in the warped extra dimension, 
the lightest KK mode increases the mass function $B(p^2)$, but the result 
yields no large difference from that for QED. 
On the $y=\pi R$ brane in the warped extra dimension and  
on the brane in the flat extra dimension, 
the results differ significantly from QED. 
In these two cases, the critical coupling $\alpha_{\rm critical}$ is less 
than that in QED. 
This is due to the large effective coupling $\alpha_1 \sim 2.0\alpha$ for 
the lightest KK mode. Comparing Fig.~\ref{Ba0} with Fig.~\ref{Ba}, 
we observe that the result for the ladder SD equation is qualitatively 
consistent with that obtained with the local four-fermion approximation in the 
previous section but the latter case overestimates 
the effect of the KK excited modes on the chiral phase transition.
We see that the propagation of the KK mode weakens the effect on the 
phase transition~\cite{Abe:2001yi}.

Next, we study how the number of KK modes, $N_{\rm KK}$, affects the chiral phase 
transition. (We recall that $N_{\rm KK}$ is the number of KK modes with mass less 
than the cutoff scale $\Lambda$.) Here we consider the SD equation in the flat 
extra dimension with $R\sim 5/\Lambda, 10/\Lambda, 15/\Lambda, 20/\Lambda$. 
Obviously, $N_{\rm KK}$ for these radii is given by $5, 10, 15$ and $20$, respectively. 
In the warped extra dimension, the KK mode mass depends on the RS 
compactification scale, $L=2\pi R$. 
We also evaluate the SD equation in the warped extra dimension, 
where $N_{\rm KK}$ is equal to $5, 10, 15$ and $20$ for the corresponding radii. 
By solving KK eigenvalue equation it is found that these values of 
$N_{\rm KK}$ are realized for $2kR\sim 1.80, 2.22, 2.47$ and $2.65$, 
respectively. Here we assume that the heaviest KK mode mass is near the 
cutoff scale. 

In models with $N_{\rm KK}=5, 10, 15$ and $20$, the gauge parameter for which
the fermion wave-function is almost unity is given by 
$\xi_{WT}\sim$ $0.285$ (flat), $0.288$ (warped, $y=\pi R$) for $N_{\rm KK}=5$, 
$\xi_{WT}\sim$ $0.275$ for $N_{\rm KK}=10$, 
$\xi_{WT}\sim$ $0.274$ for $N_{\rm KK}=15$ and 
$\xi_{WT}\sim$ $0.272$ for $N_{\rm KK}=20$. 
using the gauge parameter $\xi=\xi_{WT}$, we solved Eqs.~(\ref{eq:SDA}) 
and (\ref{eq:SDB}) numerically.
In Figs.~\ref{ADDNkk} and~\ref{RS05Nkk} we illustrate the behavior of the 
mass function $B(p^2)$ for the flat extra dimension 
and for the $y=\pi R$ brane in the warped extra dimension. 
Assuming the critical exponent is 0.5, 
the critical coupling $\alpha_{\rm critical}$ with $N_{\rm KK}=5$, $10$, $15$ 
and $20$ is evaluated, respectively, as~\cite{Abe:2001yi}
\begin{eqnarray}
\alpha_{\rm critical} &=& 0.275,\ 0.140,\ 0.094 \textrm{ and } 0.071 
\textrm{ (flat)}, \nonumber \\
\alpha_{\rm critical} &=& 0.187,\ 0.078,\ 0.048 \textrm{ and } 0.034 
\textrm{ (warped; $y=\pi R$ )}.
\nonumber 
\end{eqnarray}
In both brane models, chiral symmetry breaking
is enhanced by KK mode summation, 
and the critical coupling approaches the ordinary electromagnetic 
coupling constant $\sim 1/137$ as $N_{\rm KK}$ increases\footnote{ 
In particular on the $y=\pi R$ brane in the warped extra dimension, 
it is expected that chiral symmetry breaking is strongly enhanced by 
the mode function for larger $kR$. 
Therefore, it is possible that chiral symmetry is 
broken for the ordinary electromagnetic coupling constant 
for larger $N_{\rm KK}$ on $y=\pi R$ brane.}.

\subsection{QCD in 4D}
\label{sec:qcd}
Next we review the analysis of QCD in four-dimension. 
The main difference between the case of QCD and QED is the existence 
of the gauge boson self couplings. We incorporate it by introducing 
running coupling on the vertex in the improved SD equation. 
Taking $N_{\rm KK}=0$ in Eqs.~(\ref{eq:SDA}) and (\ref{eq:SDB}),
the SD equation for usual QCD in four dimensional space-time 
(i.e. QCD on the brane) is obtained.
In this case the integration kernels (\ref{eq:kerL}) 
and (\ref{eq:kerK}) become 
\begin{eqnarray}
L_\xi (p^2,q^2;M_0=0,\alpha_0=\alpha)
 &=& \frac{\alpha}{4\pi} \,\xi\, q^2\, f_0^2 (p^2,q^2), 
     \label{eq:QCDkernelL} \\
K_\xi (p^2,q^2;M_0=0,\alpha_0=\alpha)
 &=& \frac{\alpha}{4\pi} \,(3 +\xi)\, f_0(p^2,q^2), 
     \label{eq:QCDkernelK}
\end{eqnarray}
where 
\begin{eqnarray}
f_0 (p^2,q^2) 
 &=& \theta (p^2-q^2) \frac{1}{p^2} + 
     \theta (q^2-p^2) \frac{1}{q^2}.
\nonumber 
\end{eqnarray}
We usually use Landau gauge $\xi=0$ that results in 
$L_\xi (p^2,q^2;M_0=0,\alpha_0=\alpha)=0$, i.e. 
$A(p^2)=1$ in the SD equation. In that case the SD equation 
becomes single equation only with $B(p^2)$. 
$A(p^2)=1$ satisfies the QED like Ward-Takahashi identity.

We put the running coupling on the vertex function in the SD equation 
to compensate for neglecting three and four point gluon self 
interactions ${\cal L}_{\rm SI}$ in the ladder approximation. 
That is so called improved ladder approximation. 
We use the one-loop perturbative running coupling,
\begin{eqnarray}
\alpha \equiv \frac{\pi}{3C_2(F)} \lambda (k^2), 
\nonumber 
\end{eqnarray}
where
\begin{eqnarray}
\lambda (z) &=& \frac{\lambda_0}{1+\lambda_0B \ln \frac{z}{\mu^2}}, 
\nonumber 
\end{eqnarray}
and $B = \frac{4\pi^2}{3C_2(F)}\beta_0$. We define $\beta_0$ by 
$\beta (g) = -\beta_0 g^3 + {\cal O}(g^5)$ and 
$\beta_0 = \frac{1}{16\pi^2} \big[ \frac{11}{3}C_2(G)-\frac{4}{3}N_{\rm f} T(F) \big]$ 
where $G$ and $F$ stand for the gauge group and the fermion representation 
under the group respectively. $C_2(G)$, $C_2(F)$ and $T(F)$ are 
defined by 
$\sum_{c,d}f_{acd}f_{bcd} \equiv C_2(G) \delta_{ab}$, 
$\textrm{tr} (T^a T^b) \equiv T(F) \delta_{ab}$ and 
$\sum_a T^a T^a \equiv C_2(F)1$. 
In the case that $G=SU(N=3)$ and $F$ is fundamental representation, 
these values are given by
\begin{eqnarray}
&&C_2(G)=N=3,\quad T(F)=\frac{1}{2}, \quad C_2(F)=\frac{N^2-1}{2N}=4/3, \nonumber \\ 
&&B=\frac{1}{12C_2(F)}\left( \frac{11N-2N_{\rm f}}{3} \right)
 =\frac{9}{16}, \quad (N=3,N_{\rm f}=3). 
\nonumber 
\end{eqnarray}

We also define $\Lambda_{\rm QCD}$, the dynamical scale of QCD, as 
\begin{eqnarray}
\frac{1}{\lambda (z)} - B \ln z = 
  \frac{1}{\lambda_0} - B \ln \mu^2 
  \ \equiv \ -B \ln \Lambda_{\rm QCD}^2, 
     \qquad \left( \Lambda_{\rm QCD}^2 \equiv 
     \mu^2/e^{\frac{1}{\lambda_0 B}} \right),
\nonumber 
\end{eqnarray}
and obtain 
\begin{eqnarray}
\lambda (z) &=& \frac{16}{9}\frac{1}{\ln (z/\Lambda_{\rm QCD}^2)}. 
\label{eq:runlambda}
\end{eqnarray}
Following the analysis in Ref.~\cite{Miransky:vj} we introduce 
IR cutoff $z_{\rm IF}$ in the running coupling as 
\begin{eqnarray}
\alpha (z) &=& 
  \frac{\pi}{3C_2(F)} \left[
  \theta (z_{\rm IF}-z) \lambda (z_{\rm IF}) +
  \theta (z-z_{\rm IF}) \lambda (z)
  \right], 
\nonumber 
\end{eqnarray}
and we take the following approximation
\begin{eqnarray}
\alpha &\to& \alpha (z) \simeq \alpha (x+y),
\label{eq:HMapprox} 
\end{eqnarray}
in the integration kernel (\ref{eq:QCDkernelL}) and 
(\ref{eq:QCDkernelK}). It is called Higashijima-Miransky 
approximation. This is due to the consideration that 
the mean value of the $\theta$-dependent part in $\alpha (z)$ 
where $z=x+y-\sqrt{xy}\cos \theta$ has only a negligible effect
after the angle integration in the SD 
equation\footnote{In the usual QCD analysis we introduce 
further approximation $\alpha (x+y) \simeq \alpha (\max [x,y])$, 
but we don't use it here.}. 
It is known that the approximation (\ref{eq:HMapprox}) 
violates the chiral Ward-Takahashi identity\footnote{As suggested 
in the second paper in Ref.~[9], we can keep the chiral Ward-Takahashi 
identity to use a nonlocal gauge.}. 
In this thesis we choose the gauge parameter $\xi$ to nearly 
hold the chiral Ward-Takahashi identity. 
In the following numerical analyses we use the experimental value 
\begin{eqnarray}
\Lambda_{\rm QCD} \simeq 200 \textrm{ MeV}, 
\nonumber 
\end{eqnarray}
and choose $\ln (z_{\rm IF}/\Lambda_{\rm QCD}^2)=1$ 
for the IR cutoff\footnote{To obtain the realistic pion decay 
constant in QCD in the ladder approximation, we need more 
careful treatment for the value of 
$\Lambda_{\rm QCD}$ and $z_{\rm IF}$~\cite{Aoki:1990eq}. 
In this thesis we don't touch the detailed value of the results 
but the order of them.}. 

The composite Nambu-Goldstone field corresponds to the 
composite Higgs field in the dynamical electroweak symmetry 
breaking scenario. Its decay constant is obtained by the 
Pagels-Stokar approximation~\cite{Pagels:1979hd},
\begin{eqnarray}
f_\pi^2 
= \frac{N}{4\pi^2} \int_0^{\Lambda^2} xdx
  \frac{B(x)\left( B(x)-xB'(x)/2 \right)}{\left(A^2(x)x+B^2(x)\right)^2}. 
\label{eq:PSformula}
\end{eqnarray}

\begin{figure}[htbp]
\begin{center}
\begin{minipage}{0.48\linewidth}
   \centerline{\epsfig{figure=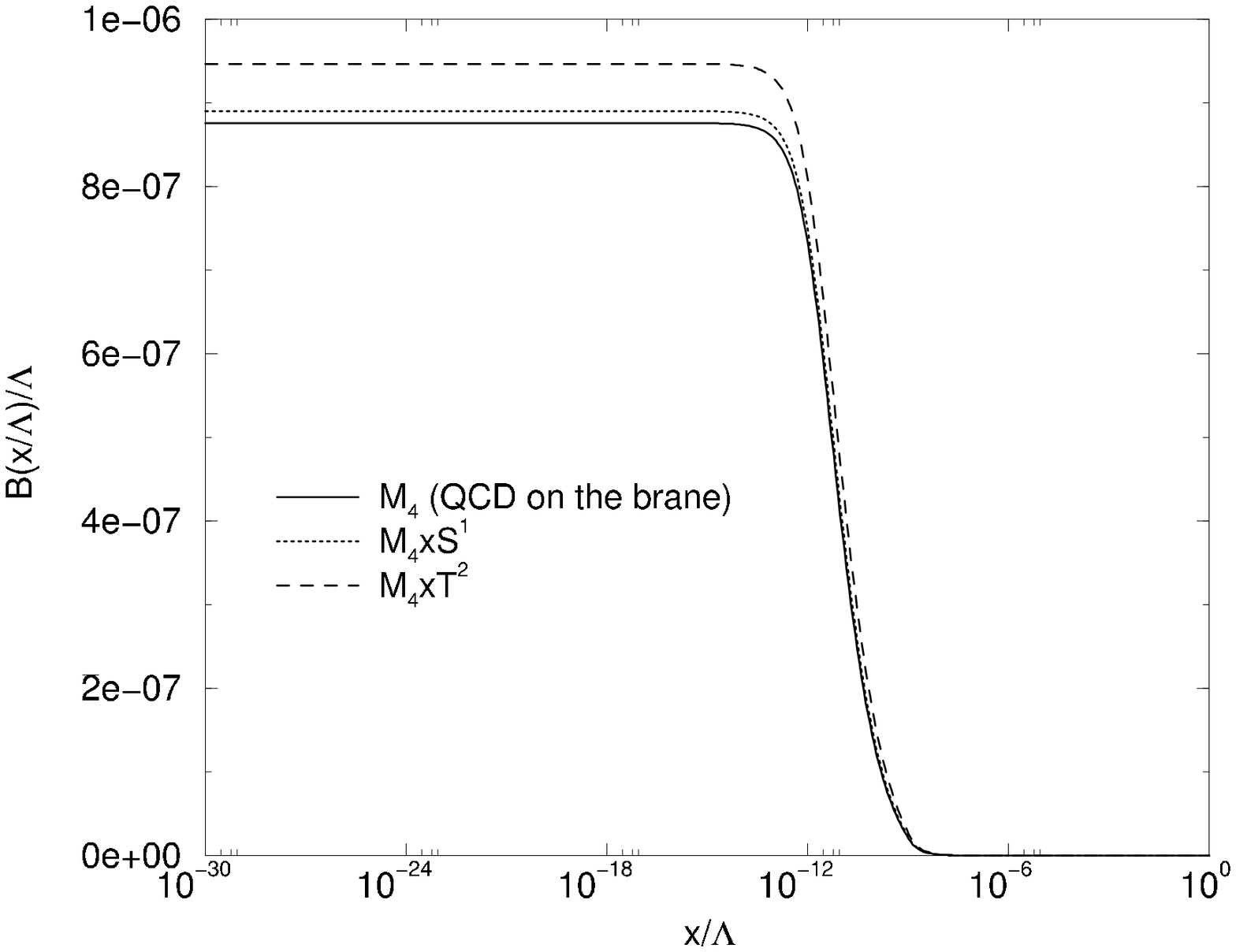,width=\linewidth}}
\end{minipage}
\hfill
\begin{minipage}{0.48\linewidth}
   \centerline{\epsfig{figure=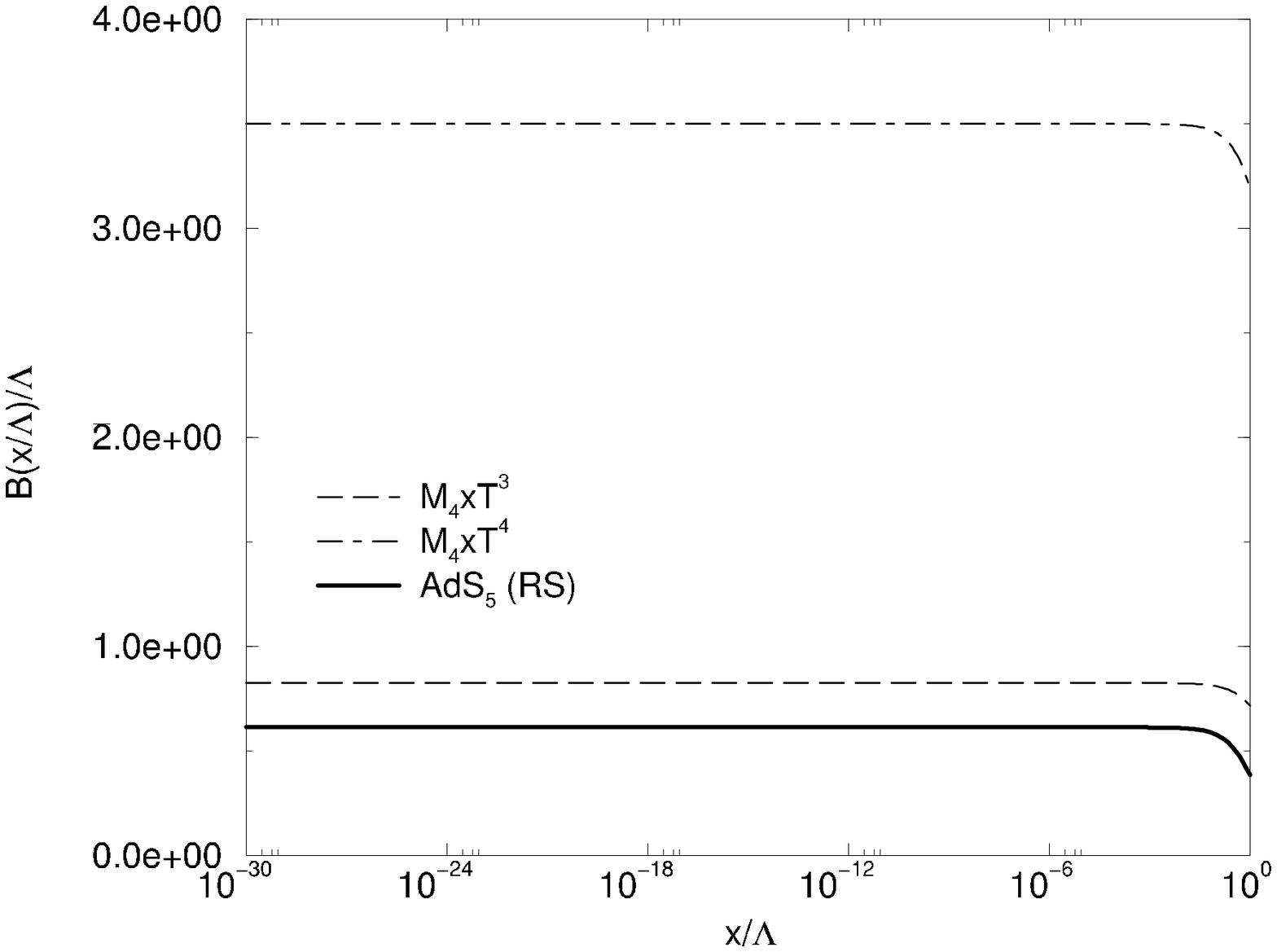,width=\linewidth}}
\end{minipage}
\end{center}
\caption{The behavior of $B(x)$ with various bulk space-time 
derived from the improved ladder SD equation. 
The case of (4D usual) QCD, bulk QCD in the flat brane world 
with $(4+\delta)$-dimensional bulk space-time ($\delta=1,2,3,4$) 
and in the RS brane world are shown. 
We set $\Lambda_{\rm QCD}=200$ MeV for all cases. 
The cutoff is taken as $\Lambda=\Lambda_{\rm (4+\delta)D}=10$ TeV 
for QCD $(\delta=0)$ and flat case $(\delta \ne 0)$, 
and for the flat brane world we take universal radii 
$R\Lambda_{\rm (4+\delta)D}=1$. 
For the RS brane world we take $k=\Lambda_{\rm 5D}=M_{\rm Pl}=10^{16}$ TeV
and $\Lambda=\pi e^{-\pi kR}\Lambda_{\rm 5D}=10$ TeV $(kR=11.35)$.}
\label{fig:SDB}
\begin{center}
\begin{minipage}{0.48\linewidth}
   \centerline{\epsfig{figure=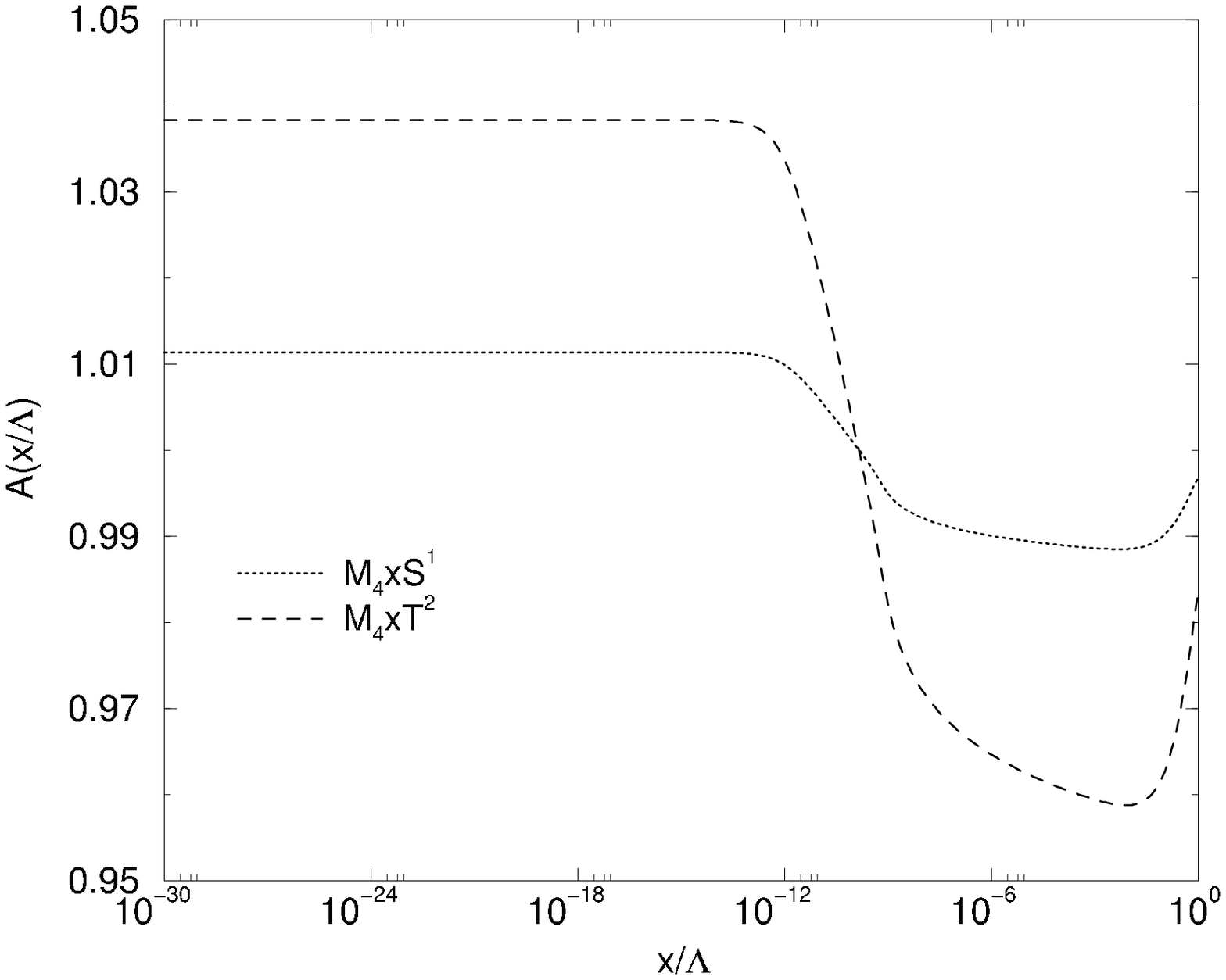,width=\linewidth}}
\end{minipage}
\hfill
\begin{minipage}{0.48\linewidth}
   \centerline{\epsfig{figure=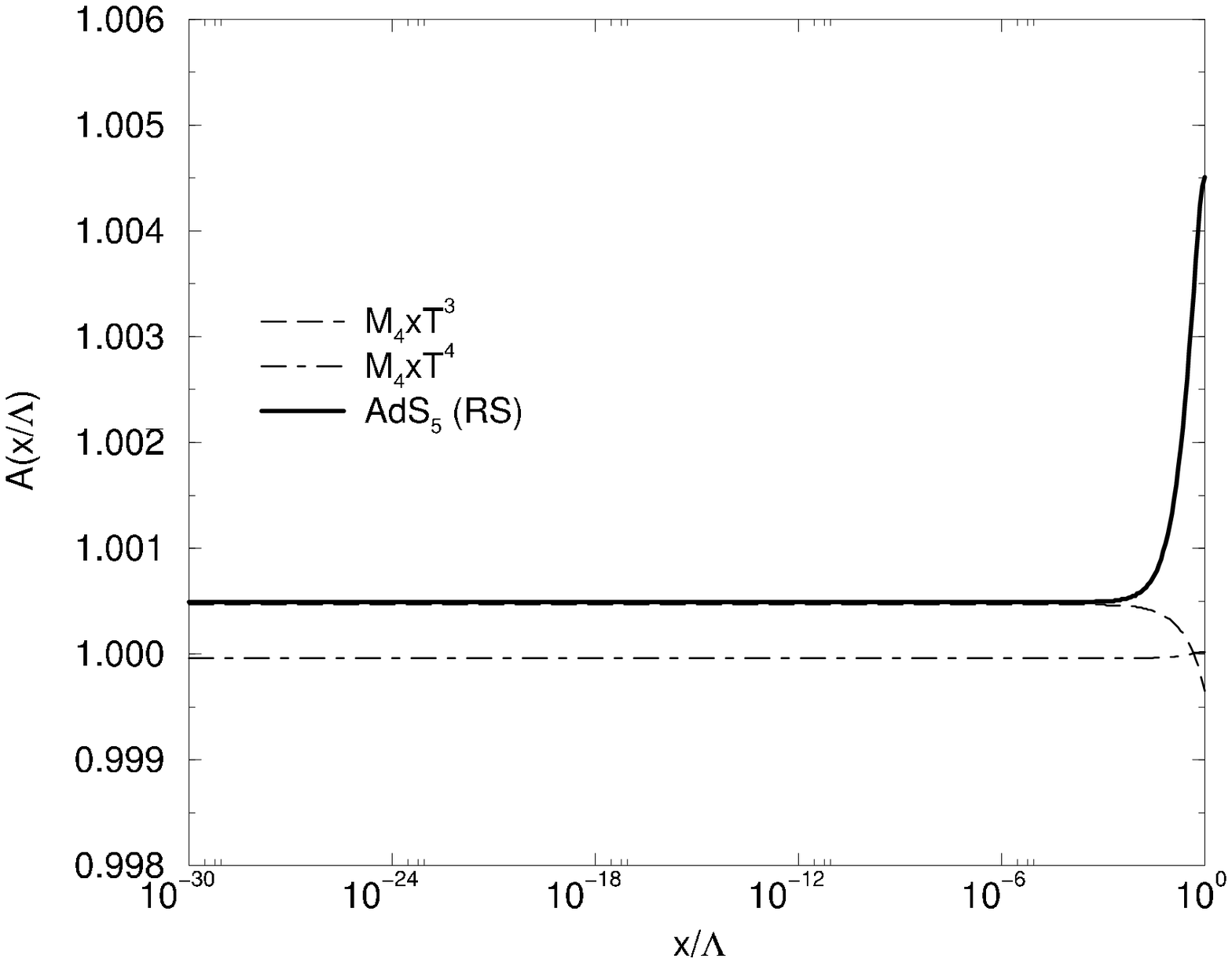,width=\linewidth}}
\end{minipage}
\end{center}
\caption{The behavior of $A(x)$ with various bulk space-time 
derived from the improved ladder SD equation. 
The case of (4D usual) QCD, bulk QCD in the flat brane world 
with $4+\delta$ dimensional bulk space-time ($\delta=1,2,3,4$) 
and in the RS brane world are shown. 
The mass parameters are taken as the same in the caption in 
Fig.~\ref{fig:SDB}. 
In order to satisfy Ward-Takahashi identity, i.e. $A(x) \simeq 1$ 
within $\pm 5$\%, we take the gauge fixing parameter $\xi=0$ for QCD, 
$\xi=0.02$, $0.07$, $0.57$ and $0.63$ for flat brane world with 
$\delta=1$, $2$, $3$ and $4$ respectively and $\xi=0.50$ for RS case.}
\label{fig:SDA}
\end{figure}

\begin{figure}[t]
\begin{center}
\begin{minipage}{0.48\linewidth}
   \centerline{\epsfig{figure=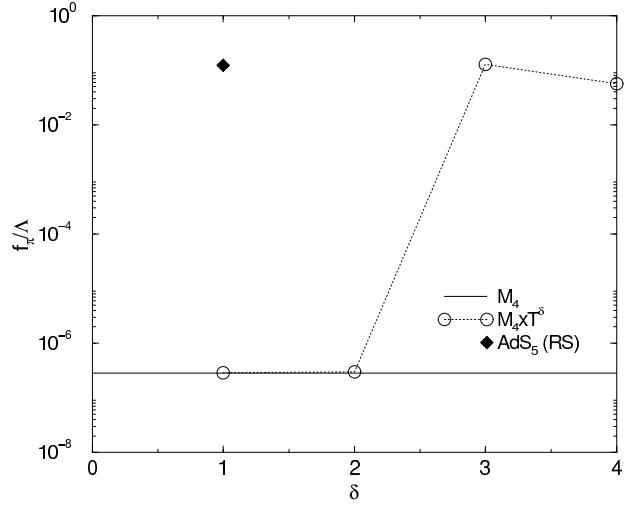,width=\linewidth}}
\end{minipage}
\end{center}
\caption{The behavior of the decay constant $f_\pi$ of the composite 
scalar field with various bulk space-time derived from the Pagels-Stokar formula. 
The case of (4D usual) QCD, bulk QCD in the flat brane world 
with $(4+\delta)$-dimensional bulk space-time ($\delta=1,2,3,4$) 
and in the RS brane world is shown together. 
The parameters are taken as the same in the caption in Fig.~\ref{fig:SDB}.}
\label{fig:fpi}
\end{figure}

Here we show the numerical results of the improved ladder SD 
equation in 4D QCD. We set the cutoff scale $\Lambda=10$ TeV 
in order to compare with the result of the brane world models. 
By using the replacement (\ref{eq:HMapprox}) 
we solve the SD equation (\ref{eq:SDA}), (\ref{eq:SDB}) 
and calculate the mass function 
$B(x)$. The behavior of the mass function $B(x)$ is shown 
as the thin solid line in Fig.~\ref{fig:SDB}. 
The scale of the fermion mass function is a little bit smaller
than the QCD scale, 
$B(x=\Lambda_{\rm QCD}^2) \sim {\cal O}(\Lambda_{\rm QCD}/10)$. 
Substituting the obtained mass function $B(x)$ to 
Eq.~(\ref{eq:PSformula}) we obtain the decay constant of the 
composite scalar field. As is shown with the thin solid line 
in Fig.~\ref{fig:fpi}, the decay constant of 4D QCD is also 
smaller than the QCD scale, $f_\pi \sim {\cal O}(\Lambda_{\rm QCD}/100)$. 
Since it is the result of the ordinary 4D QCD, the decay constant
is too small as Higgs field to break electroweak symmetry. 

%%%%%%%%%%%%%%%%%%%%%%%%%%%%%%%%%%
%%%%%%%%%%%%%%%%%%%%%%%%%%%%%%%%%%
%%%%%%%%%%%%%%%%%%%%%%%%%%%%%%%%%%
%%%%%%%%%%%%%%%%%%%%%%%%%%%%%%%%%%

\subsection{QCD in flat bulk}
\label{sec:bqcdf}
Now we analyze the dynamical fermion mass induced 
by the bulk Yang-Mills theory in flat brane world, 
$M_4 \times T^\delta$~\cite{Abe:2002yb}. For simplicity we set
the same radii for all extra directions, i, e,
 universal extra dimensions. We study $SU(N=3)$ bulk Yang-Mills 
theory (QCD) with $N_f=3$ flavor fermion as a concrete example.
For QCD extended in the compact extra dimension with the radius $R$, 
the running coupling $\alpha (z)$ has the power low 
behavior~\cite{Dienes:1998vh} in the region $1/R^2 \le z$, 
\begin{eqnarray}
\alpha^{-1} (z)
&=& \alpha^{-1}_0 
   +2\pi \beta_0 \ln \frac{z}{\mu_0^2} 
   -2\pi \tilde{\beta}_0 \ln \frac{z}{\mu_R^2}
   +2\pi \tilde{\beta}_0 \frac{2X_\delta}{\delta}
    \left[ \left(\frac{z}{\mu_R^2}\right)^{\delta/2} -1 \right], 
\nonumber 
\end{eqnarray}
where $\mu_R=1/R$ and 
\begin{eqnarray}
X_\delta 
&=& \frac{\pi^{\delta/2}}{\Gamma (1+\delta/2)} 
 =  \frac{2\pi^{\delta/2}}{\delta \Gamma (\delta/2)}, \quad 
\alpha_0 \equiv \alpha (\mu_0^2), 
\nonumber 
\end{eqnarray}
and $\tilde{\beta}_0$ is the beta-function coefficient 
stimulating from the bulk fields. Thus we obtain the running 
coupling of the bulk QCD with $N_{\rm f}=3$ flavor fermion on the brane,
\begin{eqnarray}
\lambda (z)
&=& \frac{1}
    {B \ln (z/\Lambda_{\rm QCD}^2) 
      -\widetilde{B} 
       \left[ \ln (z/\mu_R^2) 
      -\frac{2X_\delta}{\delta} 
        \left\{ (z/\mu_R^2)^{\delta/2}-1 \right\}
        \right] \theta (z-\mu_R^2)}, 
\label{eq:powerlow}
\end{eqnarray}
where 
\begin{eqnarray}
\left\{
\begin{array}{l}
B=\frac{1}{24C_2(F)}\left( \frac{11N-2N_{\rm f}}{3} \right)
 =9/16 \quad (N=3,N_{\rm f}=3) \nonumber \\
\widetilde{B} 
 =\frac{1}{24C_2(F)}\left( \frac{11N-2\tilde{N}_{\rm f}}{3} \right)
 =11/16 \quad (N=3,\tilde{N}_{\rm f}=0) 
\end{array}
\right.. \label{eq:BandtB}
\end{eqnarray}

Using the running coupling (\ref{eq:powerlow}), we numerically
solve the improved ladder SD equation and find the behaviors 
of the fermion wave-function $A(x)$ and the mass function  $B(x)$ 
on the brane. We analyze the SD equation on the 
$(y_1^\ast,\ldots,y_\delta^\ast)$ $=(0,\ldots,0)$ brane, 
with $\Lambda=\Lambda_{\rm (4+\delta)D}=10$ TeV, $\Lambda_{\rm QCD}=200$ MeV 
and $R\Lambda_{\rm (4+\delta)D}=1$ 
as typical cases, where $\Lambda_{\rm (4+\delta)D}$ and $\Lambda$ are 
the bulk fundamental scale and the reduced cutoff scale on the brane 
respectively. For $\delta=1,2,3$ and $4$, the behavior of 
$B(x)$ and $A(x)$ are drawn in Figs.~\ref{fig:SDB} and~\ref{fig:SDA} 
respectively. For $\delta<3$ the fermion mass function behaves
like 4D QCD, i.e. 
$B(x=\Lambda_{\rm QCD}^2) \sim {\cal O}(\Lambda_{\rm QCD}/10)$.
For $\delta \ge 3$ the situation is dramatically changed. 
The fermion mass function develops a value near 
the {\it reduced} cutoff scale, 
$B(x=\Lambda_{\rm QCD}^2) \sim {\cal O}(\Lambda)$.
Using the Pagels-Stokar formula (\ref{eq:PSformula}),
we calculate the decay constant of the composite scalar field. 
As is shown in Fig.~\ref{fig:fpi} the scale of the decay constant 
is summarized as $f_\pi \sim {\cal O}(\Lambda_{\rm QCD}/10)$ 
for $\delta<3$ and $f_\pi \sim {\cal O}(\Lambda/10)$ for $\delta \ge 3$. 
Therefore if the bulk space-time has seven or more dimensions, 
the bulk Yang-Mills theory induces the TeV scale decay constant on 
the brane~\cite{Abe:2002yb}. 
There is a possibility that the bulk QCD or technicolor 
can provide a composite Higgs field in the flat brane world. 
In this section we take 
$1/R = \Lambda = \Lambda_{\rm (4+\delta)D} = 10$ TeV. 
We comment that this setup merges into the scenario of 
`large extra dimension'~\cite{Antoniadis:1990ew,Arkani-Hamed:1998rs} 
in which the fundamental scale is around TeV and we assume the 
existence of a large extra dimension where only the gravity propagates.

\subsection{QCD in warped bulk}
\label{sec:bqcdw}
Here we study the bulk QCD in the RS brane world~\cite{Abe:2002yb}.
In the warped brane world a mass scale 
$\Lambda$ on the brane at $y=y^\ast$ is suppressed by 
the warp factor, i.e. 
$\Lambda \simeq W(y^\ast) \Lambda_{\rm 5D}$, 
where $\Lambda_{\rm 5D}$ is the fundamental scale of the theory 
and $W(y)=e^{-\pi k|y|}$ is the warp factor. 
As is discussed in the previous subsection, it is natural to 
consider that the effective theory of the brane fermion should be 
regularized by the {\it reduced} fundamental scale 
{\it on the brane}. 
Hence we cutoff the loop momentum in the SD equation at the {\it reduced} 
cutoff scale $\Lambda$ on the brane,
\begin{eqnarray}
\Lambda \simeq e^{-k|y^\ast|}\Lambda_{\rm 5D} 
        \simeq e^{-k|y^\ast|}k
        \qquad (y^\ast =0,\pi R). 
\nonumber 
\end{eqnarray}
For $kR \simeq 11$, UV cutoff scale $\Lambda$ is suppressed by 
$10^{-15}$ on the $y^\ast=\pi R$ brane. Randall and Sundrum
obtain the weak and Planck hierarchy form this warp suppression
factor in the original RS model~\cite{Randall:1999ee}. 

The behavior of the running coupling is nontrivial 
in the RS space-time. 
In this thesis we evaluate it in terms of the 4D effective 
theory below the cutoff scale $\Lambda$ by using the truncated 
KK approach. 
In the region $kR \gg 1$ and $M_n \ll k$, KK mass spectrum of
the gauge boson reduces to the asymptotic form (\ref{eq:KKmassasym}). 
Comparing it with the torus case (\ref{eq:KKmasstorus}), we find that 
the running coupling in the RS space-time is obtained 
by replacing $\mu_R$ with $\mu_{kR}$ and shifting the KK lowest 
threshold by $\mu_{kR}/4$. In addition we must take the orbifold 
condition into account. It drops all the $Z_2$ odd modes. 
Only a half contribution comes from the KK excited modes.
Thus we replace $\widetilde{B} \to \widetilde{B}/2$ in 
Eq.~(\ref{eq:powerlow}). The running coupling reads
\footnote{We use usual logarithmic running below the compactification scale 
and power low running above the scale derived from the truncated KK approach. 
In Ref.~\cite{Choi:2002ps} the running coupling of 
the bulk gauge theory in AdS$_5$ is investigated precisely. 
By using it we will obtain more accurate result, which is kept 
as a future work.} 
\begin{eqnarray}
\lambda (z)
  = \frac{1}
    {B \ln (z/\Lambda_{\rm QCD}^2) 
      -(\widetilde{B}/2) 
       \left[ \ln (\sqrt{z}/\mu_{kR} + 1/4)^2 
      -4\left\{ \sqrt{z}/\mu_{kR} - 3/4 \right\}
        \right] \theta (z-(3/4)^2\mu_{kR}^2)}, 
\label{eq:RSpowerlow}
\end{eqnarray}
where $B$ and $\widetilde{B}$ are given by Eq.~(\ref{eq:BandtB}). 

Applying the truncated running coupling (\ref{eq:RSpowerlow})
we evaluate the SD equation on the $y=\pi R$ brane, with 
$k=\Lambda_{\rm 5D}=M_{\rm Pl}=10^{16}$ TeV,
$\Lambda=\pi e^{-\pi kR}\Lambda_{\rm 5D}=10$ TeV and 
$\Lambda_{\rm QCD}=200$ MeV as typical cases.
The solution of $B(x)$ and $A(x)$ are plotted in Figs.~\ref{fig:SDB} 
and~\ref{fig:SDA} respectively~\cite{Abe:2002yb}. 
Calculating the Eq.~(\ref{eq:PSformula}) we obtain the decay 
constant of the composite scalar as is shown in Fig.~\ref{fig:fpi}. 
In the RS brane world both scales of the fermion mass function 
and the decay constant are near the {\it reduced} cutoff scale,
$B(x=\Lambda_{\rm QCD}^2) \sim {\cal O}(\Lambda/10)$, 
$f_\pi \sim {\cal O}(\Lambda/10)$ $\sim$ ${\cal O}(1)$ TeV. 
Therefore the bulk Yang-Mills theory induces TeV scale decay 
constant on the brane~\cite{Abe:2002yb}. 
It means that the bulk QCD has possibility
to realize the composite Higgs scenario of electroweak symmetry
breaking. 

The original theory has only the Planck scale, it has no TeV scale 
source but the fermion mass and decay constant of the composite 
scalar are generated at TeV scale through the warp suppression of 
the {\it reduced} cutoff scale on the $y=\pi R$ brane. 
KK excited modes of the gauge field are localized at the $y=\pi R$ 
brane in RS model. It enhances the contribution from the KK modes
and realize the TeV scale composite scalar on the brane with
only one extra dimension. This is the difference from the 
(4+1)-dimensional flat brane world, $\delta=1$. 
The bulk gauge theory with only Planck scale in RS brane world can 
acquire a TeV scale dynamically due to a fermion pair condensation 
on the $y=\pi R$ brane through the effect of the gauge boson KK excited 
modes localized at the brane. 
The warp suppression mechanism in the RS model is realized 
dynamically.

The extension of above results to the {\it bulk SM gauge fields} 
will make us to propose two types of primary models of dynamical 
electroweak symmetry breaking. 
We call one of them {\it (1,2)-3 model} in which the 
first and second generations in the SM are confined on the $y=0$ 
brane, and the third generation is on the $y=\pi R$ brane 
(see Fig.~\ref{fig:dewsbrs}). 
The KK excited mode is not localized on the $y=0$ brane shown in 
Fig.~\ref{fig:WF}. The coupling between the KK excited 
gluons and the brane fermion are suppressed by the mode function
$\chi_n (y=0)$ on the $y=0$ brane.
Therefore we expect that the top quark-pair is condensed on 
the $y=\pi R$ brane, while the up and charm quark are not 
condensed.
The top mode SM~\cite{Miransky:1988xi} can be realized in 
RS brane world~\cite{Rius:2001dd}. 
We note that the zero modes function, $\chi_0$, are constant, so 
that the fermions on the both brane equally 
couples to the massless gauge filed (see also Fig.~\ref{fig:WF}). 
We can construct the other model, {\it (1,2,3)-4 model}, which
is defined as follows. All the SM fermions lie in $y=0$ brane 
and the fourth generation (or the technifermion) lies 
in the $y=\pi R$ brane. In this model 
the fourth generation can develops TeV scale mass function 
and decay constant by KK excited gluons (or techni-gluons).

It is expected that these models can provide composite Higgs 
and the TeV scale physics simultaneously. Although some 
problems are remained, e.g. how to dynamically realize the 
detailed mass relation (dynamical Yukawa couplings) in SM, 
or flavor breaking etc.. 
%%%%%%%%%%%%%%%%%%%%%%%%%%%%%%%%%%
\begin{figure}[htbp]
\begin{center}
\begin{minipage}{0.48\linewidth}
   \centerline{\epsfig{figure=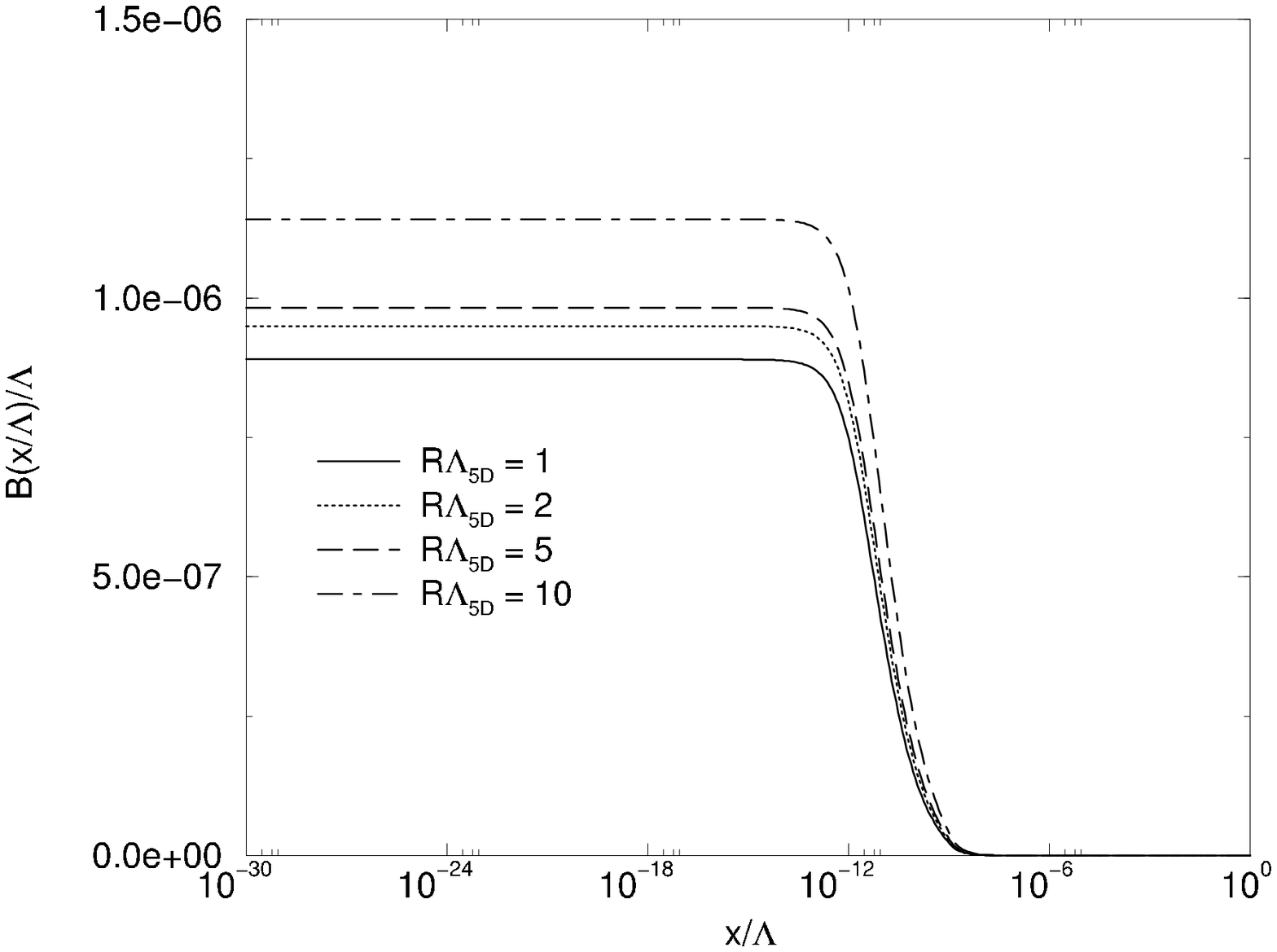,width=\linewidth}}
\end{minipage}
\hfill
\begin{minipage}{0.48\linewidth}
   \centerline{\epsfig{figure=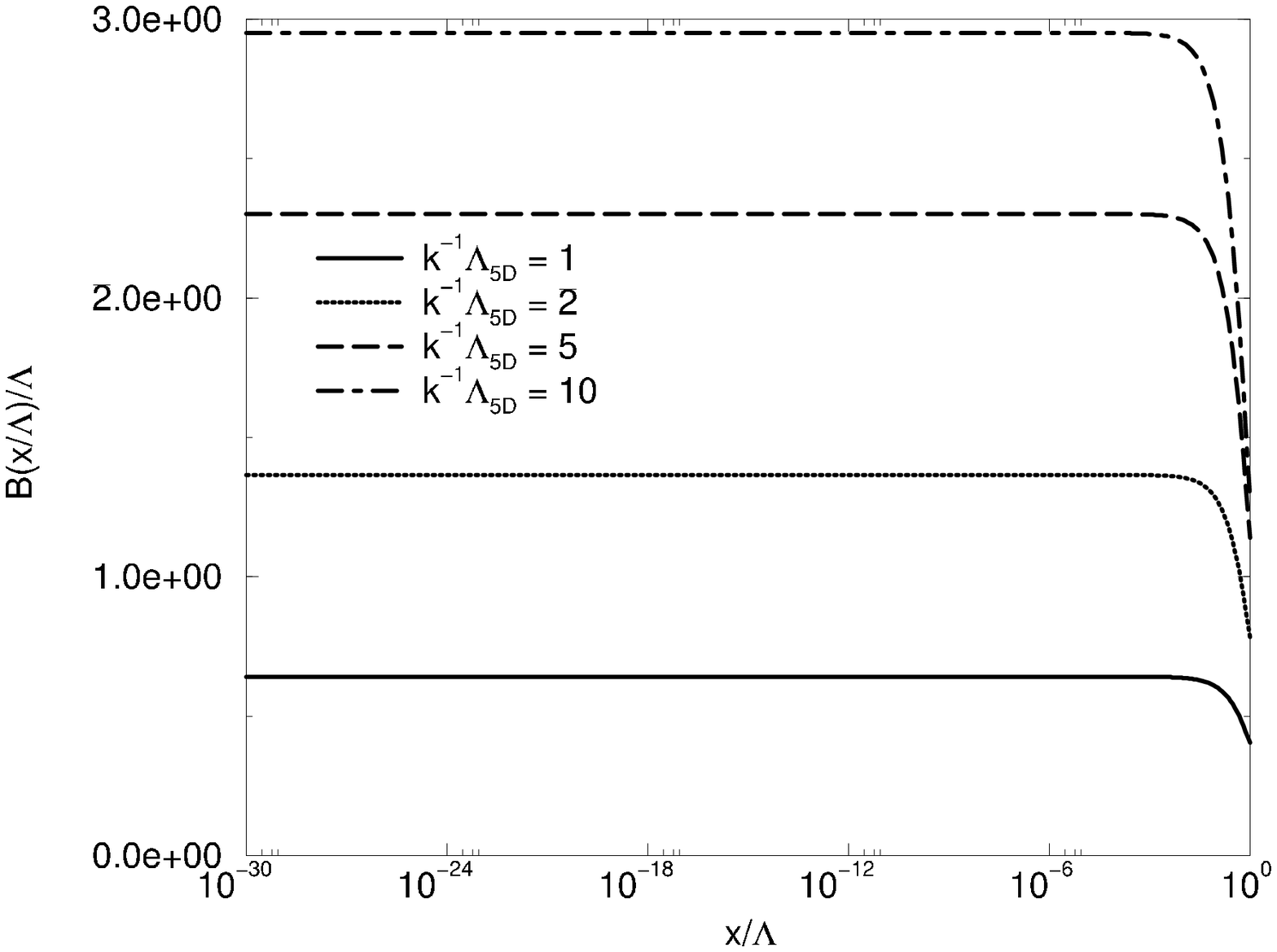,width=\linewidth}}
\end{minipage}
\end{center}
\caption{The behavior of $B(x)$ with $(4+1)$-dimensional bulk space-time 
derived from the improved ladder SD equation. 
The case of $(4+1)$-dimensional bulk QCD in the flat brane world 
with the radius $R\Lambda_{\rm 5D}$$=1,2,5$ and $10$, and the case 
in the RS brane world with the AdS$_5$ curvature radius 
$k^{-1}\Lambda_{\rm 5D}$$=1,2,5$ and $10$ are shown. 
We set $\Lambda_{\rm QCD}=200$ MeV for all cases. 
The cutoff is taken as $\Lambda=\Lambda_{\rm 5D}=10$ TeV for flat case. 
For the RS brane world we take $k=\Lambda_{\rm 5D}=M_{\rm Pl}=10^{16}$ TeV
and $\Lambda=\pi e^{-\pi kR}\Lambda_{\rm 5D}=10$ TeV $(kR=11.35)$.}
\label{fig:SDBR}
\begin{center}
\begin{minipage}{0.48\linewidth}
   \centerline{\epsfig{figure=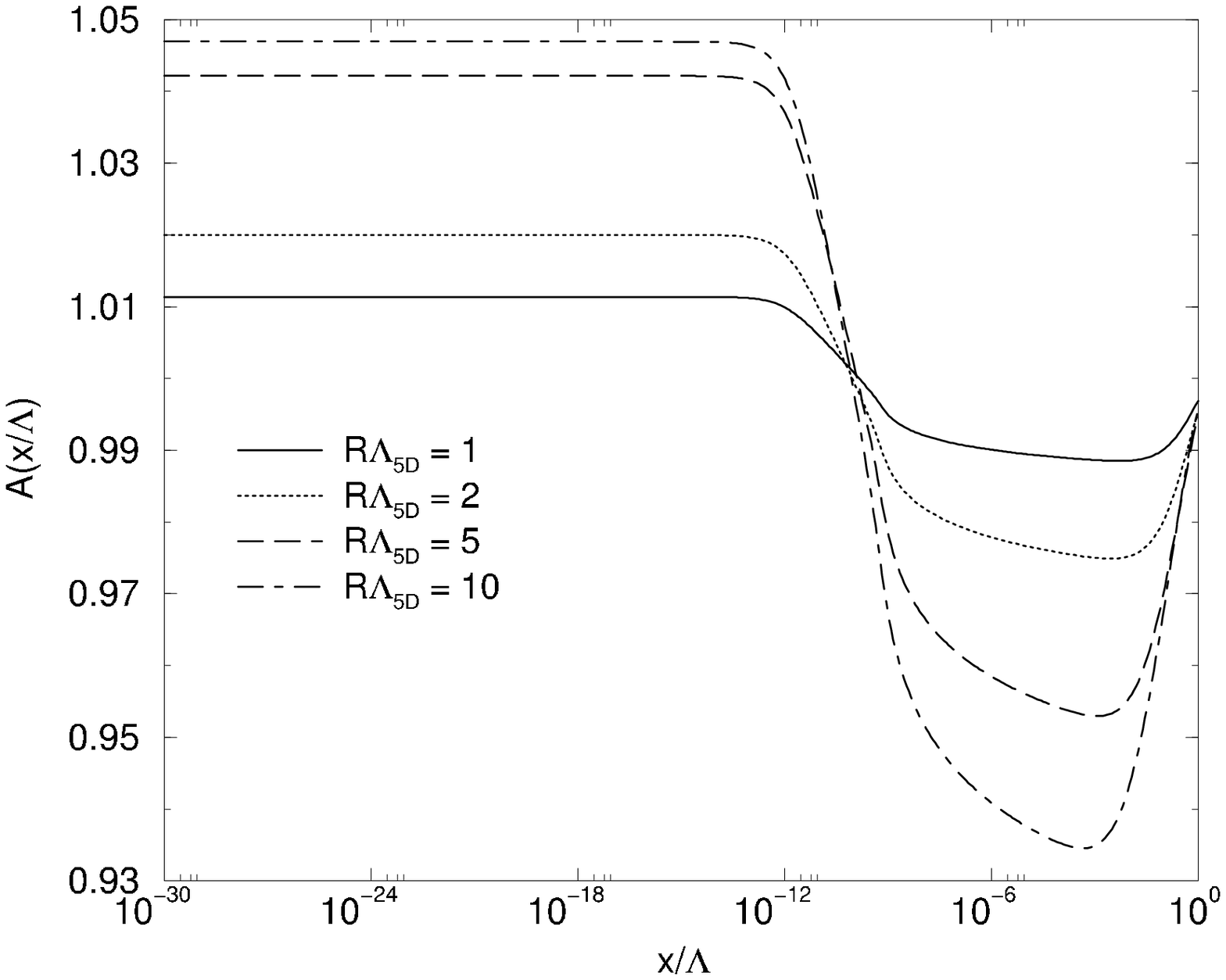,width=\linewidth}}
\end{minipage}
\hfill
\begin{minipage}{0.48\linewidth}
   \centerline{\epsfig{figure=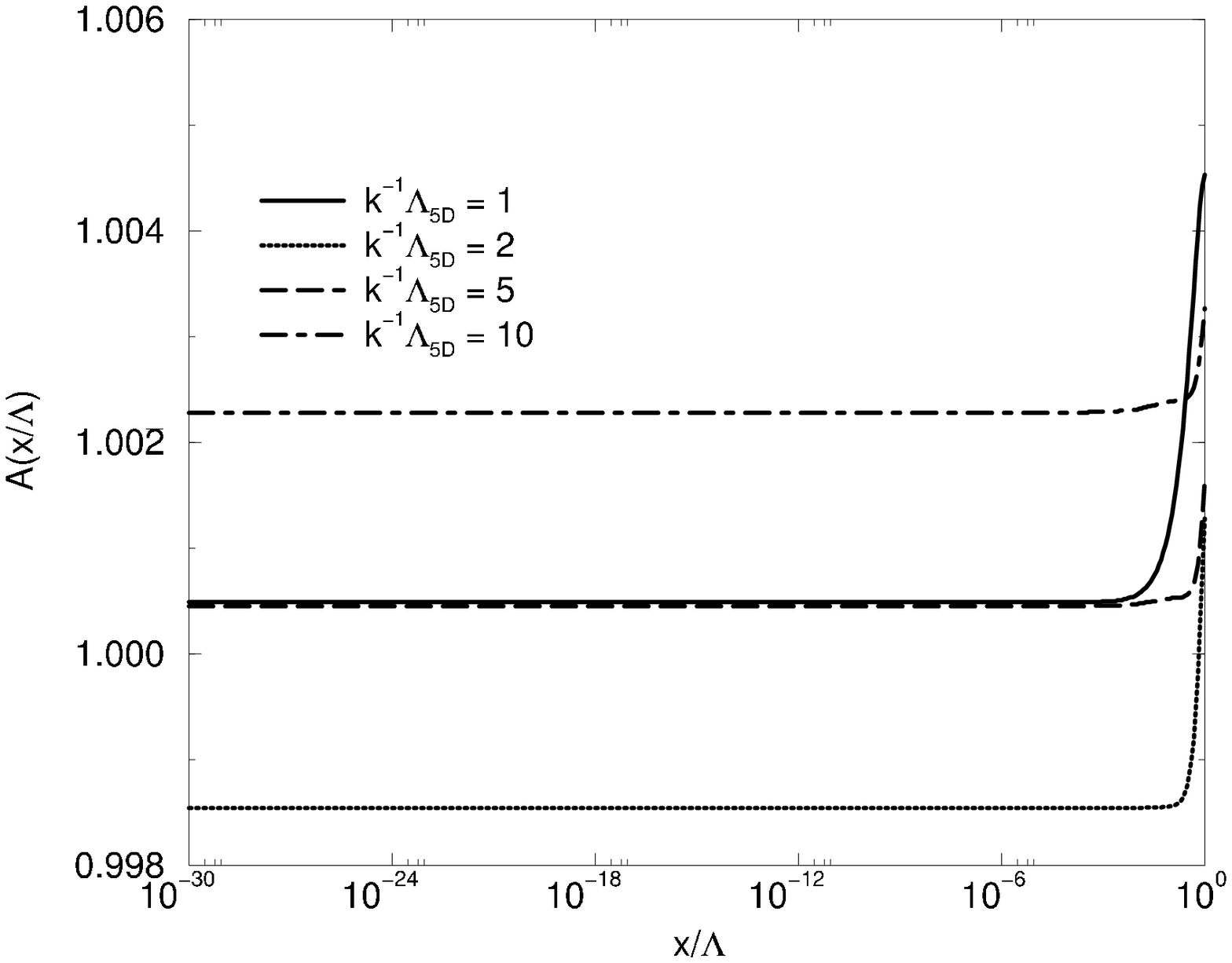,width=\linewidth}}
\end{minipage}
\end{center}
\caption{The behavior of $A(x)$ with $(4+1)$-dimensional bulk space-time 
derived from the improved ladder SD equation. 
The case of $(4+1)$-dimensional bulk QCD in the flat brane world 
with the radius $R\Lambda_{\rm 5D}$$=1,2,5$ and $10$, and the case 
in the RS brane world with the AdS$_5$ curvature radius 
$k^{-1}\Lambda_{\rm 5D}$$=1,2,5$ and $10$ are shown. 
The mass parameters are taken as the same in the caption in 
Fig.~\ref{fig:SDBR}. 
In order to satisfy Ward-Takahashi identity, i.e. $A(x) \simeq 1$ 
within $\pm 7$\%, we take the gauge fixing parameter 
$\xi=0.020$, $0.040$, $0.080$ and $0.104$ for flat brane world with 
$R\Lambda_{\rm 5D}=1$, $2$, $5$ and $10$ respectively, and 
$\xi=0.35$, $0.37$, $0.41$ and $0.50$ for RS brane world with 
$k^{-1}\Lambda_{\rm 5D}=1$, $2$, $5$ and $10$ respectively.}
\label{fig:SDAR}
\end{figure}
%%%%%%%%%%%%%%%%%%%%%%%%%%%%%%%%%%
%%%%%%%%%%%%%%%%%%%%%%%%%%%%%%%%%%
\begin{figure}[t]
\begin{center}
\begin{minipage}{0.48\linewidth}
   \centerline{\epsfig{figure=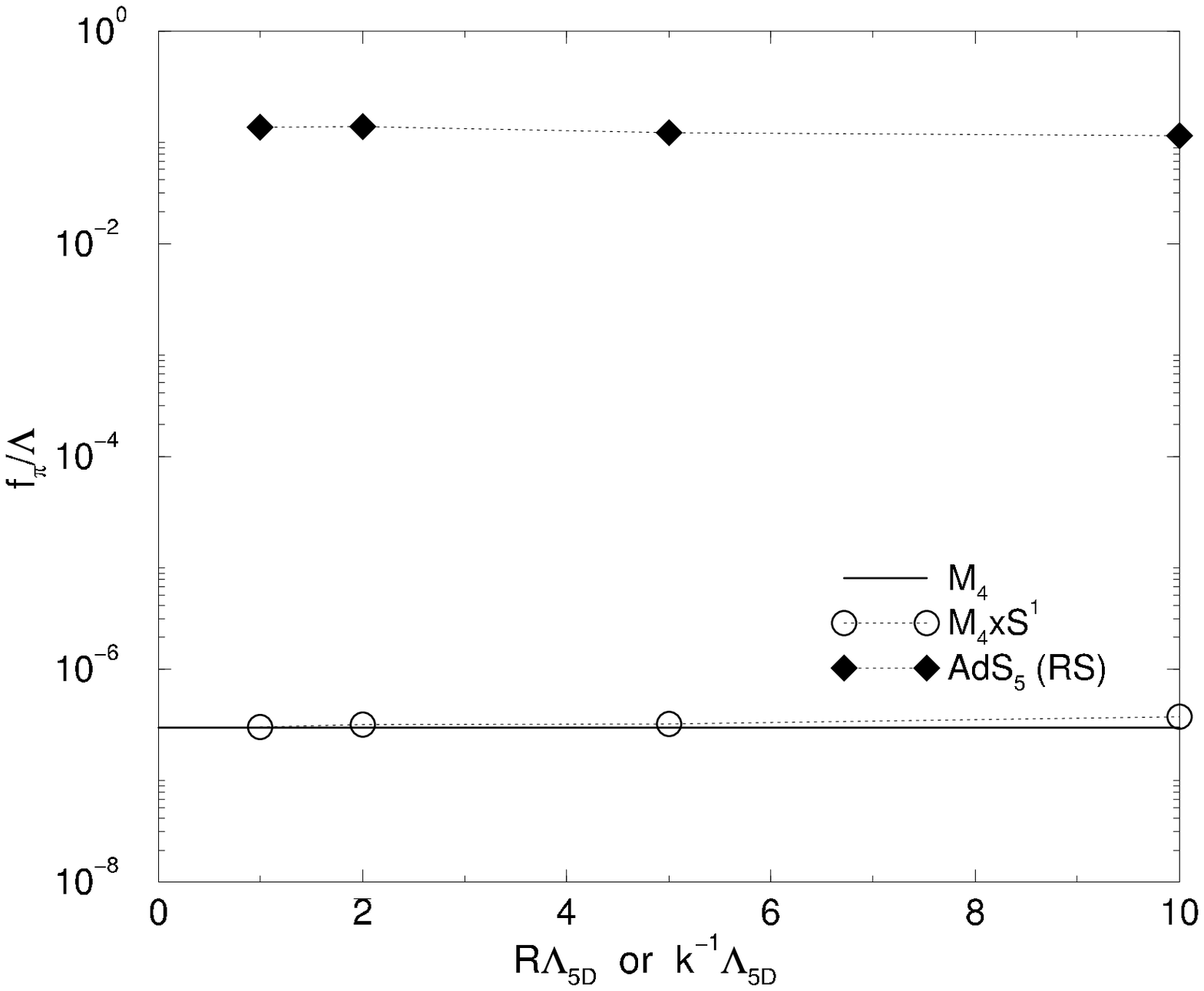,width=\linewidth}}
\end{minipage}
\end{center}
\caption{The behavior of the decay constant $f_\pi$ of the composite 
scalar field with $(4+1)$-dimensional bulk space-time 
derived from the Pagels-Stokar formula. 
The case of $(4+1)$-dimensional bulk QCD in the flat brane world 
with the radius $R\Lambda_{\rm 5D}$$=1,2,5$ and $10$, and the case 
in the RS brane world with the AdS$_5$ curvature radius 
$k^{-1}\Lambda_{\rm 5D}$$=1,2,5$ and $10$ are shown together. 
The parameters are taken as the same in the caption in Fig.~\ref{fig:SDBR}.}
\label{fig:fpiR}
\end{figure}
%%%%%%%%%%%%%%%%%%%%%%%%%%%%%%%%%%

\subsection{Behavior and classification of dynamical mass}

The behaviors of $B(x)$ shown in Fig.~\ref{fig:SDB} are divided 
into two pieces. QCD in the flat bulk space-time with $\delta=1$ and $2$ is 
classified into {\it Yang-Mills type}, i.e. the contribution from
the KK modes is extremely small and $B(x)$ has a similar profile to
the ordinary QCD (QCD on the brane). On the other hand, 
flat bulk space-time with $\delta=3$, $4$ and the RS brane world 
are {\it NJL type}, i.e. KK modes contribution dominates the 
symmetry breaking~\cite{Abe:2002yb}. 

In Figs.~\ref{fig:SDB},~\ref{fig:SDA} and~\ref{fig:fpi}, 
we consider only the case $N_{\rm KK}=1$. 
Finally we analyze the $R$- and $k$-dependence of the wave-function
$A(x)$ and the mass function $B(x)$ on the brane embedded in 
(4+1)-dimensional bulk space-time. The behaviors of $A(x)$ and $B(x)$
are shown in Figs.~\ref{fig:SDBR},~\ref{fig:SDAR} and~\ref{fig:fpiR}. 
There are little $R$- and $k$-dependence. Only a small 
$N_{\rm KK}$-dependence is observed in the behavior of $B(x)$.
Eqs.~(\ref{eq:KKmasstorus}) or (\ref{eq:KKmassasym}) show 
that the number of the KK mode $N_{\rm KK}$ increases as $R$ or 
$k^{-1}$ increases. The influence of the KK mode propagation
seems to be enhanced for larger $R$ or $k^{-1}$.
In Fig.~\ref{fig:SDBR} we clearly observe that the mass 
function $B(x)$ grows up as $R$ or $k^{-1}$ increases.
In Eqs.~(\ref{eq:powerlow}) or (\ref{eq:RSpowerlow}) the 
power low running terms contain the step function 
$\theta (z-\mu_R^2)$ or $\theta (z-(3/4)^2\mu_{kR}^2)$. 
Since the KK mode mass scale $\mu_R$ and $\mu_{kR}$ are 
proportional to $1/R$ and $k$ respectively, the contribution 
from these terms is enhanced as $R$ or $k^{-1}$ increases.
The contribution from the power low running suppresses 
dynamical symmetry breaking.
For example, if we drop the power low running term
in $(4+1)$-dimensional flat bulk space-time, $B(x)$ blows up 
from the {\it Yang-Mills type} to the {\it NJL type} as 
$N_{\rm KK}$ increases.
Thus we conclude that the bulk gluon self interactions 
${\cal L}_{\rm SI}$, i.e. the power low running coupling, 
act as a suppression factor for the dynamical mass on the 
brane~\cite{Abe:2002yb}.

\section{Discussions: pion and Higgs boson from single theory}
\label{sec:concl}

We have studied the basic structure of dynamical symmetry breaking 
on the four dimensional brane in the bulk Yang-Mills theory (QCD). 
Using the 4D effective theory of the bulk QCD and the improved 
ladder SD equation, we have found that the dynamical mass scale 
can be affected by the {\it reduced} cutoff scales on the brane. 
Starting from the Yang-Mills theory in $M_4 \times T^\delta$ space-time 
which couples to a fermion on the four dimensional brane we derive 
four dimensional effective theory by KK reduction. 
We take $A_{3+i}=0$ $(i=1,\ldots,\delta)$ gauge in extra directions 
but leave the gauge fixing parameter free parallel to the brane. 
Our system needs a proper regularization because it is a theory beyond 
four dimensional space-time where the gauge theory is nonrenormalizable. 
We impose the cutoff regularization at the UV {\it reduced} 
scale $\Lambda$ on the brane fermion lives in. For the flat brane 
world the reduced scale should be equal to the fundamental scale 
in the original bulk Yang-Mills theory, i.e. 
$\Lambda \simeq \Lambda_{\rm (4+\delta)D}$. 
The RS brane world, however, has a reduced scale suppressed 
by the warp factor, that is $\Lambda \simeq e^{-k|y^\ast|}\Lambda_{\rm 5D}$ 
which depends on the position in the extra direction. 

Based on the 4D effective theory we derive the SD equation for the fermion 
propagator on the brane. From the four dimensional point of view the
equation corresponds to the simultaneous integral equation. The loop 
integral consists of one massless and $N_{\rm KK}$ massive vector field 
where $N_{\rm KK}$ is the number of the KK modes under the reduced cutoff 
scale. Using the iteration method we numerically solve it for typical 
values of radius, $R\Lambda_{\rm (4+\delta)D}=1$, and the (AdS) curvature 
scale, $k=\Lambda_{\rm 5D}$. In order to make the analysis consistent with 
the QED like Ward-Takahashi identity, we choose the appropriate 
value of the gauge fixing parameter for each analysis. 

The results of the numerical analysis is as follows. For a flat extra 
dimension $\delta=1$ the dynamical mass of the brane fermion 
is the same scale as the usual 4D QCD result, i.e. there is little
contribution from the KK excited gauge boson. We also study the case 
of the flat brane world with more than one extra dimension. The numerical 
analysis of the SD equation shows that the number of the extra dimension
must be no less than three, $\delta \ge 3$, to generate the TeV scale 
dynamical fermion mass on the brane. Therefore it is obtained in the case 
of the Yang-Mills theory in seven or more dimensional bulk space-time. 

\begin{figure}[t]
\centerline{\epsfig{figure=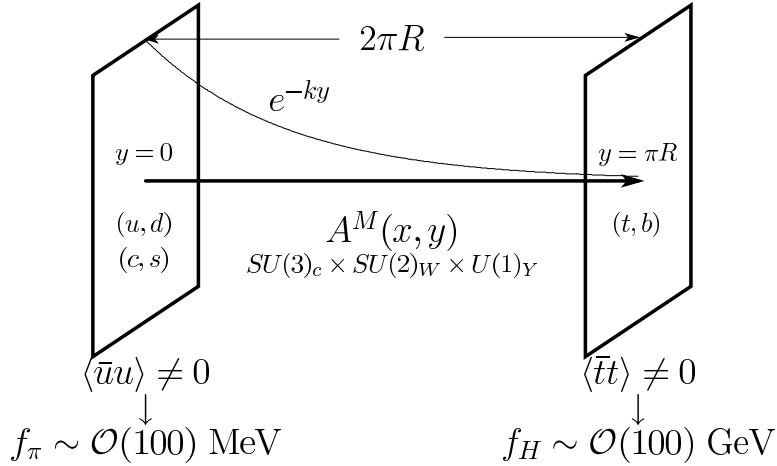,width=0.6\linewidth}}
\caption{Dynamical realization of both pion and electroweak Higgs boson 
from single theory; QCD in warped brane world. Although the RS warped 
brane world has only one fundamental scale $M_{\rm Pl}$, we might have 
both the scale of pion and electroweak Higgs boson via bulk QCD dynamics 
if we assume that the 3rd generation lives on $y=\pi R$ brane while 
the 1st and 2nd generation live on $y=0$ brane. This is because 
the gauge boson KK modes are localized at $y=\pi R$ brane while 
the gauge boson zero mode is flat in extra direction (see Fig~\ref{fig:WF}). 
Then the 1st and 2nd generation decouple from the KK gauge bosons. 
The electroweak scale is realized by the RS warp factor while the 
pion scale is given by the usual (4D) QCD dynamics via dimensional 
transmutation from the logarithmically running coupling.}
\label{fig:dewsbrs}
\end{figure}

The RS brane world has one extra dimension which corresponds to 
$\delta=1$ case but it is curved. Because the gauge boson KK excited 
modes localized at the $y=\pi R$ brane due to the curvature effect 
(see Fig.\ref{fig:WF}), we have a significant result. 
The dynamical mass of the brane fermion at the $y=\pi R$ is the order 
of the {\it reduced} cutoff scale that is warp suppressed from the 
fundamental scale~\cite{Abe:2002yb}.
We also estimate the decay constant of the composite Nambu-Goldstone scalar
field by using the Pagels-Stokar approximation~\cite{Pagels:1979hd}. 
The decay constant is also warp suppressed on the $y=\pi R$ brane
in the RS brane world. 
If the chiral symmetry is gauged as in the SM case, the composite
scalar absorbed in the longitudinal component of the gauge boson.
This decay constant is directly related to the mass of the gauge boson. 
The result in the RS brane world provides a possibility to realize 
the situation that electroweak symmetry breaking is triggered by 
the condensation of the composite Higgs field on the $y=\pi R$ brane.
Such a mechanism is caused by the bulk Yang-Mills theory, for example 
the technicolor or the QCD itself in the SM (for the latter case, 
see Fig.~\ref{fig:dewsbrs}). 
Therefore we can solve the weak and Planck hierarchy from the warp factor, 
that is realized, in a sense, dynamically. This is one of the dynamical 
realization of so called Randall-Sundrum model. 

Our analyses are based on the 4D effective theory of the bulk gauge theory 
with an explicit UV cutoff in the loop momentum integration. 
We can also check the regularization dependence 
of our model by comparing the results of the lattice regularized 
(deconstructed) version. The deconstructed bulk field theory 
on the RS background~\cite{Abe:2002rj} will be given in 
Part~\ref{part:hdnifd} of this thesis. 

The results of our analysis are also important in a theoretical sense. 
We consider that the SD equation includes certain nonperturbative effect 
of the theory. The nontrivial results obtained here implies that 
the bulk gauge theory may have rich nonperturbative dynamics 
that depends on the configurations of the extra space.
In the brane world we should take account of such effect 
in addition to the perturbative analysis. 
Our results imply that the bulk gauge dynamics can induce various types 
of the {\it dynamical} symmetry breaking in the brane world, 
depending on the bulk-brane configuration. 
For instance, we may obtain the dynamical breaking of the grand unification 
gauge symmetry~\cite{Kugo:1994qr} (or even supersymmetry?), 
in addition to the possibility of dynamical electroweak symmetry breaking 
emphasized in this thesis. 
We are also interested in the consistency of these models with the realistic 
D-brane system in the string theory.

%%%%%%%%%%%%%%%%%%%%%%%%%%%%%%%%%%%%%%%%%%%%%%%%%%%%%%%%%%%%%%%%%%%%%
%%%%%%%%%%%%%%%%%%%%%%%%%%% Part III %%%%%%%%%%%%%%%%%%%%%%%%%%%%%%%%
%%%%%%%%%%%%%%%%%%%%%%%%%%%%%%%%%%%%%%%%%%%%%%%%%%%%%%%%%%%%%%%%%%%%%
\part{Higher-dimensional nature in four-dimension}
\label{part:hdnifd}

\chapter{Warped gauge theory}
\label{chap:wgt}

In part~\ref{part:ssbb} and~\ref{part:dsbb} of this thesis 
we have studied SUSY and nonSUSY gauge theory in 
higher-dimensional space-time with some boundary terms. 
We have seen that extra dimensional configurations e.g. 
wave function localization due to boundary terms or curvature 
of extra dimension play important role in 
spontaneous or dynamical symmetry breaking in bulk and boundary 
system. However the gauge theory in more than four-dimensional 
space-time is nonrenormalizable and has less predictability. 
We should consider that the underlying theory such as string 
theory gives appropriate regularization to such higher-dimensional 
effective theory. 

In Part~\ref{part:hdnifd} of this thesis\footnote{Part~\ref{part:hdnifd} 
is based on Ref.~\cite{Abe:2002rj} with T.~Kobayashi, N.~Maru and K.~Yoshioka.} 
we give one of such regularization by latticizing the extra dimensions, 
that is so called deconstruction\footnote{
Deconstruction was originally proposed 
by~\cite{Arkani-Hamed:2001ca,Hill:2001mu} and so many 
interesting models were proposed~\cite{Cheng:2001vd}-\cite{Arkani-Hamed:2002pa} 
by utilizing it.}. 
For a finite lattice spacing the gauge theory is defined in 
four-dimensional space-time and is renormalizable. 
It may also be interesting that we treat such four-dimensional theory 
as an underlying theory just above the cut off scale of the effective 
`higher-dimensional' theory considered previously.

We formulate the four-dimensional gauge theory which induces, effectively, 
a gauge theory in (five-dimensional) general warped background\footnote{
Deconstruction of some specific curved geometries is also discussed 
in~\cite{Cheng:2001vd,Sfetsos:2001qb}}~\cite{Abe:2002rj}. 
We call such four-dimensional gauge theory 
`warped gauge theory'. It is interesting that the warped gauge theory 
gives a wave function localization in the index space of the product gauge group. 

\section{4D formulation for curved geometries}

\begin{figure}[t]
\centerline{\epsfig{figure=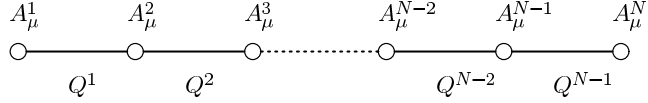,width=0.5\linewidth}}
\caption{Moose diagram for bulk vector fields
in the orbifold extra dimension.}
\label{fig:links}
\end{figure}

Following Refs.~\cite{Arkani-Hamed:2001ca,Hill:2001mu}, we introduce
$SU(n)_i$ gauge theories with gauge couplings $g_i$ ($i=1,\cdots,N$),
and scalar fields $Q_i$ ($i=1,\cdots,(N-1)$) which are in
bi-fundamental representations of $SU(n)_i\times SU(n)_{i+1}$.
The system is schematized by the segment diagram in Fig.~\ref{fig:links}.

The gauge invariant kinetic term of the scalars $Q_i$ is written by
\begin{eqnarray}
  {\cal L} &=& \sum_{i=1}^{N-1} (D_\mu Q_i)^\dagger (D^\mu Q_i),
\label{eq:Qkin}
\end{eqnarray}
where the covariant derivative is given by
$D_\mu Q_i=\partial_\mu Q_i-ig_iA^i_\mu Q_i+ig_{i+1}Q_i A^{i+1}_\mu$.
We assume that the scalar fields $Q_i$ develop their VEVs proportional
to the unit matrix, $\langle Q_i \rangle = v_i$, which break the gauge
symmetry to a diagonal $SU(n)_{diag}$. From the kinetic term
(\ref{eq:Qkin}), the mass terms for the vector fields $A_\mu^i$ are 
obtained
\begin{eqnarray}
  {\cal L}_{gm} &=& \frac{1}{2}\sum_{i=1}^{N-1}
  \Big|v_i(g_{i+1}A_\mu^{i+1}-g_iA_\mu^i)\Big|^2 \label{gKT}
  \ = \ \frac{1}{2} \sum_{i=1}^{N-1}\sum_{j=1}^N
  \left|D_{ij}^{\,\rm vec} A_\mu^j\right|^2,
\label{eq:gau-mass}
\end{eqnarray}
where the $(N-1)\times N$ matrix $D^{\,\rm vec}$ is defined as
\begin{eqnarray}
  D^{\,\rm vec} &=& \pmatrix{ v_1 & & \cr & \ddots & \cr & & v_{N-1} }
  \pmatrix{ -1 & 1 & & \cr
    & \ddots & \ddots & \cr
    & & -1 & 1 }
  \pmatrix{ g_1 & & \cr & \ddots & \cr & & g_N }.
\label{eq:diffop}
\end{eqnarray}

These mass terms result in that we have a massless gauge boson
corresponding to the unbroken gauge symmetry, which is given by the
following linear combination:
\begin{equation}
  \tilde A^{(0)}_\mu \;=\; \sum^N_{i=1}
  \biggl(\frac{g_{diag}}{g_i}\biggr) A^i_\mu,
\label{zero-A}
\end{equation}
where $g_{diag}^{-2}\equiv\sum_{i=1}^N g_i^{-2}$ and $g_{diag}$ is the
gauge coupling of the low energy gauge theory $SU(n)_{diag}$. The
profile of $\tilde A_\mu^{(0)}$ is independent of the values 
of $v_i$'s. It is found from this that the massless vector field is
`localized' at the points with smaller gauge couplings. If the gauge
couplings take a universal value $^\forall g_i=g$, the massless 
mode $\tilde A^{(0)}_\mu$ has a constant `wave-function' along 
the `index space' of gauge groups. As seen below, this direction
labeled by $i$ becomes the fifth spatial dimension in a continuum 
limit ($N\to\infty$). The localization behavior can easily be
understood from the fact that for smaller gauge coupling $g_i$,
the symmetry-breaking scale $g_iv$ of $SU(n)_i$ becomes lower,
and hence the corresponding vector field $A_\mu^i$ becomes more
dominant component in the low energy degree of
freedom $\tilde A^{(0)}_\mu$.

It is interesting to note that this vector localization mechanism
ensures the charge universality. Suppose that there is a field in a
nontrivial representation of the $SU(n)_i$ only. That is, it couples
only to $A^i_\mu$ with the strength $g_i$. This corresponds to a
four-dimensional field confined on a brane. If there are several
such fields, they generally have different values of gauge
couplings. However note that these fields couple to the massless
modes $\tilde A_\mu^{(0)}$ with an {\it universal} gauge
coupling $g_{diag}$ defined above. This is because in the presented
mechanism, vector fields are localized depending on values of gauge
couplings.

\medskip

As for massless eigenstates, the mass eigenvalues and wave-functions
are obtained by diagonalizing the mass matrix (\ref{eq:diffop}).
The simplest case is the universal couplings:
\begin{equation}
^\forall v_i \,=\, v, \qquad ^\forall g_i \,=\, g.
\end{equation}
In this case, one obtains the mass eigenvalues of $D^{\,\rm vec}$ as
\begin{eqnarray}
m_n &=& 2gv \sin\biggl(\frac{n\pi}{2N}\biggr), \qquad
(n=0,\ldots,N-1).
\nonumber 
\end{eqnarray}
In the limit $N\to\infty$ with $L\equiv 2N/gv\,$ fixed (the limit to
continuum 5D theory), the eigenvalues become
\begin{eqnarray}
m_n \,\simeq\, \frac{2n\pi}{L}.
\nonumber 
\end{eqnarray}
These are the same spectrum as that of the bulk gauge boson
in the $S^1/Z_2$ extra dimension with the radius $L/2\pi$.

\subsection{VEVs and Couplings generating AdS$_5$ background}
\begin{figure}[htbp]
\centerline{\epsfig{figure=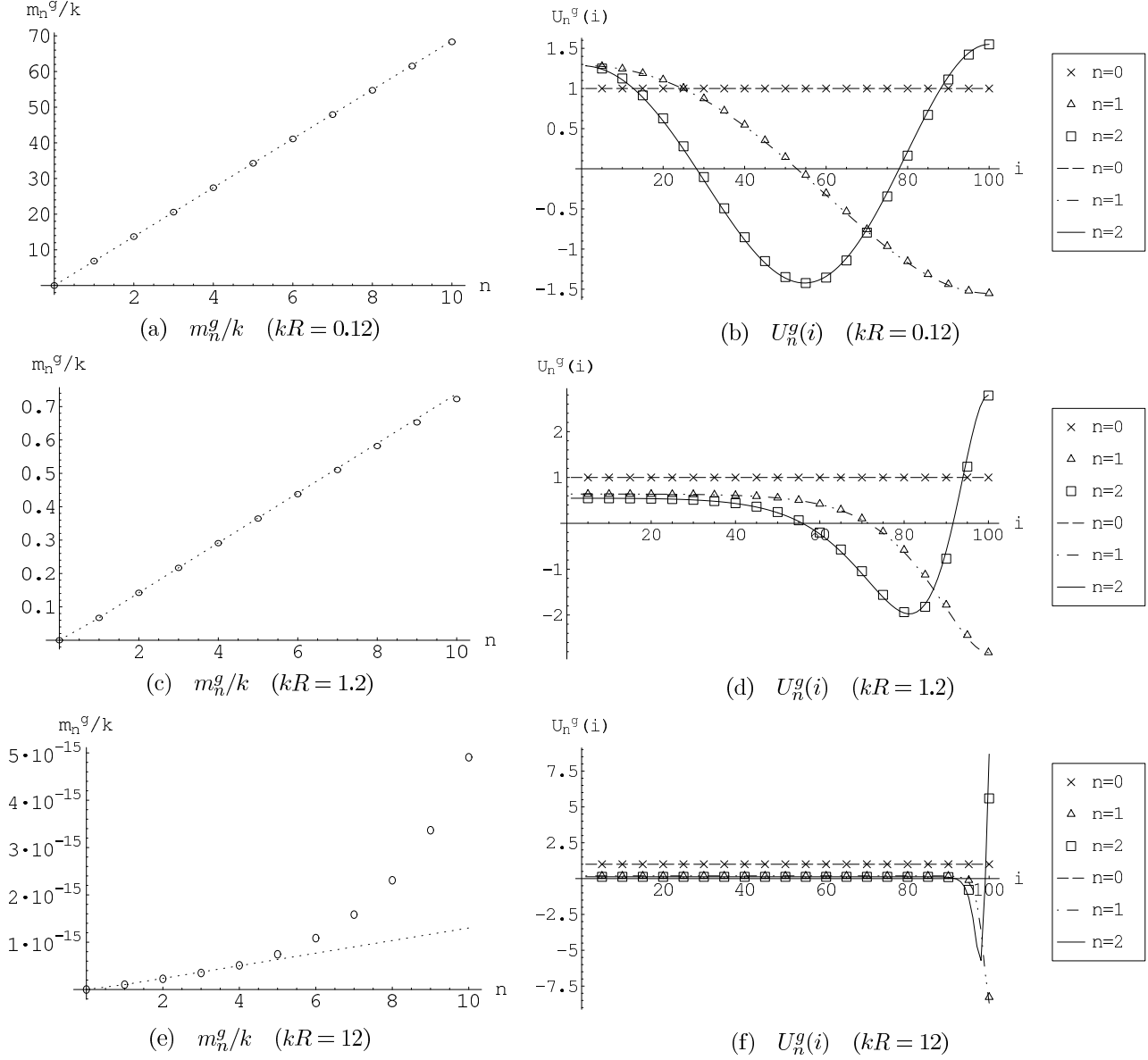,width=\linewidth}}
\caption{Numerical comparison of warped gauge theory and 
gauge theory in AdS$_5$. The mass spectra $m_n^g$ are shown 
in (a), (c) and (e), and the wave function profiles $U_n^g(i)$ 
are shown in (b), (d) and (f) for a vector field in 
warped gauge theory (depicted by symbols) 
and gauge theory in AdS$_5$ (depicted by lines). 
$n$ labels the index of KK excited mode and $i$ stands for the 
lattice site or fifth dimension $y=iL/2N$.}
\label{fig:vectores}
\end{figure}

First we consider the series of VEVs $v_i$ and couplings $g_i$ 
that generates a vector field on RS warped background, namely AdS$_5$. 
Such model can be obtained by choosing the universal $g_i$ 
and by varying $v_i$~\cite{Abe:2002rj} as 
\begin{eqnarray}
\mbox{RS} : \qquad  g_i \,=\, g, \qquad
v_i \,=\, ve^{-ki/(gv)}.
\label{eq:RSparaf}
\end{eqnarray}
Substituting this and taking the continuum limit, Eq.~(\ref{gKT})
becomes
\begin{eqnarray}
  {\cal L}_{gm} &=& \frac{1}{2} \int_0^{L/2}\! dy\,
  \Bigl[e^{-ky}\partial_y A_\mu(x,y)]^2,
\label{eq:contlimrs}
\end{eqnarray}
where $y$ represents the coordinate of extra dimension:
$y \leftrightarrow i/gv$ ($i=1,\ldots,N$) and
$A_\mu^i(x) \leftrightarrow A_\mu(x,y)$, etc. It is found that
Eq.~(\ref{eq:contlimrs}) successfully induces the kinetic energy term 
along the extra dimension and mass terms for the vector field on the 
warped background metric
\begin{equation}
  ds^2 = G_{MN}dx^M dx^N = e^{-2ky} \eta_{\mu\nu}dx^\mu dx^\nu-dy^2, 
\label{eq:rsbkg}
\end{equation} 
where $\eta_{\mu\nu}={\rm diag}(1,-1,-1,-1)$ with $\mu=0,1,2,3$.
We here conclude that we can obtain the vector field on RS warped 
background by varying only the VEVs $v_i$. In the following we will 
see that nonuniversal gauge couplings $g_i$ induce other interesting 
results beyond the effects from background metric. 
The RS warped gauge theory (\ref{eq:RSparaf}) is numerically compared 
with continuum RS one in Fig.~\ref{fig:vectores}. 

\subsection{Abelian case with nonuniversal gauge couplings}
Now let us compare the 4D model with generic couplings 
to extra dimensional ones. We define the dimensionless parameters $f_i$
and $h_i$ as~\cite{Abe:2002rj}
\begin{eqnarray}
  g_i \,=\, gf_i, \qquad v_i \,=\, vh_i.
\label{eq:deffh}
\end{eqnarray}
First we restrict ourselves to the case that the gauge group 
is $U(1)$, namely Abelian theory with {\it no vector self-couplings}. 
Similarly substituting Eq.~(\ref{eq:deffh}) and taking the continuum limit, 
Eq.~(\ref{gKT}) becomes 
\begin{eqnarray}
  {\cal L}_{gm} &=& \frac{1}{2} \int_0^{L/2}\! dy\,
  \Bigl[h(y)\partial_y\bigl(f(y)A_\mu(x,y)\bigr)\Bigr]^2. 
\label{eq:contlim}
\end{eqnarray}
Eq.~(\ref{eq:contlim}) induces the kinetic energy term along the extra
dimension and mass terms for the vector field on the warped background
metric
\begin{equation}
  ds^2 = G_{MN}dx^M dx^N = (f(y)h(y))^2 \eta_{\mu\nu}dx^\mu dx^\nu-dy^2. 
\label{eq:generalbkg}
\end{equation} 
The bulk and boundary mass terms are $y$-dependent and proportional to
the derivatives of $f(y)$.

\subsection{Non-Abelian case with nonuniversal gauge couplings}
In the above Abelian case we have discussed that the nonuniversal $f_i$ 
can be interpreted as $y$-dependent bulk/boundary masses in the 
warped extra dimension. Next we treat the non-Abelian theory with 
vector self-couplings. Since in this case we also have $y$-dependent
vector self-couplings in addition to the $y$-dependent bulk/boundary 
masses, it may be convenient and instructive to see $f_i$ 
as a $y$-dependent coefficient of the vector kinetic term~\cite{Abe:2002rj}. 
To this end, we define the four-dimensional field $A_5^i$,
\begin{eqnarray}
Q_i \equiv v_i e^{-ia (g_i A_5^i + g_{i+1} A_5^{i+1})/2} 
    = v_i \left(1-ia (g_i A_5^i + g_{i+1} A_5^{i+1})/2
                 +{\cal O}(a^2)\right),
\label{eq:defa5}
\end{eqnarray}
where $a\equiv L/(2N)$ is the lattice spacing, which goes to zero in
the continuum limit. 
Rescaling the gauge fields 
$\sqrt{N} f_i (A^i_\mu,\,A^i_5) \to (A^i_\mu,\,A^i_5)$,
the kinetic term 
$-\frac{1}{4} \sum_{i=1}^N F_{\mu \nu}^i F^{i \mu \nu}$
and Eq.~(\ref{eq:Qkin}) become 
\begin{eqnarray}
  {\cal L}^{gQ}_{kin} 
&=& -\frac{1}{4} \sum_{i=1}^N a \frac{f^{-2}_i}{L/2} 
  F^i_{\mu \nu} F^{i\mu \nu} \nonumber \\ &&
    +\sum_{i=1}^{N-1} a \frac{h^2_i}{L/2}
    \Bigg| \partial_\mu A_5^{i+\frac{1}{2}} - \frac{A^{i+1}_\mu-A^i_\mu}{a} 
           -i \hat{g} [A^i_\mu, A_5^{i+\frac{1}{2}}] 
\nonumber \\ && \qquad \qquad \qquad \qquad  \qquad \qquad  
           +i \hat{g} A_5^{i+\frac{1}{2}} (A^{i+1}_\mu - A^i_\mu) 
           + {\cal O}(a^{1/2}) \Bigg|^2, 
\label{eq:gQkin2}
\end{eqnarray}
where $\hat{g}=g/\sqrt{N}$
and $A_5^{i+\frac{1}{2}}\equiv (A_5^i+A_5^{i+1})/2$. 
In the continuum limit  
$N \to \infty$ with $L$ and $\hat{g}$ fixed,
Eq.~(\ref{eq:gQkin2}) results in 
\begin{eqnarray}
  {\cal L}^{gQ}_{kin} 
&=& -\frac{1}{4} \int_{0}^{L/2}dy \frac{f^{-2}(y)}{L/2}
      \Big[  \eta^{\mu \nu} \eta^{\rho \sigma} 
             F_{\mu \rho} F_{\nu \sigma}
            -2(h(y)f(y))^2 \eta^{\mu \nu} F_{\mu 5} F_{\nu 5} 
      \Big],
\nonumber 
\end{eqnarray}
where $F_{MN}(x,y) = 
\partial_M A_N (x,y) - \partial_N A_M (x,y) 
- - i\hat{g} [A_M (x,y), A_N (x,y)]$. 
This completely reproduces the 5D Yang-Mills 
kinetic term with a $y$-dependent coefficient
\begin{eqnarray}
  {\cal L}^{vec}_{warp} 
&=& -\frac{1}{4} \int_0^{L/2}\! dy\, f^{-2}(y) 
  \sqrt{-G} G^{MN}G^{AB} F_{MA}F_{NB}
\label{eq:5dym}
\end{eqnarray}
on the warped-background metric (\ref{eq:generalbkg}),
provided that $g_{\rm 5D} = \sqrt{L/2}\,\hat{g}$. 
This is the generic correspondence between 
the present 4D model and continuum 5D theory. From 
Eq.~(\ref{eq:5dym}), we thus find the $y$-dependent factor 
$f^{-2}(y)$ in front of the canonical Yang-Mills term, which
corresponds to a 5D dilaton VEV\@.
The factor does carry the origin of the massless
vector localization shown in Eq.~(\ref{zero-A}).\footnote{For a
continuum 5D analysis, see~\cite{Kehagias:2000au}.} 
With the constant gauge coupling $g_i \equiv g$ ($f_i$ = 1), one
obtains the bulk vector field with a constant zero mode on the 
warped metric (\ref{eq:generalbkg}).
A field redefinition $f^{-1}(y) A_M \to A_M$ in Eq.~(\ref{eq:5dym}) 
gives the previous bulk/boundary mass terms
but one then has $y$-dependent vector self-couplings in non-Abelian cases. 

\section{Toward dynamical realization}
\label{sec:DDWD}

We have shown that the 4D models with nonuniversal VEVs, gauge and 
other couplings can describe 5D physics on curved backgrounds, 
including the RS model with the exponential warp factor. In the 
continuum 5D theory, this factor is derived as a solution of the 
equation of motion for gravity. On the other hand, in the 4D 
viewpoint, warped geometries are generated by taking the couplings and 
VEVs as appropriate forms. In the previous sections, we have just 
assumed their typical forms and examined its consequences. 
If one could identify how to control these couplings by 
underlying {\it dynamics}~\cite{Abe:2002rj}, resultant 4D theories turn 
out to provide attractive schemes to discuss low energy physics such 
as tiny coupling constants. 

\subsection{Dimensional transmutation in strong coupling}

First we consider the scalar VEVs $\langle Q_i\rangle=v_i$.
A simple way to dynamically control them is to introduce additional
strongly-coupled gauge theories~\cite{Arkani-Hamed:2001ca}. Consider
the following set of asymptotically free gauge theories:
\begin{eqnarray}
A_\mu^i &:& SU(n)_i\textrm{ gauge field }(g,\Lambda) \nonumber \\
\hat A_\mu^i &:& SU(m)_i\textrm{ gauge field }
(\hat{g}_i,\hat{\Lambda}_i),
\nonumber 
\end{eqnarray}
where $\Lambda$ and $\hat{\Lambda}_i$ denote the dynamical scales. We
have, for simplicity, assumed the common values of $g$ and $\Lambda$
for all $SU(n)_i$. In addition, two types of fermions are introduced
\begin{eqnarray}
\xi^i \;:\; (n,\bar{m},1), \qquad
\bar\xi^i \;:\; (1,m,\bar{n}),
\nonumber 
\end{eqnarray}
where their representations 
under $SU(n)_i \times SU(m)_i \times SU(n)_{i+1}$ gauge groups are
shown in the parentheses. If $\hat{\Lambda}_i \gg \Lambda$, 
the $SU(m)_i$ theories confine at a higher scale than $SU(n)_i$, and
the fermion bilinear composite scalars $Q_i \sim \xi^i\bar\xi^i$
appear. Their VEVs $v_i$ are given by the dynamical 
scales $\hat{\Lambda}_i$ of the $SU(m)_i$ gauge theories through the
dimensional transmutation as
\begin{eqnarray}
\langle Q_i \rangle \,\equiv\, v_i \,\simeq\, \hat{\Lambda}_i
\,=\, \mu\, e^{\frac{1}{2\hat{\beta} \hat{g}_i^2(\mu)} },
\label{eq:DTvev}
\end{eqnarray}
where $\hat{\beta}$ is a universal one-loop gauge beta function 
for $SU(m)_i$ ($\hat\beta<0$). The gauge couplings $\hat g_i$
generally take different values and thus leads to different values 
of $v_i$. For example, a linear dependence of $1/\hat g_i^2$ on the
index $i$ is amplified to an exponential behavior of $v_i$. That is,
\begin{eqnarray}
\frac{1}{\hat{g}_i^2(\mu)}
\,=\, -2\hat\beta\biggl(\frac{k}{gv}\biggr)i
\quad\longleftrightarrow\quad v_i \,=\, ve^{-ki/gv},
\nonumber 
\end{eqnarray}
which reproduces bulk fields on the RS background as shown before.
Index dependences of gauge couplings are actually generic situations,
and may also be controlled, for example, by some mechanism fixing
dilatons or a radiatively induced kinetic terms discussed below.
A supersymmetric extension of the above scenario is achieved with
quantum-deformed moduli spaces~\cite{Csaki:2002em}.

\subsection{FI terms in supersymmetric warped $U(1)$ gauge theory}

Another mechanism which dynamically induces nonuniversal VEVs is
obtained in supersymmetric cases~\cite{Abe:2002rj}. Consider the gauge
group $\prod_i U(1)_i$ and the chiral superfields $Q_i$ with
charges $(+1,-1)$ under $U(1)_i\times U(1)_{i+1}$. It is assumed
that the scalar components $q_i$ of $Q_i$ develop their
VEVs $\langle q_i \rangle = v_i$. The $D$ term of each $U(1)_i$ is
given by
\begin{equation}
D_i \,=\, \epsilon_i + |q_i|^2 - |q_{i-1}|^2  + \cdots,
\end{equation}
where $\epsilon_i$ is the coefficient of the Fayet-Iliopoulos (FI)
term, and the ellipsis denotes contributions from other fields charged
under $U(1)_i$, which are assumed not to have VEVs. Given nonvanishing
FI terms, $\epsilon_i \neq 0$, the $D$-flatness conditions mean
\begin{equation}
v_{i-1}^2 \,=\, v_i^2 + \epsilon_i,
\end{equation}
and nonuniversal VEVs $v_i$ are indeed realized. In this case, a
dynamical origin of nonuniversal VEVs is nonvanishing FI
terms. These may be generated at loop level. Furthermore, if matter
contents are different for each gauge theory, $\epsilon_i$ themselves
have complicated forms. This supersymmetric warped $U(1)$ gauge theory 
might reveal the localization nature of 5D $S^1/Z_2$ orbifold model 
with generic FI terms on boundaries that was discussed in 
Chapter~\ref{chap:dtermsb}.

\section{Discussions: field localization in index space}
\label{sec:conclusion}

We have formulated 4D models which provide 5D field theories on
generic warped backgrounds. The warped geometries are achieved with
generic values of symmetry breaking VEVs, gauge couplings and other
couplings in the models. We have focused on field localization
behaviors along the index space of gauge theory (the fifth dimension
in the continuum limit), which is realized by taking relevant choices
of mass parameters.

As a good and simple application, we have constructed 4D models
corresponding to bulk field theories on the AdS$_5$ Randall-Sundrum
background. The localized wave-functions of massless modes are
completely reproduced with a finite number of gauge groups.
In addition, the exponentially suppressed spectrum of KK modes is
also generated. These results imply that most properties of brane
world models can be obtained within 4D gauge theories. Supersymmetric
extensions have also been investigated. In 5D warped models, bulk and
boundary mass terms of spinors and scalars satisfy complicated forms
imposed by supersymmetry on the RS background. However, we show in our
formalism that these forms of mass terms are derived from a 
4D {\it global} supersymmetric model on the {\it flat} background.

The conditions on the model parameters should be explained by some
dynamical mechanisms if one considers the models from a fully 4D
viewpoint. One of interesting ways is to include additional
strongly-coupled gauge theories. In this case, a 
small ${\cal O}(1)$ difference between gauge couplings is converted to
exponential profiles of symmetry breaking VEVs via dimensional
transmutation, and indeed generates a warp factor of the RS model.
A difference of gauge couplings is achieved by, for example, dynamics
controlling dilatons, or radiative corrections to gauge couplings.
Supersymmetrizing models provide a mechanism of dynamically realizing
nonuniversal VEVs with $D$-flatness conditions.

Our formulation makes sense not only in the 4D points of view but
also as lattice-regularized 5D theories. In this sense, such as
the AdS/CFT correspondence might be clearly seen with our
formalism. As other applications, it can be applied to construct
various types of curved backgrounds and bulk/boundary masses.
For example, we have discussed massless vector localization by varying 
the gauge couplings $g_i$. Furthermore, one might consider the models
in which some fields are charged under only some of gauge groups. That
seems not a bulk nor brane fields, but `quasi-bulk'
fields. Applications including these phenomena will be studied
elsewhere.

\chapter{Conclusions and discussions}
\label{chap:cd}

One of the most important subjects in modern theoretical particle 
physics is the unification of all fundamental interactions. 
We know few successful candidates for the theory of quantum gravity. 
The most promising one is superstring theory defined in ten-dimensional 
space-time, and we necessarily have extra dimensions to derive standard model 
in four-dimensional space-time from it. Recent interesting perspective 
is that various problems in the standard model can be solved in a 
process of compactifying nontrivially such extra dimensions, 
e.g., with orbifolding or with D-branes (orientifolding). 
In this thesis we notice such aspect, and investigate the structure of 
spontaneous or dynamical symmetry breaking in the bulk and boundary system 
as a general and most simple approximation of the effective theory. 

In Part~\ref{part:ssbb} we studied spontaneous supersymmetry breaking 
and its communication from the hidden to visible sector in the system. 
The low scale supersymmetry breaking picture preferably merges into 
the bulk and boundary system because we can sequester the supersymmetry 
breaking (hidden) sector by utilizing the branes and spatiality 
between them in extra dimensions. 
We reviewed that the $F$ term supersymmetry breaking contribution 
can be mediated to visible sector through various mechanisms like 
gravity, gauge, anomaly, gaugino, Kaluza-Klein and radion mediation. 
We obtained soft breaking parameters in the visible sector caused by 
each mechanisms. This is useful for understanding low energy 
phenomenology such as sparticle flavor structure under the 
supersymmetry breaking. 

In a case that local $U(1)$ symmetry with (nonintegrable) FI term 
exists in the brane world models, we have seen that both the field 
localization effect and the $D$ term supersymmetry breaking 
contributions also play important roles in low energy~\cite{Abe:2002ps}. 
For instance, if the $U(1)$ symmetry gives different charge per 
generation, a nontrivial VEV of the gauge scalar field induces 
different localization profile in extra dimension depending on the charge. 
That may cause a hierarchical structure in the Yukawa coupling, 
in addition to the contribution from the Froggatt-Nielsen mechanism. 
And if the charge is different between quarks and leptons, 
the proton decay may be suppressed by such localization. 
At the same time, with such localization effects, 
we have $D$ term contribution to the soft scalar 
masses which is universal everywhere in bulk up to $U(1)$ charge, 
and may affect the low energy physics. 

If the supersymmetry is broken with higher energy scale, 
we need another mechanism to stabilize electroweak Higgs mass scale. 
We have such a candidate that electroweak symmetry is broken by 
fermion bilinear condensation, that is a composite Higgs scenario. 
For such scenario we need strong binding forces between fermion and 
anti-fermion to cause such a condensation. The gauge theory with 
compact extra dimensions may give such strong forces caused by 
KK gauge bosons. We have studied in detail the dynamics 
of such gauge theory with matters on 
boundaries~\cite{Abe:2002yb,Abe:2001yi,Abe:2001ax,Abe:2000ny}. 
We have analyzed dynamical fermion masses on a boundary and 
a decay constant of fermion bilinear 
operator~\cite{Abe:2002yb} by using the improved ladder 
Schwinger-Dyson equation and the Pagels-Stoker formula. 
We obtain usual 4D Yang-Mills type or gauged NJL type strong dynamics 
from the bulk gauge theory, depending on the structure of the bulk and 
the boundary configuration, i.e., the position of the brane, the number of 
extra dimensions and its curvature. The dynamical mass scale, induced 
on the boundaries by the former Yang-Mills type and the latter NJL 
type dynamics, is usual dimensional transmutational scale $\Lambda_{\rm QCD}$ 
and a reduced scale from bulk fundamental scale, respectively. 
The reduced scale is warp suppressed on $y^\ast=\pi R$ brane 
in warped brane world and in this case the dynamical mass scale can 
be electroweak scale, though the fundamental scale in the bulk 
is near the Planck scale. This is, in a sense, dynamical 
realization of so called Randall-Sundrum scenario\footnote{
Even in the usual elementary Higgs scenario in RS background, 
our result also insists that we need large corrections to the masses of 
$W$, $Z$ bosons and some quarks, which come from the quark bilinear 
condensations on the $y=\pi R$ brane, if we put SM (QCD) gauge fields 
in bulk and the quarks on that brane.}. 

These results are important for applying our system to phenomenological 
models such as top mode SM. In this thesis we suggest a possibility to 
realize a simple primary model of top quark condensation in RS brane 
world that is summarized in Fig.~\ref{fig:dewsbrs}. 
Our analysis here might be useful to construct any other dynamical 
models in the bulk and boundary system. 
Also in a theoretical sense it is concluded that theories in 
the bulk and boundary system have a nontrivial and interesting 
vacuum structure and/or a nonperturbative dynamics. 
Technically it remains as a future work that we analyze the system 
adopting another way of gauge fixing in order to satisfy Ward-Takahashi 
identity in more accuracy than one utilized in this thesis. 
It is because the slight violation of the identity, namely 
the shift of the value of wave function renormalization 
factor $A(x)$ from $1$, also deviates a bit the value of the dynamical 
mass function $B(x)$ from the correct one. This may be done 
with nonlocal gauge fixing formulated in appendix~\ref{sec:nlgf}. 

In both SUSY and nonSUSY case it is understood that the wave function 
localization in extra dimensions gives important effects on the 
spontaneous or dynamical symmetry breaking in the bulk and boundary system. 
Such nontrivial field profiles that lead to the hierarchical 
couplings, strong-weak phase splitting in extra dimensions 
are generated by some boundary terms, e.g., FI terms and curvature 
effects of extra dimensions. This can be the geometrical origin of 
the hierarchical structures between parameters in SM. 
We have also studied one of the regularization of higher-dimensional 
theory in order to obtain a renormalizable theory. 
We formulated warped gauge theory as the regularized theory, 
and showed that field localization 
is realized in the index space of the product gauge group~\cite{Abe:2002rj}. 
By using warped gauge theory, we obtain more predictive results 
from the bulk and boundary theories. Inversely, in the warped gauge theory, 
we understand that we need fine-tuned couplings in four-dimensional 
theory to realize, in a purely four-dimensional framework, 
the same nontrivial results as in the higher-dimensional theories. 
In this sense, we can say that the higher-dimensional theories have 
more rich structure than the four-dimensional ones. 

Phenomenologically interesting features investigated in this thesis are 
based on the nontrivial compactification of the higher-dimensional 
field theory. So the future task is to check that such interesting 
models can be derived from the compactification of just superstring theory. 
It may be interesting to reanalyze the orbifold compactification of heterotic 
string, D-branes in type I superstring, and their intersection~\cite{Ibanez:2001nd} 
in terms of such phenomenological sense.

\chapter*{Acknowledgements}

First I would like to thank Profs. Taichiro~Kugo (Kyoto U.), 
Taizo~Muta (Hiroshima U.) and Masanori~Okawa (Hiroshima U.) 
very much for their hospitality as my supervisors. 

I greatly thank Tomohiro~Inagaki (Hiroshima U.) and 
Tatsuo~Kobayashi (Kyoto U.) for many precious discussions and 
important comments, especially during our collaboration. 

I would also like to thank Kenji~Fukazawa (Kure N. C. T.), 
Michio~Hashimoto (KEK), Tetsutaro~Higaki (Kyoto U.), 
Yasuhiko~Katsuki (Kure U.), Nobuhito~Maru (U. of Tokyo), 
Hironori~Miguchi (SONY LSI, Fukuoka), Seiji~Mukaigawa (Iwate U.), 
Hiroaki~Nakano (Niigata U.), Kensaku~Ohkura (Hiroshima U.), 
Masaharu~Tanabashi (Tohoku U.) and Koichi~Yoshioka (Kyushu U.) 
for fruitful discussions and correspondences. 

I thank all of the other members in theoretical 
particle physics groups in Hiroshima and Kyoto university 
for their hospitality and encouraging me.

\appendix

\chapter{Wave-function renormalization and soft parameters}
\label{app:wrsp}

We can determine soft terms of an observable field via its 
wave function renormalization if the supersymmetry breaking 
is communicated by perturbatively renormalizable interactions. 
That means that we can obtain all the information about the 
soft terms from an anomalous dimension and a beta function. 
For example, we should calculate one-loop and two-loop 
diagrams to obtain the form of a gaugino and a sfermion soft 
mass respectively in the minimal gauge-mediated scenario. 
It is, however, easier to determine anomalous dimensions and 
beta functions than such loop calculations. In this appendix 
we review the calculus how to determine soft parameters 
from the wave function renormalization following G.~F.~Giudice 
and R.~Rattazzi~\cite{Giudice:1997ni}. 

For such purpose we are interested in the case that the 
supersymmetry-breaking scale and the mediation scale are 
completely determined by the VEVs of a scalar and an auxiliary 
component of a chiral superfield $X$: 
\[ \langle X \rangle = M + \theta^2 F \]
In a conventional gauge-mediated scenario a messenger field 
receives a mass via a superpotential $X\bar{\Phi}\Phi$. 
In the case $F \ll M^2$ supersymmetry is broken only by 
the soft terms below the messenger scale, and we can utilize 
explicitly supersymmetric formulation to chase the supersymmetry 
breaking effects. 

Thanks to the nonrenormalization theorem of the superpotential 
the gaugino and the sfermion masses are generated by $X$-dependent 
renormalization of the kinetic terms (wave-function) of them. 
That is, only the gauge coupling $g^2$ and the matter wave function 
renormalization $Z_i$ is the source of the soft masses. 
Therefore in the case with only one (spurion) field $X$, 
we calculate $g^2$ and $Z_i$ in the supersymmetric limit $X=M$ 
($M$ is a c-number), and then we replace $M$ with the superfield 
$X$ in the final results. This is the most important point, however, 
we should note that for $Z_i(M,\mu)$ the replacement which 
is consistent with a chiral reparameterization $X \to e^{i\phi}X$ is 
$M \to \sqrt{X^\dagger X}$, while $M \to X$ should be performed for 
the gauge coupling $g^2(M,\mu)$.

\subsubsection{Gaugino mass}

The gaugino mass renormalized at scale $\mu$ is extracted from the 
$X$-dependent wave function renormalization: 
\begin{eqnarray}
{\cal L} = \int d^2 \theta \, S(X) W^{a \alpha}W^a_{\alpha} + {\rm h.c.}, 
\label{eq:gaugeKT}
\end{eqnarray}
where $S(X)$ is expanded around $\theta, \bar{\theta}=0$ as 
\begin{eqnarray}
S(X) 
 &=& S(X(0,0))
    +\left( \theta \frac{\partial}{\partial \theta} + 
     \bar{\theta} \frac{\partial}{\partial \bar{\theta}} \right)
     S(X(0,0)) 
    +\frac{1}{2!} 
     \left( \theta \frac{\partial}{\partial \theta} + 
     \bar{\theta} \frac{\partial}{\partial \bar{\theta}} \right)^2
     S(X(0,0)) +\ldots \nonumber \\
 &=& \left[ 
     S + \frac{1}{2} \frac{\partial S}{\partial X} 
         \frac{\partial X}{\partial (\theta^2)} \theta^2 
       + (\theta,\ \bar{\theta},\ \bar{\theta}^2,\ 
          \theta^2 \bar{\theta}^2 {\rm term})
     \right]_{X=X(0,0)=M} \nonumber \\
 &=& S(M) \left[
     1 + \frac{1}{2} \frac{1}{S} \frac{\partial S}{\partial X} 
         \frac{\partial X}{\partial (\theta^2)} \theta^2 
       + (\theta,\ \bar{\theta},\ \bar{\theta}^2,\ 
          \theta^2 \bar{\theta}^2 {\rm term})
     \right]_{X=M}. 
\label{eq:gaugeKFexp}
\end{eqnarray}
Canonically normalizing the gaugino kinetic term in (\ref{eq:gaugeKT}) 
and using $\partial X/\partial (\theta^2) =F$, 
we obtain the gaugino mass in the form 
\begin{eqnarray}
\bar{M}_g(\mu) 
&=& -\frac{1}{2} \left. \frac{1}{S}
     \frac{\partial S}{\partial X} \right|_{X=M} \!\!\! F \quad
 = -\frac{1}{2} \left. \frac{1}{X} 
     \frac{\partial \ln S(X,\mu)}{\ln X}
     \right|_{X=M} F. 
\label{eq:gauginomassf}
\end{eqnarray}
Then we know that we can calculate gaugino mass from a function $S(X)$. 
Because the scalar component $S(M,\mu)$ of $S(X)$ is a coefficient function 
of the gauge kinetic term, we can determine it: 
\begin{eqnarray}
S(M,\mu) = \frac{\alpha^{-1}(M,\mu)}{16\pi}-\frac{i\Theta}{32\pi^2}, 
\nonumber 
\end{eqnarray}
where $\alpha = g^2/4\pi$ is a gauge coupling and $\Theta$ is 
a topological vacuum angle. 

The function $\alpha (M,\mu)$ is 
given by integrating the renormalization group (RG) evolution 
from the ultra-violet (UV) scale $\Lambda_{\rm UV}$ to the 
observation scale $\mu$ over the messenger threshold $M$. 
From the one-loop RG equation we find 
\begin{eqnarray}
\frac{d}{dt}\alpha^{-1} = \frac{b}{2\pi} \qquad (t=\ln \mu),
\label{1loopRGE}
\end{eqnarray}
where the beta-function coefficient $b$ is given by $b=3N_c-N_f$ 
for the $SU(N_c)$ gauge theory with $N_f$ flavor. 
Above the messenger scale $M$ the running of the gauge coupling 
suffers from messenger fields $\Phi_i$ and $\bar{\Phi}_i$ ($i=1,\ldots,N$), 
which transforms as $N_c+\bar{N}_c$ of $SU(N_c)$, and the beta-function 
coefficient becomes $b'=b-N$. Therefore the one-loop running of the 
gauge coupling is given by 
\begin{eqnarray}
{\rm Re}S(M,\mu) 
&=& \frac{\alpha^{-1}(M,\mu)}{16 \pi} \nonumber \\
&=& \frac{\alpha^{-1}(\Lambda_{\rm UV},\Lambda_{\rm UV})}{16 \pi} 
   +\frac{b'}{32\pi^2} \ln \frac{|M|}{\Lambda_{\rm UV}}
   +\frac{b}{32\pi^2} \ln \frac{\mu}{|M|}. 
\label{eq:gaugerunning}
\end{eqnarray}
This equation and the {\it holomorphic property} determines the 
form of $S(X,\mu)$. In this equation we replace $M$ as $M \to X$ 
and obtain 
\begin{eqnarray}
S(X,\mu) 
&=& S(\Lambda_{\rm UV},\Lambda_{\rm UV})
   +\frac{b'}{32\pi^2} \ln \frac{X}{\Lambda_{\rm UV}}
   +\frac{b}{32\pi^2} \ln \frac{\mu}{X}. 
\label{eq:KF}
\end{eqnarray}
The imaginary part of this equation reproduces 
renormalization $\Theta \to (b-b') \arg (X)$ 
that shows the chiral anomaly. 

By differentiating Eq.~(\ref{eq:KF}) by $\ln X$ 
for fixed $S(\Lambda_{\rm UV})$ we finally obtain 
the form of the gaugino mass 
\begin{eqnarray}
\bar{M}_g(\mu) 
 &=& -\frac{1}{2} \left.
      \frac{1}{S(X,\mu)} \frac{b'-b}{32\pi^2} 
      \frac{F}{X} \right|_{X=M} \nonumber \\
 &=& -\frac{1}{2} 16\pi \,\alpha(M,\mu)\, \frac{-N}{32\pi} 
      \frac{F}{M}
      \nonumber \\
 &=&  \frac{\alpha(\mu)}{4\pi}N \frac{F}{M}. 
\label{eq:gauginomass}
\end{eqnarray}
This agrees with the result from the direct loop 
integration of some diagrams. 

\subsubsection{Sfermion mass}

We also define the wave function renormalization $Z_Q$ of a chiral 
superfield $Q$ by 
\begin{eqnarray}
{\cal L} = \int d^4 \theta Z_Q(X,X^\dagger) Q^\dagger Q. 
\label{eq:matterKF}
\end{eqnarray}
Different from the gaugino case here we should take account of 
$D$ term renormalization and $Z_Q$ is {\it real} and is a 
function of {\it both} $X$ and $X^\dagger$. 

We expand $Z_Q$ around $\theta^2,\bar{\theta}^2=0$ as 
\begin{eqnarray}
\lefteqn{
Z_Q(X(\theta,\bar{\theta}),X^\dagger (\theta,\bar{\theta}))} 
\hspace{-1cm} \nonumber \\
&=& Z_Q(X(0,0),X^\dagger (0,0)) 
  + \frac{\partial Z_Q(X(0,0),X^\dagger (0,0))}
         {\partial X} F \theta^2 
    \nonumber \\ &&
  + \frac{\partial Z_Q(X(0,0),X^\dagger (0,0))}
         {\partial X^\dagger} 
    F^\dagger \bar{\theta}^2 
  + \frac{\partial^2 Z_Q(X(0,0),X^\dagger (0,0))}
         {\partial X \partial X^\dagger} 
    F F^\dagger \theta^2 \bar{\theta}^2
    \nonumber \\
&=& \left[ Z_Q
  + \frac{\partial Z_Q}
         {\partial X} F \theta^2 
  + \frac{\partial Z_Q}
         {\partial X^\dagger} 
    F^\dagger \bar{\theta}^2 
  + \frac{\partial^2 Z_Q}
         {\partial X \partial X^\dagger} 
    F F^\dagger \theta^2 \bar{\theta}^2
    \right]_{\tiny \matrix{X=X(0,0)=M \cr 
    X^\dagger = X^\dagger (0,0) = M^\dagger}}
    \nonumber \\
&=& \left[ Z_Q \left( 1
  + \frac{\partial \ln Z_Q}
         {\partial X} F \theta^2 
  + \frac{\partial \ln Z_Q}
         {\partial X^\dagger} 
    F^\dagger \bar{\theta}^2 
  + \frac{1}{Z_Q} \frac{\partial^2 Z_Q}
         {\partial X \partial X^\dagger} 
    F F^\dagger \theta^2 \bar{\theta}^2
    \right)
    \right]_{\tiny \matrix{X=M \cr X^\dagger = M^\dagger}}
    \nonumber \\
&=& \left[ Z_Q \left\{
    \left( 1+\frac{\partial \ln Z_Q}{\partial X} 
           F \theta^2 \right)^\dagger
    \left( 1+\frac{\partial \ln Z_Q}{\partial X} 
           F \theta^2 \right)
    +\frac{1}{X^\dagger X} 
     \frac{\partial \ln Z_Q}
          {\partial \ln X \partial \ln X^\dagger}
     FF^\dagger \theta^2 \bar{\theta}^2
    \right\}
    \right]_{\tiny \matrix{X=M \cr X^\dagger = M^\dagger}}
    \nonumber \\
&=& \left[ Z_Q 
    \left( 1+\frac{\partial \ln Z_Q}{\partial X} 
           F \theta^2 \right)^\dagger
    \left( 1+\frac{\partial \ln Z_Q}{\partial X} 
           F \theta^2 \right)
    \left\{ 1+ \frac{
     \frac{1}{X^\dagger X} 
     \frac{\partial \ln Z_Q}
          {\partial \ln X \partial \ln X^\dagger}
     FF^\dagger \theta^2 \bar{\theta}^2}
    {\left( 1+\frac{\partial \ln Z_Q}{\partial X} 
           F \theta^2 \right)^\dagger
    \left( 1+\frac{\partial \ln Z_Q}{\partial X} 
           F \theta^2 \right)}
    \right\}
    \right]_{\tiny \matrix{X=M \cr X^\dagger = M^\dagger}}. 
\label{matterKFexp}
\end{eqnarray}
In Eq.~(\ref{eq:matterKF}) we canonically normalize the 
matter superfield 
\begin{eqnarray}
{\cal Z}Q \equiv \left.
\sqrt{Z_Q} 
\left( 1+\frac{\partial \ln Z_Q}{\partial X} 
       F \theta^2 \right)
\right|_{\tiny \matrix{\ \cr X=M \cr X^\dagger=M^\dagger}} 
\hspace{-1cm}Q 
\qquad \quad \to Q
\label{eq:mattercanor}
\end{eqnarray}
and the mass of a scalar component of $Q$ results in 
\begin{eqnarray}
\tilde{m}_Q^2 (\mu) 
&=& -\left. \left[
    \frac{
     \frac{1}{X^\dagger X} 
     \frac{\partial \ln Z_Q}
          {\partial \ln X \partial \ln X^\dagger}
     FF^\dagger \theta^2 \bar{\theta}^2}
    {\left( 1+\frac{\partial \ln Z_Q}{\partial X} 
           F \theta^2 \right)^\dagger
    \left( 1+\frac{\partial \ln Z_Q}{\partial X} 
           F \theta^2 \right)}
    \right]_{D}
     \right|_{\tiny \matrix{X=M \cr X^\dagger = M^\dagger}}
    \nonumber \\
&=& -\left. \frac{1}{X^\dagger X} 
     \frac{\partial \ln Z_Q}
          {\partial \ln X \partial \ln X^\dagger}
     FF^\dagger 
     \right|_{\tiny \matrix{X=M \cr X^\dagger = M^\dagger}}
    \nonumber \\
&=& -\left. 
     \frac{\partial \ln Z_Q(X,X^\dagger,\mu)}
          {\partial \ln X \partial \ln X^\dagger}
     \right|_{\tiny \matrix{X=M \cr X^\dagger = M^\dagger}}
     \frac{FF^\dagger}{MM^\dagger}. 
\label{eq:sfermionmassf}
\end{eqnarray}
By following the similar way we can obtain a corresponding 
formula of $A$ term. 

The one-loop wave function renormalization of the chiral 
superfield $Q$ is given by 
\begin{eqnarray}
\frac{d}{dt} \ln Z_Q = \frac{c}{\pi} \alpha, 
\label{eq:matterRGE}
\end{eqnarray}
where $c$ is the second Casimir of the gauge group 
that is given by $c=(N^2-1)/2N$ for $SU(N)$ fundamental 
representation. The function $Z_Q(M,\mu)$ is determined 
by integrating Eq.~(\ref{eq:matterRGE}) from $\Lambda_{\rm UV}$ 
to $\mu$ and matching it at the threshold $M$ with the tree 
level one. Namely, 
\begin{eqnarray}
Z_Q(X,X^\dagger,\mu) 
= Z_Q(\Lambda_{\rm UV}) 
  \left[ \frac{\alpha (\Lambda_{\rm UV})}
              {\alpha (X)} \right]^{2c/b'}
  \left[ \frac{\alpha (X)}
              {\alpha (\mu)} \right]^{2c/b}, 
\label{eq:matterRGrun}
\end{eqnarray}
where we define 
\begin{eqnarray}
\alpha^{-1} (\mu)
 &\equiv& 16\pi {\rm Re}S(\sqrt{XX^\dagger},\mu)
 \ = \ \alpha^{-1}(X) 
    +\frac{b}{4\pi} \ln \frac{\mu^2}{XX^\dagger}
    \nonumber \\
\alpha^{-1} (X)
 &\equiv& 16\pi {\rm Re}
          S(\sqrt{XX^\dagger},\sqrt{XX^\dagger})
 \ = \ \alpha^{-1}(\Lambda_{\rm UV}) 
    +\frac{b'}{4\pi} \ln 
     \frac{XX^\dagger}{\Lambda_{\rm UV}^2},
\label{eq:defcouprun}
\end{eqnarray}
by performing $M \to \sqrt{XX^\dagger}$ in the 
function $S(M,\mu)$ in Eq.~(\ref{eq:gaugerunning}). 
We note that $\alpha$, i.e., the real part of $S(X,\mu)$ 
is nonchiral though $S(X,\mu)$ is chiral. 

By differentiating Eq.~(\ref{eq:sfermionmassf}) by $X$ 
for fixed UV parameters $Z_Q(\Lambda_{\rm UV})$ and 
$\alpha (\Lambda_{\rm UV})$, we obtain the soft parameters: 
\begin{eqnarray}
\tilde{m}_Q^2(\mu) 
 &=& 2c\frac{\alpha^2(\mu)}{(4\pi^2)}N
     \left[ \xi^2+\frac{N}{b}(1-\xi^2) \right]
     \left( \frac{F}{M} \right)^2
     \label{eq:sfermionmass} \\
A_i (\mu)
 &=& \frac{2c_i}{b} \frac{\alpha(\mu)}{4\pi}
     N(\xi-1)\frac{F}{M}
     \label{eq:softAterm} \\
\xi &\equiv& \frac{\alpha (M)}{\alpha (\mu)}
     = \left[ 1+\frac{b}{2\pi} \alpha (\mu) 
              \ln \frac{M}{\mu} \right]^{-1}. 
     \nonumber
\end{eqnarray}
These coincide with the well-known soft terms 
given by taking the effect of leading-log in the 
gauge-mediated scenario.

\chapter{SUSY formula in 5D $S^1/Z_2$}
\label{app:osform}

In this appendix we review a formula of off-shell 
supersymmetric vector and hyper multiplets in 5D 
following Ref.~\cite{GrootNibbelink:2002qp}. 
We neglect the gravity (Weyl) multiplet. 

\section{Symplectic reality condition}
Here we consider one vector multiplet 
$V=\{ A_M,\lambda_i, \Sigma, \vec{D} \cdot \vec{\sigma}_{ij} \}$ 
and a set of hyper multiplets 
$H^\alpha = \{ \Phi_i^{\,\alpha}, \Psi^\alpha, F_i^{\,\alpha} \}$ 
where $i,j=1,2$ are $SU(2)_R$ indices and $\alpha=1,\ldots,2n$. 
$\lambda_i$ and $\Psi^\alpha$ are $USp(2)$ and $USp(2n)$ Majorana spinors, 
respectively with reality conditions, 
\begin{eqnarray}
\lambda_i = \varepsilon_i^{\,j} \lambda^C_j, \quad 
\Psi^\alpha = ({\mathbf 1} \otimes \varepsilon)^\alpha_{\,\beta} (\Psi^\beta)^C, 
\label{eq:symplecticrcf} 
\end{eqnarray}
where 
$\varepsilon=\left( \begin{array}{cc} 0 & 1 \cr -1 & 0 \end{array} \right)$, 
${\mathbf 1} \otimes \varepsilon = 
\left( \begin{array}{cc} 0 & {\mathbf 1}_n  \cr 
-{\mathbf 1}_n & 0 \end{array} \right)$, 
$\Psi^C \equiv C \bar\Psi^T$, and the charge conjugation matrix $C$ is 
defined as 
\begin{eqnarray}
C^{-1}\gamma^MC=(\gamma^M)^T, \quad 
C^T=-C, \quad C^\dagger = C^{-1}. \nonumber 
\end{eqnarray}
$\Phi_i^{\,\alpha}$ also satisfies symplectic reality condition, 
\begin{eqnarray}
\Phi^{\,\alpha}_i = 
({\mathbf 1} \otimes \varepsilon)^\alpha_{\,\beta} 
\Phi^{\,\beta}_j (\varepsilon^{-1})^{\,j}_i. 
\label{eq:symplecticrcb} 
\end{eqnarray}

\section{Lagrangian and SUSY transformation}
The Lagrangian is given by 
\begin{eqnarray}
{\cal L} &=& {\cal L}_V + {\cal L}_H, \nonumber \\
{\cal L}_V &=& {\rm tr} \left[ -\frac{1}{4}F_{MN}F^{MN} 
+ \frac{1}{2}(\partial_M \Sigma)^2 
+ \frac{1}{2}(\vec{D}\cdot \vec{\sigma}_{ij})^2 
+ i\bar\lambda_i \gamma^M D_M \lambda^i \right], \nonumber \\
{\cal L}_H &=& \frac{1}{2}|D_M \Phi_i^{\,\alpha}|^2 
+ \frac{1}{2}|F_i^{\,\alpha}|^2
-\frac{1}{2}g^2\Sigma_a \Sigma_b 
\Phi_i^{\,\alpha \dagger} (t^a t^b)_{\alpha \beta} \Phi^{i \beta}
-\frac{1}{2}g\Phi_i^{\,\alpha \dagger} (t^a)_{\alpha \beta} \Phi_j^{\,\beta} 
\vec{D}_a \cdot \vec{\sigma}^{ji} \nonumber \\ &&
+ \frac{i}{2}\bar\Psi^\alpha \gamma^M D_M \Psi_\alpha 
+ \frac{1}{2}g \Sigma_a \bar\Psi^\alpha (t^a)_{\alpha \beta} \Psi^{\beta}
+ 2g\bar\Psi^\alpha (t^a)_{\alpha \beta} \Phi^{\beta}_i \lambda^i_a, 
\nonumber  
\end{eqnarray}
where $D_M=\partial_M+igA_M^at^a$ is a covariant derivative, 
$(t^a)_{\alpha \beta} \equiv -(T^a \otimes \sigma_3)_{\alpha \beta} = 
\left( \begin{array}{cc} -T^a & 0 \\ 0 & T^a \end{array} \right)_{\alpha \beta}$, 
and $T^a$ is a generator of the gauge group. This Lagrangian is invariant 
under the (global) SUSY transformation, 
\begin{eqnarray}
&& \left\{
\begin{array}{rcl}
\delta A_M &=& i\bar\epsilon^i \gamma_M \lambda_i, \nonumber \\ 
\delta \lambda_i &=& F^{MN}\gamma_{MN}\epsilon_i/4 
         - \gamma^M \partial_M \Sigma \epsilon_i/2 
         - i\vec{\sigma}_i^{\,j} \cdot \vec{D} \epsilon_j/2, \nonumber \\ 
\delta \Sigma &=& i\bar\epsilon^i \lambda_i, \nonumber \\ 
\delta \vec{D} &=& \bar{\epsilon}_i \vec{\sigma}^{ij} 
         \gamma^M \partial_M \lambda_j, 
\end{array}
\right. \nonumber \\ 
&& \left\{
\begin{array}{rcl}
\delta \Phi_i^\alpha &=& i\bar\epsilon_i \Psi^\alpha, \\ 
\delta \Psi^\alpha &=& -\gamma^M D_M \Phi_i^\alpha \epsilon^i 
         - F_i^\alpha \epsilon^i 
         -ig\Sigma (T^a)^\alpha_{\ \beta} \Phi_i^\beta \epsilon^i \\ 
\delta F_i^\alpha &=& i\bar\epsilon_i \gamma^M D_M \Psi^\alpha 
         + 2g\bar\epsilon_j (T^a)^\alpha_{\,\beta}\Phi^{j\beta} \lambda_i^a 
         + \bar\epsilon_i \Sigma_a (T^a)^\alpha_{\,\beta} \Psi^\beta 
\end{array}
\right. \nonumber 
\end{eqnarray}
where $\epsilon_i$ is a SUSY transformation parameter satisfying 
a $USp(2)$ Majorana condition $\epsilon_i = \varepsilon_i^{\,j} \epsilon_j^C$. 

\section{Orbifold condition}
In order to obtain 4D ${\cal N}=1$ SUSY, we perform $S^1/Z_2$ orbifolding, 
\begin{eqnarray}
\epsilon_i = \epsilon_i(y) = \epsilon_i(-y) = 
(\sigma_3)_i^{\,j}i\gamma^5 \epsilon_j(y) = 
(\sigma_3)_i^{\,j}i\gamma^5 \epsilon_j, \quad 
i\gamma^5= \left( \begin{array}{cc} {\mathbf 1} & 0 \\ 
0 & -{\mathbf 1} \end{array} \right). 
\label{eq:ocpara}
\end{eqnarray}
In addition we require that the gauge symmetry is not broken. 
Then we have 
\begin{eqnarray}
&& \left\{
\begin{array}{rcl}
\lambda_i(-y) &=& (\sigma_3)_i^{\,j}i\gamma^5 \lambda_j(y), \\ 
\Sigma(-y) &=& -\Sigma(y), \\
\vec{\sigma}_{ij} \cdot \vec{D} (-y) &=& 
(\sigma_3 T^a) \vec{\sigma}_i^{\,k} \cdot \vec{D}_a (y) (\sigma_3)_{kj}, 
\end{array}
\right. \label{eq:ocvec} \\
&& \left\{
\begin{array}{rcl}
\Phi_i^{\,\alpha}(-y) &=& 
-({\mathbf 1} \otimes \sigma_3)^\alpha_{\,\beta} \Phi_j^{\,\beta}(y) 
(\sigma_3)^j_{\,i}, \\
\Psi^\alpha(-y) &=& ({\mathbf 1} \otimes \sigma_3)^\alpha_{\,\beta} 
i\gamma^5 \Psi^\beta(y), \\ 
F_i^{\,\alpha}(-y) &=& 
({\mathbf 1} \otimes \sigma_3)^\alpha_{\,\beta} F_j^{\,\beta}(y) 
(\sigma_3)^j_{\,i},
\end{array}
\right. \label{eq:ochyp} 
\end{eqnarray}
These orbifold condition (\ref{eq:ocpara})-(\ref{eq:ochyp}) 
and reality conditions (\ref{eq:symplecticrcf})-(\ref{eq:symplecticrcb}) 
are solved simultaneously and result in 
\begin{eqnarray}
\epsilon_i = 
\left( \begin{array}{c} \epsilon \\ -\epsilon^C \end{array} \right), \quad 
\lambda_i = 
\left( \begin{array}{c} \lambda \\ -\lambda^C \end{array} \right), \nonumber 
\end{eqnarray}
\begin{eqnarray}
(\Phi_i^{\,\bar\alpha} | \Phi_i^{\,\bar\alpha+n}) &=& 
\left( \begin{array}{c|c} (\Phi_-^{\,\bar\alpha})^* & \Phi_+^{\,\bar\alpha} \\ 
-(\Phi_+^{\,\bar\alpha})^* & \Phi_-^{\,\bar\alpha} \end{array} \right), 
\nonumber \\ 
(\Psi^{\bar\alpha}|\Psi^{\bar\alpha+n}) &=& 
(\Psi^{\bar\alpha}|-(\Psi^{\bar\alpha})^C), \nonumber \\
(F_i^{\,\bar\alpha} | F_i^{\,\bar\alpha+n}) &=& 
\left( \begin{array}{c|c} F_+^{\,\bar\alpha} & -(F_-^{\,\bar\alpha})^* \\ 
F_-^{\,\bar\alpha} & (F_+^{\,\bar\alpha})^* \end{array} \right), 
\nonumber 
\end{eqnarray}
where the row of the matrices corresponds to the $SU(2)_R$ index $i$, 
$\epsilon=\epsilon_{+L}-\epsilon_{-R}$, $\epsilon^C=\epsilon_{+R}+\epsilon_{-L}$, 
$\lambda=\lambda_{+L}-\lambda_{-R}$, $\lambda^C=\lambda_{+R}+\lambda_{-L}$, 
$\Psi^{\bar\alpha}=\Psi^{\bar\alpha}_{+L}-\Psi^{\bar\alpha}_{-R}$, 
$(\Psi^{\bar\alpha})^C=\Psi^{\bar\alpha}_{+R}+\Psi^{\bar\alpha}_{-L}$, 
the subscript $\pm$ represents the eigenvalues of the parity operator 
$\sigma_3i\gamma^5$, and $\bar\alpha=1,\ldots,n$. 
The parities of all the independent fields are summarized as follows: 
\[V:~
\begin{array}{l|c|c|c|c|c|c|c|}
\mbox{state} &A_\mu & A_5 & \Sigma 
& \lambda_{\pm L} & \lambda_{\pm R} & D_3 & D_{1,2}
\\\hline
\mbox{parity} & + & - & - & \pm & \pm & + & -
\end{array}\,,\]
\[\Phi:~ 
\begin{array}{l|c|c|c|c|}
\mbox{state} &\Phi_\pm & \Psi_{\pm L} & \Psi_{\pm R} & F_\pm
\\\hline
\mbox{parity} &\pm & \pm & \pm & \pm 
\end{array}\,.\]

The SUSY transformations of even components are in the form of 
\begin{eqnarray}
&& \left\{ \begin{array}{rcl}
\delta A_\mu &=& 
i\bar\epsilon_+ \gamma_\mu \lambda_+, \\
\delta \lambda_+ &=& 
F^{\mu \nu}\gamma_{\mu \nu}\epsilon_+/4 
-i\widetilde{D_3}\epsilon_+/2, \\
\delta \widetilde{D_3} &=& 
\bar\epsilon_+ \gamma^\mu \partial_\mu \lambda_+, 
\end{array} \right. \nonumber \\
&& \left\{ \begin{array}{rcl}
\delta \Phi_+^{\bar\alpha} &=& 
i\bar\epsilon_{+R}\Psi_{+L}^{\bar\alpha}, \\
\delta \Psi_{+L}^{\bar\alpha} &=& 
-\gamma^\mu D_\mu \Phi_+^{\bar\alpha} \epsilon_{+R} - 
\widetilde{F}_+^{\bar\alpha} \epsilon_{+L}, \\
\delta \widetilde{F}_{+}^{\bar\alpha} &=& 
\bar\epsilon_{+L} (i\gamma^\mu D_\mu \Psi_{+L}^{\bar\alpha} + 
2g (T^a)^{\bar\alpha}_{\,\bar\beta} \Phi_+^{\,\bar\beta} \lambda_{+R,a}), 
\end{array} \right.
\nonumber 
\end{eqnarray}
where $D_\mu=\partial_\mu+igA_\mu^aT^a$, 
$\widetilde{D_3}=D_3 - \partial_y \Sigma$ and 
$\widetilde{F}_+=F_+ - \partial_y \Phi_-^*$. 
Now we see that, in the 4D ${\cal N}=1$ language, we have 
a massless vector multiplet $\widetilde{V}=
\{ A_\mu, \lambda_+, \widetilde{D_3} \}$ 
and chiral multiplets $\Phi_+^{\bar\alpha}=
\{ \Phi_+^{\bar\alpha},\Psi_{+L}^{\bar\alpha}, \widetilde{F}_+^{\bar\alpha} \}$. 

\section{Bulk and boundary coupling}
So the SUSY is 4D ${\cal N}=1$ because of the above orbifolding, 
we can also write down the coupling between the above bulk 
multiplets and some 4D boundary multiplets. 
We introduce chiral multiplets, 
$\phi_0^{\,\bar\alpha_0} = 
\{ \phi_0^{\,\bar\alpha_0}, 
\psi_{0L}^{\,\bar\alpha_0}, 
f_0^{\,\bar\alpha_0} \}$ and 
$\phi_\pi^{\,\bar\alpha_\pi} = 
\{ \phi_\pi^{\,\bar\alpha_\pi}, 
\psi_{\pi L}^{\,\bar\alpha_\pi}, 
f_\pi^{\,\bar\alpha_\pi} \}$, 
on both boundary-fixed planes of $S^1/Z_2$ orbifold, 
where $\bar\alpha_0=1,\ldots,n_0$ and $\bar\alpha_\pi=1,\ldots,n_\pi$. 
These chiral multiplets obey the same SUSY transformation rules 
as the above $\Phi_+$. Then the bulk and boundary coupling 
can be written as 
\begin{eqnarray}
{\cal L}_{\rm boundary} &=& 
\sum_{I=0,\pi} \delta (y-IR) \Bigg[ 
D_\mu \phi_I^{\,\bar\alpha_I \dagger} D^\mu \phi_{I \bar\alpha_I} 
+i \bar\psi_{IL}^{\,\bar\alpha_I} D\!\!\!\!/ \psi_{IL \bar\alpha_I} 
+f_I^{\,\bar\alpha_I \dagger} f_{I \bar\alpha_I} 
\nonumber \\ && 
-g \phi_I^{\,\bar\alpha_I \dagger} (T^a)_{\bar\alpha_I \bar\beta_I} 
\phi_I^{\,\bar\beta_I} \widetilde{D_3}_{,a} 
- 2g \left( \bar\psi_{IL}^{\,\bar\alpha_I}
\lambda_{+R,a} (T^a)_{\bar\alpha_I \bar\beta_I} 
\phi_I^{\,\bar\beta_I} + {\rm h.c.} \right) \Bigg]
\nonumber 
\end{eqnarray}

\section{Abelian vector multiplet}
If the gauge symmetry is $U(1)$, the Fayet-Iliopoulos term 
is invariant under the gauge symmetry and 4D ${\cal N}=1$ SUSY. 
This means that the FI term can be written without conflicting 
any symmetry in 5D $S^1/Z_2$ orbifold. In the 5D language, 
the FI terms on the boundary fixed planes 
\begin{eqnarray}
{\cal L}_{\rm FI} 
= \widetilde{D_3} \xi = (D_3-\partial_y \Sigma)\, \xi, \quad 
\xi = \sum_{I=0,\pi} \xi_I \delta (y-IR) \nonumber
\end{eqnarray}
are invariant. 

\subsection{4D scalar potential}
From the Lagrangian 
${\cal L}_{\rm total}={\cal L}+{\cal L}_{\rm boundary}+{\cal L}_{\rm FI}$, 
we can write the effective 4D potential as~\cite{GrootNibbelink:2002qp} 
\begin{eqnarray}
V_4 &=& \int_0^{\pi R} dy 
\Bigg[ 
\frac{1}{2} \Bigg( -\partial_y \Sigma + \xi 
+\frac{1}{2} g \Phi_i^{\,\alpha \dagger} 
 (q)_{\alpha \beta} \Phi_j^\beta (\sigma_3)^{ji} 
+g \sum_{I=0,\pi} \delta (y-IR) \phi_I^{\,\bar\alpha_I \dagger} 
Q_{\bar\alpha_I \bar\beta_I} \phi_I^{\,\bar\beta_I} \Bigg)^2 
\nonumber \\ && \hspace{1.5cm} 
-\frac{1}{2} \Bigg( -D_3 + \xi 
+\frac{1}{2} g \Phi_i^{\,\alpha \dagger} 
 (q)_{\alpha \beta} \Phi_j^\beta (\sigma_3)^{ji} 
+g \sum_{I=0,\pi} \delta (y-IR) \phi_I^{\,\bar\alpha_I \dagger} 
Q_{\bar\alpha_I \bar\beta_I} \phi_I^{\,\bar\beta_I} \Bigg)^2 
\nonumber \\ && \hspace{1.5cm} 
+ \frac{1}{8} g^2 \sum_{s=1,2} \left( 
\Phi_i^{\,\alpha \dagger} (q)_{\alpha \beta} \Phi_j^\beta (\sigma_s)^{ji} 
\right)^2 
- \frac{1}{2} \sum_{s=1,2} \left( -D_s + \frac{1}{2} g 
\Phi_i^{\,\alpha \dagger} (q)_{\alpha \beta} \Phi_j^\beta (\sigma_s)^{ji} 
\right)^2 
\nonumber \\ && \hspace{1.5cm} 
+ \frac{1}{2} \Big( \partial_y \Phi_i^{\,\alpha \dagger} 
-igA_5 \Phi_i^{\,\beta \dagger} (q)_\beta^{\,\alpha} 
- g\Sigma (\sigma_3)_i^{\,j} \Phi_j^{\,\beta \dagger} (q)_\beta^{\,\alpha} 
\Big) \nonumber \\ && \hspace{3cm} \times
\Big( \partial_y \Phi_{\,\alpha}^i 
+igA_5 (q)_{\alpha \gamma} \Phi^{i \gamma}  
- g\Sigma (q)_{\alpha \gamma} \Phi_j^{\,\gamma} (\sigma_3)^{ji} 
\Big) \Bigg], 
\nonumber 
\end{eqnarray}
where 
\begin{eqnarray}
\Phi_i^{\,\alpha \dagger} (q)_{\alpha \beta} \Phi_j^\beta (\sigma_1)^{ji} &=& 
-2q_{\bar\alpha} (\Phi_+^{\,\bar\alpha} \Phi_-^{\,\bar\alpha} + {\rm h.c.}), 
\nonumber \\
\Phi_i^{\,\alpha \dagger} (q)_{\alpha \beta} \Phi_j^\beta (\sigma_2)^{ji} &=& 
-2iq_{\bar\alpha} (\Phi_+^{\,\bar\alpha} \Phi_-^{\,\bar\alpha} - {\rm h.c.}), 
\nonumber \\
\Phi_i^{\,\alpha \dagger} (q)_{\alpha \beta} \Phi_j^\beta (\sigma_3)^{ji} &=& 
2q_{\bar\alpha} (|\Phi_+^{\,\bar\alpha}|^2 - |\Phi_-^{\,\bar\alpha}|^2), 
\nonumber \\
(q)_{\alpha \beta} &=& -(Q \otimes \sigma_3) \ = \ 
\left(
\begin{array}{cc}
-Q & 0 \\ 0 & Q 
\end{array}
\right)_{\alpha \beta}, 
\nonumber 
\end{eqnarray}
in the representation of independent bulk fields, and 
$Q_{\bar\alpha \bar\beta} 
= q_{\bar\alpha} \delta_{\bar\alpha \bar\beta}$ 
and $Q_{\bar\alpha_I \bar\beta_I} 
= q_{\bar\alpha_I} \delta_{\bar\alpha_I \bar\beta_I}$ 
are the generator of the $U(1)$ gauge transformation. 
$q_{\bar\alpha}$ and $q_{\bar\alpha_I}$ are the $U(1)$ charge of 
the bulk hyper multiplet and of the $y=IR$ brane chiral multiplet, 
respectively. 

We choose $A_5=0$ gauge and integrate out $D_{1,2,3}$ auxiliary fields. 
The above potential finally takes the form 
\begin{eqnarray}
V_4 &=& \int_0^{\pi R} dy 
\Bigg[ 
\frac{1}{2} \Bigg( -\partial_y \Sigma + \xi 
+gq_{\bar\alpha} (|\Phi_+^{\,\bar\alpha}|^2 - |\Phi_-^{\,\bar\alpha}|^2)  
+\sum_{I=0,\pi} \delta (y-IR) g q_{\bar\alpha_I} |\phi_I^{\,\bar\alpha_I}|^2 \Bigg)^2 
\nonumber \\ && \hspace{1.5cm} 
+ |gq_{\bar\alpha} \Phi_+^{\bar\alpha} \Phi_-^{\bar\alpha}|^2 
+ |\partial_y \Phi_+^{\,\bar\alpha} - gq^{\bar\alpha}\Sigma \Phi_+^{\,\bar\alpha}|^2 
+ |\partial_y \Phi_-^{\,\bar\alpha} + gq^{\bar\alpha}\Sigma \Phi_-^{\,\bar\alpha}|^2 
\Bigg], 
\label{eq:appsp}
\end{eqnarray}
where the same indices are contracted only between upper and lower ones.

\section{One-loop induced FI terms on boundaries}
\label{app:olidfi}

One-loop calculations have been done to show the FI terms 
are generated at the loop level in 
Ref.~\cite{Ghilencea:2001bw,Barbieri:2001cz,Scrucca:2001eb,Barbieri:2002ic}.
The one-loop tadpole diagrams for $D_3$ and $\Sigma$ 
propagated by the bulk hypermultiplet in the loop result in 
\begin{eqnarray}
\Xi_{D_3} 
&=& gq \frac{\pi R}{4} \sum_{n,n',n"} 
    \eta_{n}\eta_{n'}\eta_{n"} 2 \delta_{n,n'+n"}
    \int \frac{d^4p}{(2\pi^4)}\frac{\delta_{n,n"}}{p^2-n'^2/R^2}
    D_n, \nonumber \\
\Xi_{\Sigma} 
&=& -2gq \frac{\pi R}{4} \sum_{n,n',n"} 
    \eta_{n}\eta_{n'}\eta_{n"} 2 \delta_{n,n'+n"}
    \int \frac{d^4p}{(2\pi^4)}\frac{\delta_{n,n"}}{p^2-n'^2/R^2}
    \left( -\frac{n"}{R} \right) \Sigma_n, 
\nonumber 
\end{eqnarray}
where $\eta_0 = 1/\sqrt{\pi R}$, $\eta_{n>0} = 1/\sqrt{2\pi R}$, 
and $D_n$ and $\Sigma_n$ are the $n$-th KK excited modes of 
$D_3$ and $\Sigma$ respectively. Therefore we have 
\begin{eqnarray}
\Xi_{\rm bulk} 
&=& \Xi_{D_3}+\Xi_{\Sigma} \nonumber \\
&=& \sum_n g \textrm{tr}(q) \int \frac{d^4p}{(2\pi^4)} 
    \frac{1}{p^2-n^2/R^2} \eta_{2n} 
    \left( -D_{2n}+(\partial_y \Sigma)_{2n} \right) \nonumber \\ 
&=& \xi_{\rm bulk}(y) (-D_3+\partial_y \Sigma) \nonumber \\ 
\xi_{\rm bulk}(y) 
&=& g\frac{\textrm{tr}(q)}{32\pi^2}
    \left( \Lambda^2+\frac{1}{4}(\ln \lambda^2)\partial_y^2 \right) 
    \left[ \delta(y)+\delta(y-\pi R) \right], 
\label{eq:dfibulk}
\end{eqnarray}
where we regulate the divergent part of four-dimensional integral 
by using cut-off $\Lambda$ as 
\begin{eqnarray}
\left. \int \frac{d^4p}{(2\pi^4)} 
\frac{1}{p^2-n^2/R^2} \right|_{\rm div} 
= \frac{1}{16\pi^2} \left( \Lambda^2
    -\frac{n^2}{R^2}\ln \Lambda^2 \right).
\nonumber 
\end{eqnarray}

By performing the similar calculations we have 
\begin{eqnarray}
\Xi_{\rm brane} 
&=& \xi_{\rm brane}(y) (-D_3+\partial_y \Sigma) \nonumber \\ 
\xi_{\rm brane}(y) 
&=& g\frac{\Lambda^2}{16\pi^2} \sum_{I=0,\pi} 
    \textrm{tr}(q_I)\delta(y-IR), 
\label{eq:dfibrane}
\end{eqnarray}
from the tadpole diagram propagated by brane chiral multiplets. 

Eqs.~(\ref{eq:dfibulk}) and (\ref{eq:dfibrane}) tells us that 
even if we have no FI terms in the tree-level, we have nonzero 
FI terms that is dynamically generated. Furthermore we should note 
that even if we have no chiral multiplets on the branes, FI terms 
are dynamically generated on the boundary due to the orbifolding. 
This is rather nontrivial and results in spontaneous localization 
of bulk fields at the boundary due to the orbifolding. 

In section~\ref{sec:2}, however, we consider the {\it generic}\footnote{
As is calculated above it was shown~\cite{Scrucca:2001eb,
GrootNibbelink:2002qp,Marti:2002ar} that the one-loop generated FI terms 
include the second derivative of delta-functions, $\delta''(y)$ and 
$\delta''(y-\pi R)$ as well as delta-function form like Eq.~(\ref{5D-FI}). 
In this thesis we do not consider such FI term, but we can extend our analysis 
into the case with $\delta''(y)$ and $\delta''(y-\pi R)$.} FI terms on 
the boundaries and we are not concerned with whether it is generated 
dynamically at the loop-level or not. We also consider the case with 
bulk $U(1)$ symmetry is anomalous in general, which also gives 
interesting results.

\chapter{Higher modes with FI terms}
\label{app:higher1}

Following \cite{Abe:2002ps} here we study in detail 
profiles and mass eigenvalues of higher modes with 
the nontrivial VEV of $\Sigma$, which is obtained 
in section~\ref{sec:lifi} as well as~\ref{sec:2.2.3}. 

\subsection*{Case with $^\forall q, q_I>0$}
\label{app:higher1}
First we study the case of section~\ref{sec:lifi}.
{}From the potential (\ref{potential}) we have the $\Phi_\pm$ square 
term after $\Sigma$ develops the VEV $\langle \Sigma \rangle$. 
It is written as 
\begin{eqnarray}
-{\cal L}_{\rm m}^{\Phi_\pm} &=& 
   \pm \left( \xi(y)-\partial_y \langle \Sigma \rangle \right) 
   gq |\Phi_\pm|^2 
 + |\partial_y \Phi_\pm \mp gq \langle \Sigma \rangle \Phi_\pm|^2
\nonumber \\ &=& 
-\Phi_\pm^\dagger \left( \partial_y^2 \mp gq \xi (y) 
 - (gq)^2 \langle \Sigma \rangle^2 \right) \Phi_\pm 
\equiv 
-\Phi_\pm^\dagger \Delta_\pm \Phi_\pm, 
\label{KKop}
\end{eqnarray}
where $\Delta_\pm \equiv \partial_y^2 \mp gq \xi(y) 
- (gq)^2 \langle \Sigma \rangle^2$. The eigenvalues of the operator 
$\Delta_\pm$ correspond to the KK spectrum of $\Phi_\pm$. 

We solve the eigenvalue equation 
\begin{eqnarray}
\Delta_\pm \Phi_\pm + \lambda \Phi_\pm =0, 
\label{PhiEE}
\end{eqnarray}
where $\lambda$ is the eigenvalue of the operator $\Delta_\pm$. 
Following~\cite{GrootNibbelink:2002qp} we redefine $\Phi_\pm$ as 
\begin{eqnarray}
\Phi_\pm = 
\widetilde\Phi_\pm e^{\pm gq \int_0^y dy' \langle \Sigma(y') \rangle},
\label{redeftp}
\end{eqnarray}
and put this into Eq.~(\ref{PhiEE}) which results in the eigenvalue 
equation for $\widetilde\Phi_\pm$
\begin{eqnarray}
\left( \partial_y^2 \pm 2gq \langle \Sigma \rangle \partial_y 
\pm gq \left\{ \partial_y \langle \Sigma \rangle - \xi(y) \right\} 
+ \lambda \right) \widetilde\Phi_\pm =0, 
\label{tPhiEE}
\end{eqnarray}
with the boundary conditions 
\begin{eqnarray}
\partial_y \widetilde\Phi_+(y)\Big|_{y=0,\pi R}=
\widetilde\Phi_-(y)\Big|_{y=0,\pi R}=0. 
\label{tPhiBC}
\end{eqnarray}

The VEVs $\partial_y \langle \Sigma \rangle$ and 
$\langle \Sigma \rangle$ are given by 
Eqs.~(\ref{vevphi}) and (\ref{VEVSigma-1}) respectively. 
In the bulk ($0<y<\pi R$) eigenvalue equation (\ref{tPhiEE}) becomes 
\begin{eqnarray}
\partial_y^2 \widetilde\Phi_\pm 
\pm 2gq \left( Cy+{\xi_0 \over 2}\right) \partial_y \widetilde\Phi_\pm 
 +  \left( \lambda \pm gqC \right) \widetilde\Phi_\pm =0.
\label{bulkEE}
\end{eqnarray}

In the case with $\xi_0 + \xi_\pi = C =0$, 
Eq.~(\ref{bulkEE})  has been solved already in Ref.~\cite{GrootNibbelink:2002qp}.
(Mass)$^2$ eigenvalues $\lambda_k$ and wave functions 
of higher modes for $Z_2$ even bulk fields are obtained 
\begin{eqnarray}
\lambda_k &=& {k^2 \over R^2} + {(gq\xi_0)^2 \over 4}, \label{eq:czerokk} \\
\tilde \Phi^{(k)}_+ &=& \alpha_- e^{\alpha_+ y} 
- \alpha_+ e^{\alpha_- y}, \label{eq:czerowf}
\end{eqnarray}
up to normalization, where 
\begin{equation}
\alpha_\pm = - {1 \over 2}gq\xi_0 \pm {k \over R }i.
\end{equation}
$Z_2$ odd bulk fields have the same (mass)$^2$ eigenvalues 
$\lambda_k$ and the corresponding wave functions are obtained as 
\begin{eqnarray}
\lambda_k &=& {k^2 \over R^2} + {(gq\xi_0)^2 \over 4}, \nonumber \\
\tilde \Phi^{(k)}_- &=& e^{-\alpha_+ y} - e^{-\alpha_- y},
\nonumber 
\end{eqnarray}
up to normalization.

\begin{figure}[t]
\begin{center}
\begin{minipage}{0.48\linewidth}
   \centerline{\epsfig{figure=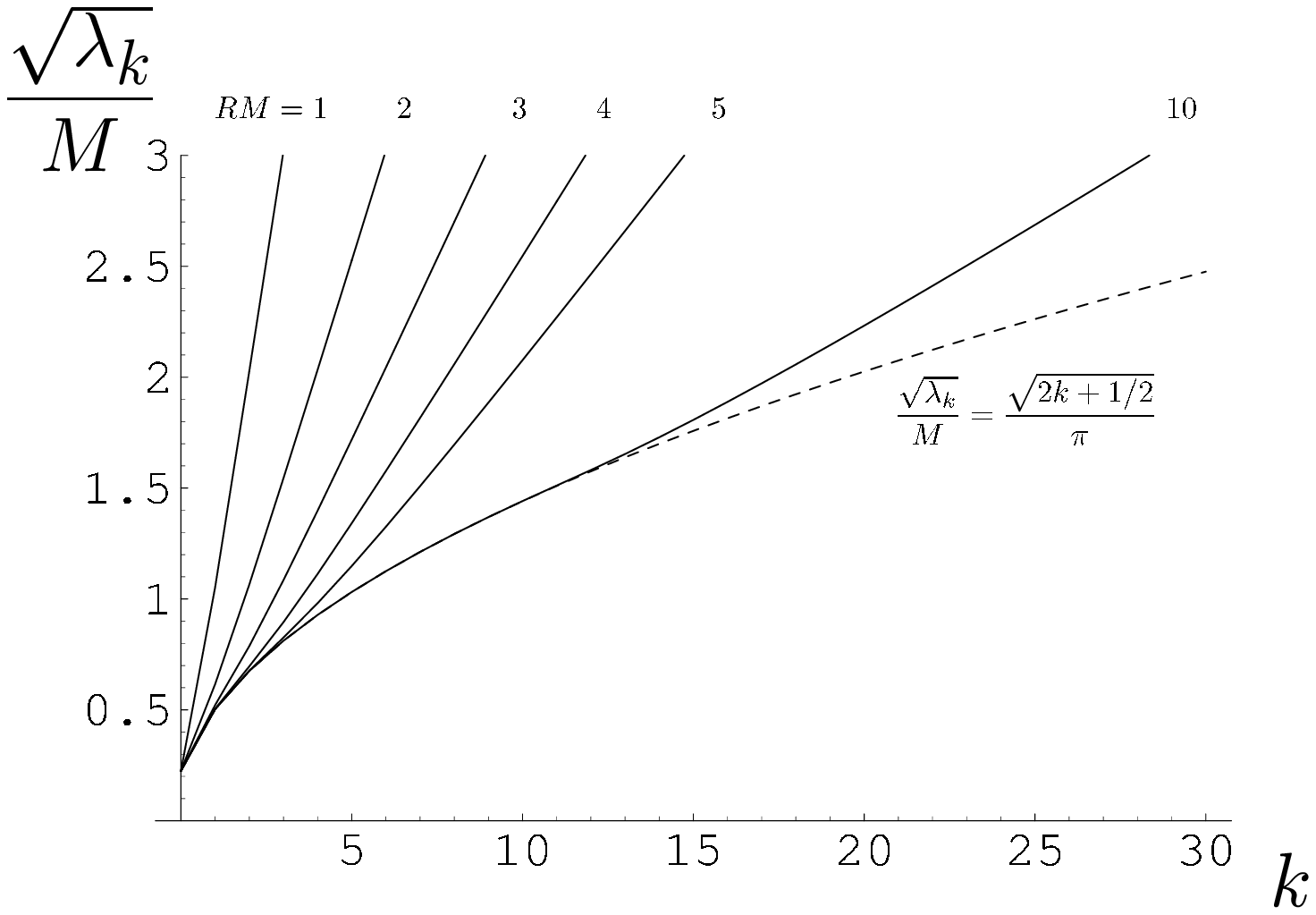,width=\linewidth}}
   \centerline{(a) $\Phi_+$, $z_0=RM$, $z_\pi=0$}
\end{minipage}
\hfill
\begin{minipage}{0.48\linewidth}
   \centerline{\epsfig{figure=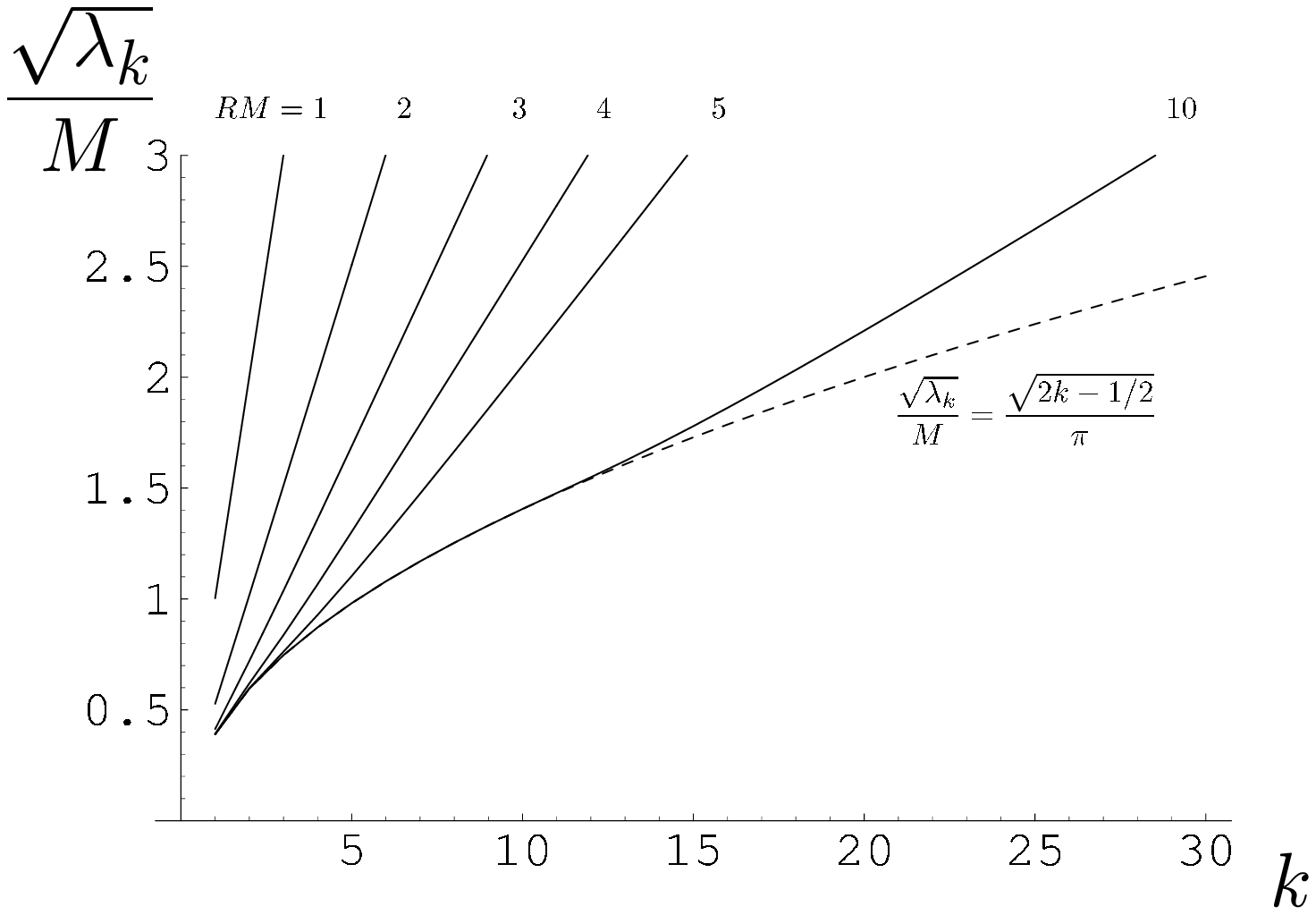,width=\linewidth}}
   \centerline{(b) $\Phi_-$, $z_0=RM$, $z_\pi=0$}
\end{minipage}
\end{center}
\begin{center}
\begin{minipage}{0.48\linewidth}
   \centerline{\epsfig{figure=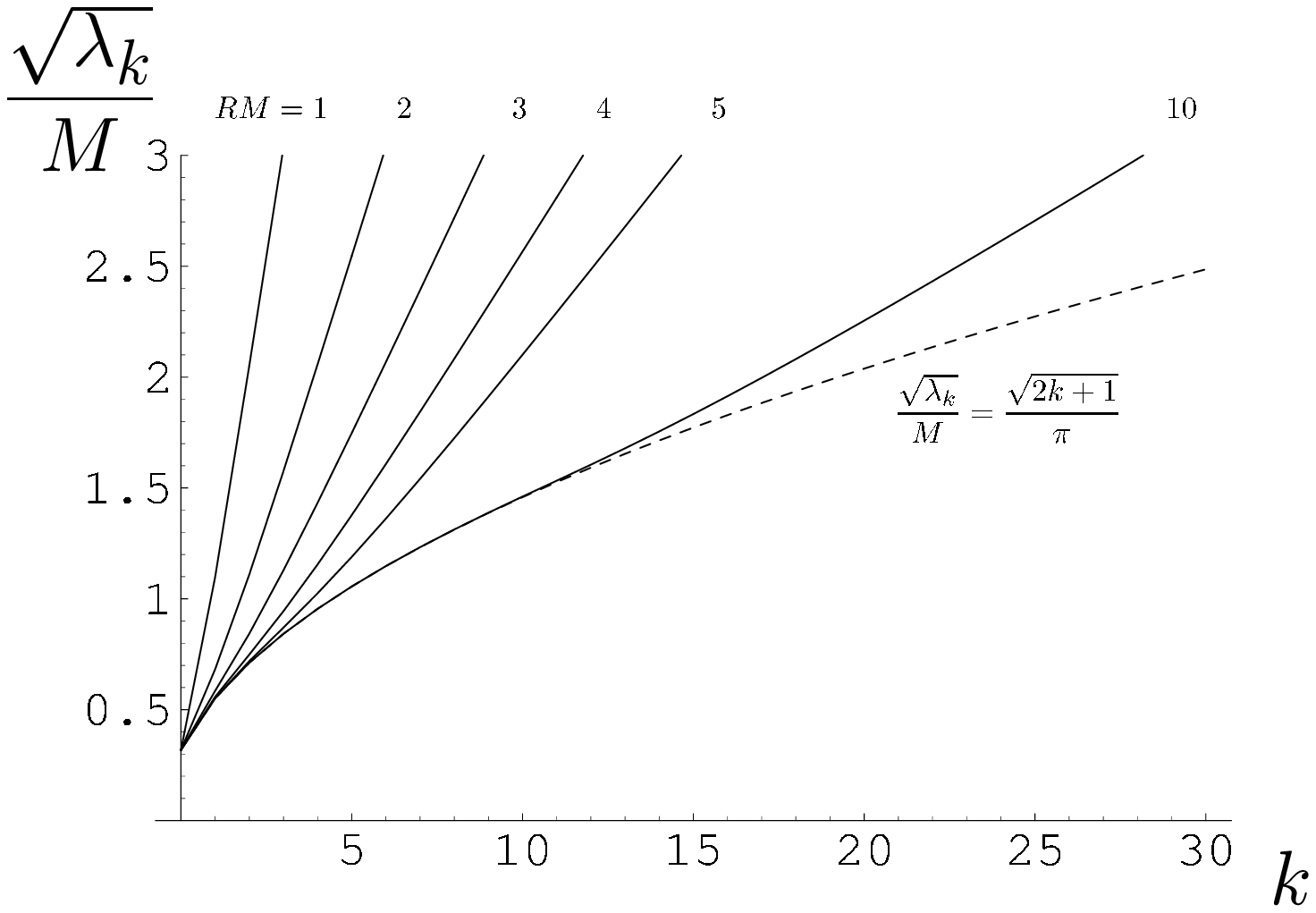,width=\linewidth}}
   \centerline{(c) $\Phi_+$, $z_0=z_\pi=RM/\sqrt{2}$}
\end{minipage}
\hfill
\begin{minipage}{0.48\linewidth}
   \centerline{\epsfig{figure=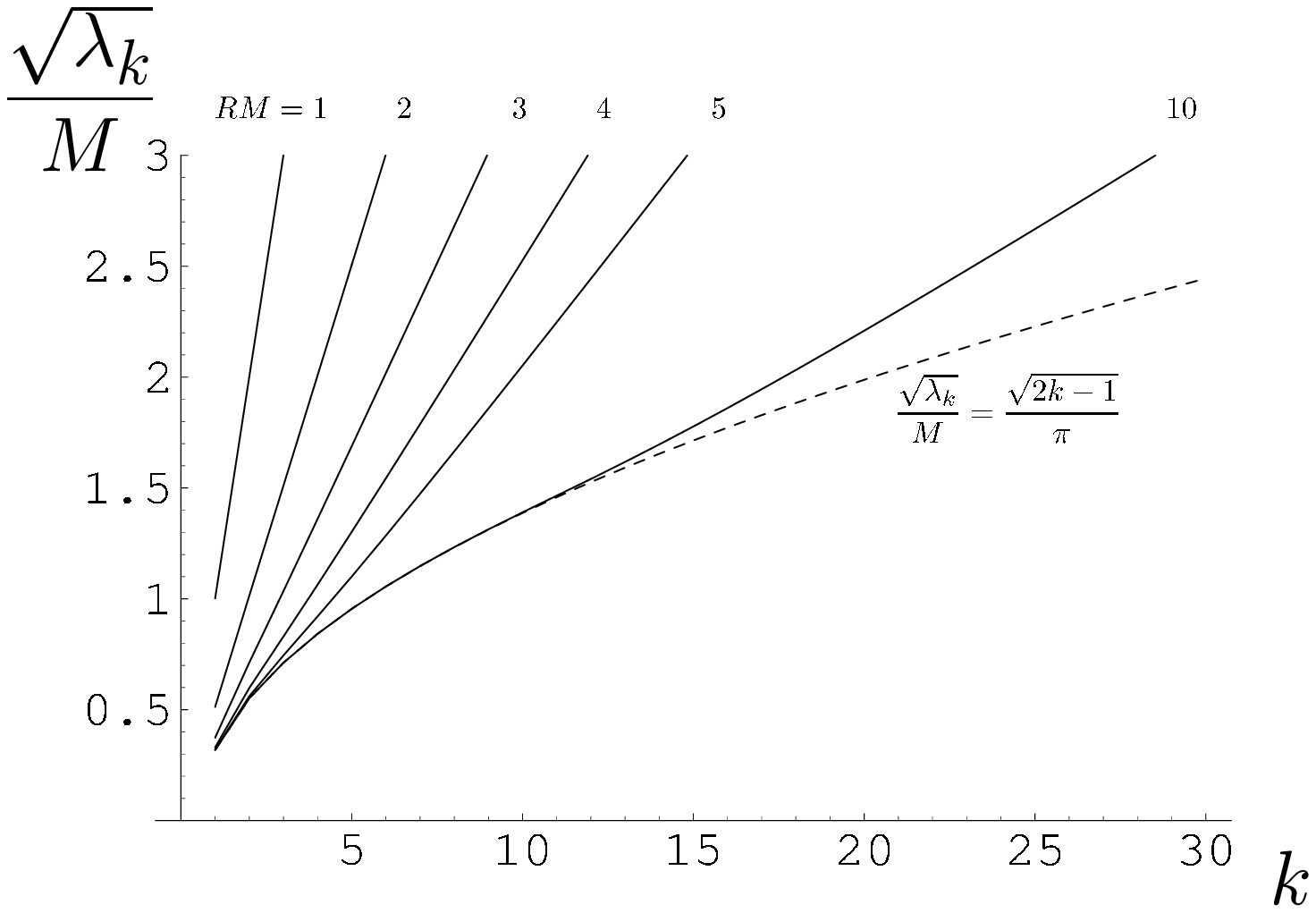,width=\linewidth}}
   \centerline{(d) $\Phi_-$, $z_0=z_\pi=RM/\sqrt{2}$}
\end{minipage}
\end{center}
\caption{KK spectrum $\sqrt{\lambda_k}/M$ up to 30th excited mode, 
for the case (a), (b) $z_0=RM$ and $z_\pi=0$ 
($l_{\xi_0}=1/(4\pi^2g_{\rm 4D}^2q)$ and 
$l_{\xi_\pi}=0$), 
and for the case (c), (d) $z_0=z_\pi=RM/\sqrt{2}$ 
($l_{\xi_0}=l_{\xi_\pi}=1/(4\pi^2g_{\rm 4D}^2q)$). 
The discrete eigenvalues are joined by solid lines. 
{}From the figures (a) and (b) we see that 
$\lambda_k$ 
$= \left( \frac{M}{\pi} \right)^2 
\left\{ \left( \frac{\pi k}{RM} \right)^2 \pm \frac{1}{2} \right\}$ 
($k \gg RM$), and 
$\lambda_k$ 
$= \left( \frac{M}{\pi} \right)^2 \left(2k\pm \frac{1}{2} \right)$ 
($k \ll RM$). 
{}From the figures (c) and (d) we see that 
$\lambda_k$ 
$= \left( \frac{M}{\pi} \right)^2 
\left\{ \left( \frac{\pi k}{RM} \right)^2 \pm 1 \right\}$ 
($k \gg RM$), and 
$\lambda_k$ 
$= \left( \frac{M}{\pi} \right)^2 (2k\pm 1)$ 
($k \ll RM$). 
The case with $k \ll RM$ is 
rather nontrivial because $\lambda_k$ is linear in $k$ and 
it does not depend on $R$, the size of extra dimension. 
This is understood from the fact that Eq.~(\ref{tPhiEEz}) 
is Hermite differential equation (harmonic oscillator) 
for $R \to \infty$. 
We also note that the first excited mode of $\Phi_-$ 
does not become tachyonic because 
$\lambda_k/M^2 \ge (2k-1)/\pi^2$ 
from the figure (b) and (d).}
\label{fig:sp1}
\end{figure}

In the case with $\xi_0 + \xi_\pi = C \ne 0$, 
the differential equation (\ref{bulkEE}) can be simplified as 
\begin{eqnarray}
\partial_z^2 \widetilde\Phi_\pm (z) 
\mp z \partial_z \widetilde\Phi_\pm (z)
+ \nu \widetilde\Phi_\pm (z) = 0, 
\label{tPhiEEz}
\end{eqnarray}
where $z=(Cy+{\xi_0 \over 2})\sqrt{\frac{2gq}{-C}}$ and 
$\nu=\frac{\lambda}{-2gqC}\mp \frac{1}{2}$. 
The boundary condition (\ref{tPhiBC}) becomes as 
\begin{eqnarray}
\partial_z 
\widetilde\Phi_+(z)\Big|_{z=z_0,z_\pi}= 
\widetilde\Phi_-(z)\Big|_{z=z_0,z_\pi}=0, 
\label{tPhiBCz}
\end{eqnarray}
where $z_0={\xi_0 \over 2}\sqrt{\frac{2gq}{-C}}$ and 
$z_\pi=-\frac{\xi_\pi}{2}\sqrt{\frac{2gq}{-C}}$. 
By parameterizing 
\begin{eqnarray}
\xi_I = 2g  l_{\xi_I} M^2, 
\nonumber 
\end{eqnarray}
where $I=(0,\pi)$, 
$l_{\xi_I}$ is the dimensionless parameter  
and $M$ is a representative scale for FI terms, 
we obtain the boundary parameters $z_0$ and $z_\pi$ as 
\begin{eqnarray}
z_0 = 
\frac{2\pi g_{\rm 4D} \sqrt{q}  l_{\xi_0}}
{\sqrt{ l_{\xi_0}+ l_{\xi_\pi}}}
RM, \quad 
z_\pi =  -r^\pi_0 z_0,
\label{defz}
\end{eqnarray}
where $r^\pi_0 =  l_{\xi_\pi} /  l_{\xi_0} 
=\xi_\pi / \xi_0$. 
The solution of Eq.~(\ref{tPhiEEz}) 
is given by 
\begin{eqnarray}
\widetilde\Phi_\pm (z) = f_1^\pm F_1^\pm (z,\nu) + f_2^\pm F_2^\pm (z,\nu), 
\nonumber 
\end{eqnarray}
where $f_1^\pm$ and  $f_2^\pm$ are arbitrary constants and 
\begin{eqnarray}
F_1^\pm (z,\nu) &=& F(\mp \nu/2\,;\,1/2\,;\,\pm z^2/2), \nonumber \\
F_2^\pm (z,\nu) &=& z F(1/2 \mp \nu/2\,;\,3/2\,;\,\pm z^2/2). 
\nonumber 
\end{eqnarray}
Here, $F(a\,;\,b\,;\,c)$ is the Kummer's hypergeometric function. 
Requiring that both of $f_1^\pm$ and $f_2^\pm$ can not be zero 
at the same time with the boundary condition 
(\ref{tPhiBCz}), the following equations should be satisfied: 
\begin{eqnarray}
\partial_z F_1^{+}(z,\nu)\Big|_{z=z_0} 
\partial_z F_2^{+}(z,\nu)\Big|_{z=z_\pi} 
&=& 
\partial_z F_1^{+}(z,\nu)\Big|_{z=z_\pi}
\partial_z F_2^{+}(z,\nu)\Big|_{z=z_0}, 
\nonumber \\
F_1^-(z,\nu)\Big|_{z=z_0}F_2^-(z,\nu)\Big|_{z=z_\pi} 
&=& F_1^-(z,\nu)\Big|_{z=z_\pi}F_2^-(z,\nu)\Big|_{z=z_0}. 
\label{KKEE}
\end{eqnarray}
These determine the eigenvalue $\nu$ which also gives $\lambda$. 
We numerically solve Eq.~(\ref{KKEE}) and estimate the eigenvalue 
for the typical case which is shown in Fig.~\ref{fig:sp1}. 
{}From the figure, in the case $r^\pi_0=0$ we see
$m_k^2 \equiv \lambda_k$ for $\Phi_{\pm}$ as 
\begin{eqnarray}
\begin{array}{llll}
\lambda_k & \sim & \left( \frac{M}{\pi} \right)^2 
\left\{ \left( \frac{\pi k}{RM} \right)^2 \pm \frac{1}{2} \right\} 
& (k \gg RM), \\
\lambda_k & \sim & \left( \frac{M}{\pi} \right)^2 (2k\pm \frac{1}{2}) 
& (k \ll RM), 
\end{array} \label{m-eigen1}
\end{eqnarray}
and in the case $r^\pi_0=1$, we obtain for $\Phi_{\pm}$ 
\begin{eqnarray}
\begin{array}{llll}
\lambda_k & \sim & \left( \frac{M}{\pi} \right)^2 
\left\{ \left( \frac{\pi k}{RM} \right)^2 \pm 1 \right\} & (k \gg RM), \\
\lambda_k & \sim & \left( \frac{M}{\pi} \right)^2 (2k\pm 1) & (k \ll RM), 
\end{array} \label{m-eigen2}
\end{eqnarray}
all with the fixed parameter 
$\xi_0/\sqrt{\pi R} \equiv M^2/(\sqrt{2}\pi^2 g_{\rm 4D} q)$. 
The case with $k \ll RM$ is 
rather nontrivial because $\lambda_k$ is linear in $k$ and 
it does not depend on $R$, the size of extra dimension. 
This behavior can be understood from the following explanation.
Eq.~(\ref{tPhiEEz}) 
is Hermite differential equation (harmonic oscillator) 
for $R \to \infty$. 
Thus, we have the spectrum like a harmonic oscillator for $k\ll RM$.
On the other hand, when $\lambda$ is large compared with 
$g\xi_0/R$ and $gC$, the term with 
$\partial_y \tilde \Phi_{\pm}$ in Eq.~(\ref{bulkEE}) 
can be neglected.
Thus, we would have $\lambda \sim (k/R)^2$.
We also note that the first excited mode ($k=1$) of $\Phi_-$ 
does not become tachyonic because 
$\lambda_k/M^2 \ge (2k-1)/\pi^2$ 
from Eqs.~(\ref{m-eigen1}) and (\ref{m-eigen2}).
Thus, the above vacuum is a local minimum.

The SUSY breaking scalar mass terms are constants along 
the $y$ direction.
Hence, fermionic higher modes corresponding to  $\Phi_{\pm}$ have the 
same profiles as their scalar partners.
In Eqs.~(\ref{m-eigen1}) and (\ref{m-eigen2}), 
the terms $\pm M^2/(2\pi^2)$ and  $\pm M^2/(\pi^2)$ are 
the contributions from the SUSY breaking scalar masses, respectively.
The corresponding fermionic modes do not have such contributions, 
that is, for example, for both $r^\pi_0 =0$ and $1$, we have 
\begin{eqnarray}
\begin{array}{llll}
\lambda_k & \sim & \left( \frac{M}{\pi} \right)^2 
 \left( \frac{\pi k}{RM} \right)^2  & 
(k \gg RM), \\
\lambda_k & \sim & \left( \frac{M}{\pi} \right)^2 2k & 
(k \ll RM). 
\end{array}
\label{fm-eigen1}
\end{eqnarray}

On the other hand, in the case that $C$ is small compared with $\xi_0/R$, 
we can consider that 
the term with $C$ in Eq.~(\ref{bulkEE}) 
is a perturbation from the case with $C=0$.
For example, the model which will be discussed in section~\ref{sec:3} 
would correspond to such a case.
In the first order of such perturbation, (mass)$^2$ 
eigenvalues $\lambda_{k}$ are shifted from (\ref{eq:czerokk}) as 
\begin{eqnarray}
\displaystyle
\begin{array}{ll}
\lambda_{k} = 
\displaystyle
  {k^2 \over R^2} + {(gq\xi_0)^2 \over 4} 
+ C \, 
\frac{\pi R(gq)^2 \xi_0}{e^{gq\xi_0 \pi R}-1} \times 
\frac{(k/R)^2+(gq\xi_0)^2/4}{(k/R)^2+5(gq\xi_0)^2/4}
& \textrm{for } \Phi_+, \\
\displaystyle
\lambda_{k} = 
  {k^2 \over R^2} + {(gq\xi_0)^2 \over 4} 
&  \textrm{for } \Phi_-.   
\end{array}
\nonumber 
\end{eqnarray}
Note that for $\Phi_-$ the first order perturbation in $C$ 
exactly cancels the $D$ term contribution (\ref{Dterm-1}).
There is no tachyonic mode for $\Phi_-$.
The mass eigenvalues of the corresponding fermionic modes are 
obtained by adding $\pm gqC$ for $\Phi_\pm$.

\subsection*{Case with $^\exists q<0$}
\label{app:higher2}
Next we analyze the case with $q<0$ for only one bulk field $\Phi_+$ 
as studied in section~\ref{sec:2.2.3}. 

First we show the simple case that argument value 
inside the $\tan$ function in (\ref{vevSigma-3}) is small enough, 
such that we can approximate Eq.~(\ref{vevSigma-3}) as 
\begin{equation}
\langle \Sigma (y) \rangle \approx - 
{\xi_0 + \xi_\pi \over 2 \pi R}y  + 
{\xi_0 \over 2}{\rm sgn}(y) 
+ {\xi_\pi \over 2}({\rm sgn }(y-\pi R)-1),
\label{approx-1}
\end{equation}
where we have taken $m=0$.
This VEV is exactly the same as Eq.~(\ref{VEVSigma-1}).
Note that in the present case SUSY is unbroken.
Actually, the VEV of $\Phi_+$ (\ref{vevphi-3}) is 
approximated as 
\begin{equation}
\langle |\Phi_+ (y)|^2 \rangle \approx  -
{\xi_0 + \xi_\pi \over 2 \pi R},
\end{equation}
at the same level of approximation as Eq.~(\ref{approx-1}).
That cancels the VEV $\langle \Sigma \rangle$.
Thus, mass eigenvalues and profiles of higher modes 
are obtained in the same way as in section~\ref{sec:lifi}, 
but in this case there is no SUSY breaking scalar mass terms.
Namely, both bosonic and fermionic modes have the same 
spectrum, e.g. Eq.~(\ref{fm-eigen1}).

Actually for more details let us consider another bulk field 
$\Phi'_\pm$ with charge $\pm q'$. 
{}From the potential (\ref{potential}) we have the $\Phi'_\pm$ square 
term after $\Sigma$ and $\Phi_+$ develop VEVs $\langle \Sigma \rangle$ 
and $\langle \Phi_+ \rangle$. 
It is written by 
\begin{eqnarray}
-{\cal L}_{\rm m}^{\Phi'_\pm} &=& 
|\partial_y \Phi'_\pm \mp gq' \langle \Sigma \rangle \Phi'_\pm|^2
\nonumber \\ &=& 
-\Phi_\pm'^\dagger \left( \partial_y^2 
 \mp gq' (\partial_y \langle \Sigma \rangle) 
 - (gq')^2 \langle \Sigma \rangle^2 
\right) \Phi'_\pm 
\equiv 
-\Phi_\pm'^\dagger \Delta_\pm \Phi'_\pm, 
\label{KKop}
\end{eqnarray}
where $\Delta_\pm \equiv \partial_y^2 
 \mp gq' (\partial_y \langle \Sigma \rangle) 
 - (gq')^2 \langle \Sigma \rangle^2$. 
The eigenvalues of the operator 
$\Delta_\pm$ correspond to the KK spectrum of $\Phi'_\pm$. 

We solve the same eigenvalue equation as (\ref{PhiEE}). 
The VEVs $\partial_y \langle \Sigma \rangle$ and 
$\langle \Sigma \rangle$ in $\Delta_\pm$ are derived from 
Eqs.~(\ref{D-flat3}), (\ref{vevphi-3}) and (\ref{vevSigma-3}). 
By defining $z \equiv \tan (ay+b+c_0)$, in the bulk ($0<y<\pi R$) 
the eigenvalue equation becomes 
\begin{eqnarray}
(1+z^2) \partial_z^2 \Phi'_\pm (z) 
+ 2z \partial_z \Phi'_\pm (z) 
-  \left( n_\pm(n_\pm+1) - \frac{\nu_\pm}{1+z^2} \right) 
\Phi'_\pm (z) =0, 
\nonumber 
\end{eqnarray}
where $n_\pm(n_\pm+1)=q'(q'\pm q)/q^2$ and 
$\nu_\pm=\lambda/a^2 \pm q'/q +n_\pm(n_\pm+1)$. 
Thus we obtain 
\begin{eqnarray}
\Phi'_\pm (z) = 
  A_\pm P_{n_\pm}^{\nu_\pm} (iz) 
+ B_\pm Q_{n_\pm}^{\nu_\pm} (iz),
\nonumber 
\end{eqnarray}
where $P_n^\nu (x)$ ($Q_n^\nu (x)$) is the first (second) class 
associated Legendre function. 
$A_\pm$ and $B_\pm$ are constants. 
The mass eigenvalue $\lambda$ can be obtained 
by requiring that both of $A_\pm$ and $B_\pm$ can not be zero 
at the same time with the boundary conditions 
in a way similar to the previous case.

%%%%%%%%%%%%%%%%%%%%%%%%%%%%%%%%%%
%%%%%%%%%%%%%%%%%%%%%%%%%%%%%%%%%%

\chapter{Schwinger-Dyson equation for boundary propagator}

\section{Local gauge fixing}
The Schwinger-Dyson equation for a propagator of (Dirac) fermion on a 
$D$-dimensional brane coupled to a Yang-Mills gauge field in 
($D+d$)-dimensional bulk space-time is written as 
\begin{eqnarray}
\lefteqn{iS^{-1}(p)} \hspace{15.2cm} \nonumber \\
=  iS_0^{-1}(p) 
 + \sum_n \int \frac{d^Dq}{(2\pi)^Di} 
   \left[ -ig_n(p-q)T^a \Gamma^M \right] S(q) 
   \left[ -ig_n(p-q)T^a \Gamma^N \right] D_{MN}^{(n)}(p-q), 
\label{apeq:appSD}
\end{eqnarray}
where 
\begin{eqnarray}
iS^{-1}(p) &=& A(-p^2)p\!\!\!/-B(p^2), \nonumber \\
iS_0^{-1}(p) &=& p\!\!\!/, \nonumber \\
D_{MN}^{(n)}(k) &=& \frac{-i}{k^2-M_n^2} 
 \left[ \eta_{MN}-\left( 1-\xi \right) 
\frac{k_Mk_N}{k^2-\xi M_n^2} \right]. 
\nonumber 
\end{eqnarray}

For $D=4$ the SD equation (\ref{apeq:appSD}) becomes the 
simultaneous integral equation of $A$ and $B$,
\begin{eqnarray}
A(p^2) &=& 1+ \int_0^{\Lambda^2} dq^2\, 
            \frac{q^2A(q^2)}{A^2(q^2)q^2 +B^2(q^2)} 
            \sum_{n=0}^{N_{\rm KK}} L_\xi (p^2,q^2;M_n,\alpha_n),  
\label{apeq:SDA} \\
B(p^2) &=& \int_0^{\Lambda^2} dq^2\, 
         \frac{q^2 B(q^2)}{A^2(q^2)q^2 +B^2(q^2)} 
         \sum_{n=0}^{N_{\rm KK}} K_\xi (p^2,q^2;M_n,\alpha_n), 
\label{apeq:SDB}
\end{eqnarray}
where 
\begin{eqnarray}
L_\xi (p^2,q^2;M_n,\alpha_n)
&=& \frac{\alpha_n}{4\pi} 
    \Bigg[ q^2f_{M_n^2}^2 (p^2,q^2) 
         + \frac{2q^2}{M_n^2} \left\{ f_{M_n^2} (p^2,q^2)-
                             f_{\xi M_n^2} (p^2,q^2) \right\} 
           \nonumber \\  && \hspace{1cm} 
         - \frac{(q^2)^2+p^2q^2}{2M_n^2} \left\{ f_{M_n^2}^2 (p^2,q^2)-
                             f_{\xi M_n^2}^2 (p^2,q^2) \right\} 
    \Bigg], \label{apeq:kerL} \\
K_\xi (p^2,q^2;M_n,\alpha_n)
&=& \frac{\alpha_n}{4\pi} 
    \bigg[ 4f_{M_n^2} (p^2,q^2) 
         + \frac{p^2+q^2}{M_n^2} \left\{ f_{M_n^2} (p^2,q^2)-
                             f_{\xi M_n^2} (p^2,q^2) \right\} 
           \nonumber \\  && \hspace{1cm} 
         - \frac{p^2q^2}{M_n^2} \left\{ f_{M_n^2}^2 (p^2,q^2)-
                             f_{\xi M_n^2}^2 (p^2,q^2) \right\} 
    \Bigg], \label{apeq:kerK}
\end{eqnarray}
\begin{eqnarray}
f_M (p^2,q^2) 
&=& \frac{2}{p^2+q^2+M+\sqrt{(p^2+q^2+M)^2 -4p^2q^2}}, 
\nonumber 
\end{eqnarray}
\begin{eqnarray}
\alpha_n &\equiv& \alpha \chi_n^2(y^\ast)
\qquad (n \ne 0), \nonumber \\ 
\alpha_0 &=& \alpha = g^2/4\pi. 
\nonumber 
\end{eqnarray}

\section{Non-local gauge fixing}
\label{sec:nlgf}
By introducing nonlocal gauge fixing parameter $\xi = \xi(k)$ 
and running coupling $g = g(k)$ we now rewrite SD equation 
(\ref{apeq:appSD}) as 
\begin{eqnarray}
iS^{-1}(p)
=  iS_0^{-1}(p)-C_F \int \frac{d^Dq}{(2\pi)^Di} 
               \Gamma^M S(q) \Gamma^N D_{MN}(p-q), 
\label{apeq:appSDnl}
\end{eqnarray}
where $C_F=T^aT^a$ and 
\begin{eqnarray}
S(q) &=& i\frac{A(-q^2)q\!\!\!/-B(-p^2)}{A(-q^2)q^2-B(-p^2)}, \nonumber \\
D_{MN}(k) &=& -i D(k) \left[ \eta_{MN}-\eta(k) \frac{k_Mk_N}{k^2} \right], \nonumber \\
\eta(k) &=& 1-\bar\xi(k), \qquad \bar\xi(k) \ =\ D^{-1}(k)D_\xi(k), \nonumber \\
D(k) &=& \sum_n \frac{g_n^2(k)}{k^2+M_n^2}, \qquad 
D_\xi(k) \ = \ \sum_n \frac{g_n^2(k)}{\xi^{-1}(k)k^2+M_n^2}.
\nonumber 
\end{eqnarray}

By adopting the following parameter 
\begin{eqnarray}
\bar\xi(z) &=& 
D-1-\frac{(D-1)(D-2)}{z^{D-1}D(z)}\int_0^z dt\,t^{D-2} D(t), 
\nonumber 
\end{eqnarray}
we obtain the SD equations for $A$ and $B$ as 
\begin{eqnarray}
A(x) &=& 1, \nonumber \\
B(x) &=& 
(D-1)C_F C_D \int_0^{\Lambda^2} dy 
\frac{y^{\frac{D-2}{2}}B(y)}{A^2(y)y+B^2(y)} K(x,y), \nonumber \\
K(x,y) &=& \int_0^\pi d\theta \sin^{D-2} \theta 
\left[ 2D(z)-(D-2)z^{1-D}\int_0^z dt\,t^{D-2}D(t) \right], \nonumber \\
D(z) &=& \sum_n \frac{g_n^2(z)}{z+M_n^2}, 
\nonumber 
\end{eqnarray}
where $z=x+y-2\sqrt{xy}\cos \theta$ and 
\begin{eqnarray}
C_D = \frac{1}{2^D\pi^{\frac{D+1}{2}\Gamma \left(\frac{D-1}{2}\right)}}.
\nonumber 
\end{eqnarray}
Therefore we know that the function $D(z)$ is important for 
analyzing the SD equation. We rewrite $D(z)$ as 
\begin{eqnarray}
D(z) 
&=& g^2(z) \sum_n \frac{\hat\chi_n^2}{z+M_n^2} \nonumber \\
&=& -g^2(z) \int_0^\infty dw\,w\, \partial_w 
    \left( e^{-zw}\sum_n \hat\chi_n^2 e^{-M_n^2w} \right), 
\label{apeq:fundz}
\end{eqnarray}
where 
$\hat\chi_n \equiv \Lambda^{\frac{4-D}{2}} \chi_n(y^\ast)/\chi_0(y^\ast)$. 
In the truncated KK theory the $1$-loop running coupling $g(z)$ 
is calculated as 
\begin{eqnarray}
g^2(z) = 
\left( g^{-2}(z_0)
+\beta_0 \ln \frac{z}{z_0} 
-\tilde\beta_0 \left\{ \ln \frac{z}{z_r} - \frac{2X_d}{d} 
\left[ \left(\frac{z}{z_r} \right)^{d/2}-1 \right] \right\} 
\Theta \left(\frac{z}{z_r}-1 \right) 
\right)^{-1}, 
\nonumber 
\end{eqnarray}
where $z_r$ is $(\textrm{mass difference})^2$ between the nearest 
two KK modes, $\Theta(x)$ is a step-function and 
\begin{eqnarray}
X_d = \frac{2\pi^{d/2}}{d\Gamma(d/2)}. 
\nonumber 
\end{eqnarray}
$\beta_0$ and $\tilde\beta_0$ is the one-loop beta-function coefficient with 
full fields and only bulk fields respectively. 
$z_0$ is given by the dynamical scale $\Lambda_{\rm dyn}$ of the Yang-Mills 
theory as 
\begin{eqnarray}
\frac{1}{g^2(z_0)}-\beta_0 \ln z_0 
\equiv -\beta_0 \ln \Lambda_{\rm dyn}^2.
\nonumber 
\end{eqnarray}

\subsection{Flat extra dimensions}
If the extra dimensions are flat and compact with universal radii $R$, 
we have $M_n=n/R$ ($n=\sqrt{n_1^2+\cdots+n_d^2}$) and 
$\hat\chi_n=\Lambda^{\frac{4-D}{2}}$. Eq.~(\ref{apeq:fundz}) becomes 
\begin{eqnarray}
\Lambda^{D-4}D(z) 
&=& -g^2(z) \int_0^\infty dw\,w\, \partial_w 
    \left[ e^{-zw} \left\{ \vartheta_3 \left( 
    \frac{iw}{\pi R^2} \right) \right\}^d \right] \nonumber \\
&=&  g^2(z) \int_0^\infty dw\,
    e^{-zw} \left\{ \vartheta_3 \left( 
    \frac{iw}{\pi R^2} \right) \right\}^d 
    \ -\ \left[ w e^{-zw} \left\{ \vartheta_3 \left( 
    \frac{iw}{\pi R^2} \right) \right\}^d \right]_{w=0}^{w=\infty} 
    \nonumber \\
&=& (\pi R^2)^{d/2} z^{d/2-1} g^2(z) \, 
\Gamma \left( 1-\frac{d}{2}\,,\,0 \right), 
\nonumber 
\end{eqnarray}
where we use Jacobi $\vartheta_3$ function defined by 
$\vartheta_3(\tau)=\sum_{n=-\infty}^\infty e^{i\pi \tau n^2}$ 
and $r \equiv \pi \left( X_d \right)^{2/d}$. 
In the third line we have used the asymptotic form 
$\vartheta_3 \left( \frac{iw}{\pi R^2} \right) \simeq R\sqrt{\frac{\pi}{w}}$ 
for $w/R^2 \ll 1$. 

If we introduce infrared and ultraviolet cut-off 
$\mu_0=\sqrt{z_0}$ and $\Lambda$ respectively 
as in the truncated KK calculations, we obtain 
\begin{eqnarray}
\Lambda^{D-4}D(z) &=&  
    g^2(z) \int_{r\Lambda^{-2}}^{r\mu_0^{-2}} dw\, 
    e^{-zw} \left\{ \vartheta_3 \left( 
    \frac{iw}{\pi R^2} \right) \right\}^d, \nonumber \\
&=& (\pi R^2)^{d/2} z^{d/2-1} g^2(z) 
\left[ \Gamma \left( 1-\frac{d}{2}\,,\,r\frac{z}{\Lambda^2} \right) 
- \Gamma \left( 1-\frac{d}{2}\,,\,r\frac{z}{\mu_0^2} \right) \right], 
\nonumber 
\end{eqnarray}
$\Gamma(a,z)=\int_z^\infty dt\,t^{a-1} e^{-t}$ is the 
Legendre second class imperfect gamma-function.

\chapter{Bulk fields in AdS$_5$}

Here we briefly review the field theory on the RS background,
following Ref.~\cite{Gherghetta:2000qt}. One of the original
motivations to introduce a warped extra dimension by Randall and
Sundrum is to provide the weak and Planck mass hierarchy by the
exponential factor in the space-time metric. This factor is called
`warp factor', and the bulk space `a warped extra dimension'. Such a
nonfactorizable geometry with the warp factor distinguishes the RS
brane world from others.

Consider the fifth dimension $y$ compactified on an orbifold
$S^1/{\bf Z}_2$ with radius $R$ and two three-branes at the orbifold
fixed points, $y=0$ and $y=L/2 \equiv \pi R$. 
The Einstein equation for this five-dimensional setup leads to 
the solution~\cite{Randall:1999ee},
\begin{equation}
ds^2 \,=\, e^{-2\sigma}\eta_{\mu\nu}dx^\mu dx^\nu-dy^2,
\qquad \sigma \,=\, k|y|,
\label{metric}
\end{equation}
where $k$ is a constant with mass dimension one. Let us study a vector
field $A_M$, a Dirac fermion $\Psi$ and a complex scalar $\phi$ in the
bulk specified by the background metric (\ref{metric}). The 5D action
is given by
\begin{equation}
  S_5 \,=\, \int d^4x\int_0^{\pi R} dy\sqrt{-g}\,
    \Bigg[\frac{-1}{4g^2_5}F^2_{MN}+
    \left|\partial_M\phi\right|^2+i\bar{\Psi}\gamma^MD_M\Psi
    -m^2_\phi|\phi|^2-m_\Psi\bar{\Psi}\Psi\Bigg],
\label{kin}
\end{equation}
where $\gamma_M=(\gamma_\mu,i\gamma_5)$ and the covariant derivative
is $D_M=\partial_M+\Gamma_M$ where $\Gamma_M$ is the spin connection
given by $\Gamma_\mu= i\gamma_5\gamma_\mu \sigma'/2$ 
and $\Gamma_4=0$. From the transformation properties under 
the ${\bf Z}_2$ parity, the mass parameters of scalar and fermion
fields are parameterized as\footnote{In Ref.~\cite{Gherghetta:2000qt},
the integral range with respect to $y$ is taken
as $-\pi R\leq y\leq\pi R$. Here we adopt $0\leq y\leq \pi R$, and
then the boundary mass parameter $b$ in Eq.~(\ref{eq:smass}) is
different from that in Ref.~\cite{Gherghetta:2000qt}
by the factor $1/2$.}
\begin{eqnarray}
m^2_\phi &=& ak^2+\frac{b}{2}\sigma'',\label{eq:smass} \\
m_\Psi &=& c\sigma' \label{eq:fmass},
\end{eqnarray}
where $a,b$ and $c$ are dimensionless parameters.

Referring to~\cite{Gherghetta:2000qt}, the vector, scalar and spinor
fields are cited together by the single notation
$\Phi=\{A_\mu,\ \phi,\ e^{-2\sigma}\Psi_{L,R}\}$.
The KK mode expansion is performed as
\begin{equation}
\Phi(x^\mu,y) \,=\, {1\over\sqrt{2\pi R}}
\sum_{n=0}^\infty \Phi^{(n)}(x^\mu)f_n(y).
\end{equation}
By solving the equations of motion, the eigenfunction $f_n$ is given by 
\begin{eqnarray}
f_n(y)
&=& \frac{e^{s\sigma/2}}{N_n}\left[J_\alpha(\frac{m_n}{k}e^{\sigma})
  +b_{\alpha}(m_n)\, Y_\alpha(\frac{m_n}{k}e^{\sigma})\right],
  \label{eq:modefunction}
\end{eqnarray}
where $\alpha = \sqrt{(s/2)^2+M^2_\Phi/k^2}$,
$s=\{2,\ 4,\ 1\}$, $M^2_\Phi=\{0,\ ak^2,\ c(c\pm 1)k^2\}$
for each component in $\Phi$. $N_n$ is the normalization factor 
and $J_\alpha$ and $Y_\alpha$ are the Bessel functions. The
corresponding KK spectrum $m_n$ is obtained by solving
\begin{eqnarray}
b_{\alpha}(m_n) \,=\, b_{\alpha}(m_ne^{\pi kR}).
\label{eq:KKequation}
\end{eqnarray}

A supersymmetric extension of this scenario was discussed
in~\cite{Gherghetta:2000qt,Gherghetta:2001kr}. The on-shell field
content of a vector supermultiplet is $(A_M,\lambda_i,\Sigma)$ where
$A_M$, $\lambda_i~(i=1,2)$ and $\Sigma$ are the vector, two Majorana
spinors and a real scalar in the adjoint representation,
respectively. Also a hypermultiplet consists of $(H_i,\Psi)$, 
where $H_i$ $(i=1,2)$ are two complex scalars and $\Psi$ is a Dirac
fermion. By requiring the action (\ref{kin}) to be invariant under
supersymmetric transformation on the warped background, one finds that
the five-dimensional masses of the scalar and spinor fields have to
satisfy
\begin{eqnarray}
m^2_\Sigma&=&-4k^2+2\sigma'', \label{eq:Sig} \\
m_\lambda&=&\frac{1}{2}\sigma', \label{eq:lam} \\
m^2_{H^{1,2}}&=&\left(c^2\pm c-\frac{15}{4}\right)k^2
 +\left(\frac{3}{2}\mp c\right)\sigma'',\label{eq:H} \\
m_\Psi&=&c\sigma', \label{eq:Psi}
\end{eqnarray}
where $c$ remains as an arbitrary dimensionless parameter.
That is, $a=-4$, $b=2$ and $c=1/2$ for vector multiplets and
$a=c^2\pm c-15/4$ and $b=3/2 \mp c$ for hypermultiplets. There is no
freedom to choose bulk masses for vector supermultiplets and
only one freedom parameterized by $c$ for bulk hypermultiplets.
It should be noted that in the warped 5D models, fields contained in
the same supermultiplet have different bulk/boundary masses.
That is in contrast with the flat case.

The $Z_2$ even components in supermultiplets have massless modes
with the following wave-functions
\begin{eqnarray}
\frac{1}{\sqrt{2\pi R}} ~~&&
\textrm{ for } \;V^{(0)}_\mu \,
\textrm{ and } \,\lambda^{1\,(0)}_L, \nonumber \\[1mm]
\frac{e^{(1/2- c)\sigma}}{N_0\sqrt{2\pi R}} &&
\textrm{ for } \;H^{1\, (0)} \,
\textrm{ and } \,\Psi^{(0)}_L. \label{eq:hzero}
\end{eqnarray}
The subscript $L$ means the left-handed ($Z_2$ even) component.
The massless vector multiplet has a flat wave-function in the extra
dimension. On the other hand, the wave-function for massless chiral
multiplets involve the $y$-dependent contribution from the space-time
metric, which induces a localization of the zero modes. The zero modes
with masses $c>1/2$ and $c<1/2$ localize at $y=0$ and $y=\pi R$,
respectively. The case with $c=1/2$ corresponds to the conformal limit
where the kinetic terms of zero modes are independent of $y$.

\pagebreak[0]

\end{document}